\documentclass[11pt]{article}
\usepackage{epsfig}
\usepackage{amsfonts}
\usepackage{amsmath}
\usepackage{bbm,bm}
\usepackage{booktabs}
\usepackage{iftex}
\ifPDFTeX
  \PackageError{Thesis}{Chinese text requires XeLaTeX or LuaLaTeX}{Please compile this thesis with xelatex instead of pdflatex.}
\fi
\usepackage[UTF8,fontset=fandol]{ctex}

\usepackage{float}
\usepackage{tcolorbox}
\usepackage{slashed}
\usepackage{pdfpages}
\usepackage{fancyhdr}
\allowdisplaybreaks[1]

\usepackage{amsthm}
\usepackage{amssymb}
\usepackage{graphicx}
\usepackage{tikz}
\usetikzlibrary{shapes.misc}
\usepackage{tikz-feynman}
\usepackage{pgfplots}
\usepackage{chngcntr}
\usepackage{xcolor}

\usepackage[
  colorlinks=true,
  linkcolor=blue,
  citecolor=blue,
  filecolor=black,
  urlcolor=blue,
  breaklinks=true]{hyperref}

\makeatletter
\DeclareRobustCommand{\captionfootnotemark}{%
  \begingroup
    \protected@xdef\@thefnmark{\thefootnote}%
    \@footnotemark
  \endgroup
}
\newcommand{\captionfootnotetext}[1]{%
  \begingroup
    \protected@xdef\@thefnmark{\thefootnote}%
    \@footnotetext{#1}%
  \endgroup
}
\makeatother

\usepackage[
    backend=biber, 
    style=numeric-comp, 
    giveninits=true,
    sorting=none, 
]{biblatex}
\newcommand{\sumint}{\mathop{\ooalign{\hfil$\displaystyle\sum$\hfil\cr\kern0.15ex$\displaystyle\int$\cr}}}

\usepackage{ulem}
\newcommand{\flag}{\textcolor{red}{}}
\newcommand{\thesisversion}{1} 
\newcommand{\iffullversion}[2]{\ifnum\thesisversion=1 #1\else #2\fi}
\newcommand{\blindfield}[1]{\iffullversion{#1}{}}
\newcommand{\englishmonthname}{%
  \ifcase\month\or January\or February\or March\or April\or May\or June\or July\or August\or September\or October\or November\or December\fi
}

\newcommand{\thesisenglishdate}{\englishmonthname\ \number\day, \number\year}
\newcommand{\reproducedfromref}[1]{\iffullversion{ This figure is reproduced from Ref.~\cite{#1}.}{}}
\newcommand{\publicationentry}[4]{%
  \item \textbf{#1}\par
  \iffullversion{#2}{#3}\par
  {\itshape #4}\par
}

\def\lsi{\raise0.3ex\hbox{$<$\kern-0.75em\raise-1.1ex\hbox{$\sim$}}}

\renewcommand{\vec}[1]{{\bf #1}}

\makeatletter \@addtoreset{equation}{section} \makeatother

\makeatletter
\renewcommand\section{\@startsection{section}{1}{\z@}%
  {-5.5ex \@plus -1ex \@minus -.2ex}
  {2.3ex \@plus.2ex}%
  {\normalfont\large\bfseries}}
\renewcommand\subsection{\@startsection{subsection}{2}{\z@}%
  {-3.25ex\@plus -1ex \@minus -.2ex}%
  {1.5ex \@plus .2ex}%
  {\normalfont\normalsize\bfseries}}
\renewcommand\thesection{\@arabic\c@section}
\renewcommand\thesubsection{\thesection.\@arabic\c@subsection}
\renewcommand{\@seccntformat}[1]{%
  \csname the#1\endcsname.\hspace{1.0em}}
\makeatother

\graphicspath{
  {./figures/}
}

\newcounter{savedtitlepagepage}

\begin{document}






\begin{titlepage}
\setcounter{page}{1}

\begin{flushright}

\end{flushright}
\begin{centering}

\includegraphics[width=4cm]{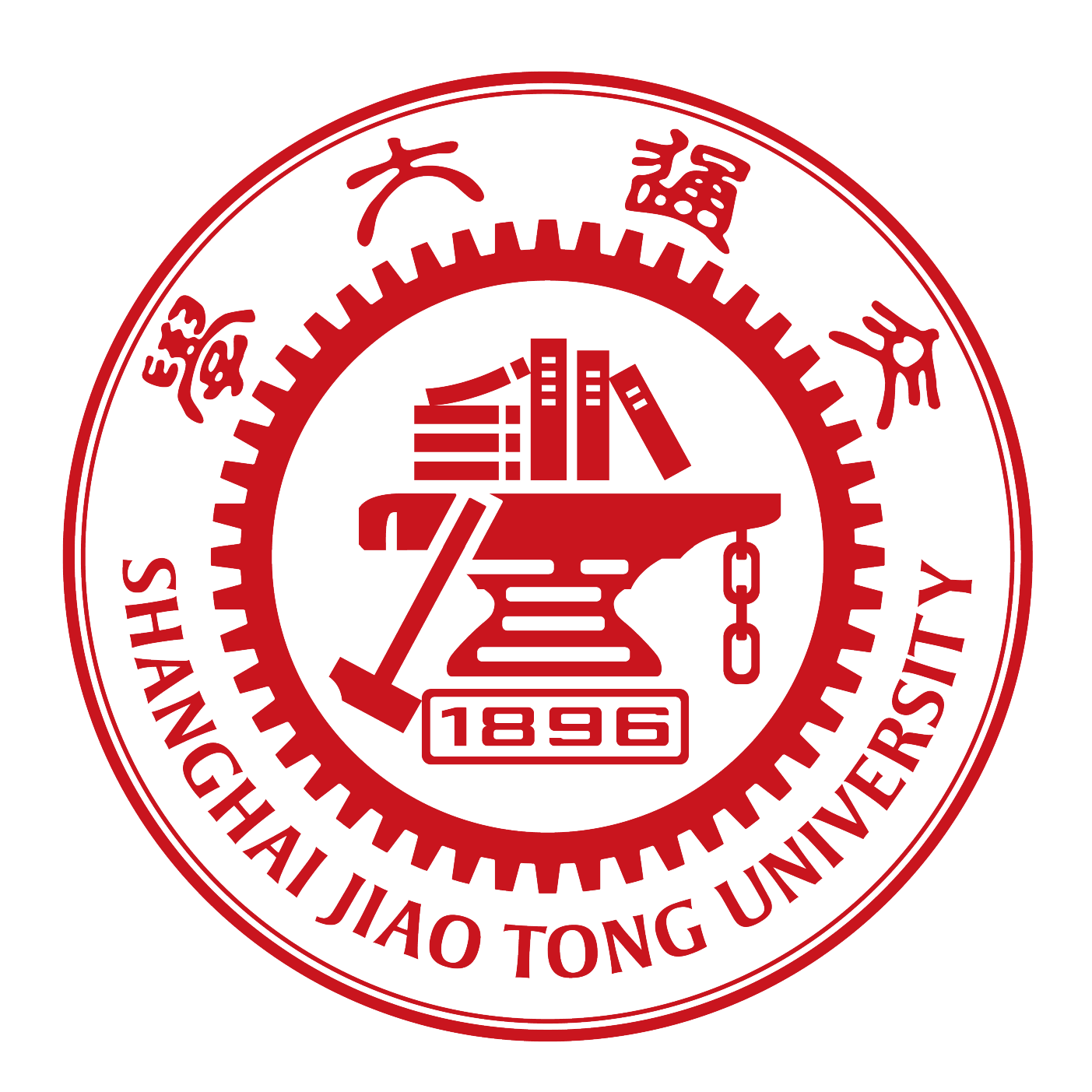}

{\fontsize{14pt}{20pt}\selectfont A Dissertation Submitted to
Shanghai Jiao Tong University\\[0.2em]
for the Degree of Doctor of Philosophy}

\vspace{3cm}

\begin{center}
\songti \zihao{3} \textbf{\parbox{0.92\textwidth}{\centering
Electroweak Baryogenesis: Advances in \\ Sphaleron Rate Calculations
and \\ Implications of  Thermal Phase Transitions}}
\end{center}

\vspace{1.2cm}

\renewcommand{\thefootnote}{\fnsymbol{footnote}}

{\fontsize{15pt}{18pt}\selectfont
\begin{tabular}{@{}c@{}}
Author: \blindfield{Yanda Wu} \\
Supervisor: \blindfield{Michael Ramsey-Musolf}
\end{tabular}
}

\vspace{2.0cm}

\includegraphics[width=6cm]{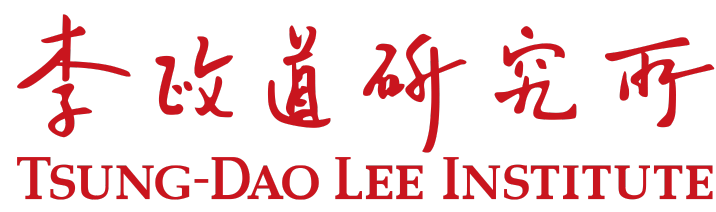}

{\fontsize{14pt}{20pt}\selectfont
\renewcommand{\arraystretch}{1.0}
\begin{tabular}{@{}c@{}}
\thesisenglishdate \\
\end{tabular}
}

\vspace{0.8cm}
\vspace{0.8cm}


\newpage
\mbox{\bf Abstract}

\end{centering}

\vspace{0.3cm}

\noindent
The origin of baryon asymmetry and the nature of dark matter remain two of the central puzzles in modern cosmology and particle physics. This thesis focuses on baryon asymmetry, with particular emphasis on recent developments in electroweak baryogenesis (EWBG), while also exploring connections to dark matter with candidates of
weakly interacting massive particle and primordial black holes. In the EWBG framework, the Sakharov conditions require baryon number violation (BNV), additional C and CP violation, and a first-order phase transition. This thesis mainly advances our understanding of the first and third ingredients.

We study how new scalar particles modify baryon number violating processes. For general \(SU(2)\) scalar multiplets, we classify the associated topological field configurations and construct the corresponding sphaleron and monopole solutions. Both sphaleron and monopole configurations can induce BNV, though their detailed dynamics differ. The sphaleron energy and monopole mass are computed in different phase transition scenarios, particularly when the Standard Model Higgs sector is extended by an additional scalar multiplet.
These monopole solutions are also related to minimal dark matter phenomenology and can contribute to the relic density. At finite temperature, we analyze how new scalar particles affect EWBG and revisit the traditional baryon-preservation condition based on the sphaleron rate. We formulate a new framework to compute the sphaleron rate within three-dimensional thermal effective field theory, where issues such as gauge dependence and quantum fluctuations can be treated more systematically. We also discuss broader implications of first-order phase transitions, including the possibility that a delayed transition produces primordial black holes, as well as the prospects for probing the relevant new scalar particles at future lepton colliders such as CEPC.

To make the thesis self-contained, we also provide detailed introductions to sphaleron, baryon number violation, homotopy theory, instanton, Adler-Bell-Jackiw (ABJ) anomaly, vacuum structure of non-abelian gauge theories, and thermal field theory.

\vfill
\newpage
\thispagestyle{plain}

\begin{center}
  {\Large\bfseries Suggestion on the Chinese Name of Sphaleron\par}
\end{center}

\vspace{0.5cm}

\noindent
The term \emph{sphaleron} was first introduced by Klinkhamer and Manton in Ref.~\cite{Klinkhamer:1984di}\footnote{Although the sphaleron solution had already been constructed by Manton in an earlier work \cite{Manton:1983nd}.}. It is derived from the classical Greek adjective ``$\sigma\phi\alpha\lambda\epsilon\rho\acute{o}\varsigma$'', meaning ``ready to fall.'' As illustrated below, the sphaleron is a static but unstable saddle-point solution of the energy functional. If one constructs classical paths interpolating between vacua with Chern-Simons numbers $N_{\rm CS}=1$ and $N_{\rm CS}=2$, then among the maximal field energies attained along all such paths, the sphaleron corresponds to the configuration of minimal energy. It is characterized by a half-integer value of $N_{\rm CS}$. 

For this reason, we suggest the Chinese rendering ``鞍子'' for \emph{sphaleron} \footnote{On Wikipedia, \emph{sphaleron} is translated into Chinese as ``滑子,'' emphasizing its etymological meaning of ``ready to fall.'' However, we believe that ``鞍子'' is a more appropriate rendering, as it better reflects the saddle-point nature of the sphaleron configuration.}.
The character ``鞍'' emphasizes the saddle-point nature of the configuration, while the suffix ``子'' is used only as a concise physics term for a non-perturbative field configuration and is not meant to imply an asymptotic particle state. Similarly, instanton are already referred to in Chinese as ``瞬子''.

\vspace{0.1cm}

\begin{center}
  \includegraphics[width=0.82\textwidth,height=0.42\textheight,keepaspectratio]{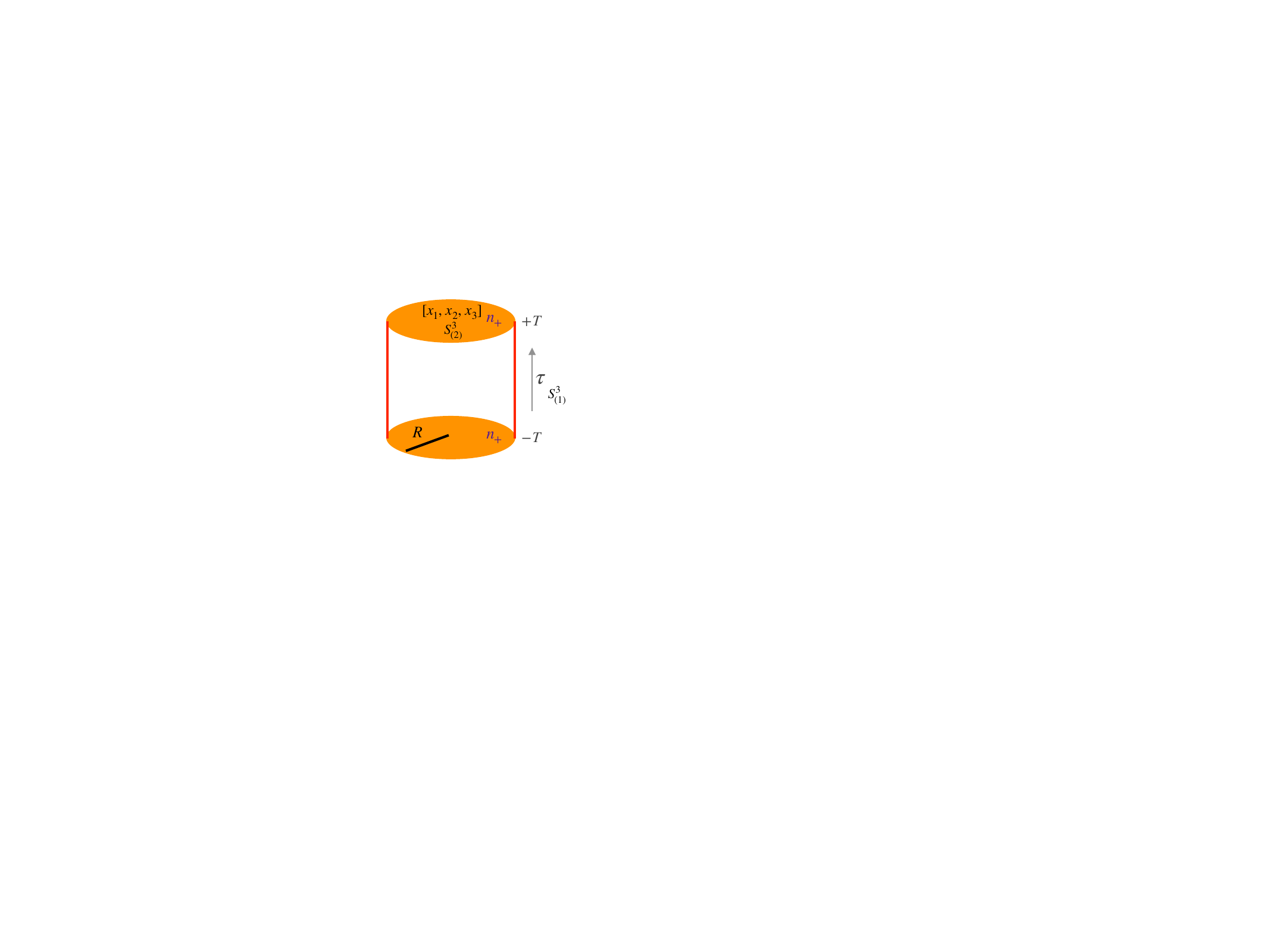}
\end{center}

\vfill
\setcounter{savedtitlepagepage}{\value{page}}
\end{titlepage}
\setcounter{page}{\numexpr\value{savedtitlepagepage}+1\relax}

\tableofcontents
\renewcommand{\thefootnote}{\fnsymbol{footnote}}
\iffullversion{\footnotetext[1]{yanda.wu7@sjtu.edu.cn}}{}
\clearpage

\renewcommand{\thefootnote}{\arabic{footnote}}
\setcounter{footnote}{0}

\section{Introduction}
\label{sec:intro}

In this introduction, we summarize the key contributions and main novelties of this thesis. We also outline the background material introduced throughout the thesis that is necessary for understanding these contributions and novelties. The overall motivation is to address major puzzles of our universe, such as baryon asymmetry and dark matter, which will be introduced in detail in Sec.~\ref{sec:puzzles}. For baryon asymmetry, we focus on electroweak baryogenesis, whose dynamics will be discussed in Sec.~\ref{sec:EWBG_four_processes}. The advantages of electroweak baryogenesis and its experimental tests are deferred to Sec.~\ref{sec:baryon_asymmetry}. For dark matter, we focus on the weakly interacting massive particle (WIMP) and primordial black hole (PBH) candidates introduced in Sec.~\ref{sec:dark_matter}.

The computation of sphaleron rates, together with the analysis of thermal phase transitions and their implications, constitutes the main technical core of this thesis. These two themes are not unique to electroweak baryogenesis; they also appear in a wide range of baryogenesis scenarios. Therefore, the methods and results developed here can be applied more broadly to other baryogenesis mechanisms, as well as to cosmological phenomena driven by thermal phase transitions.

Around these two topics, the thesis is written to be self-contained for readers with basic knowledge of quantum field theory and cosmology, without requiring prior familiarity with homotopy theory, topological field configurations, the ABJ anomaly, vacuum structure of non-abelian gauge theories, or thermal effective potentials. These topics are introduced throughout the thesis. For readers already familiar with them, we highlight below the key contributions and novelties that can be read independently. Since many background topics are presented at an almost pedagogical level, we also list the sections in which these materials are developed in detail.

\subsection{Contributions and Novelties of This Thesis}
\label{sec:contributions_and_novelty}

Below, we summarize the key contributions and main novelties of this thesis, together with the sections where the details are provided.
\begin{itemize}
  \item For an \(SU(2)\) scalar multiplet in a general representation undergoing the electroweak phase transition, in Sec.~\ref{sec:topology_and_construction_for_general_multiplet} we classify the resulting topological field configurations (TFCs) by hypercharge \(Y\): if \(Y\neq 0\) (as for the Standard Model Higgs), the resulting TFC is a sphaleron; if \(Y=0\), the resulting TFC is a monopole (as for a real triplet, here extended to higher-dimensional representations). In addition, zero hypercharge can contribute to part of the WIMP relic density. We also construct, for the first time, the sphaleron matrix for a general-dimensional representation and establish its connection to the monopole matrix. In Sec.~\ref{sec:mass_and_eom_for_general_multiplet}, we compute the general-dimensional sphaleron energy and monopole mass, and discuss how the situation changes in multi-step phase transitions.
  
  \item In Sec.~\ref{sec:sphaleron_power_counting} and Sec.~\ref{sec:sphaleron_scaling_properties_and_numerical_fittings}, we develop a gauge-invariant formalism for the sphaleron rate in \(SU(2)\)+Higgs theory within the three-dimensional thermal effective field theory (3D EFT). This formalism is gauge invariant under power counting up to \(\mathcal{O}(g^4)\), where \(g\) is the \(SU(2)\) gauge coupling. As a benchmark, we compare our result with the SM sphaleron rate from lattice simulations and find good agreement; see Fig.~\ref{fig:sph_SM_vev_scaling_fit} in Sec.~\ref{sec:sphaleron_rate_crossover}. Using this gauge-invariant formalism, we compute the baryon-preservation condition for electroweak baryogenesis, and replace the previous approximate criterion \(v_c/T_c \gtrsim 1\) with the new gauge-invariant criterion on \(x=\lambda_3/g_3^2\); see Fig.~\ref{fig:yplot} and the discussion at the end of Sec.~\ref{sec:gauge_invariant_BPC}. An important implication of this formalism is that, even when CP violation overproduces the baryon asymmetry, the washout can be computed precisely to recover the observed value today.
  The previous condition can be gauge dependent if gauge dependence is not handled consistently. We apply this formalism to the real triplet extension of the SM in Sec.~\ref{sec:sphaleron_rate_and_BPC_in_RSM}, and find that the baryon washout rate can be very large in much of the parameter space, placing strong constraints on beyond the Standard Model (BSM) theories. Using the real triplet model, we also show that two-loop thermal corrections are necessary for a consistent perturbative expansion of both phase-transition properties and sphaleron rates, which can be seen from the comparison of the one-loop and two-loop results in Fig.~\ref{fig:sigmaSM_x_a2_Msigma_one_loop} and Fig.~\ref{fig:sigmaSM_x_a2_Msigma_two_loop}, respectively.
  
  \item In Sec.~\ref{sec:PBH_from_delayed_FOPT} and Sec.~\ref{sec:collider_probe_FOPT_light_singlet}, we explore broader cosmological and phenomenological implications of a first-order phase transition. On the cosmology side, we study primordial-black-hole production from a delayed first-order phase transition in a simple scalar model, and show in Eq.~\eqref{eq:fpbh_approx} that the resulting relic abundance is super-exponentially sensitive to the phase-transition parameters, so that even small parameter variations can lead to huge variations in the PBH relic abundance. On the phenomenology side, using a light real-singlet extension of the SM as an example, we show in Fig.~\ref{fig:experimental_limits_on_scaled_branching_ratio} that collider searches for BSM scalars---especially at future lepton colliders such as CEPC---can indirectly probe the first-order phase transition required by electroweak baryogenesis through exotic Higgs decays, and can cover a large portion of the parameter space yielding a strong first-order phase transition.
\end{itemize}

\subsection{Background and Prerequisite Knowledge}

Below, we list the background topics introduced in this thesis, together with the sections where details are provided. We do not include all possible background material, for example primordial black hole in detail, because primordial black hole appear here mainly as a consequence of delayed first-order phase transitions rather than as a central topic. We therefore cite the relevant literature for interested readers. The topics listed below are directly connected to the key contributions and novelties of this thesis and are necessary for understanding them.
\begin{itemize}
    \item Homotopy theory is the mathematical framework used throughout this thesis to understand topological field configurations (TFCs), including their existence, classification, and stability. It can also be used to illuminate the vacuum structure of gauge theory. We provide a pedagogical introduction to homotopy theory in Sec.~\ref{sub:homotopy-theory}, discuss the general properties of TFCs and the role of homotopy theory in Sec.~\ref{sec:roles_of_topology}, and present detailed mathematical formulations of common TFCs (cosmic string, monopole, sphaleron, and instanton) in Sec.~\ref{sec:Examples of topological field configurations}.
    
    \item The ABJ anomaly is introduced in Appendix~\ref{sec:ABJ_anomaly} using a simple sigma model with the diagrammatic approach. This anomaly is fundamental to understanding baryon-number violation in the SM, a key ingredient in baryogenesis mechanisms.

    \item The sphaleron and monopole mechanisms for baryon-number violation are introduced in Sec.~\ref{sec:sphaleron_baryon_violation} and Sec.~\ref{sec:monopole_induced_bnv}, respectively. For pedagogical purposes, we also introduce instanton and tunneling in quantum mechanics and the Standard Model instanton mechanism for baryon-number violation in Appendices~\ref{sec:instanton_QM} and~\ref{sec:instanton_baryon_violation}. These quantum-mechanical examples provide visualizations that aid understanding of the more complex topological field configurations.
    
    \item Matsubara frequencies and the one-loop thermal effective potential in a general $R_\xi$ gauge are introduced in Appendices~\ref{app:matsubara_frequencies} and \ref{app:effective_potential_computation}, respectively. Dimensional reduction in thermal effective field theory is discussed in Sec.~\ref{sec:thermal_effective_potential_DR}. Gauge cancellation in the dimensionally reduced effective theory is illustrated with the Higgs self-coupling example in Sec.~\ref{sec:gauge_dependence_effective_potential}.
\end{itemize}

\newpage
\section{Puzzles of Our Universe}
\label{sec:puzzles}

\subsection{Baryon Asymmetry}
\label{sec:baryon_asymmetry}

There is strong cosmological and astrophysical evidence for the baryon asymmetry (matter--antimatter asymmetry) of our Universe. The origin of this asymmetry remains an open question, and many mechanisms have been proposed to explain it. In this section, we first introduce the observational evidence, then discuss the Sakharov conditions for baryogenesis, and finally explain how topological field configurations, such as sphaleron, play a central role. We also briefly review representative baryogenesis scenarios, with a focus on electroweak baryogenesis, including its advantages and experimental tests. For general reviews of this subject, see Refs.~\cite{Riotto:1999yt,Dine:2003ax,Bodeker:2020ghk}.

Two key cosmological probes of baryon asymmetry are the Cosmic Microwave Background (CMB) \cite{Planck:2018vyg} and Big Bang Nucleosynthesis (BBN) \cite{Fields:2019pfx}. The asymmetry is commonly parameterized by the baryon-to-photon ratio \footnote{
Alternatively, one may use the baryon-to-entropy ratio,
\(Y_B \equiv n_B/s = (8.70\pm 0.04)\times 10^{-11}\)
\cite{Planck:2018vyg,Fields:2019pfx}.}:
\begin{align}
\eta = \frac{n_B - n_{\bar{B}}}{n_\gamma} \approx (6.10\pm 0.04) \times 10^{-10} \ .
\end{align}
Here \(n_B\), \(n_{\bar{B}}\), and \(n_\gamma\) denote the number densities of baryons, antibaryons, and photons, respectively. In addition, the Universe cannot be a patchwork of matter and antimatter domains, because the resulting diffuse gamma-ray background would exceed observational limits \cite{Cohen:1997ac}.

\paragraph{Sakharov Conditions and Baryogenesis Mechanisms}
Any successful baryogenesis mechanism must satisfy three necessary conditions, known as the Sakharov conditions \cite{Sakharov:1967dj}:  
(1) baryon-number violation,  
(2) C and CP violation,  
(3) departure from thermal equilibrium.  

Here we briefly comment on the third condition. Assuming CPT invariance, particles and antiparticles have equal masses, so in thermal equilibrium their distributions are identical. Therefore, an out-of-equilibrium environment is needed to bias baryon-number-violating processes and produce an excess of baryons over antibaryons. Such a departure can arise from a first-order phase transition, where bubble nucleation and expansion provide non-equilibrium dynamics, as in electroweak baryogenesis \cite{Kuzmin:1985mm}.   It can also arise from out-of-equilibrium decays of heavy particles, as in leptogenesis \cite{Fukugita:1986hr}, or through topological defects, such as superconducting strings \cite{Brandenberger:1998mj}.

One may alternatively consider CPT violation, in which case thermal equilibrium can be maintained while still generating an asymmetry. Since CPT is generally assumed to hold at zero temperature, such violation is typically associated with the expanding early Universe. In this picture, CPT violation acts like an effective chemical potential, biasing baryons and antibaryons even in equilibrium. This is the idea behind spontaneous baryogenesis \cite{Cohen:1987vi,Cohen:1988kt}\footnote{We note that if the probabilities of generating positive and negative chemical potentials are equal, the final baryon asymmetry may still vanish. This can be avoided, for example, during inflation.} and, later, gravitational baryogenesis \cite{Davoudiasl:2004gf}. Other baryogenesis scenarios include Affleck--Dine baryogenesis \cite{Affleck:1984fy}, GUT baryogenesis \cite{Kolb:1990vq}, resonant leptogenesis \cite{Pilaftsis:2003gt}, axiogenesis \cite{Co:2019wyp}, baryogenesis from Hawking radiation \cite{Hook:2014mla}, mesogenesis \cite{Elor:2018twp}, and Peccei--Quinn genesis \cite{Chun:2025brc}; see also the summary in Ref.~\cite{Elor:2022hpa}. In this thesis, we focus on electroweak baryogenesis, especially recent developments in sphaleron-rate calculations and thermal phase-transition analyses. These techniques are not unique to electroweak baryogenesis and can also be applied to other baryogenesis scenarios, as well as to cosmological phenomena driven by thermal phase transitions. For reviews of electroweak baryogenesis, from classic treatments to recent syntheses, see Refs.~\cite{Morrissey:2012db,Cohen:1993nk,Cline:2006ts,Konstandin:2013caa,Garbrecht:2018mrp,White:2016nbo,vandeVis:2025efm,Barni:2025ifb}. Closely related ideas have also been explored, where electroweak-symmetric regions are provided by cosmic strings, domain walls, and seeded phase transitions \cite{Brandenberger:1992ys,Brandenberger:1994bx,Trodden:1994ve,Brandenberger:1994mq,Abel:1995uc,Dasgupta:1996ys,Brandenberger:2005bx,Blasi:2022woz,Sassi:2023cqp,Sassi:2024cyb,Schroder:2024gsi,Wei:2024qpy,Azzola:2024pzq,Azzola:2026mbq,Azzola:2026cwa,Bai:2026udq}; other catalyst-like objects and exploding primordial black holes provide additional analogues \cite{Bai:2021xyf,Klipfel:2026nzx}. For recent baryogenesis scenarios operating during the electroweak crossover through sphaleron dynamics, see Refs.~\cite{Kharzeev:2019rsy,Hong:2023zrf,Tanaka:2025cpw,Ogawa:2026olw} \footnote{However, the physical properties of the ``sphaleron-like configurations'' discussed in these references remain under debate.}.

\paragraph{Why the Standard Model Is Not Enough: Electroweak Baryogenesis in Context}
Before discussing the details of electroweak baryogenesis, we briefly explain why the pure Standard Model (SM) cannot account for the observed baryon asymmetry, and why new physics is needed. The SM does provide one Sakharov ingredient: baryon-number violation through electroweak sphaleron processes at high temperature. However, it fails in the other two respects:
(i) For the measured Higgs mass (\(m_H\approx125\,\mathrm{GeV}\)), lattice studies show that the electroweak transition is a smooth crossover rather than a strong first-order phase transition \cite{Kajantie:1995kf,Kajantie:1996mn,Kajantie:1996qd,Csikor:1998ge,Aoki:1999fi}. Without bubble nucleation and expanding phase boundaries, the required out-of-equilibrium dynamics is insufficient.
(ii) Even if the SM transition were first order, CP violation from the Cabibbo--Kobayashi--Maskawa (CKM) matrix is not sufficient to generate the observed baryon asymmetry \cite{Gavela:1993ts,Gavela:1994dt,Huet:1994jb}.

In electroweak baryogenesis, sphaleron still provide baryon-number violation, while new BSM physics supplies stronger C \& CP violation and a first-order phase transition. For example, non-equilibrium dynamics may be induced by additional scalar fields coupled to the Higgs sector, and CP violation may arise from new complex phases in scalar or fermion couplings. Candidate extensions include singlet scalars \cite{Profumo:2007wc,Espinosa:1993bs,Barger:2007im,Damgaard:2015con,Kotwal:2016tex,Brauner:2016fla,Cline:2017qpe,Gould:2019qek,Carena:2019une,Niemi:2021qvp,Wang:2022dkz,Niemi:2024axp,Ellis:2022lft,Giovanakis:2024rvg,Chen:2025ksr,Ahriche:2007jp,Espinosa:2011ax,Curtin:2014jma,Profumo:2014opa,Vaskonen:2016yiu,Beniwal:2017eik,Kurup:2017dzf,Alves:2018jsw,Kozaczuk:2019pet,Harigaya:2022ptp,Bosse:2026bdk}, new SU(2) multiplets \cite{FileviezPerez:2008bj,Chowdhury:2011ga,Patel:2012pi,Niemi:2018asa,Chao:2018xwz,Niemi:2020hto,Wu:2023mjb,Inoue:2015pza,Chala:2018opy,Zhou:2022mlz}, two-Higgs-doublet models (2HDM) and its extensions \cite{Hammerschmitt:1994fn,Turok:1991uc,Davies:1994id,Cline:1996mga,Dorsch:2013wja,Bernon:2017jgv,Kainulainen:2019kyp,Biekotter:2025fjx,Aiko:2025tbk,Lee:2025hgb,Garbrecht:2025jgy,Gent:2025csq,Fromme:2006cm,Ginzburg:2009dp,Cline:2011mm,Tranberg:2012jp,Dorsch:2014qja,Basler:2016obg,Dorsch:2016nrg,Su:2020pjw,Fabian:2020hny,Zhou:2020irf,Aoki:2021oez,Goncalves:2021egx,Biekotter:2021ysx,Basler:2021kgq,Wang:2022yhm,Biekotter:2022kgf}, models with effective operators \cite{Ham:2004zs,Bodeker:2004ws,Grojean:2004xa,Delaunay:2007wb,Cai:2017tmh,Chala:2018ari,Ellis:2018mja,Phong:2020ybr,Camargo-Molina:2021zgz,Cai:2022bcf,Kanemura:2022txx,Qin:2024idc,Chala:2024xll,Oikonomou:2024jms}, and the MSSM and its extensions \cite{Carena:1996wj,Delepine:1996vn,Cline:1996cr,Funakubo:2005pu,Curtin:2012aa,Katz:2015uja,Huang:2014ifa,Chatterjee:2022pxf,Ning:2025zfh,Huet:1995sh,Worah:1997ni,Cline:1998hy,Cline:2000nw,Ham:2004nv,Huber:2006wf,Funakubo:2009eg,Carena:2011jy,Bi:2015qva,Baum:2020vfl}. These new particles can be searched for at colliders \cite{Curtin:2014jma,Kozaczuk:2019pet,Wagner:2023vqw,Ramsey-Musolf:2024ykk,Biekotter:2025fjx,Bosse:2026bdk}, providing a direct test window for electroweak baryogenesis; see the collider target \cite{Ramsey-Musolf:2019lsf}. The bubble-wall velocity during a first-order phase transition plays an important role both in determining the generated baryon asymmetry and in predicting the resulting gravitational-wave signal; focused studies of bubble-wall velocities can be found in Refs.~\cite{Steinhardt:1981ct,Ignatius:1993qn,Arnold:1993wc,Moore:1995ua,Moore:1995si,Espinosa:2010hh,Konstandin:2014zta,Bodeker:2017cim,Lewicki:2021pgr,Ai:2021kak,Azatov:2021ifm,Gouttenoire:2021kjv,Laurent:2022jrs,DeCurtis:2022hlx,Azatov:2023xem,Ai:2023suz,Li:2023xto,Wang:2023lam,Ai:2023see,Long:2024sqg,Branchina:2025jou,Carena:2025flp,Ramsey-Musolf:2025jyk,Ai:2025bjw,Si:2025vdt,vandeVis:2025plm,Branchina:2025adj,Munzenberg:2025nwa}. The electroweak-baryogenesis sequence is discussed in Sec.~\ref{sec:EWBG_four_processes}. In parallel, stochastic gravitational waves from electroweak-scale first-order phase transitions may be probed by future space-based detectors such as LISA \cite{LISA:2017pwj}, Taiji \cite{Ruan:2018tsw}, and TianQin \cite{TianQin:2015yph}. For relevant gravitational-wave studies, see Refs.~\cite{Huber:2008hg,Caprini:2009yp,Caprini:2009fx,Hindmarsh:2013xza,Hindmarsh:2015qta,Jinno:2016vai,Cai:2017cbj,Chao:2017vrq,Hindmarsh:2017gnf,Ellis:2018mja,Chala:2018ari,Caprini:2019egz,Dev:2019njv,Guo:2020grp,Wang:2020jrd,Bian:2021ini,Tenkanen:2022tly,Ellis:2022lft,Kierkla:2023von,Ramsey-Musolf:2024ykk,Caprini:2024gyk,Guan:2025idx,Biekotter:2025fjx}.

\paragraph{Role of Sphaleron in Baryogenesis and Their Computation}
Although electroweak sphaleron transitions already exist in the SM, BSM particles can strongly affect the sphaleron rate. Beyond electroweak baryogenesis \cite{Kuzmin:1985mm}, sphaleron are also central in leptogenesis \cite{Fukugita:1986hr}, where a lepton asymmetry from heavy-neutrino decays is partially converted into baryon asymmetry, and in axiogenesis \cite{Co:2019wyp}, where PQ charge is converted through sphaleron processes\footnote{Sphaleron are not required in all baryogenesis mechanisms; for instance, baryon-number violation can arise differently in supersymmetric settings \cite{Nilles:1983ge,Haber:1984rc}.}. The SM sphaleron solution was first constructed by Manton and Klinkhamer \cite{Manton:1983nd,Klinkhamer:1984di}. An equivalent construction appears in Ref.~\cite{Klinkhamer:1990fi} and is discussed in Sec.~\ref{sec:manton_sphaleron} (see also Ref.~\cite{Wu:2023mjb}). Alternative ans\"atze were proposed in Refs.~\cite{Akiba:1988ay,Akiba:1989xu}, with equivalent physical properties. Many sphaleron-focused studies are available, e.g.\ Refs.~\cite{Ahriche:2014jna,Fuyuto:2014yia,Fuyuto:2015jha,Tye:2015tva,Gan:2017mcv,Qiu:2018wfb,Kharzeev:2019rsy,Zhou:2019uzq,Hamada:2020rnp,Phong:2020ybr,Kanemura:2020yyr,BarrosoMancha:2022mbj,Wu:2023mjb,Drewes:2023khq,Hong:2023zrf,Hu:2023gbp,Qin:2024idc,Matchev:2025ivr,Matchev:2025irm,Tanaka:2025cpw,Li:2025kyo,Annala:2025aci}\footnote{This thesis mainly focuses on electroweak sphalerons; for sphaleron studies in QCD or general SU(N) theories, see Refs.~\cite{Moore:2010jd,Klinkhamer:2017fqi,Altenkort:2020axj,Bonanno:2023xfv,Bonanno:2023thi,Gao:2023djs,Bodeker:2025ahg,Guin:2026kbp}.}. Currently, there is no experimental evidence for sphaleron-induced baryon- and lepton-number-violating processes \cite{Mattis:1991bj,Mattis:1992nt,Ellis:2016ast,CMS:2018ozv,Ringwald:2018gpv,Papaefstathiou:2019djz}; however, their mathematical foundation in homotopy theory is well understood. A detailed introduction to homotopy theory is given in Sec.~\ref{sub:homotopy-theory}. If the weak mixing angle is not set to zero, the sphaleron acquires a dipole moment~\cite{Klinkhamer:1984di,James:1992re,Klinkhamer:1990fi,Kunz:1992uh}. As a result, the presence of a magnetic field can affect the sphaleron energy. For studies on the impact of magnetic fields on sphaleron properties, see Refs.~\cite{Comelli:1999gt,Ho:2020ltr,Annala:2023jvr,Patel:2023ybi,Di:2025ncl}. For studies of magnetic-field helicity from sphalerons, see Ref.~\cite{Cornwall:1997ms,Vachaspati:2001nb,Copi:2008he}. More generally, the influence of magnetic fields on the vacuum structure is discussed in Refs.~\cite{Chernodub:2022ywg,Adhikari:2024bfa}.

Because the physically relevant quantity is the finite-temperature sphaleron rate, thermal field theory is essential. In this thesis, we use dimensionally reduced three-dimensional effective field theory (3D EFT), which enables gauge-invariant power counting and direct comparison with lattice simulations, which are also formulated in 3D EFT. This method systematically integrates out non-zero Matsubara modes and heavy modes, retaining the scales most relevant for sphaleron dynamics. This framework was developed in Refs.~\cite{Ginsparg:1980ef,Appelquist:1981vg,Kajantie:1995dw,Braaten:1995cm} and extended to many BSM scenarios; see, for example, Refs.~\cite{Andersen:2017ika,Niemi:2018asa,Croon:2020cgk,Niemi:2021qvp,Niemi:2020hto,Gould:2021ccf,Hirvonen:2021zej,Ekstedt:2022bff,Ekstedt:2022zro,Gould:2023ovu,Ekstedt:2024etx,Li:2025kyo,Liu:2026ask}.

For SU(2)+Higgs theory, one-loop determinant calculations of the sphaleron rate were carried out in Refs.~\cite{Carson:1989rf,Carson:1990jm,Baacke:1993aj,Baacke:1994ix}, building on the zero-mode analysis \footnote{Here we note that the perturbative formalism used to compute sphaleron rates shares many features with that used for bubble-nucleation rates; early developments of the latter include Refs.~\cite{Langer:1967ax,Langer:1969bc,Coleman:1977py,Callan:1977pt,Linde:1980tt,Affleck:1980ac}. On the one hand, both are semiclassical barrier-crossing problems governed by a classical field solution, a unstable mode, and fluctuation determinants around the classical field configuration. On the other hand, in the context of first-order phase transitions, this formalism has been developed considerably, from early analyses of gauge independence \cite{Nielsen:1975fs,Fukuda:1975di,Metaxas:1995ab,Patel:2011th,Garny:2012cg,Andreassen:2014gha} to recent gauge-invariant 3D EFT treatments, higher-order computations, and improved nucleation-rate calculations \cite{Gould:2021ccf,Lofgren:2021ogg,Hirvonen:2021zej,Ekstedt:2022zro,Ekstedt:2024etx,Kierkla:2025qyz,Hallfors:2025key,Liu:2026ask,Navarrete:2025yxy}. 
Similar analyses also arise in the sphaleron case and will be discussed in later sections. In this sense, the study of sphaleron transitions and bubble nucleation is closely related at the level of formalism. For recent developments in vacuum decay and finite-temperature tunneling, see also \cite{Witten:2010cx,Tanizaki:2014xba,Andreassen:2016cff,Andreassen:2017rzq,Espinosa:2018szu,Espinosa:2019hbm,Mou:2019gyl,Hertzberg:2019wgx,Ai:2019fri,Khoury:2021zao,Espinosa:2022ofv,Nishimura:2023dky,Ai:2023yce,Steingasser:2023gde,Steingasser:2024ikl,Garbrecht:2024end,Garbrecht:2025alb,Barni:2026dhc}. For quantum simulation of ``sphaleron-like" dynamics, see Ref.~\cite{Huang:2025nlo}.} of Ref.~\cite{Arnold:1987mh}. Later perturbative improvements were developed in Ref.~\cite{Burnier:2005hp}, mainly for the SM crossover regime. Lattice studies include Refs.~\cite{Moore:1998swa,Moore:1999fs,Moore:2000mx,Moore:2010jd,DOnofrio:2014rug,Altenkort:2020axj,BarrosoMancha:2022mbj,Annala:2023jvr,Annala:2025aci}. While lattice methods can access first-order transitions, their simulations are computationally intensive and thus impractical for extensive parameter scans in beyond-the-Standard-Model theories. For this reason, perturbative analyses remain necessary to facilitate comprehensive parameter space exploration. A perturbative three-dimensional effective-field-theory formalism for computing the sphaleron rate during first-order phase transitions was developed only recently in Ref.~\cite{Li:2025kyo}. This development is one of the key contributions of this thesis, and the corresponding technical advances are summarized in Sec.~\ref{sec:contributions_and_novelty}.

\subsection{Dark Matter}
\label{sec:dark_matter}

Observational evidence for dark matter spans multiple length scales: galactic scales (e.g.\ galaxy rotation curves \cite{Begeman:1991iy}), galaxy-cluster scales (from the early velocity-dispersion measurements in the Coma cluster \cite{Zwicky:1933gu} to later analyses of the Bullet Cluster \cite{Robertson:2016xjh}), and cosmological scales (especially the Cosmic Microwave Background, CMB \cite{Planck:2018vyg}). These observations indicate that dark matter contributes about 27\% of the total energy density of the Universe \cite{Planck:2018vyg}, roughly five times the abundance of ordinary baryonic matter. Many dark-matter candidates have been proposed; for broad reviews, see Refs.~\cite{Bertone:2004pz,Feng:2010gw,Bertone:2016nfn,Battaglieri:2017aum,Arbey:2021gdg,Bozorgnia:2024pwk}.

In this thesis, we focus on two well-motivated dark-matter candidates: WIMP \cite{Jungman:1995df,Cirelli:2005uq} and PBH \cite{Zeldovich:1967lct,Hawking:1971ei,Carr:2016drx,Villanueva-Domingo:2021spv,Escriva:2022duf,Bartolo:2018evs,Carr:2020gox}. Among them, only PBH will be discussed in detail. Other important candidates, such as axions and axion-like particles \cite{Preskill:1982cy,Abbott:1982af,Dine:1982ah}, are not treated in detail here. Still, axion physics is closely related to vacuum structure of non-Abelian gauge theories and instantons, topics that are central to this thesis (see Sec.~\ref{sec:BNV_in_gauge_theory}).

Although dark matter is not the main subject of this thesis, it appears naturally as a byproduct of our study of baryogenesis and thermal phase transitions. For example, an SU(2) multiplet with monopole topology during the electroweak phase transition can contribute to the WIMP relic abundance. We do not perform a detailed relic-density analysis for WIMP; extensive studies already exist in the literature \cite{Cirelli:2005uq,Arcadi:2017kky,Roszkowski:2017nbc,Chao:2018xwz,Chiang:2020rcv}. On the other hand, delayed first-order phase transitions can produce PBH, which may constitute part or all of dark matter. For this reason, we include dark matter as an independent section, parallel to baryon asymmetry, while keeping its scope narrower. We also emphasize that baryogenesis and dark matter are not fully independent topics; they can share a common origin, as in asymmetric dark matter scenarios \cite{Petraki:2013wwa}.

If a first-order phase transition proceeds slowly, the Universe can undergo supercooling, and some patches may remain in the false vacuum for a long period even after most patches have transitioned into the new phase. In that case, sufficiently large inhomogeneities between true- and false-vacuum regions can collapse into black holes \cite{Crawford:1982yz,Hawking:1982ga,Kodama:1982sf}. 
This possibility has been studied extensively in recent years
\cite{Kodama:1982sf,Hall:1989hr,Khlopov:1998nm,Jedamzik:1999am,Wang:2020jrd,
Liu:2021svg,Hashino:2021qoq,Hashino:2022tcs,He:2022amv,Kawana:2022olo,
Lewicki:2023ioy,Gouttenoire:2023naa,Banerjee:2023brn,Gouttenoire:2023pxh,
Banerjee:2023qya,Baldes:2023rqv,Salvio:2023blb,Conaci:2024tlc,
Lewicki:2024ghw,Flores:2024lng,Kanemura:2024pae,Cai:2024nln,
Goncalves:2024vkj,Banerjee:2024fam,Arteaga:2024vde,Banerjee:2024cwv,
Wu:2024lrp,Hashino:2025fse,Zou:2025sow,Franciolini:2025ztf,Cao:2025jwb,
Kierkla:2025vwp,Zhang:2025kbu,Huang:2025hos,An:2026hiq,Wang:2026zvz,
Ning:2026nfs,Ai:2026zrs} \footnote{For studies of the sphaleron energy outside the PBH, see Ref.~\cite{DeLuca:2021oer}}. See also Refs.~\cite{Garriga:2015fdk,Ai:2024cka} for mechanisms involving phase transitions.
For the delayed first-order phase transition mechanism, the PBH relic abundance is highly sensitive to model parameters; the perturbative analysis in Ref.~\cite{Wu:2024lrp} \footnote{We note that the analysis of Ref.~\cite{Wu:2024lrp} has since been significantly improved, especially in the treatment of gauge dependence \cite{Franciolini:2025ztf,Wang:2026zvz} and in the analysis of the matter-dominated era after a slow phase transition \cite{Ai:2026zrs}. However, to keep this PhD thesis self-contained, we still present the original analysis here and direct the reader to these later developments for more accurate results.} shows a super-exponential dependence of the relic density on those parameters. Therefore, if the predicted PBH abundance exceeds the observed dark-matter density, the corresponding parameter region is excluded. This provides a complementary probe of BSM theories that realize first-order phase transitions, alongside collider and gravitational-wave searches.

\newpage
\section{Aspects of Topological Field Configurations}
\label{sec:aspects-TPC}

Why do we need to study topological field configurations (TFCs), and what are they?  
The purpose of this section is to answer these two questions. One important role of TFCs is in realizing baryon-number violation in gauge theories: TFCs are non-trivial solutions of the classical equations of motion, and quantum fluctuations around these classical solutions can induce effective interactions that violate baryon and lepton numbers. This mechanism will be discussed in detail in the next section. 

\flag One of the main novelties of this thesis is the discussion of TFCs for a general-dimensional \(SU(2)\) scalar multiplet undergoing the electroweak phase transition, since this is the scenario of interest in many BSM theories, where either sphaleron or monopole solutions may appear depending on the hypercharge; see Sec.~\ref{sec:topology_and_construction_for_general_multiplet}.

In the present section, we first give a pedagogical introduction to TFCs, including homotopy theory and its application to field theory, equations of motion, and field energy. Reviews of TFCs can be found in Refs.~\cite{Coleman:1978ae,Tong:2005un}. We note that homotopy theory provides the mathematical tools needed to study TFCs, including their existence, classification, and stability. Homotopy theory is also important for understanding gauge-vacuum structure and the sphaleron mechanism for baryon-number violation, which will be discussed in Sec.~\ref{sec:BNV_in_gauge_theory}. It is therefore important to understand homotopy theory before studying TFCs. For this reason, we first provide a pedagogical introduction to homotopy theory.

\subsection{Homotopy Theory}
\label{sub:homotopy-theory}

Suppose \(X\) and \(Y\) are two manifolds, and consider continuous maps from the domain \(X\) to the target \(Y\), \(f: X \to Y\). Two maps \(f\) and \(g\) are said to be \emph{homotopic} if there exists a continuous deformation from \(f\) to \(g\), i.e.\ a continuous map \(H: X \times [0,1] \to Y\) such that \(H(x,0)=f(x)\) and \(H(x,1)=g(x)\) for all \(x\in X\). The second parameter in \(H\) can be viewed as a ``time'' parameter that continuously deforms \(f\) into \(g\). The homotopy relation is an equivalence relation: it is reflexive (\(f\) is homotopic to itself), symmetric (if \(f\) is homotopic to \(g\), then \(g\) is homotopic to \(f\)), and transitive (if \(f\) is homotopic to \(g\) and \(g\) is homotopic to \(h\), then \(f\) is homotopic to \(h\)). A based map is defined by first choosing fixed points \(x_0\in X\) and \(y_0\in Y\), and then requiring \(f(x_0)=y_0\) for all maps \(f:X\to Y\).

The set of all maps from \(X\) to \(Y\) can be classified into equivalence classes under the homotopy relation; these are called \emph{homotopy classes}. For example, one such class consists of maps that can be continuously deformed to a constant map, i.e.\ maps that are continuously connected to a map satisfying \(f(x)=y_0\) for all \(x\in X\). This class is called the \emph{trivial homotopy class}. In general, there are also non-trivial homotopy maps that cannot be continuously deformed to a constant map, and these form non-trivial homotopy classes. Following Ref.~\cite{Manton:2004tk}, it is useful to study homotopy classes with \(X\) chosen to be an \(n\)-sphere \(S^n\), since this is also the situation of interest for the topological field configurations discussed later.

The \(n\)-sphere \(S^n\) is defined as the set of points in \(\mathbb{R}^{n+1}\) that are at a fixed distance (the radius) from a given point (the center). For example, \(S^1\) is a circle, \(S^2\) is the surface of an ordinary sphere, and \(S^3\) is the three-dimensional surface of a four-dimensional ball. If we choose the center to be the origin, then \(S^n\) can be defined as the set of points \((x_1,x_2,\ldots,x_{n+1})\) in \(\mathbb{R}^{n+1}\) such that
\begin{align}
  \label{eq:definition_Sn}
  x_1^2+x_2^2+\ldots+x_{n+1}^2=1 \ .
\end{align}

The homotopy classes of based maps from \(S^n\) to a target space \(Y\) are denoted by \(\pi_n(Y)\). The base point may be chosen as the north pole of \(S^n\), \((0,0,\ldots,1)\in\mathbb{R}^{n+1}\), together with a given point \(y_0\in Y\). It can be proved that, for \(n\geqslant 1\), \(\pi_n(Y)\) has a group structure and is called the \(n\)-th \emph{homotopy group} of \(Y\). It has an associative group operation, an identity element (the homotopy class of the constant map), and inverses (represented by traversing the same map in the opposite orientation) \cite{Manton:2004tk}. Thus, the trivial homotopy class of constant maps is the identity element of the homotopy group. To further explore the other elements of homotopy groups, we now consider the case in which the target space \(Y\) is also a sphere, \(S^m\).

The \(U(1)\) and \(SU(2)\) gauge groups that will be discussed later are isomorphic to \(S^1\) and \(S^3\), respectively. Here \emph{isomorphic} means that there exists a bijective map between the two groups that preserves the group operation. Suppose we have two groups \(A\) and \(B\), together with a map \(\phi:A\to B\). The bijection means that for every element \(b\in B\), there exists a unique element \(a\in A\) such that \(\phi(a)=b\), and for every element \(a\in A\), there exists a unique element \(b\in B\) such that \(\phi^{-1}(b)=a\). Preservation of the group operation means that for any two elements \(a_1,a_2\in A\), we have \(\phi(a_1*a_2)=\phi(a_1)*'\phi(a_2)\), where \(*\) and \(*'\) are the group operations in \(A\) and \(B\), respectively. Thus, an isomorphism can be understood as a one-to-one correspondence between the elements of two groups that preserves their algebraic structure. If the bijective condition does not hold, the map is called a \emph{homomorphism}, and the two groups are not isomorphic. In other words, isomorphism is a stronger condition than homomorphism.
Sometimes we encounter the term \emph{diffeomorphism}, it is defined as an isomorphism of differentiable manifolds. Even though the $n-$sphere $S^n$ that we considered is a differentiable manifold, we will use the term isomorphism in the following discussion, since this concept is more commonly used in the literature on topological field configurations. We denote isomorphism by the symbol \(\cong\). For example, the \(U(1)\) group can be represented as the set of complex numbers of unit magnitude, which can be visualized as points on a circle, i.e.\ \(S^1\). A one-to-one correspondence can be established between the elements of \(U(1)\) and the points on \(S^1\), such that the group operation in \(U(1)\) corresponds to the addition of angles on the circle. Therefore, we have \(U(1)\cong S^1\). Similarly, an element of \(SU(2)\) can be written as
\begin{align}
  \label{eq:SU(2)_element}
  g=a_0 \mathbb{I}+i a_1 \sigma_1+i a_2 \sigma_2+i a_3 \sigma_3 \ ,
\end{align}
where \(\mathbb{I}\) is the \(2\times 2\) identity matrix, \(\sigma_i\) are the Pauli matrices, and \(a_0^2+a_1^2+a_2^2+a_3^2=1\), which is the definition of \(S^3\). A rigorous mathematical proof of the isomorphism between $SU(2)$ and $S^3$ is beyond the scope of this thesis, which can be found in Ref.~\cite{wiki:special_unitary_group} (note that the terminology used there is diffeomorphism).

\begin{figure}[t]
\center
\includegraphics[width=12cm]{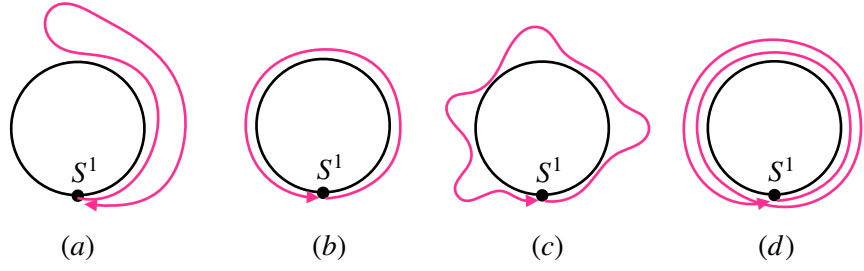}
\caption{Visualization of the homotopy group \(\pi_1(S^1)\). The black circle represents the target space \(S^1\), and the pink curve represents a map from the domain \(S^1\) to the target \(S^1\). The black dot at the bottom of the circle denotes the base point. Panel (a) is homotopic to the trivial map and therefore corresponds to the identity element \(0\) in \(\pi_1(S^1)\). Panels (b) and (c) are homotopic to each other and correspond to the element \(1\), representing one anticlockwise winding of the circle. Panel (d) corresponds to the element \(2\), representing two anticlockwise windings.}
\label{fig:Pi_1_S_1}
\end{figure}

\begin{figure}[t]
\center
\includegraphics[width=12cm]{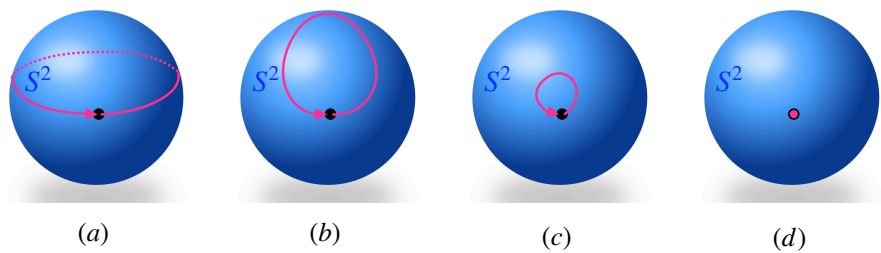}
\caption{Visualization of the homotopy group \(\pi_1(S^2)\). The blue sphere represents the target space \(S^2\), and the pink curve represents a map from \(S^1\) to \(S^2\). The black dot on the sphere denotes the base point. All panels (a), (b), (c), and (d) are homotopic to each other, since each pink curve can be continuously deformed to the trivial map. They therefore all correspond to the same element \(0\) in the group \(\pi_1(S^2)\). Hence, \(\pi_1(S^2)\cong 0\).}
\label{fig:Pi_1_S_2}
\end{figure}

Below, we discuss the homotopy groups of spheres, \(\pi_n(S^m)\), especially for \((n,m)=(1,2,3)\), which are relevant for the topological field configurations introduced later. One can show that \(\pi_1(S^1)\cong \mathbb{Z}\) and \(\pi_1(S^2)\cong 0\). Since both \(\pi_1(S^1)\) and \(\mathbb{Z}\) are groups, we use the isomorphism symbol \(\cong\) to denote their equivalence. The group \(\mathbb{Z}\) is the set of integers. Some intuition for these results is provided by Figs.~\ref{fig:Pi_1_S_1} and \ref{fig:Pi_1_S_2}. For \(\pi_1(S^1)\), panel (a) represents the trivial map, since it can be continuously deformed to a constant map. Thus, panel (a) represents the identity element \(0\) of \(\pi_1(S^1)\). Panels (b) and (c) represent two maps that are homotopic to each other, and therefore correspond to the same element, denoted by \(1\), which represents one anticlockwise wrapping of the circle. Panel (d) represents a map that is not homotopic to the maps in panels (a), (b), and (c), and thus corresponds to a different element, denoted by \(2\). The wrapping can be either clockwise or anticlockwise, so the elements \(-1\) and \(-2\) also exist in the group \(\pi_1(S^1)\). More generally, any integer \(n\) can be represented by a map that wraps around the circle \(n\) times, and therefore \(\pi_1(S^1)\cong \mathbb{Z}\). By contrast, for \(\pi_1(S^2)\), Fig.~\ref{fig:Pi_1_S_2} shows that all the maps in panels (a), (b), (c), and (d) are homotopic to each other. Hence they all represent the same element, and therefore \(\pi_1(S^2)\cong 0\). More generally \cite{hatcher_algebraic_2001},
\begin{align}
  \pi_n(S^n) &\cong \mathbb{Z}, \quad \text{for } n\geqslant 1 \ ,\\
  \pi_n(S^m) &\cong 0, \quad \text{for } n<m \ .
\end{align}
Here \(n\) and \(m\) are positive integers.

Applying these results to the manifolds of the gauge groups, we obtain the relations that will be used later in the discussion of topological field configurations:
\begin{align}
  \pi_1(U(1)) &\cong \pi_1(S^1)\cong\mathbb{Z}\ ,\\
  \pi_2(SU(2)) &\cong \pi_2(S^3)\cong 0 \ ,\\
  \pi_3(SU(2)) &\cong \pi_3(S^3)\cong\mathbb{Z}\ .
\end{align}
The homotopy group \(\pi_1(U(1))\cong \mathbb{Z}\) will be relevant for the cosmic-string solution discussed later, while \(\pi_3(SU(2))\cong \mathbb{Z}\) will be relevant for the instanton solution. As mentioned above, only for \(n\geqslant 1\) does \(\pi_n(Y)\) have a group structure. For \(n=0\), the homotopy classes of maps from \(S^0\) to \(S^0\) are denoted by \(\pi_0(S^0)\), and for \(S^0=\{+1,-1\}\), one typically obtains two classes if one point is chosen as the base point. Although \(\pi_0(S^0)\) does not have a group structure, it still provides a topological classification of maps from \(S^0\) to \(S^0\), as in the case of the kink solution in one spatial dimension.

Finally, we discuss homotopy groups of coset spaces. A \emph{coset space} is defined as the quotient of a group \(G\) by a subgroup \(H\), denoted by \(G/H\). This structure appears frequently in gauge theories with spontaneous symmetry breaking, where \(G\) is the original gauge group and \(H\) is the unbroken subgroup after symmetry breaking. According to homotopy theory \cite{Hilton_1953}, the homotopy groups of the coset space \(G/H\) satisfy\footnote{The proof of this statement involves the exact homotopy sequence, which lies beyond the scope of this thesis. We refer the reader to Ref.~\cite{Hilton_1953} for details.}
\begin{align}
  \label{eq:pi_2_G/H}
  \pi_2(G/H)=\pi_1(H) \ ,
\end{align}
provided that \(G\) is connected, i.e.\ \(\pi_0(G)=0\), and simply connected, i.e.\ \(\pi_1(G)=0\). Note that \(SU(2)\) is connected and simply connected, while \(SO(3)\) is connected but not simply connected. For example, for the symmetry-breaking pattern \(SU(2)\to U(1)\), we have \(G/H=SU(2)/U(1)\), and therefore \(\pi_2(SU(2)/U(1))=\pi_1(U(1))\cong \mathbb{Z}\). Recalling that \(\pi_2(S^2)\cong \mathbb{Z}\), \(SU(2)\cong S^3\), and \(U(1)\cong S^1\), one can understand this result by visualizing\footnote{A rigorous proof of this statement involves the Hopf fibration, which expresses \(S^3\) in terms of \(S^1\) and \(S^2\). See Ref.~\cite{wiki:hopf_fibration} for more details.} the coset space \(SU(2)/U(1)\) as being isomorphic to the two-sphere \(S^2\). Later we will see that the 't Hooft--Polyakov monopole is a topological field configuration associated with the non-trivial homotopy group \(\pi_2(SU(2)/U(1))\cong \mathbb{Z}\) \footnote{Strictly speaking, for the scalar field transforms in the adjoint representation, one may instead regard the monopole vacuum manifold as \(SO(3)/U(1)\). This does not change the conclusion, however, since \(SO(3)/U(1)\cong S^2\). Indeed, \(SO(3)\cong SU(2)/Z_2\), where \(Z_2=\{\pm \mathbb{I}\}\). Since the relevant \(U(1)\) subgroup already contains this \(Z_2\), namely the elements \(e^{i0}\) and \(e^{i\pi}\), the double covering \(SU(2)\to SO(3)\) does not affect the resulting coset space. A more detailed discussion of coset spaces for various groups can be found in Ref.~\cite{Lee2012}.}.

\begin{figure}[t]
\center
\includegraphics[width=5cm]{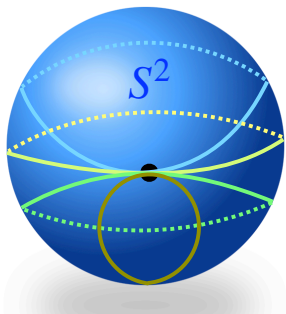}
\caption{Visualization of the homotopy group \(\pi_1(\text{Maps}_0(S^1 \to S^2))\). Here the target space is the space of maps from \(S^1\) to \(S^2\), represented by the colored loops on the blue sphere. Although each individual loop is contractible to the base point, all loops together need not be simultaneously contractible. This configuration corresponds to the element \(1\) in the homotopy group \(\pi_1(\text{Maps}_0(S^1 \to S^2))\).}
\label{fig:Pi_1_Maps_S1_S2}
\end{figure}

Finally, we consider another kind of homotopy group that will be relevant for the sphaleron solution. Previously, the target space was taken to be an \(m\)-sphere \(S^m\). We now consider the case in which the target space is itself a space of maps, denoted by \(\text{Maps}_0(X\to Y)\), namely the set of all maps from \(X\) to \(Y\), where the subscript \(0\) indicates that we include the constant map that sends all points in \(X\) to a given point in \(Y\). Specifically, we consider the spaces \(\text{Maps}_0(S^q \to S^m)\), and study the homotopy classes of maps from \(S^n\) to \(\text{Maps}_0(S^q \to S^m)\), denoted by \(\pi_n(\text{Maps}_0(S^q \to S^m))\). Following Ref.~\cite{Manton:2004tk}, the result is
\begin{align}
  \pi_n(\text{Maps}_0(S^q \to S^m))\cong \pi_{n+q}(S^m) \ .
\end{align}

To build intuition for this relation, consider \(\pi_1(\text{Maps}_0(S^1 \to S^2))\). Since \(\pi_1(S^2)\cong 0\), every individual map in \(\text{Maps}_0(S^1 \to S^2)\) can be continuously contracted to the trivial map. However, \(\pi_1(\text{Maps}_0(S^1 \to S^2))\) need not be trivial. A non-trivial element of \(\pi_1(\text{Maps}_0(S^1 \to S^2))\) can be visualized in Fig.~\ref{fig:Pi_1_Maps_S1_S2}. In that figure, the target space is the space of maps from \(S^1\) to \(S^2\), represented by the various colored loops on the blue sphere. Although each of these loops is contractible to the base point, all loops together are not \emph{simultaneously} contractible, and this corresponds to the element \(1\) in \(\pi_1(\text{Maps}_0(S^1 \to S^2))\). Here ``all loops together'' means that the colored loops are labeled, or parameterized, by another parameter \(\mu\), coming from the domain \(S^1\) of the homotopy group \(\pi_1(\text{Maps}_0(S^1 \to S^2))\). This leads to the concept of a non-contractible loop (NCL), which will be discussed in the following example.

A more relevant example for the existence of the sphaleron solution is
\begin{align}
  \label{eq:Pi_1_Maps_S2_S3}
  \pi_1(\text{Maps}_0(S^2 \to S^3))\cong \pi_3(S^3)\cong\mathbb{Z}\ .
\end{align}
In \(\text{Maps}_0(S^2 \to S^3)\), the domain \(S^2\) is the spatial boundary of three-dimensional space, while the target \(S^3\) is the manifold of the \(SU(2)\) gauge group. Since \(\pi_2(S^3)\cong 0\), every individual map in \(\text{Maps}_0(S^2 \to S^3)\) is homotopic to the trivial map. However, \(\pi_1(\text{Maps}_0(S^2 \to S^3))\cong \mathbb{Z}\) means that there exists an additional parameter, coming from the domain \(S^1\), that labels different maps in \(\text{Maps}_0(S^2 \to S^3)\), and that all of these labeled maps together can be non-contractible to the trivial map. This additional parameter is called the NCL parameter, and is denoted by an angle \(\mu\) running from \(0\) to \(\pi\) \cite{Manton:1983nd}. A detailed discussion of the \(\mu\) parameter will be given later in the introduction to the sphaleron. Here we emphasize one difference between the homotopy group \(\pi_1(\text{Maps}_0(S^2 \to S^3))\) and the homotopy group \(\pi_3(S^3)\). The latter classifies maps from \(S^3\) to \(S^3\), and will be relevant for the instanton solution discussed later. For the instanton case, the domain \(S^3\) is the boundary of four-dimensional Euclidean space, which differs from the sphaleron case, where the domain is \(S^2\) supplemented by one NCL parameter. Thus, although both groups are isomorphic to \(\mathbb{Z}\), they are constructed in different ways.

\subsection{General features of Topological Field Configurations}
\label{sec:roles_of_topology}

In this subsection, we first introduce the general types of topological field configurations and then discuss the role of homotopy groups in the existence of TFCs. Specific examples of TFCs, including cosmic string, monopole, sphaleron, and instanton, will be discussed in the next subsection.

\paragraph{Note on terminology.} As we will see in Sec.~\ref{sec:homotopy_and_TFCs}, the field solution, or field configuration, at the spatial or Euclidean-spacetime boundary defines a map from an \(n\)-sphere to the vacuum manifold of the gauge group. As introduced in the previous subsection, the homotopy group of this map provides a topological classification of field configurations. Field configurations that belong to different elements of the homotopy group cannot be continuously deformed into one another and are therefore topologically distinct. It is this property that gives these field configurations their topological character.

\subsubsection{types of topological field configurations}

As summarized in Ref.~\cite{Jackiw:1977yn}, topological field configurations can be divided into three kinds\footnote{Note on terminology: topological field configurations are sometimes called classical field solutions in the literature.}: (1) static solutions, (2) time-dependent solutions, and (3) imaginary-time solutions. In this section, we mainly focus on the first and third kinds, which are relevant for the TFCs that we will study in this thesis. Examples of the first kind include cosmic string, monopole, and sphaleron; the main example of the third kind is the instanton. The second kind of time-dependent solutions, although is also interesting in field theory, is not the focus of this thesis. We refer the reader to Refs.~\cite{Jackiw:1977yn} for examples of time-dependent solutions. \flag

We first use a \(1+1\)-dimensional example to illustrate the differences among these three kinds of solutions. Consider a Lagrangian for a real scalar field \(\phi\),
\begin{align} \label{eq:real_scalar_lagrangian}
  \mathcal{L} = \frac{1}{2} (\partial_\mu \phi)^2 - V(\phi) \ ,
\end{align}
and use the metric convention \(g_{\mu\nu}=\text{diag}(+,-)\). The equation of motion is
\begin{align}
  \partial_\mu \partial^\mu \phi + \frac{dV}{d\phi}
  = \frac{\partial^2}{\partial t^2}\phi - \frac{\partial^2}{\partial x^2}\phi + \frac{dV}{d\phi}
  = 0 \ .
\end{align}
A time-dependent solution \(\phi(t,x)\) satisfies the full equation above. A static solution \(\phi(x)\) satisfies
\begin{align}
  -\frac{d^2}{dx^2}\phi + \frac{dV}{d\phi} = 0 \ .
\end{align}
An imaginary-time solution \(\phi_E(\tau,x)\) satisfies
\begin{align}
  -\frac{d^2}{d\tau^2}\phi_E - \frac{d^2}{dx^2}\phi_E + \frac{dV}{d\phi_E} = 0 \ .
\end{align}

Here we comment on the Hamiltonian for the static solution \(\phi(x)\) using this simple example. This discussion is useful for understanding the field energy in non-Abelian gauge theory later. The Hamiltonian density for a static solution is
\begin{align} \label{eq:Hamiltonian_scalar_theory}
  \mathcal{H} = \frac{1}{2} \left(\frac{d\phi}{dx}\right)^2 + V(\phi) \ .
\end{align}
Therefore, for static configurations, the Hamiltonian density \(\mathcal{H}\) equals minus the static Lagrangian density \(\mathcal{L}\). This result can be generalized to gauge theories under the static gauge condition \(A_0=0\), which is the gauge used in our discussion of static solutions in SU(2) gauge--Higgs theory.\footnote{Even in static gauge, the relation \(\mathcal{H}=-\mathcal{L}\) does not hold for interaction terms such as \(a\,F_{\mu\nu}\tilde{F}^{\mu\nu}\). However, such terms do not appear in our discussion of static solutions in SU(2) gauge--Higgs theory.} One implication is that the equation of motion for a static solution can be derived by applying the Euler--Lagrange equation to the Hamiltonian density. This approach is frequently used in the literature to derive equations of motion for static solutions in gauge--Higgs theory, although this point is often not stated explicitly.

\subsubsection{application of homotopy theory to topological field configurations.}
\label{sec:homotopy_and_TFCs}

A necessary condition for the existence of topological field configurations (TFCs) is finite classical action, or finite classical energy for static configurations. We now explain how topology, through homotopy groups, enters this condition.

The starting point is the principle of stationary action. In the present context, it is sufficient to focus on the classical action; quantum and thermal corrections will be discussed in later sections. In ordinary perturbative QFT, quantization is usually performed around the vacuum, with no classical background field, for which \(S_{\text{classical}}=0\), and one studies only fluctuations.

Now return to the purely classical theory and restrict attention to static field configurations. The requirement is that the classical action be finite. As discussed below Eq.~\eqref{eq:Hamiltonian_scalar_theory}, for static configurations \(\mathcal{H}=-\mathcal{L}\), so
\(
L=\int d^n x\,\mathcal{L}=-\int d^n x\,\mathcal{H} \ .
\)
Here $n$ is the spatial dimension. Therefore, for static TFCs, a necessary condition for the action to be finite is that the energy be finite,
\(
E=\int d^n x\,\mathcal{H}<\infty \ .
\)
This finite-action condition applies not only to classical field theory but also to quantum field theory, where, as pointed out in Ref.~\cite{Coleman:1978ae}, the requirement of finite action is motivated by the semiclassical approximation to the path integral.

Consider the classical energy in a general theory,
 a necessary condition for finite \(E\) is that \(\mathcal{H}\to 0\) at spatial infinity, i.e.\ as \(|x|\to\infty\). The spatial boundary is a sphere \(S^{n-1}\), for example, \(S^2\) is the boundary of three-dimensional space \footnote{The definition of $n-$sphere is given previously in Eq.~\eqref{eq:definition_Sn}.}. The condition \(\mathcal{H}=0\) implies that the fields asymptotically approach vacuum configurations at the boundary \footnote{For gauge field, this corresponds to the pure gauge configuration. For the example of $U(1)$ gauge field, see later discussion in Eq.~\eqref{eq:vortex_configuration}.}. The asymptotical condition is required for the derivative terms in the Hamiltonian density to vanish at infinity, which can be checked from the explicit numerical solutions of the radial profiles of the field equations of motions, which will be shown in later subsections. For a pure scalar theory, the vacua are the minima of the scalar potential. For example, with
\begin{align}
  V(\phi)=\lambda(\phi^2-a^2)^2 \ ,
\end{align}
the vacua are \(\phi=\pm a\). A nontrivial stable solution is the kink connecting \(\phi=-a\) at \(x=-\infty\) and \(\phi=+a\) at \(x=\infty\). However, such a kink is stable only in \(1+1\) dimensions, where the spatial boundary consists of two disconnected points, matching the two discrete vacua.

More generally, Ref.~\cite{Jackiw:1977yn} proves that, for a spinless scalar theory described by Eq.~\eqref{eq:real_scalar_lagrangian}, stable static finite-energy solutions exist only in one plus one dimension. In higher dimensions, one must introduce gauge fields or other internal structure so that the vacuum manifold is sufficiently nontrivial. If the vacuum manifold is \(M\), stability requires a nontrivial map \(S^{n-1}\to M\), i.e.\ a map that cannot be continuously deformed to a point. As we discussed in the previous subsection, such maps are classified by \(\pi_{n-1}(M)\). If \(\pi_{n-1}(M)\) is nontrivial, stable TFCs can exist. Therefore:
\begin{tcolorbox}[colback=yellow!10!white, colframe=yellow!50!black]
The homotopy group \(\pi_{n-1}(M)\) gives the topological condition for nontrivial finite-energy field configurations: \(\mathcal{H}\) must vanish at spatial or Euclidean spacetime infinity.
\end{tcolorbox}

Table~\ref{tab:topological-defects} summarizes common TFCs and their associated homotopy groups. We first give a brief explanation here, and then present the detailed constructions in Sec.~\ref{sec:Examples of topological field configurations}.

\refstepcounter{footnote}
\begin{table}[h]
\centering
\caption[Classification of common topological field configurations by homotopy group]{Classification of common topological field configurations by homotopy group\captionfootnotemark. Here \(M\) denotes the vacuum manifold. The monopole listed here is the 't Hooft--Polyakov monopole, not the Dirac monopole. The term ``imaginary'' for the instanton means that it is defined in imaginary time. The symbol $\cong$ denotes isomorphism (see definition above Eq.~\eqref{eq:SU(2)_element}), which means that there exists a one-to-one correspondence between the two groups that preserves the group operation. The homotopy group for the sphaleron is discussed previously in Eq.~\eqref{eq:Pi_1_Maps_S2_S3}.}
\begin{tabular}{|c|c|c|c|}
\hline
\textbf{Topological Defect} & \textbf{Homotopy Group} & \textbf{Vacuum Manifold} & \textbf{Properties} \\
\hline
Kink; Domain Wall & \(\pi_0(M)\) & \(M\cong \{+1,-1\}\) & static, stable \\
\hline
Vortex; Cosmic String & \(\pi_1(M)\) & \(M\cong S^1\) & static, stable \\
\hline
Monopole & \(\pi_2(M)\) & \(M\cong  S^2\) & static, stable \\
\hline
Sphaleron & \(\pi_1(\text{Maps}_0(S^2\to M))\) & \(M\cong S^3\) & static, unstable \\
\hline
Instanton & \(\pi_3(M)\) & \(M\cong S^3\) & imaginary \\
\hline
\end{tabular}
\label{tab:topological-defects}
\end{table}
\captionfootnotetext{Note that there is another kind of field configuration, ``texture'' \cite{Turok:1989ai,Turok:1990zg}, which arises from $\pi_3(M)=\mathbb{Z}$ and is not discussed in this thesis. Such a scenario is also of interest in studies of Chern-Simons number violation during Higgs bubble collisions \cite{Bhusal:2025lvm}.}

The configurations listed in Table.~\ref{tab:topological-defects} are central to this thesis. The second column shows \(\pi_{n-1}(M)\), which classifies maps from the boundary sphere \(S^{n-1}\) to \(M\). Here \(n\) is the characteristic spatial dimension of the configuration. In the last column, ``static'' means time independent, ``stable'' means a local minimum of the energy functional, and ``unstable'' means a saddle point. Equivalently, fluctuations around stable TFCs have only positive and zero eigenvalues (modes), while unstable TFCs have at least one negative eigenvalue (mode).

For concreteness, a kink in one spatial dimension interpolates between two discrete vacua: \(\phi=v_-\) as \(x\to-\infty\) and \(\phi=v_+\) as \(x\to+\infty\). In three spatial dimensions, this generalizes to a domain wall, a codimension-1 \footnote{The codimension of a subspace \(A\) embedded in a space \(D\) is defined as
  \(\mathrm{codim}(A)=\dim(D)-\dim(A)\).
  In the present case, a domain wall is a two-dimensional surface embedded in ordinary three-dimensional space, so its codimension is \(3-2=1\).} object separating regions with different vacuum values. Thus, although a domain wall exists in three spatial dimensions, its characteristic dimension is one, and it therefore belongs to the \(\pi_0(M)\) class. During a first-order phase transition, domain walls can form when bubbles with opposite-sign VEVs collide.

Likewise, a two-dimensional vortex generalizes to a cosmic string in three dimensions: a codimension-2 object localized in two directions and extended along one. Cosmic strings may be either infinite or closed loops. Their vacuum manifold is typically \(S^1\), for example, from spontaneous breaking of a \(U(1)\) gauge symmetry. In general,
\begin{align}
  M \;=\; G/H \ ,
\end{align}
where \(G\) is the original symmetry group and \(H\) is the unbroken subgroup after spontaneous symmetry breaking.

For monopoles, \(M\cong S^2\), as in \(SU(2)\to U(1)\) \footnote{If one considers the breaking chain $SU(2)\to U(1)\to 1$, string topology can also appear in the second step of symmetry breaking \cite{Shifman:2002yi,Blasi:2026iyq}.}. In the Standard Model, however, \(SU(2)\times U(1)\to U(1)_{\text{em}}\), so \(M\cong S^3\). The map \(S^2\to S^3\) is topologically trivial, so there is no stable monopole in the Standard Model. However, as we discussed in Eq.~\eqref{eq:Pi_1_Maps_S2_S3}, the following homotopy group is nontrivial:
\begin{align}
  \label{eq:Pi_1_Maps_S2_S3_repeated}
  \pi_1({\rm Maps}(S^2\rightarrow S^3)) \;=\; \pi_3(S^3) \;\cong\;\mathbb{Z}\ .
\end{align}
By introducing an additional parameter, one can construct a map from \(S^3\)—formed by the spatial boundary together with that extra parameter—to the vacuum manifold \(S^3\). In Manton's construction, this parameter is denoted by \(\mu\) \cite{Manton:1983nd} and is called the NCL parameter. The sphaleron corresponds to the configuration at \(\mu=\pi/2\). It is static but unstable, since small fluctuations can drive it toward either of the neighboring vacuum configurations. \flag The parameter \(\mu\) runs from \(0\) to \(\pi\), and the configurations at \(\mu=0\) and \(\mu=\pi\) are both vacua. For intermediate values \(0<\mu<\pi\), the field energy is higher than the vacuum energy and reaches its maximum at \(\mu=\pi/2\), namely at the sphaleron. One may regard \(\mu\) as a time-dependent parameter: as \(t\to -\infty\) and \(t\to +\infty\), the field configuration approaches vacuum (\(\mu=0,\pi\)), while at \(t=t_0\) it passes through the sphaleron configuration with \(\mu=\pi/2\). This interpretation is used later in the computation of the sphaleron baryonic charge; see Eq.~\eqref{eq:Q_B sphaleron definition simplified}, where the time integral is taken from \(t=-\infty\) to \(t=t_0\). Extending the integral to \(t=+\infty\) gives a total baryonic charge of \(-1\), corresponding to a winding number \(-1\) for the map from \(S^3\) to \(S^3\). For the definition of winding number, see Eq.~\eqref{eq:winding_number_SU(2)_equivalent_definitions}.

For instanton, the vacuum manifold is also \(S^3\), associated with \(SU(2)\). The instanton is an imaginary-time solution, and the boundary of four-dimensional Euclidean space is a \(S^3\). The map \(S^3\to S^3\) is nontrivial, leading to instanton solutions. However, as we mentioned at the end of the previous subsection, although the final homotopy group for the sphaleron and instanton is the same, \(\pi_3(S^3)\cong \mathbb{Z}\), the construction of the two homotopy groups is different. For the sphaleron, the domain \(S^3\) is formed by the spatial boundary \(S^2\) together with the NCL parameter \(\mu\). For the instanton, the domain \(S^3\) is simply the boundary of four-dimensional Euclidean space. Both sphaleron and instanton are relevant to baryon-number violation in gauge theories, but they play different physical roles. A detailed discussion is given in the next section.

\subsection{Examples of Topological Field Configurations}
\label{sec:Examples of topological field configurations}

In the previous subsection, we discussed the role of homotopy groups in the existence of TFCs and gave general comments on several common TFCs. In this subsection, we present the mathematical details of several TFCs, including their field ans\"atze, equations of motion, and field energies. The examples include cosmic string, monopole, sphaleron, and instanton. These examples will be used in later sections to discuss their cosmological implications. Since all these TFCs are related to gauge theories, we first summarize the gauge choices used in these examples.

For static TFCs, we always use static gauge, i.e.\ \(A_0=0\) for Abelian gauge theory and \(A_0^a=0\) for non-Abelian gauge theory, where \(a\) is the generator index. The benefits of this gauge choice are: (1) it simplifies the equations of motion in polar or spherical coordinates, and (2) it makes the Hamiltonian density equal to minus the Lagrangian density, i.e.\ \(\mathcal{H}=-\mathcal{L}\), which simplifies the derivation of equations of motion from the Hamiltonian density. The second point was discussed below Eq.~\eqref{eq:Hamiltonian_scalar_theory}. Gauge redundancy is not completely removed by setting \(A_0=0\), since one can still perform gauge rotations of the spatial components. In polar or spherical coordinates, we further impose radial gauge, \(A_r=0\) or \(A_r^a=0\), after which no further local gauge freedom remains. \flag The benefit of radial gauge is that it simplifies the equations of motion in polar or spherical coordinates. For example, for the cosmic string solution, the ansatz for the gauge field in polar coordinates is \(A_\theta\propto 1/r\), which would be more complicated if \(A_r\) were nonzero.
For imaginary-time TFCs, however, static gauge is not used, as will be seen in the instanton example.

\subsubsection{cosmic string (Nielsen--Olesen vortex)}

We focus on local, or gauge, cosmic strings, which arise from spontaneous breaking of a local \(U(1)\) gauge symmetry in the Abelian--Higgs model \footnote{The global $U(1)$ symmetry breaking can also leads to string solution, which we refer to Ref.~\cite{Hindmarsh:1994re} for more details.}. This solution was first constructed by Nielsen and Olesen \cite{Nielsen:1973cs}. Although they called it a vortex, the field configuration is the same as that of a cosmic string lying in the \(x\)-\(y\) plane and extending along the \(z\) direction. We therefore use the term cosmic string throughout this thesis. Reviews can be found in Ref.~\cite{Hindmarsh:1994re}. The simplest model that gives local cosmic strings is the Abelian--Higgs model, with Lagrangian density
\begin{align}
  \mathcal{L} = -\frac{1}{4} F_{\mu\nu} F^{\mu\nu} + |D_\mu \phi|^2 - V(\phi) \ ,
\end{align}
where \(F_{\mu\nu}=\partial_\mu A_\nu - \partial_\nu A_\mu\) is the field-strength tensor of the \(U(1)\) gauge field \(A_\mu\), \(D_\mu = \partial_\mu - i g A_\mu\) is the covariant derivative, and
\begin{align}
  V(\phi)=\frac{\lambda}{2}\left(|\phi|^2 - v^2\right)^2 \ ,
\end{align}
is the scalar potential with a nonzero vacuum expectation value (VEV) \(v\). As noted above, we use static gauge, \(A_0=0\). The Hamiltonian density is then
\begin{align}
  \mathcal{H}=\frac{1}{4} F_{ij}F_{ij} + (D_i \phi)^* D_i \phi + V(\phi) \ ,
\end{align}
which has mass dimension four. We define the string tension \(\mu\), which has mass dimension two, by integrating \(\mathcal{H}\) over the transverse plane:
\begin{align}
  \mu = \int d\theta \, dr\ r \mathcal{H} \ .
\end{align}
We seek nontrivial static solutions with finite \(\mu\). Finite string tension requires \(\mathcal{H}\) to decrease faster than $1/r$ as \(r\to \infty\). This implies
\begin{align}
  & F_{ij} = 0 \ , \\
  & D_i \phi = 0 \ , \\
  & V(\phi) = 0 \Rightarrow |\phi|=v \ . 
\end{align}
Field configurations satisfying these conditions form the vacuum manifold \(\mathcal{M}\). This manifold is topologically \(S^1\), since the phase of \(\phi\) can take any value from \(0\) to \(2\pi\). The spatial boundary of the \(x\)-\(y\) plane at infinity is also topologically \(S^1\). The map \(S^1 \to S^1\) is characterized by \(\pi_1(S^1)=\mathbb{Z}\), which is nontrivial. Therefore, stable cosmic-string solutions exist in the Abelian--Higgs model. In cylindrical coordinates \((r,\theta,z)\), the ansatz with winding number \(n\in\mathbb{Z}\) is
\begin{align} \label{eq:vortex_configuration}
  &A_0 = 0\ ,\quad A_r = 0\ ,\quad A_\theta = \frac{n}{g r} f(\tilde{r}),\quad A_z = 0 \ , \quad \phi = v h(\tilde{r}) e^{i n \theta} \ .
\end{align}
Here \(A_0=A_r=0\) are the gauge choices discussed above, \(\tilde{r}=\sqrt{\lambda}vr\) is the dimensionless radius, and \(f(\tilde r)\) and \(h(\tilde r)\) are dimensionless profile functions determined by the numerical equations of motion. \flag This local string contains a quantized magnetic flux
\begin{align}
  \Phi = \int d^2 x\, \vec{B}\cdot \hat{z} = \int_{S^1} \vec{A}\cdot d\vec{l}=\int_{S^1} A_\theta r d\theta = \frac{2\pi n}{g} \ ,
\end{align}
where \(\vec{B}=\nabla\times \vec{A}\) is the magnetic field. The $S^1$ in the second integral is the spatial boundary of the \(x\)-\(y\) plane at infinity, where the gauge field approaches a pure gauge configuration and $f(\tilde{r})\to 1$.
Since the symmetry breaking pattern is \(U(1)\to 1\), the vacuum manifold is \(M=G/H=U(1)\cong S^1\). The number $n$ in Eq.~\eqref{eq:vortex_configuration} and the quantized magnetic flux can be understood as element $n$ in the homotopy group \(\pi_1(G/H)\cong \mathbb{Z}\), which classifies maps from the spatial boundary \(S^1\) to the vacuum manifold \(S^1\).  The quantized magnetic flux is a topological physical quantity, which has been verified in experiments on superconductors \cite{Deaver:1961zz, PhysRevLett.7.51}. 

The field equations are
\begin{align}
  &[D^\mu D_\mu + \lambda (|\phi|^2 - v^2)]\phi = 0 \ , \\
  &\partial_\nu F^{\mu \nu} + ig \left(\phi^* D^\mu \phi- (D^\mu\phi)^* \phi \right) = 0 \ .
\end{align}
From these equations, one derives the profile equations for \(f(\tilde r)\) and \(h(\tilde r)\). The full dimensionless system depends only on
\begin{align}
  \beta=\frac{\lambda}{g^2} \ .
\end{align}
The radial profile equations are
\begin{align}
\label{eq:master}
&h''(\tilde{r})+\frac{h'(\tilde{r})}{\tilde{r}}-n^2\frac{h(\tilde{r})}{\tilde{r}^2}+2n^2\,\frac{f(\tilde{r})}{\tilde{r}^2}\,h(\tilde{r})-n^2\frac{f(\tilde{r})^2}{\tilde{r}^2}\,h(\tilde{r})-\left(h(\tilde{r})^2-1\right)\,h(\tilde{r})=0 \ ,\\
&f''(\tilde{r})-\frac{f'(\tilde{r})}{\tilde{r}}+2\,\frac{g^2}{\lambda}\left(1-f(\tilde{r})\right)\,h(\tilde{r})^2=0 \ .
\end{align}
The string tension can be written as
\begin{align}
  \mu &= \int d\theta \, dr\ r \mathcal{H} \nonumber \\
  &=2\pi\,v^2\,\int_0^\infty\,\tilde{r}\,d\tilde{r}\,\Bigg[h'(\tilde{r})^2+n^2\frac{h(\tilde{r})^2}{\tilde{r}^2}\left(1-f(\tilde{r})\right)^2
+\frac{1}{2}\left(h(\tilde{r})^2-1\right)^2+n^2 \frac{\beta}{2}\frac{f'(\tilde{r})^2}{\tilde{r}^2}\Bigg] \ .
\end{align}
The boundary conditions are
\begin{align}
  &h(0)=0,\ h(\infty)=1 \ ,\\
  &f(0)=0,\ f(\infty)=1 \ .
\end{align}
Numerically, one finds that \(f'(\tilde r)\to 0\) at large \(\tilde r\), which ensures that the string tension is finite.

\subsubsection{'t Hooft--Polyakov monopole}

The 't Hooft--Polyakov monopole \cite{tHooft:1974kcl,Polyakov:1974ek} is a static, spherically symmetric solution in the \(SU(2)\) gauge--Higgs theory \footnote{In addition to gauged monopole, global monopole also exist; see Ref.~\cite{Barriola:1989hx} and more recent studies \cite{Nakano:2026zme}.}, where the \(SU(2)\) gauge symmetry is spontaneously broken to \(U(1)\) by an adjoint real scalar field \(\phi^a\), with \(a=1,2,3\). The Lagrangian density is
\begin{align} \label{eq:SU(2)_adjoint_model}
  \mathcal{L} = -\frac{1}{4} F_{\mu\nu}^a F^{a\mu\nu} + \frac{1}{2} (D_\mu \phi^a)(D^\mu \phi^a) - V(\phi^a) \ ,
\end{align}
where \(F_{\mu\nu}^a = \partial_\mu A_\nu^a - \partial_\nu A_\mu^a + g \epsilon^{abc} A_\mu^b A_\nu^c\) is the field-strength tensor of the \(SU(2)\) gauge field \(A_\mu^a\), \(D_\mu \phi^a = \partial_\mu \phi^a + g \epsilon^{abc} A_\mu^b \phi^c\) is the covariant derivative of the adjoint scalar field, and
\begin{align}
  V(\phi^a) = \frac{\lambda}{4} \left( \phi^a \phi^a - v^2 \right)^2 \ ,
\end{align}
is the scalar potential. Adopting static gauge \(A_0^a=0\), the Hamiltonian density reads
\begin{align}
  \mathcal{H} = \frac{1}{4} F_{ij}^a F_{ij}^a + \frac{1}{2} (D_i \phi^a)(D_i \phi^a) + V(\phi^a) \ .
\end{align}
We define the monopole mass \(M\) by integrating the Hamiltonian density over all space:
\begin{align}
  M = \int d\theta\, d\phi\, dr\ r^2 \sin\theta\ \mathcal{H} \ .
\end{align}
In analogy with the cosmic-string case, finite monopole mass requires each term in \(\mathcal{H}\) to vanish at spatial infinity, i.e.\ as \(r\to \infty\). Following Ref.~\cite{Rubakov:2002fi}, field configurations satisfying \(\mathcal{H}=0\) at spatial infinity are
\begin{align}
  \label{eq:monopole_vacuum_confg}
  A_i^a(\vec{x})=\frac{1}{gr}\epsilon^{aij} n^j \ , \phi^a(\vec{n}) = v n^a \ .
\end{align}
Here \(\vec{n}=\vec{x}/r\) is the unit radial vector. The vacuum manifold is isomorphism to \(S^2\). This follows from the symmetry-breaking pattern \(SU(2)\to U(1)\), for which \(M=SU(2)/U(1)\cong S^2\), as we explained above in the paragraph after Eq.~\eqref{eq:pi_2_G/H}. The spatial boundary at infinity is a $2-$sphere \(S^2\). The map from the boundary \(S^2\) to the vacuum manifold \(S^2\) is classified by \(\pi_2(S^2)\cong \mathbb{Z}\), which is nontrivial. \flag
The monopole configuration in Eq.~\eqref{eq:monopole_vacuum_confg} corresponds to the element \(1\) in \(\pi_2(S^2)\cong \mathbb{Z}\), as can be seen from the fact that the unit radial vector \(\vec{n}\) wraps around the vacuum manifold \(S^2\) once as one goes around the spatial boundary \(S^2\).
Therefore, stable monopole solutions exist in the \(SU(2)\) gauge--Higgs theory. A monopole configuration can then be written as
\begin{align} \label{eq:monopole_confg}
  A_i^a=\frac{1}{gr}\epsilon^{aij} n^j (1-F(\tilde{r})) \ , \phi^a = v n^a (1-H(\tilde{r})) \ ,
\end{align}
where \(H(\tilde{r})\) and \(F(\tilde{r})\) are dimensionless profile functions determined by the equations of motion, which we do not repeat here. The boundary conditions for \(H(\tilde{r})\) and \(F(\tilde{r})\) are
\begin{align}
  &F(0)=1,\ F(\infty)=0 \ , \\
  &H(0)=1,\ H(\infty)=0 \ .
\end{align}
One can check that the gauge configuration in Eq.~\eqref{eq:monopole_confg} satisfies the radial gauge condition, \(A_r^a=0\).

We now comment on the term ``monopole'', i.e.\ why this configuration has magnetic-monopole properties. First, unlike the Abelian--Higgs model for cosmic strings, the model in Eq.~\eqref{eq:SU(2)_adjoint_model} contains a massless vector field after spontaneous symmetry breaking. The SU(2) field strength tensor \(F_{\mu\nu}^a\) contains a non-Abelian part, so it cannot be directly identified with the electromagnetic field strength. It is convenient to define an effective gauge-invariant electromagnetic field strength. 't Hooft \cite{tHooft:1974kcl} constructed a gauge invariant field strength:
\begin{align} \label{eq:gauge invariant strengh for monopole}
  \mathcal{F}_{\mu\nu} = \frac{1}{v}\phi^a F_{\mu\nu}^a - \frac{1}{g v^3}\epsilon_{abc}\phi^a(D_\mu \phi^b)(D_\nu \phi^c) \ .
\end{align}
If one imposes unitary gauge, \(\phi^a=v\delta^{a3}\), then Eq.~\eqref{eq:gauge invariant strengh for monopole} reduces to \(\mathcal{F}_{\mu\nu}\rightarrow F_{\mu\nu}=\partial_\mu A_\nu^3-\partial_\nu A_\mu^3\), the usual electromagnetic field strength. Specifically, for this scalar configuration, the second term in Eq.~\eqref{eq:gauge invariant strengh for monopole} cancels the non-Abelian part of the first term. Because this definition is gauge invariant, it can be used to study the electromagnetic properties of the monopole solution in Eq.~\eqref{eq:monopole_confg}. One can verify that, far from the monopole center,
\begin{align}
  \mathcal{E}_i = \mathcal{F}_{0i} = 0 \ , \\
  \mathcal{B}_i = -\frac{1}{2} \epsilon_{ijk} \mathcal{F}_{jk} =\frac{1}{g r^2}  n_i \ .
\end{align}
Thus, the configuration in Eq.~\eqref{eq:monopole_confg} has a non-zero magnetic field far from the center, while the electric field vanishes. Furthermore, the magnetic field carries monopole charge
\begin{align}
  g_M=\frac{1}{g} \ .
\end{align}
The monopole mass and its profile equations of motion share many similarities with the sphaleron case, which we discuss in detail in the following subsection.

\subsubsection{Manton sphaleron}
\label{sec:manton_sphaleron}

In the SM, the Higgs field coupled to \(SU(2)_L\) is a complex doublet, in contrast to the real adjoint field considered in the monopole case. In addition, there is an extra gauge group, \(U(1)_Y\). Therefore, the symmetry pattern and homotopy structure are different from those in the monopole case. Working in the static gauge, \(A_0^a=0\), we start from the Hamiltonian density
\begin{align} \label{eq:Hamiltonian_density}
  \mathcal{H} = \frac{1}{4} F_{ij}^a F_{ij}^a + \frac{1}{4}f_{ij} f_{ij} + (D_i \Phi)^\dagger (D_i \Phi) + V(\Phi) \ ,
\end{align}
where \(\Phi\) is the complex Higgs doublet, and \(D_i = \partial_i - i g A_i^a \sigma^a / 2 - i g' a_i Y\) is the covariant derivative, with \(Y\) the hypercharge. $A_i^a$ and $a_i$ are the \(SU(2)_L\) and \(U(1)_Y\) gauge fields, respectively. $\sigma^a$ are the Pauli matrices. The \(SU(2)\) and \(U(1)\) field strengths are \(F_{ij}^a\) and \(f_{ij}\), respectively. The potential is
\begin{align}
  V(\Phi) = \lambda \left( |\Phi|^2 - \frac{v^2}{2} \right)^2 \ .
\end{align}
In this case, the vacuum manifold is
\begin{align}
  M=\frac{SU(2)_L\times U(1)_Y}{U(1)_{\rm em}}\cong S^3 \ ,
\end{align}
rather than \(S^2\). Since \(\pi_2(S^3)=0\), there is no stable monopole solution in the SM. However, as first constructed by Manton \cite{Manton:1983nd}, the mapping
\begin{align}
  \pi_1({\rm Maps}(S^2\rightarrow S^3)) \;=\; \pi_3(S^3) \;\cong\;\mathbb{Z}\ ,
\end{align}
exists. By introducing an additional topological parameter \(\mu\), one can construct a topological field configuration, called the sphaleron. In the following discussion, for simplicity we focus on the \(SU(2)\) gauge field and the Higgs doublet, i.e.\ the so-called zero-mixing limit, because the \(U(1)_Y\) gauge field has only a small effect on the sphaleron energy; its effect will be discussed later in Sec.~\ref{sec:mass_and_eom_for_general_multiplet}. The sphaleron configurations for the gauge and scalar fields are
\begin{align}
\label{eq:sphaleron_multiplet_confg_manton}
H &= \frac{v}{\sqrt{2}}h(\xi)U_{\text{sph}}(\mu,\theta,\phi)\begin{pmatrix}
    0\\
    1
\end{pmatrix} \ ,\\
\label{eq:sphaleron_gauge_confg_manton}
A_i^a T^a dx^i &= -\frac{i}{g}f(\xi)(\partial_i U_{\text{sph}})U_{\text{sph}}^{-1}dx^i \ .
\end{align}
Here, \(v\) denotes the Higgs VEV, while \(h(\xi)\) and \(f(\xi)\) are the radial profile functions for the Higgs and gauge fields, respectively, with \(\xi\equiv g v r\) the dimensionless radial coordinate. In the gauge-field configuration, the index \(i\) in \(A_i^a\), \(\partial_i\), and \(dx^i\) is contracted, with \(i\in[\theta,\phi]\) and \(dx^i\in[d\theta,d\phi]\). The matrix \(U_{\text{sph}}\) is
\begin{equation}
\label{eq::Manton_Umatrix_fundamental}
\begin{aligned}
U_{\text{sph}}(\mu,\theta,\phi)= \left[\begin{array}{cc}
e^{i \mu } (\cos \mu -i \cos \theta \sin \mu ) & e^{i \varphi } \sin \theta \sin \mu \\
-e^{-i \varphi } \sin \theta \sin \mu & e^{-i \mu } (\cos \mu +i \cos \theta \sin \mu)
\end{array}\right] \ .
\end{aligned}
\end{equation}
Again, one can easily verify that the sphaleron gauge configuration in Eq.~\eqref{eq:sphaleron_gauge_confg_manton} satisfies the radial gauge condition, \(A_r^a=0\). As noted in the previous subsection, the sphaleron is a static field configuration; here this refers to \(\mu=\pi/2\). 
As explained in the paragraph after Eq.~\eqref{eq:Pi_1_Maps_S2_S3_repeated}, the parameter \(\mu\) may be viewed as a time-dependent parameter along the NCL. As \(\mu\) varies from \(0\) to \(\pi\), the field energy vanishes at \(\mu=0\) and \(\mu=\pi\), corresponding to vacuum configurations, and reaches its maximum at \(\mu=\pi/2\), which defines the sphaleron. Thus, the field configuration starts from the vacuum at \(\mu=0\), builds up to the sphaleron at \(\mu=\pi/2\), and then returns to the vacuum at \(\mu=\pi\). At \(\mu=\pi/2\), the sphaleron matrix becomes
\begin{align}
  U_{\text{sph}}(\theta,\phi)= \left[\begin{array}{cc}
\cos \theta & e^{i \varphi } \sin \theta \\
-e^{-i \varphi } \sin \theta & \cos \theta
\end{array}\right]\rightarrow \frac{1}{r}\left[\begin{array}{cc}
z & x+iy \\
-x+iy & z
\end{array}\right] \ .
\end{align}
The latter form is the one used by Klinkhamer and Manton \cite{Klinkhamer:1984di}, which is more widely cited in the literature, although the essential sphaleron construction was already completed in Ref.~\cite{Manton:1983nd}. As discussed previously, the Hamiltonian density in Eq.~\eqref{eq:Hamiltonian_density} is minus the Lagrangian density for static configurations. Inserting the sphaleron ansatz into \(\mathcal{H}\), and defining the sphaleron energy by
\begin{align}
  E = \int d^3 x\,\mathcal{H} \ ,
\end{align}
one obtains
\begin{align}
  E=\frac{4\pi v}{g}\int_0^{\infty}d\xi\ \left[4 f^{\prime 2} + \frac{8}{\xi^2}[f(1-f)]^2 + \frac{1}{2}\xi^2 h^{\prime 2} + [h(1-f)]^2
+ \frac14 \frac{\lambda}{g^2}\xi^2 (h^2-1)^2\right] \ ,
\end{align}
where \(f^\prime=df/d\xi\) and \(h^\prime=dh/d\xi\). The equations of motion for the radial profiles are
\begin{align}
  \xi^2 \frac{d^2 f}{d\xi^2}&=2f(1-f)(1-2f)-\frac{\xi^2}{4}h^2(1-f) \ , \\
  \frac{d}{d\xi} \left(\xi^2\frac{dh}{d\xi} \right)&=2h(1-f)^2+\frac{\lambda}{g^2}\xi^2(h^2-1)h \ .
\end{align}
The boundary conditions are
\begin{align}
  f(0)=h(0)=0,\ \ f(\infty)=h(\infty)=1 \ \ .
\end{align}

Note that the sphaleron configurations in Eqs.~\eqref{eq:sphaleron_multiplet_confg_manton} and \eqref{eq:sphaleron_gauge_confg_manton} are not unique. Klinkhamer and Laterveer \cite{Klinkhamer:1990fi} proposed another sphaleron construction, in which the NCL, parameterized by \(\mu\), is divided into three stages:
\begin{itemize}
    \item I, \(\mu \in [-\pi/2,0]\): builds up the scalar-field configuration,
    \item II, \(\mu \in [0,\pi]\): builds up and then destroys the gauge-field configuration,
    \item III, \(\mu \in [\pi, 3\pi/2]\): destroys the scalar-field configuration \ .
\end{itemize}
Here \(\mu=-\pi/2,\ 3\pi/2\) represent vacuum states, while \(\mu=\pi/2\) denotes the sphaleron state. The field ans\"atze differ in the three stages. In phases I and III, the profile functions are
\begin{align} \label{eq::Klinkharmer_gound_gauge_new}
    &A_i^aT^a=0 \ , \\
\label{eq::Klinkharmer_gound_scalar_new}
    &H = \frac{v\left(\sin ^2 \mu+h(\xi) \cos ^2 \mu\right)}{\sqrt{2}}\left(\begin{array}{c}
     0 \\
     1
\end{array} \right)\ \ \ .
\end{align}
In phase II, the field configurations are
\begin{align}
    \label{eq:multiplet_sphaleron_confg_KL_scalar}
        &H=\frac{v}{\sqrt{2}}h(\xi)\left(\begin{array}{c}
             0 \\
            1
        \end{array} \right) \ ,\\
        \label{eq:multiplet_sphaleron_confg_KL_gauge}
        &A_i^a T^a dx^i = \frac{i}{g}(1-f(\xi))U_{\rm sph}^{-1}\partial_i U_{\rm sph} dx^i \ ,
\end{align}
where we manually switch off the \(U(1)\) mixing.

It was pointed out in Ref.~\cite{Wu:2023mjb} that these two sphaleron conventions are equivalent up to a gauge rotation. Applying a \(U_{\rm sph}^{-1}\) transformation to Eq.~\eqref{eq:sphaleron_multiplet_confg_manton} in the Manton--Klinkhamer convention gives Eq.~\eqref{eq:multiplet_sphaleron_confg_KL_scalar} in the Klinkhamer--Laterveer convention for the Higgs field. In other works, we choose a specialized gauge transformation \(U=U_{\rm sph}^{-1}\) to show the equivalence of the gauge-field configurations in the two conventions. \flag Here, ``equivalence'' means that the two conventions are related by a mathematical transformation. We note, however, that \(U_{\rm sph}^{-1}\) is itself a large gauge transformation, which shifts the winding number by one unit. Equivalently, the associated baryonic charge differs by one unit, as shown explicitly in Ref.~\cite{Wu:2023mjb}; where the baryonic charge definition is given in Eq.~\eqref{eq:Q_B sphaleron definition simplified}. Since the gauge field transforms as
\begin{align}
A_\mu^a T^a\rightarrow U A_\mu^a T^a U^{-1}-\frac{i}{g}(\partial_{\mu}U) U^{-1} \ ,
\end{align}
the sphaleron gauge field in the Manton--Klinkhamer convention transforms as
\begin{align}
\label{eq::gauge_trans_Gauge}
A_i^a T^a&\rightarrow U^{-1}\left(-\frac{i}{g}f(\xi)(\partial_{i}U) U^{-1}\right)U  -\frac{i}{g}(\partial_{i}U^{-1})U \nonumber \\
&=\frac{i}{g}(1-f(\xi))U^{-1} \partial_{i}U\ \ \ .
\end{align}
Here we omit the subscript ``sph'' for notational clarity. Multiplying by \(dx^i\), this is exactly the convention used in Eq.~\eqref{eq:multiplet_sphaleron_confg_KL_gauge}. This completes the proof that the two sphaleron conventions are equivalent.

Other sphaleron ans\"atze also exist in the literature, for example the one proposed by Akiba, Kikuchi, and Yanagida (AKY) \cite{Akiba:1988ay}, which uses a more general ansatz for both gauge and scalar fields. Their construction is physically equivalent to the Manton--Klinkhamer convention, and we do not repeat it here. The sphaleron is one of the central topics of this thesis. Other key aspects, such as its half-integer Chern--Simons number and its relation to the finite-temperature baryon-number-violation rate, will be discussed in later sections.

\subsubsection{BPST instanton}
\label{sec:instanton}

The instanton is an imaginary-time solution of Yang--Mills theory, first constructed by Belavin, Polyakov, Schwartz, and Tyupkin (BPST) \cite{Belavin:1975fg}. 't Hooft \cite{tHooft:1976rip} further studied fermion fluctuations in the instanton background and showed that instantons induce effective interactions that violate both baryon and lepton numbers. A detailed discussion of instanton-induced baryon-number violation is given in Appendix.~\ref{sec:instanton_baryon_violation}. In this subsection, we briefly introduce the instanton solution in pure \(SU(2)\) gauge theory. We start from the Euclidean action
\begin{align} \label{eq:instanton_edulidean_action}
  S_E = \int d^4 x\ \frac{1}{4} F_{\mu\nu}^a F_{\mu\nu}^a \ .
\end{align}
Here a Wick rotation \(t \to -i\tau\) has been performed. At the Euclidean boundary, i.e.\ for \(|x|\to\infty\), finite action requires \(F_{\mu\nu}^a\to 0\). The field configurations satisfying this condition form the vacuum manifold, which is topologically equivalent to \(S^3\) in \(SU(2)\) gauge theory. The Euclidean boundary at infinity is also topologically \(S^3\), in contrast to the static spatial boundary discussed earlier, which is topologically \(S^2\). The mapping from the Euclidean boundary \(S^3\) to the vacuum manifold \(S^3\) is classified by \(\pi_3(S^3)=\mathbb{Z}\), which is nontrivial. Therefore, nontrivial field solutions exist in pure \(SU(2)\) gauge theory; these are called instanton. In fact, this conclusion is not limited to pure \(SU(2)\): it generalizes to any non-Abelian simple Lie group \(G\), since \(\pi_3(G)=\mathbb{Z}\) \cite{Belavin:1975fg}.

Following the convention in Ref.~\cite{tHooft:1976rip}, a one-instanton (BPST) solution in \(SU(2)\) Yang--Mills theory is
\begin{align}
  \label{eq:BPST_instanton_confg}
  A_\mu^a(x)_{\mathrm{cl}}=\frac{2}{g}\,\frac{\eta_{a\mu\nu}\,(x-x_0)^\nu}{(x-x_0)^2+\rho^2}\ ,
\end{align}
where \(x_0\) is the instanton center, reflecting translational invariance, and \(\rho\) is the instanton size. The parameter \(\eta_{a\mu\nu}\) is the 't Hooft symbol, defined as
\begin{align}
  \eta_{a\mu\nu}=\begin{cases}
    \epsilon_{a\mu\nu} & \text{for } \mu,\nu=1,2,3 \ , \\
    \delta_{a\mu} & \text{for } \nu=4 \ , \\
    -\delta_{a\nu} & \text{for } \mu=4 \ , \\
    0 & \text{for } \mu=\nu=4 \ .
  \end{cases}
\end{align}
For the instanton solution, the Euclidean action in Eq.~\eqref{eq:instanton_edulidean_action} is
\begin{align} \label{eq:instanton_action_value}
  S_E = \frac{8\pi^2}{g^2} \ .
\end{align}
This action value plays an important role in the leading-order instanton-induced baryon-number-violation rate, which is proportional to \(\left|e^{-S_E}\right|^2=e^{-16\pi^2/g^2}\). A more detailed discussion of instanton-induced baryon-number violation is given later in Appendix.~\ref{sec:instanton_baryon_violation}.

\subsection{Extension of Sphaleron and Monopole to a General SU(2) Multiplet}

The discussions of monopoles and sphalerons in the previous sections were limited to cases in which the scalar field is in either the real adjoint representation or the complex fundamental representation of the \(SU(2)\) gauge group. However, in many BSM theories, scalar multiplets appear in more general \(SU(2)\) representations. For example, in minimal dark matter models \cite{Cirelli:2005uq}, scalar multiplets can be in higher representations, such as quintuplets or septuplets. It is therefore important to generalize the monopole and sphaleron solutions to a general \(SU(2)\) multiplet. In this subsection, we provide such a generalization and find that the hypercharge of the new multiplet plays a vital role. The discussion in this subsection thus represents one of the main novelties of this thesis.

This discussion is divided into two parts. In the first part, we discuss the homotopy structure for a general \(SU(2)\) multiplet undergoing electroweak symmetry breaking and construct the monopole and sphaleron configurations for a general \(SU(2)\) multiplet. In the second part, we discuss the masses and equations of motion of the monopole and sphaleron for a general \(SU(2)\) multiplet.

\subsubsection{topology and general construction of sphaleron and monopole}
\label{sec:topology_and_construction_for_general_multiplet}

We restrict our discussion to a single \(SU(2)\) multiplet \(\Phi\) with isospin \(J\) and hypercharge \(Y\), which undergoes electroweak symmetry breaking (EWSB). As noted in the previous subsection, the homotopy group \(\pi_{n-1}(M)\) plays a central role in determining the properties of topological field configurations. For EWSB, the vacuum manifold is
\begin{equation} \label{eq:cosetspace}
\mathcal{M}=\frac{G}{H}=\frac{SU(2)_L\times U(1)_Y}{H}\ .
\end{equation}
Here, \(G\) and \(H\) denote the symmetry groups before and after EWSB, respectively. The unbroken subgroup \(H\) is fixed by the multiplet representation through its hypercharge \(Y\). If \(Y=0\), the electric charge is determined solely by \(T_3\), and the residual symmetry is
\[
H = U(1)_Y \times U(1)_{\mathrm{em}},
\]
leaving two unbroken generators. If \(Y\neq 0\), only the electromagnetic \(U(1)\), generated by \(Q=T_3+Y\), remains unbroken, so
\[
H = U(1)_{\mathrm{em}},
\]
with one unbroken generator.

The classification of sphaleron and monopole solutions is determined by the multiplet hypercharge \(Y\) \cite{Wu:2023mjb}.  
(i) For a multiplet with \(Y\neq 0\), the resulting topological field solution is a sphaleron:
\begin{align}
{\rm Sphaleron:}\ \ \ \mathcal{M}&=\frac{SU(2)_L\times U(1)_Y}{U(1)_{\text{em}}}\cong S^3,\ \ \pi_3(\mathcal{M})=\mathbb{Z}\ .
\end{align}
Here, the symbol \(\cong\) denotes an isomorphism, and \(\pi_3(\mathcal{M})\) describes mappings of \(S^3\) into \(\mathcal{M}\).

(ii) For a multiplet with \(Y=0\), the resulting topological field solution is a monopole:
\begin{align} \label{eq:topology_for_monopole}
{\rm Monopole:}\ \ \ \mathcal{M}&=\frac{SU(2)_L\times U(1)_Y}{U(1)_{\text{em}}\times U(1)_Y}\cong S^2,\ \pi_2(\mathcal{M})=\mathbb{Z}\ \ .
\end{align}
For the monopole case, the corresponding neutral scalar can contribute to the WIMP dark-matter abundance \cite{Cirelli:2005uq,Chao:2018xwz}, since it does not couple to the \(Z\)-boson current due to zero hypercharge. However, we note that the mass range relevant for electroweak baryogenesis is typically below 1 TeV \cite{Ramsey-Musolf:2019lsf}. In contrast, if the multiplet is to account for the entire dark matter relic abundance, as shown in Ref.~\cite{Chao:2018xwz}, its required mass is much larger than 1 TeV. Specifically, for the septuplet considered later, the mass needed to explain the total dark matter relic density is around 10 TeV. Therefore, in the mass range of interest for a first-order phase transition (less than 1 TeV), the multiplet can contribute only a small fraction of the dark matter relic density.

The scalar and gauge field configurations for sphalerons (monopoles) can be parameterized as
\begin{align}
\label{eq:sphaleron_multiplet_confg_general}
\Phi &= \frac{v_\phi}{\sqrt{2}}\phi(\xi)U_{\text{sph (mon)}}(\mu,\theta,\phi)\begin{pmatrix}
  0\\
  \cdots\\
  1 \\
  \cdots \\
  0
\end{pmatrix},\\ 
\label{eq:sphaleron_gauge_confg_general}
A_i^a T^a dx^i &= -\frac{i}{g}f(\xi) (\partial_i U_{\text{sph (mon)}} )U_{\text{sph (mon)}}^{-1}dx^i\ ,
\end{align}
where \(v_\phi\) is the multiplet vev, while \(\phi(\xi)\) and \(f(\xi)\) are the scalar and gauge radial profile functions, respectively.

To proceed, we must specify the matrices \(U_{\text{sph}}\) and \(U_{\text{mon}}\) for an arbitrary \(SU(2)\) multiplet. Although sphalerons and monopoles share the same schematic ansatz, their parameter dependence differs because of the distinct topological requirements discussed above. Concretely, \(U_{\text{sph}}\) depends on \((\theta,\phi)\) and the NCL parameter \(\mu\), whereas \(U_{\text{mon}}\) does not depend on \(\mu\). In both cases, we define
\begin{align} \label{eq:one_form_Fa_definition}
iU^{-1} \partial_i U dx^i = \sum_{a=1}^3 F_a T^a\ ,
\end{align}
with \(T^a\) the \(SU(2)\) generators and \(F_a\) scalar one-forms built from the relevant angular variables.

We first focus on \(U_{\text{sph}}\). The monopole matrix \(U_{\text{mon}}\) can then be obtained from \(U_{\text{sph}}\), as shown below.

For the sphaleron, \(U=U_{\text{sph}}\), and in the fundamental representation the components \(F_a\) are \footnote{As in Eq.~(\ref{eq::Manton_Umatrix_fundamental}), the sphaleron matrix is usually specified at \(\mu=\pi/2\)\ .}
\begin{equation}
  \begin{aligned} \label{eq:one_form_definition}
  &F_1=-(2\sin^2\mu\cos(\mu-\phi)-\sin2\mu\cos\theta\sin(\mu-\phi))d\theta \\
  &\quad -(\sin2\mu\cos(\mu-\phi)\sin\theta+\sin^2\mu\sin2\theta\sin(\mu-\phi))d\phi \\
  &F_2=-(2\sin^2\mu\sin(\mu-\phi)+\sin2\mu\cos\theta\cos(\mu-\phi))d\theta \\
  &\quad +(-\sin2\mu\sin(\mu-\phi)\sin\theta+\sin^2\mu\sin2\theta\cos(\mu-\phi))d\phi \\
  &F_3=-\sin2\mu\sin\theta d\theta+2\sin^2\theta \sin^2\mu d\phi\ .
  \end{aligned}
\end{equation}

It is useful to construct \(U_{\text{sph}}\) and \(F_a\) in a representation-independent way for an arbitrary \(SU(2)\) multiplet with isospin \(J\) \footnote{Ahriche et al. \cite{Ahriche:2014jna} evaluated sphaleron energies for higher \(SU(2)\) scalar representations and used, without proof, that the one-form \(F_a\) is representation independent. Below, we make that invariance explicit.}. A natural starting point is to parametrize a general \(SU(2)\) element using Wigner \(D\)-matrices:
\begin{align} \label{eq:U_sph_general_single_wigner_D_attempt}
U_\text{sph}(\alpha,\beta,\gamma)=\mathcal{D}^{1/2}\big(\alpha,\beta,\gamma \big)=e^{-i\alpha T^3} e^{-i\beta T^2} e^{-i\gamma T^3}\ .
\end{align}
The Euler angles \(\alpha,\beta,\gamma\) do not depend on the representation. To relate them to sphaleron parameters, compare with the fundamental representation:
\begin{align}\label{eq::su2_2dim_wigner}
\mathcal{D}^{1/2}\big(\alpha,\beta,\gamma \big)
&=e^{-i\alpha \frac{\sigma_3}{2}} e^{-i\beta \frac{\sigma_2}{2}} e^{-i\gamma \frac{\sigma_3}{2}} \\
&=\left(
  \begin{matrix}
  e^{-i\frac{\alpha + \gamma}{2}}\cos\frac{\beta}{2}  & -e^{-i\frac{\alpha - \gamma}{2}} \sin\frac{\beta}{2}  \\
  e^{i\frac{\alpha - \gamma}{2}} \sin\frac{\theta}{2}  & e^{i\frac{\alpha + \gamma}{2}} \cos\frac{\beta}{2}
  \end{matrix}
  \right)\ .
\end{align}
Matching this expression to the sphaleron matrix in Eq.~\eqref{eq::Manton_Umatrix_fundamental} gives
\begin{equation}
\begin{aligned} \label{eula_parameter}
&\cos \left(\frac{\beta }{2}\right) \cos \left(\frac{\alpha }{2}+\frac{\gamma }{2} \right)=1+\sin^{2}\mu(\cos\theta-1) \\
&\sin \left(\frac{\beta }{2}\right) \sin \left(\frac{\alpha }{2}-\frac{\gamma }{2}\right)=\sin\phi \sin\theta \sin\mu \\
&\sin \left(\frac{\beta }{2}\right) \cos \left(\frac{\alpha }{2}-\frac{\gamma }{2}\right)=-\cos\phi \sin\theta \sin\mu \\
&\cos \left(\frac{\beta }{2}\right) \sin \left(\frac{\alpha }{2}+\frac{\gamma }{2}\right)=\sin\mu \cos\mu (\cos\theta-1)\ \ .
\end{aligned}
\end{equation}
These relations follow by expanding both matrices in the basis \(I_{2\times 2},\sigma_1,\sigma_2,\sigma_3\) and matching coefficients. Although one can solve \((\alpha,\beta,\gamma)\) in terms of \((\mu,\theta,\phi)\), the expressions are cumbersome and non-linear, which makes direct evaluation of \(F_a\) inconvenient.

A more practical approach is to construct \(U_{\text{sph}}\) from products of Wigner-\(D\) matrices. For isospin \(J\) (matrix dimension \(2J+1\)), we use
\begin{equation}
   \begin{aligned}
  \label{eq::general_u} 
   U_{\text{sph}}\big(\mu, \theta, \phi\big) 
   &= D^{J}\big(\omega_{-}, -\theta, \mu \big) D^{J}\big(\mu, \theta, \omega_+\big)\\
   &=e^{-i\omega_-T^3}e^{i\theta T^2}e^{-2i\mu T^3}e^{-i\theta T^2} e^{-i\omega_+ T^3}\ ,
\end{aligned} 
\end{equation}
where
\begin{equation}
  \omega_{\pm}=-\mu\pm(\phi-\frac{\pi}{2})\ \ \ .
\end{equation}
For \(J=1/2\), this reproduces Eq.~\eqref{eq::Manton_Umatrix_fundamental}. For larger \(J\), one promotes \(T^i\) to the \((2J+1)\)-dimensional generators. For example, for \(J=1\),
\begin{align} \label{eq:T2_T3_generators_J_1}
T^2=\frac{i}{\sqrt{2}}\begin{pmatrix}
  0 & -1 & 0 \\
  1 & 0 & -1 \\
  0 & 1 & 0
\end{pmatrix},\quad T^3=\begin{pmatrix}
  1 & 0 & 0 \\
  0 & 0 & 0 \\
  0 & 0 & -1
\end{pmatrix}\ .
\end{align}
This parametrization simplifies evaluation of the one-form because the Euler-like parameters enter linearly in the product form. The components \(F_a\) follow from
\begin{equation}
  F_a = \left[\text{Tr}(T^3)^2\right]^{-1} \text{Tr}[iU_{\text{sph}}^{-1}\partial_iU_{\text{sph}}.T^a]dx^i\ .
\end{equation}
Using this method, we explicitly checked that the same \(F_a\) is obtained for \(J\in\{1,3/2,2,5/2,3\}\), using standard \(SU(2)\) generators \cite{Fonseca:2020vke}. Therefore, the one-form \(F_a\) in Eq.~\eqref{eq:one_form_definition} is representation independent.

We now turn to the monopole. A convenient construction of \(U_{\text{mon}}\) for a general multiplet starts from \(U_{\text{sph}}\) at \(\mu=\pi/2\), multiplied on the right by a constant matrix \(U_R\) (independent of \(\theta,\phi\)):
\begin{equation} \label{eq:sph_to_mon}
  U_{\text{mon}}(\theta,\phi)=U_{\text{sph}}(\mu=\frac{\pi}{2},\theta,\phi).U_R\ .
\end{equation}
To verify that this gives the standard monopole gauge field, substitute \(U_{\text{mon}}\) into the Klinkhamer-Laterveer form gauge ansatz. From Eq.~\eqref{eq::gauge_trans_Gauge},
\begin{equation}
\begin{aligned} \label{eq:Klinkhamer_gauge_f3_equal_f_monopole}
  A_{i}^aT^adx^i = \frac{1-f(\xi)}{g}iU_{\text{mon}}^{-1} (\partial_i U_{\text{mon}})dx^i\ .
\end{aligned}
\end{equation}
Using Eq.~\eqref{eq:sph_to_mon}, this becomes
\begin{equation} \label{eq:monopole_gauge_from_rotation_UR}
\begin{aligned}
A_{i}^aT^adx^i
&= \frac{1-f(\xi)}{g}U_R^{-1}.\left[iU_{\text{sph}}^{-1} (\partial_i U_{\text{sph}})dx^i\right].U_R\\
&= \frac{1-f(\xi)}{g}U_R^{-1}\left[\sum_{i=1}^3F_aT^a \right]U_R\ \ \ .   
\end{aligned}
\end{equation}
We do not include an additional left multiplication \(U_L\) in Eq.~\eqref{eq:sph_to_mon}, because such a factor cancels in Eq.~\eqref{eq:Klinkhamer_gauge_f3_equal_f_monopole}. A convenient choice is \cite{Chao:2018xwz}
\begin{equation} \label{eq:UR matrix}
  U_R=\left(\begin{array}{ccccc}
0 & 0 & \cdots & 0 & 1 \\
0 & 0 & \cdots & -1 & 0 \\
\cdots & \cdots & \cdots & \cdots & \cdots \\
0 & (-1)^{2 J-1} & \cdots & 0 & 0 \\
(-1)^{2 J} & 0 & \ldots & 0 & 0
\end{array}\right)\ ,
\end{equation}
which implements \(T^{a*}=-U_R^{-1}T^a U_R\) (see the appendix of Ref.~\cite{Chao:2018xwz}). For \(J\in\{1,3/2,2,5/2,3\}\), Eq.~\eqref{eq:monopole_gauge_from_rotation_UR} reproduces the standard ’t Hooft--Polyakov form
\begin{equation}
  A_i^a=\epsilon^{iab}x^b F\ \ \ ,
\end{equation}
with
\begin{equation}
  F=\frac{2(-1+f(\xi))}{gr^2}\ \ \ .
\end{equation}
This confirms that the rotated sphaleron matrix yields the standard monopole gauge profile.

\subsubsection{mass and equations of motion for sphaleron and monopole}
\label{sec:mass_and_eom_for_general_multiplet}

With the general sphaleron and monopole matrix constructions established, we now evaluate their masses and equations of motion. Unlike the previous subsection, we include both the SM Higgs doublet and a new general \(SU(2)\) multiplet, since this is the most relevant setup in many BSM scenarios.

As shown in Eq.~\eqref{eq:sph_to_mon}, the monopole configuration can be obtained from the sphaleron configuration by a simple rotation. We focus on the sphaleron case; the monopole case follows analogously and is detailed in \cite{Wu:2023mjb}. To improve accuracy, we include the \(U(1)_Y\) gauge contribution to sphaleron and monopole masses. Since this contribution is well studied in \cite{Klinkhamer:1990fi}, we adopt their convention.

There are two Higgs multiplets in this setup: the SM doublet \(H\), and a new general \(SU(2)\) multiplet \(\Phi\). The gauge fields are \(A_\mu^a\) for \(SU(2)_L\) and \(a_\mu\) for \(U(1)_Y\). The field ansatz is
\begin{equation}\label{eq:multiplet_sphaleron_confg_KL_scalar_general}
\begin{aligned}
&H=\frac{v}{\sqrt{2}}h(\xi)U_{\text{sph}}(\mu,\theta,\phi)\begin{pmatrix}0\\1\end{pmatrix},\qquad
\Phi=\frac{v_\phi}{\sqrt{2}}\phi(\xi)\begin{pmatrix}0\\\vdots\\1\\\vdots\\0\end{pmatrix},\\
&A_i^a T^a dx^i = \frac{1}{g}\,(1-f(\xi))\big(F_1 T^1+F_2 T^2\big)
+ \frac{1}{g}\,(1-f_3(\xi))\,F_3 T^3,\\
&a_i dx^i = \frac{1}{g^\prime}\,(1-f_0(\xi))F_3\ .
\end{aligned}
\end{equation}
Here, \(v\) and \(v_\phi\) are the vevs of the SM Higgs doublet and the new multiplet, respectively, while \(h(\xi)\), \(\phi(\xi)\), \(f(\xi)\), \(f_3(\xi)\), and \(f_0(\xi)\) are radial profiles. The one-forms \(F_a\) (\(a=1,2,3\)) are defined in Eq.~\eqref{eq:one_form_definition}. The generators \(T^a\) are in the fundamental representation for \((D_iH)^\dagger (D_iH)\), and in the general representation for \((D_i\Phi)^\dagger (D_i\Phi)\). Since we proved in the previous subsection that \(F_a\) is representation independent, the same \(F_a\) can be used for both scalars. The field energy is
\begin{equation}
\begin{aligned} \label{eq::formal_sphaleron_formula}
    E=\frac{4\pi \Omega}{g}\int d\xi \bigg[ \frac{1}{4}F^a_{ij}F^a_{ij}+\frac{1}{4}f^a_{ij}f^a_{ij}+ (D_iH)^{\dagger}(D_iH)  + (D_i\Phi)^{\dagger}(D_i\Phi) + V(H,\Phi) \bigg]\ .
\end{aligned}
\end{equation}
Substituting Eq.~\eqref{eq:multiplet_sphaleron_confg_KL_scalar_general} into Eq.~\eqref{eq::formal_sphaleron_formula}, we obtain
\begin{equation} \label{eq:sphaleron_energy_density_terms}
\begin{aligned}
\frac{1}{4}F_{ij}^{a}F_{ij}^{a}(\xi,\mu) &= \sin ^{2} \mu\left(\frac{8}{3} f^{\prime 2}+\frac{4}{3} f_{3}^{\prime 2}\right)+\frac{8}{\xi^{2}} \sin ^{4} \mu\left\{\frac{2}{3} f_{3}^{2}(1-f)^{2}+\frac{1}{3}\left\{f(2-f)-f_{3}\right\}^{2}\right\} \\
\frac{1}{4}f_{ij}f_{ij}(\xi,\mu)&=\frac{4}{3}\left(\frac{g}{g^{\prime}}\right)^{2}\left\{f_{0}^{\prime 2} \sin ^{2} \mu +\frac{2}{\xi^{2}} \sin ^{4} \mu\left(1-f_{0}\right)^{2}\right\} \\
(D_{i}\Phi)^{\dagger}(D_{i}\Phi)(\xi,\mu)&=\frac{v^{2}_\phi}{\Omega^{2}}\left\{\frac{1}{2} \xi^{2} \phi^{\prime 2}+\frac{4}{3} \phi^{2} \sin ^{2} \mu \left\{\left(J(J+1)-J_{3}^{2}\right)(1-f)^{2}+J_{3}^{2}\left(f_{0}-f_{3}\right)^{2}\right\}\right\}\ \ \ .
\end{aligned}
\end{equation}
The Higgs doublet term \((D_iH)^{\dagger}(D_iH)\) is obtained from \((D_i\Phi)^{\dagger}(D_i\Phi)\) by setting \(J=1/2\), \(J_3=1/2\), \(\phi(\xi)\to h(\xi)\), and \(v_\phi\to v\). The scalar potential \(V(H,\Phi)\) depends on the model and is left in general form here. Varying the energy functional gives the radial equations of motion \cite{Ahriche:2014jna,Wu:2023mjb}:
\begin{equation} \label{eq::EOMs}
\begin{aligned}
f^{\prime \prime}+\frac{2}{\xi^{2}}(1-f)\left[f(f-2)+f_{3}\left(1+f_{3}\right)\right]+(1-f)(\frac{v^{2}h^{2}}{4\Omega^{2}}+\alpha \phi^{2})=0, \\
f_{3}^{\prime \prime}-\frac{2}{\xi^{2}}\left[3 f_{3}+f(f-2)\left(1+2 f_{3}\right)\right]+(\frac{v^{2}}{4\Omega^{2}}h^{2}+\beta \phi^{2})(f_{0}-f_{3})=0, \\
f_{0}^{\prime \prime}+\frac{2}{\xi^{2}}\left(1-f_{0}\right)-\frac{g^{\prime 2}}{g^{2}}(\frac{v^{2}}{4\Omega^{2}}h^{2}+\beta \phi^{2})(f_{0}-f_{3})=0, \\
h^{\prime \prime}+\frac{2}{\xi}h^{\prime}-\frac{2}{3\xi^{2}}h[2(1-f)^{2}+(f_{0}-f_{3})^{2}]-\frac{1}{g^{2} v^{2} \Omega^{2}}\frac{\partial V[h,\phi]}{\partial h}=0,\\
\phi^{\prime \prime}+\frac{2}{\xi}\phi^{\prime}-\frac{8\Omega^{2}\phi}{3 v_{\phi}^{2} \xi^{2}}[2\alpha(1-f)^{2}+\beta (f_{0}-f_{3})^{2}]-\frac{1}{g^{2} v_{\phi}^{2} \Omega^{2}}\frac{\partial V[h,\phi]}{\partial \phi}=0\ .
\end{aligned}
\end{equation}
Here, \(\alpha=4J(J+1)-4J_3^2\), \(\beta=4J_3^2\), and prime denotes \(d/d\xi\). Before solving Eq.~\eqref{eq::EOMs}, one must specify boundary conditions, obtained from finite-energy requirements \cite{Wu:2023mjb}:
\begin{equation}
 \begin{aligned} \label{eq:sphaleron_boundary_conditions}
&\xi\rightarrow0,\quad f=f_3=\phi=0,\ f_0=1 \\
&\xi\rightarrow \infty,\quad f=f_3=f_0=\phi=1\ .
\end{aligned}   
\end{equation}
For the pure SM sphaleron, the numerical radial profiles are shown in Fig.~\ref{fig:SM_sphaleron_profiles}. The coupled equations in Eq.~\eqref{eq::EOMs} were solved using \texttt{scipy.integrate.solve\_bvp}. It is useful to define
\begin{equation} \label{eq:sphaleron_energy_final_definition}
E_{\text{sph (mon)}}=\frac{4\pi \Omega}{g} \mathcal{B}_{\text{sph (mon)}}\ \ \ .
\end{equation}
For the SM sphaleron in Fig.~\ref{fig:SM_sphaleron_profiles}, one finds \(\mathcal{B}_{\text{sph}}\approx 1.9\).

\begin{figure}[t]
\center
\includegraphics[width=8cm]{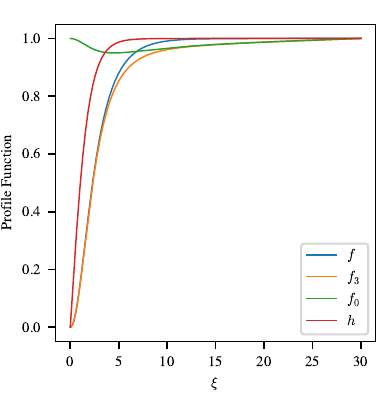}
\caption{Sphaleron radial-profile solutions for the pure Standard Model. The horizontal axis denotes the dimensionless radial coordinate \(\xi\equiv g\Omega r\), with \(\Omega=246\) GeV and \(g\) the \(SU(2)\) gauge coupling.\reproducedfromref{Wu:2023mjb}}
\label{fig:SM_sphaleron_profiles}
\end{figure}

We now consider a specific example for the additional scalar multiplet: a complex septuplet with \(Y=0\). 
We first review the motivations for choosing this representation.
In general, for an electroweak multiplet having isospin $J$, theoretical constraints restrict the maximum allowed value. When $J\geq 5$, the perturbative regime of the theory breaks down at an unacceptably low scale, around $\Lambda_{\text{Landau}}\leq 10$ TeV \cite{AbdusSalam:2013eya}. Additionally, consistency with tree-level scattering amplitude unitarity imposes upper bounds: $J\leq 7/2$ for complex scalar multiplets and $J \leq 4$ for real ones \cite{Hally:2012pu,Earl:2013jsa}. We further impose the requirement that the neutral component can contribute to dark matter. To satisfy stringent direct detection constraints from experiments, this neutral state must not interact with the $Z$ boson, necessitating $Y=0$. Only multiplets with integer $J$ possess a $J_3=0$ component, restricting our choice accordingly. The multiplet with the largest dimension satisfying both the unitarity constraint and viable dark matter criterion is the complex septuplet with $J=3$ \cite{Cirelli:2005uq}. This multiplet naturally emerges in the minimal dark matter framework \cite{Cirelli:2005uq}, where the neutral septuplet can serve as a dark matter candidate without requiring an additional discrete symmetry.

Returning to Eq.~\eqref{eq::formal_sphaleron_formula}, the only model-dependent ingredient is the scalar potential \(V(H,\Phi)\). For the SM Higgs doublet plus a complex septuplet \(\Phi\), the potential is \cite{Chao:2018xwz}
\begin{equation} \label{eq::potential_higgs_with_multiplet}
\begin{aligned}
V=&M_A^2\left(\Phi^{\dagger} \Phi\right)+\left\{M_B^2(\Phi \Phi)_0+\text { h.c. }\right\}  -\mu^2 H^{\dagger} H+\lambda\left(H^{\dagger} H\right)^2+\lambda_1\left(H^{\dagger} H\right)\left(\Phi^{\dagger} \Phi\right) \\
&+\lambda_2\left((\overline{H} H)_1(\overline{\Phi} \Phi)_1\right)_0+\left[\lambda_3(\overline{H} H)_0(\Phi \Phi)_0+\text { h.c. }\right] +V_{\text{self}}(\Phi,\overline{\Phi})\ ,
\end{aligned}
\end{equation}
with
\begin{equation}
  \begin{aligned}
    V_{\text{self}}(\Phi,\overline{\Phi}) =& \sum_{J=0}^{2 J} \kappa_k\left((\Phi \Phi)_k(\overline{\Phi}\,\overline{\Phi})_k\right)_0 +\sum_{k=0}^{2 J}\left\{\kappa_k^{\prime}\left((\Phi \Phi)_k(\Phi \Phi)_k\right)_0  +\kappa_k^{\prime \prime}\left((\overline{\Phi} \Phi)_k(\Phi \Phi)_k\right)_0+\text { h.c. }\right\}\ .
  \end{aligned}
\end{equation}
Here, \(J\) is the \(SU(2)\) isospin of the multiplet (\(J=3\) for a complex septuplet). \(\overline{H}\) and \(\overline{\Phi}\) denote the conjugate \(SU(2)\) representations of \(H\) and \(\Phi\). By Clebsch--Gordan selection rules \cite{Chao:2018xwz}, \((\Phi\Phi)_1\), \((\Phi\Phi)_3\), and \((\Phi\Phi)_5\) vanish, so only \(k\in\{0,2,4,6\}\) contribute for the septuplet. In addition, for \(J=3\), only the \(k=0\) and \(k=2\) operators are independent \cite{Chao:2018xwz,Cao:2022ocg}, reducing the number of independent self couplings.

\begin{figure}[t]
\center
\includegraphics[width=13cm]{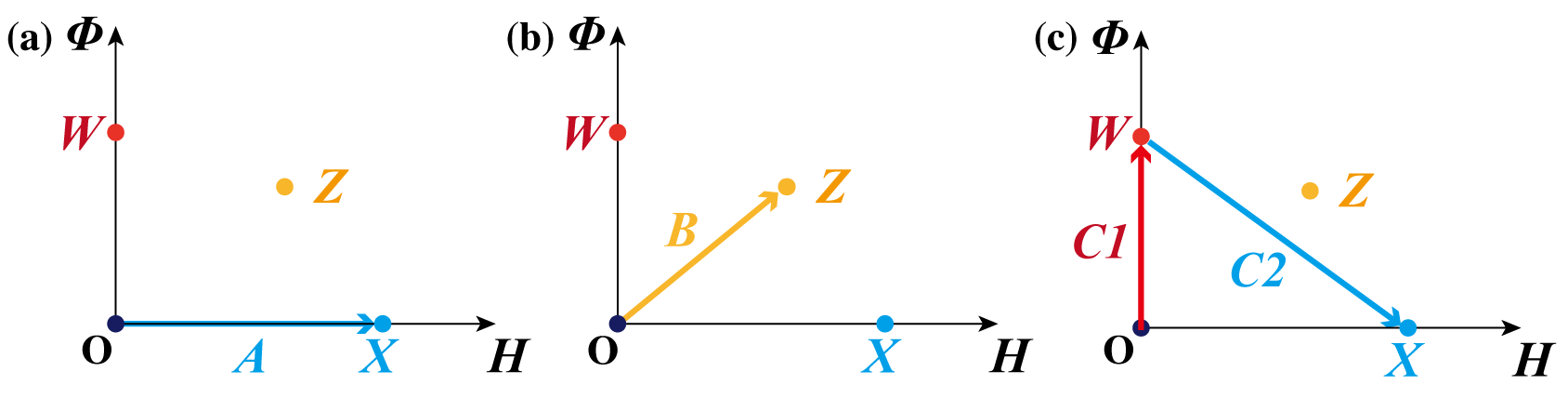}
\caption{Schematic illustration of the three scenarios for the electroweak phase transition. In each panel, the horizontal axis represents the Higgs field, while the vertical axis represents the septuplet field. Panel (a) represents the one-step phase transition from the symmetric phase $O$ to the electroweak minimum $X$; (b) represents a one-step phase transition to the mixed minimum $Z$, where both fields acquire nonzero vacuum expectation values; (c) represents a two-step phase transition from the symmetric phase $O$ to the septuplet minimum $W$, followed by a second step to the electroweak minimum $X$.\reproducedfromref{Wu:2023mjb}}
\label{fig:Phase_Transition_Three_Scenarios}
\end{figure}

\begin{figure*}[t!]
\center
\includegraphics[width=8cm]{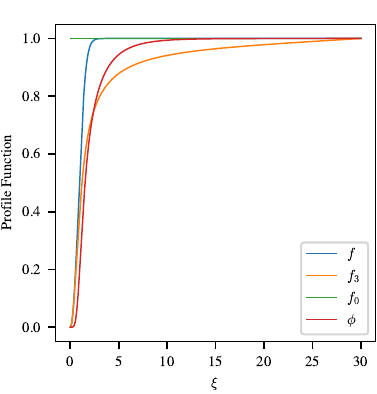}
\caption{Radial profile functions when only the septuplet obtains a vev during a two-step EWPT, for \(v=0\) GeV, \(v_\phi=500\) GeV, \(\lambda_{13}=0.05\), and \(\lambda_s=0.005\). At this stage, the Higgs profile does not appear because its vev vanishes, and the monopole-mass value is \(B_{\text{mon}}=5.001145\).\reproducedfromref{Wu:2023mjb}}
\label{fig:Two_EWPT_profile_function}
\end{figure*}

\begin{figure}[t!]
\center
\includegraphics[width=8cm]{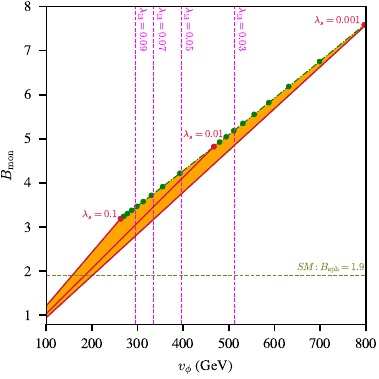}
\caption{Monopole mass \(B_{\text{mon}}\) when only the septuplet obtains a vev during a two-step EWPT. The horizontal axis is the septuplet vev \(v_\phi\) in GeV, and the vertical axis is \(B_{\text{mon}}\). The viable monopole-mass region is the part of the orange band to the right of the dashed vertical line at \(\lambda_{13}\). The olive-drab horizontal dashed line indicates the SM sphaleron benchmark \(B_{\text{sph}}\approx1.9\). The orange band shows \(B_{\text{mon}}\) across scanned values of \(v_\phi\) and \(\lambda_s\). The lower and upper red curves correspond to \(\lambda_s=0.001\) and \(\lambda_s=0.1\), respectively, with \(\lambda_s\) increasing counterclockwise. Colored dots mark points removed by the \(v_\phi\) constraint.\reproducedfromref{Wu:2023mjb}}
\label{fig:two_step_monopole_mass}
\end{figure}

When the SM is extended with the new septuplet, different phase transition patters may exist, as illustrated in Fig.~\ref{fig:Phase_Transition_Three_Scenarios}. In the figure, the horizontal axis represents the Higgs field, while the vertical axis represents the septuplet field. Panel (a) represents the one-step phase transition from the symmetric phase $O$ to the electroweak minimum $X$; panel (b) represents a one-step phase transition to the mixed minimum $Z$, where both fields acquire nonzero vacuum expectation values; panel (c) represents a two-step phase transition from the symmetric phase $O$ to the septuplet minimum $W$, followed by a second step to the electroweak minimum $X$. Two scenarios were studied in Ref.~\cite{Wu:2023mjb}:  
(i) both the SM Higgs doublet and septuplet obtain vevs (panel (b) in Fig.~\ref{fig:Phase_Transition_Three_Scenarios}), and  
(ii) only the septuplet obtains a vev (panel (c) at point $W$ in Fig.~\ref{fig:Phase_Transition_Three_Scenarios}).  
In the first case, the topological field configuration is still a sphaleron. In the second case, it is a monopole. The second case can be interpreted as an intermediate stage of a two-step phase transition, followed by a second step where only the SM Higgs doublet obtains a vev. For this intermediate phase, the relevant homotopy group is the monopole one, as shown in Eq.~\eqref{eq:topology_for_monopole}.

Each case requires detailed analysis of viable parameter space and phenomenological constraints, which is presented in \cite{Wu:2023mjb}. We do not repeat those details here. We define two combined couplings that frequently appear in sphaleron and monopole energy expressions:
\begin{equation} \label{eq::lam13}
  \lambda_{13}=\frac{1}{2}\lambda_1-\frac{1}{\sqrt{14}}\lambda_3\ ,
\end{equation}
and
\begin{equation} \label{eq::lams}
\begin{aligned}
\lambda_{s}=&\frac{1}{7}[\kappa_{0}+2\text{Re}(\kappa_{0}^{\prime})-2\text{Re}(\kappa_{0}^{\prime \prime})] +\frac{4}{21\sqrt{5}}[\kappa_{2}+2\text{Re}(\kappa_{2}^{\prime})-2\text{Re}(\kappa_{2}^{\prime \prime})]\ .
\end{aligned}
\end{equation}
It was shown in \cite{Wu:2023mjb} that, in scenario (i), septuplet contributions to the sphaleron energy are negligible, mainly due to the \(\rho\)-parameter constraint \cite{ParticleDataGroup:2024cfk} requiring a very small septuplet vev. In scenario (ii), however, the monopole mass can be significantly affected by septuplet contributions. We therefore focus on scenario (ii). In this case, the septuplet vev is \(v_\phi\), while the SM Higgs vev is zero. The numerical monopole radial profiles are shown in Fig.~\ref{fig:Two_EWPT_profile_function}, and the monopole mass is shown in Fig.~\ref{fig:two_step_monopole_mass}. We see that the monopole mass can be significantly larger than the SM sphaleron mass, depending on \(v_\phi\), \(\lambda_{13}\), and \(\lambda_s\). This can have important implications for baryon-number violation during a two-step EWPT. If the first step of the two-step EWPT generates a baryon asymmetry, that asymmetry can be well preserved when the monopole mass is sufficiently large to suppress monopole-induced BNV. However, it may still be washed out by sphaleron-induced BNV during the second step of the EWPT, which requires a careful analysis of the sphaleron energy at that stage. This Higgs-phase sphaleron-induced BNV, and the associated baryon-washout calculation, are studied in detail in Sec.~\ref{sec:electroweak_baryogenesis_sphaleron}.

\newpage
\section{Baryon Number Violation in Gauge Theories} \label{sec:BNV_in_gauge_theory}
In this section, we discuss the mechanisms of BNV in the Standard Model, and explain how the topological field configurations introduced in the previous section enter this problem. This is a broad subject, covering topics such as the ABJ anomaly, vacuum structure of non-Abelian gauge theories, instantons, and extensive use of semiclassical methods in the Euclidean path integral. It is neither realistic nor necessary to cover all of these aspects in full detail in this thesis. Indeed, many excellent review articles and books already exist; for example, Coleman's classic lectures \cite{Coleman:1978ae,Coleman:1985rnk} provide a clear introduction to instantons and vacuum structure, while the review by Rubakov and Shaposhnikov \cite{Rubakov:1996vz} is a standard reference for BNV in the Standard Model. We therefore proceed as follows. We mainly emphasize the logical chain by which BNV is understood from basic principles, while also commenting on implications and recent developments. We do not present every intermediate calculation; however, for some key steps, or for steps that are not easily found in the literature, we give more detail.

The understanding of baryon-number violation in gauge theories can be organized into three stages:
\begin{enumerate}
  \item \textbf{Tunneling in the pure-gauge vacuum sector.} Early work (for example, \cite{Callan:1977gz}) analyzed the nontrivial vacuum structure and tunneling in pure gauge theory, using Euclidean (instanton) methods without fermions.
  \item \textbf{Tunneling in the gauge and fermion sectors (instantons with fermions).} 't Hooft \cite{tHooft:1976rip} showed that fermion fluctuations in an instanton background induce effective interactions that violate baryon and lepton numbers. At zero temperature, these processes are exponentially suppressed by the instanton action, yielding rates proportional to $e^{-16\pi^2/g^2}$.
  \item \textbf{Tunneling in the gauge, fermion, and Higgs sectors.} Including the Higgs sector leads to static saddle-point solutions of the energy functional, namely sphalerons, which sit at the top of the barrier between topologically distinct vacua \cite{Manton:1983nd}. At finite temperature, the relevant Boltzmann-suppressed rate is controlled by the sphaleron energy rather than the pure-gauge instanton action. The role of fermions is similar to that in stage 2, namely to induce BNV. The new feature of stage 3 is that the Higgs sector modifies the field energy through the sphaleron configuration, which can significantly affect the BNV rate at finite temperature.
\end{enumerate}

The main focus of this thesis is stage 3: how BSM Higgs sectors modify sphaleron energies and monopole masses, and what the implications are for baryon-number violation at leading and next-to-leading (determinant) order. \flag 
Although the previous section showed that different scalar multiplets can give rise to different topological field configurations---namely, sphaleron or monopole---that conclusion applies only to theories with a single scalar multiplet undergoing symmetry breaking. In the Standard Model, the Higgs doublet already determines the vacuum structure of the theory, i.e.\ the coset space \(G/H\). Consequently, adding an extra scalar multiplet does not alter the vacuum structure \cite{Wu:2023mjb}.
Therefore, instanton-induced BNV remains a useful starting point for understanding the underlying mechanism. The common principle underlying both instanton- and sphaleron-induced BNV is the ABJ anomaly. \flag In appendix~\ref{sec:ABJ_anomaly}, we give a detailed review of the ABJ anomaly, including its derivation using a simple $\sigma$ model, its application to the Standard Model $SU(2)_L$ chiral gauge theory.

Below, we first discuss gauge-vacuum structure, winding number and Chern--Simons number in Sec.~\ref{sec:winding_number_gauge_vacuum_structure}; we then discuss sphaleron and monopole-induced baryon-number violation and its relation to instantons in Sec.~\ref{sec:sphaleron_baryon_violation} and \ref{sec:monopole_induced_bnv}.

In Appendix.~\ref{sec:instanton_QM} and \ref{sec:instanton_baryon_violation}, we present a pedagogical analysis of instanton tunneling in quantum-mechanics examples and the instanton-induced baryon-number violation in gauge theories with fermions. Since tunneling in the vacuum structure of gauge theories is highly nontrivial, it is useful to introduce these quantum-mechanics examples. We will see that these examples already capture many essential features of tunneling in gauge theories.

\subsection{Vacuum Structure of non-Abelian Gauge Theories, Winding Number and Chern--Simons Number}
\label{sec:winding_number_gauge_vacuum_structure}
\flag
We first discuss two kinds of mappings from \(S^3\) to the vacuum manifold of non-Abelian group, using \(SU(2)\) as the main example. We will see that one of these mappings generalizes to the concept of winding number, while the other generalizes to the concept of Chern--Simons (CS) number. Later in this subsection, we discuss the relation between the winding number and the CS number. For a homotopy group labeled by an integer, the winding number can be identified with the homotopy class of the mapping, as illustrated below for the \(U(1)\) example in Eq.~\eqref{eq:homotopy_group_U(1)}. We will also see later that the winding number plays an important role in understanding sphaleron- and monopole-induced baryon-number violation. Therefore, the discussion of these mappings provides an important bridge between the homotopy groups introduced previously and the baryon-number-violation mechanism to be discussed in the following subsections. We clarify the two kinds of \(S^3\) that appear here:
\begin{itemize}
  \item The first \(S^3\) is the boundary at infinity of Euclidean spacetime \cite{Belavin:1975fg,Coleman:1978ae}, i.e.\ \(|x|\to\infty\). This is the setup used in Sec.~\ref{sec:aspects-TPC} when discussing the finite-energy condition for topological field configurations. This \(S^3_{(1)}\) is illustrated in Fig.~\ref{fig:cylinder} as the boundary surface of the cylinder. The spatial boundary at infinity (\(|\vec{x}|=R\) with $R\rightarrow \infty$) together with the imaginary-time direction forms the cylindrical boundary shown in the figure, which represents the \(S^3\) boundary of Euclidean spacetime. The vertical direction represents imaginary time, while each horizontal slice represents a spatial slice at fixed time. The mapping of this \(S^3_{(1)}\) to the gauge vacuum manifold is characterized by an integer winding number, which is related to tunneling between different vacua and to baryon-number violation.
\item The second \(S^3\) is the compactified spatial slice at fixed time for a vacuum configuration \cite{Treiman:1986ep}. \flag We consider a time-independent gauge transformation \(U(\vec{x})\), under which the Lagrangian and the Hamiltonian remain invariant. When acting on a vacuum configuration, such a transformation generates another gauge-equivalent vacuum configuration. Following Jackiw, we assume that \(U(\vec{x})\) approaches a constant element at spatial infinity \cite{Treiman:1986ep}. Under this assumption, physical space \(\mathbb{R}^3\) can be regarded as compactified to the three-sphere \(S^3\), where ``compactified'' means that all points at spatial infinity are identified with a single point. The function \(U(\vec{x})\) therefore defines a map from \(S^3\) to the gauge group. Such maps are classified into distinct homotopy classes, labeled by an integer number. For vacuum gauge configurations, this integer is identified with the Chern--Simons number. This \(S^3_{(2)}\) is illustrated in Fig.~\ref{fig:cylinder} as each horizontal slice at \(\tau=\pm T\).
\end{itemize}

\begin{figure}[t!]
\center
\includegraphics[width=5cm]{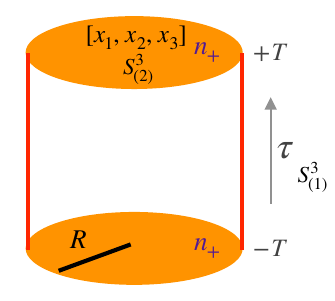}
\caption{Illustration of the spacetime structure as a cylinder. Each horizontal slice represents a spatial slice at fixed time, while the vertical direction corresponds to imaginary time. The spatial boundary at infinity (\(|\vec{x}|=R\) with $R\rightarrow \infty$), together with the imaginary-time direction, forms the surface of the cylinder, which has the topology of \(S^3_{(1)}\). At \(\tau=-T\) and \(\tau=+T\), the fields are in vacuum configurations. If the vacuum spatial slice is compactified, \flag then the vacuum configuration on each horizontal slice has the topology of \(S^3_{(2)}\), where ``compactified'' means that all points at spatial infinity are identified with a single point. The subscripts \((1)\) and \((2)\) distinguish the two kinds of \(S^3\) discussed in the main text.}
\label{fig:cylinder}
\end{figure}

In the Standard Model, the gauge groups are Abelian \(U(1)_Y\) and non-Abelian \(SU(2)_L\) and \(SU(3)_C\). The mapping from \(S^3\) to the gauge group is classified by the homotopy group \(\pi_3(G)\), where \(G\) is the gauge group. Here, the \(S^3\) can be either \(S^3_{(1)}\) or \(S^3_{(2)}\), as discussed above. The homotopy group \(\pi_3(U(1)_Y)\) is trivial, while \(\pi_3(SU(2)_L)\) and \(\pi_3(SU(3)_C)\) are both nontrivial, namely \(\mathbb{Z}\). Therefore, for the gauge-vacuum structure and instanton tunneling, we only need to consider the non-Abelian gauge groups. For the example of \(SU(2)_L\), both mappings are classified by \(\pi_3(SU(2))=\mathbb{Z}\). Below, we discuss the mathematical details of these mappings.

\flag
Since the first kind of mapping, from \(S^3_{(1)}\) to the vacuum manifold of \(SU(2)\), is directly related to tunneling between different vacua, it is the primary focus here. As shown later in Eq.~\eqref{eq:winding_number_SU(2)_equivalent_definitions}, the winding number associated with this mapping is related to the spacetime integral of \(F\tilde{F}\), where \(F\) is the field-strength tensor and \(\tilde{F}\) is its dual, defined below. As shown in Eq.~\eqref{eq:ABJ_anomaly_non_Abelian} of Appendix~\ref{sec:ABJ_anomaly}, the integral of \(F\tilde{F}\) is directly related to changes \footnote{Strictly speaking, the relation to changes in baryon and lepton number is obtained after analytically continuing from Euclidean time to Minkowski spacetime, so that the result can be interpreted in terms of real-time evolution.} in baryon and lepton number for for chiral $SU(2)$ gauge theory. Since this connection is central to the main themes of this thesis, we concentrate primarily on this case. By contrast, the second kind of mapping, from \(S^3_{(2)}\) to \(SU(2)\), characterizes the vacuum structure of the gauge theory, and the associated integer is identified with the Chern--Simons number. This Chern--Simons number is related to the winding number through Eq.~\eqref{eq:relation_winding_number_Chern_Simons}. Since the vacuum structure itself is not the main focus of this thesis, we comment on it only briefly here without going into full mathematical detail. Further mathematical details of the second kind of mapping can be found in Refs.~\cite{Treiman:1986ep,Srednicki:2007qs}.

Below, we omit the subscript \((1)\) and simply denote the first kind of \(S^3_{(1)}\) by \(S^3\). For pedagogical purposes, we first consider the simpler case of a mapping from \(S^1\) to the \(U(1)\) group. Suppose \(U\) is an element of \(U(1)\). It can be parameterized as
\begin{align}
  U^{(0)}(\theta) & = 1 \ , \\
  U^{(1)}(\theta) &= e^{i\theta} \ , \\
  U^{(n)}(\theta) &= e^{i n \theta} \ ,
\end{align}
where \(\theta\in[0,2\pi]\) is the angular variable parameterizing \(S^1\). \(U^{(0)}\) represents the trivial mapping, while \(U^{(1)}\) and \(U^{(n)}\) represent the identity mapping and the \(n\)-winding mapping from \(S^1\) to \(U(1)\), respectively. We define the winding number \(\nu\) as
\begin{align} \label{eq:winding_number_U(1)}
  \nu = \frac{i}{2\pi} \int_0^{2\pi} d\theta\ U \frac{dU^{-1}}{d\theta} \ .
\end{align}
By direct calculation, one finds \(\nu=0\) for \(U^{(0)}\), \(\nu=1\) for \(U^{(1)}\), and \(\nu=n\) for \(U^{(n)}\). Moreover, \(\nu\) is invariant under a continuous infinitesimal deformation of \(U(\theta)\), i.e.\ it is a topological invariant. To verify this, consider an infinitesimal deformation of \(U(\theta)\),
\begin{align}
  \delta U = i (\delta \lambda) U \ ,
\end{align}
where \(\delta \lambda\) is an infinitesimal real function. The variation of \(\nu\) is
\begin{align}
  \delta \nu &= \frac{i}{2\pi} \int_0^{2\pi} d\theta\ \left[ \delta U \frac{dU^{-1}}{d\theta} + U \frac{d(\delta U^{-1})}{d\theta} \right] \nonumber \\
  &= \frac{i}{2\pi} \int_0^{2\pi} d\theta\ \left[ i(\delta \lambda)\, U \frac{dU^{-1}}{d\theta} - i\,U \frac{d\bigl((\delta\lambda)U^{-1}\bigr)}{d\theta} \right] \nonumber \\
  &= \frac{1}{2\pi} \int_0^{2\pi} d\theta\ \frac{d(\delta \lambda)}{d\theta} = 0 \ ,
\end{align}
where we used the condition \(\delta \lambda=0\) at the boundaries. Thus, \(\nu\) is indeed a topological invariant. The definition of \(\nu\) in Eq.~\eqref{eq:winding_number_U(1)} is therefore a realization of the homotopy-group statement used above Eq.~\eqref{eq:vortex_configuration},
\begin{align}
  \label{eq:homotopy_group_U(1)}
  \nu = \pi_1(U(1)) \cong \mathbb{Z} \ .
\end{align}

We now generalize the above discussion to mappings from \(S^3\) to \(SU(2)\). Consider an element \(U\) of \(SU(2)\), parameterized as
\begin{align}
  U^{(0)}(x) & = \mathbb{I} \ , \\
  U^{(1)}(x) & = \frac{1}{r} (x_4 \mathbb{I} + i x_i \sigma^i) \ , \\
  U^{(n)}(x) & = \left[ U^{(1)}(x) \right]^n \ ,
\end{align}
where \(x_\mu\), with \(\mu=1,2,3,4\), are the coordinates parameterizing the Euclidean boundary \(S^3\), and \(r=\sqrt{x_\mu x_\mu}\). Here \(\mathbb{I}\) is the identity matrix, and \(\sigma^i\), with \(i=1,2,3\), are the Pauli matrices. \(U^{(0)}\) represents the trivial mapping, while \(U^{(1)}\) and \(U^{(n)}\) represent the identity mapping and the \(n\)-winding mapping from \(S^3\) to \(SU(2)\), respectively. In analogy with the \(U(1)\) case, we define the winding number \(\nu\) for the mapping \(S^3\to SU(2)\) as
\begin{align} \label{eq:winding_number_SU(2)}
  \nu = \frac{1}{24\pi^2} \int_{S^3} dS_\mu \epsilon^{\mu \nu \rho \sigma} \mathrm{Tr}\left( U\partial_\nu U^{-1} U\partial_\rho U^{-1} U\partial_\sigma U^{-1} \right) \ .
\end{align}
Here $dS^\mu=n^\mu dS$, where \(n^\mu\) is the outward-pointing unit normal vector on the Euclidean boundary \(S^3\), and \(dS^\mu\) is the oriented surface element. This definition can be understood as follows. (i) The integral is performed over the Euclidean boundary \(S^3\), which can be parameterized by three angles, e.g.\ three Euler angles. The particular choice of angles is irrelevant for the definition of the winding number, because the Jacobian from changing coordinates is canceled by the Jacobian carried by the \(\epsilon\)-symbol \cite{Coleman:1978ae}. (ii) The product contains three factors of \(U \partial_\nu U^{-1}\), since the number of generators of \(SU(2)\) is three; this is the natural generalization of the single-derivative term in Eq.~\eqref{eq:winding_number_U(1)} for the \(S^1\) case. (iii) We use the four-dimensional \(\epsilon\)-symbol, which is Lorentz invariant, to contract the Lorentz indices \cite{Belavin:1975fg}. (iv) The definition of \(\nu\) is invariant under continuous deformations, i.e.\ it is a homotopy-invariant quantity. (v) For the identity mapping \(U^{(1)}\), one can verify by direct calculation that \(\nu=1\).

We now carry out some additional algebra on Eq.~\eqref{eq:winding_number_SU(2)} to obtain a more useful form. First, define
\begin{align} \label{eq:K_mu_definition}
    K^{\mu}\equiv \frac{g^{2}}{32\pi^{2}}\epsilon^{\mu \nu \rho \sigma} \left(F_{\nu \rho}^{a}A_{\sigma}^{a}-\frac{g}{3}f^{abc}A_{\nu}^{a}A_{\rho}^{b}A_{\sigma}^{c}\right) \ ,
\end{align}
where \(F_{\nu \rho}^{a}\) is the field-strength tensor, \(A_{\sigma}^{a}\) is the gauge field, and \(f^{abc}\) is the structure constant of \(SU(2)\). One can verify that the divergence of \(K^\mu\) is
\begin{align} \label{eq:divergence_K_mu}
  \partial_\mu K^\mu = \frac{g^2}{32\pi^2} F_{\mu\nu}^a \tilde{F}^{a\mu\nu} \ ,
\end{align}
where \(\tilde{F}^{a\mu\nu}=\frac{1}{2}\epsilon^{\mu\nu\rho\sigma}F_{\rho\sigma}^a\) is the dual field-strength tensor. Performing a four-dimensional integral of both sides of Eq.~\eqref{eq:divergence_K_mu} over Euclidean space, and applying Gauss's theorem to the left-hand side, gives
\begin{align} \label{eq:surface_integral}
   \int_{S^3} dS_\mu K^\mu = \frac{g^2}{32\pi^2}\int d^4 x\  F_{\mu\nu}^a \tilde{F}^{a\mu\nu} \ .
\end{align}
Since finite action requires \(F_{\mu\nu}^a\to 0\) at the Euclidean boundary, only the second term in Eq.~\eqref{eq:K_mu_definition} contributes on the left-hand side. Thus,
\begin{align}
  \int_{S^3} dS_\mu K^\mu &= -\frac{g^3}{96\pi^2} \int_{S^3} dS_\mu \epsilon^{\mu \nu \rho \sigma} f^{abc} A_{\nu}^{a} A_{\rho}^{b} A_{\sigma}^{c} \nonumber \\
  &=  i\frac{g^3}{24\pi^2} \int_{S^3} dS_\mu \epsilon^{\mu \nu \rho \sigma} \mathrm{Tr}\left( A_{\nu}^a\tau^a A_{\rho}^b \tau^b A_{\sigma}^c \tau^c \right) \ ,
\end{align}
where \(\tau^a\) are the \(SU(2)\) generators, and we used \(\mathrm{Tr}(\tau^a\tau^b\tau^c)=if^{abc}/4\). Since the vacuum gauge configuration approaches a pure gauge at the Euclidean boundary, we can make the following replacement for \(A_\nu^a\tau^a\) in the above integral,
\begin{align} \label{eq:gauge_ansatz_boundary}
  A_{\nu}^a\tau^a \rightarrow \frac{i}{g} U\partial_\nu U^{-1} \ ,
\end{align}
noting that \((\partial_i U) U^{-1}=-U\partial_i U^{-1}\). Then
\begin{align}
  \int_{S^3} dS_\mu K^\mu &= \frac{1}{24\pi^2} \int_{S^3} dS_\mu \epsilon^{\mu \nu \rho \sigma} \mathrm{Tr}\left( U\partial_\nu U^{-1} U\partial_\rho U^{-1} U\partial_\sigma U^{-1} \right) \ .
\end{align}
Therefore, we arrive at the following two equivalent definitions of the winding number \(\nu\) for the mapping \(S^3\to SU(2)\) \footnote{In some definitions of the winding number, there is an overall minus-sign difference. Such a sign difference can be removed by changing the sign in Eq.~\eqref{eq:gauge_ansatz_boundary}, which does not affect the following discussion.},
\begin{align}
  \label{eq:winding_number_SU(2)_equivalent_definitions}
  \nu = \pi_3(SU(2)) &= \frac{1}{24\pi^2} \int_{S^3} dS_\mu \epsilon^{\mu \nu \rho \sigma} \mathrm{Tr}\left( U\partial_\nu U^{-1} U\partial_\rho U^{-1} U\partial_\sigma U^{-1} \right) \nonumber \\
  &= \frac{g^2}{32\pi^2} \int d^4 x\ F_{\mu\nu}^a \tilde{F}^{a\mu\nu} \ .
\end{align}
This winding number \(\nu\) is also known in the literature as the Pontryagin number or the second Chern class \cite{Belavin:1975fg,Manton:2004tk,Tong:2005un}.

We now briefly comment on the second kind of mapping, from \(S^3_{(2)}\) to \(SU(2)\), which usually leads to the definition of the Chern--Simons number. The Chern--Simons number is defined as \cite{Treiman:1986ep}
\begin{align}
  \label{eq:Chern_Simons_number_definition}
  N_{\rm CS} = \frac{1}{24\pi^2} \int d^3 x\ \epsilon^{ijk} \mathrm{Tr}\left( U\partial_i U^{-1} U\partial_j U^{-1} U\partial_k U^{-1} \right) \ ,
\end{align}
where the matrix \(U\) is now a function of the spatial coordinates at $\tau=\pm T$ with $T\rightarrow \infty$. The integral is performed over the compactified spatial slice, which has the topology of \(S^3_{(2)}\).

We now investigate its relation to the winding number \(\nu\) defined in Eq.~\eqref{eq:winding_number_SU(2)}. The integral in Eq.~\eqref{eq:surface_integral} can be decomposed into two parts,
\begin{align}
  \frac{g^2}{32\pi^2}\int d^4 x\ F_{\mu\nu}^a \tilde{F}^{a\mu\nu} &\equiv \int_{-\infty}^{+\infty} dt \int_{-\infty}^{+\infty} d^3 x\ \partial_\mu K^\mu \nonumber \\
  &=\int d^{3} x\, K^{0}\bigg\vert^{t=+ \infty}_{ t=-\infty}+\int_{-\infty}^{+\infty}dt \int_{S} \vec{K}\cdot \vec{dS} \ .
  \label{eq:decomposition_surface_integral}
\end{align}
Another useful definition is
\begin{align}
  \label{eq:definition_P_ta_tb}
  \mathcal{P}(t_a,t_b) = \int_{t_a}^{t_b} dt \int d^3 x\ \partial_\mu K^\mu \ ,
\end{align}
where the previous definition of winding number $\nu$ in Eq.~\eqref{eq:winding_number_SU(2)} is recovered by taking \(t_a=-\infty\) and \(t_b=+\infty\). We will later see that \(\mathcal{P}(t_a,t_b)\) is useful for the computation of the sphaleron baryonic charge, as discussed in detail in Sec.~\ref{sec:instanton_sphaleron_difference}. As mentioned in Ref.~\cite{Treiman:1986ep}, for vacuum configurations at \(t=\pm\infty\), we assume the gauge matrix \(U\) to approach a constant matrix at spatial infinity\footnote{This is not the case for non-vacuum configurations. For example, at \(t=t_0\) with finite \(t_0\), the second term on the right-hand side of Eq.~\eqref{eq:decomposition_surface_integral} does not vanish. This appears in the computation of the sphaleron baryonic charge, as discussed in detail in Ref.~\cite{Wu:2023mjb} and Sec.~\ref{sec:instanton_sphaleron_difference}.}. \flag 
Then \(\partial_i U=0\), and the pure-gauge form in Eq.~\eqref{eq:gauge_ansatz_boundary} implies \(A_i\to 0\) as \(|\vec x|\to\infty\). At the same time, \(F_{ij}\to 0\). Since \(K^\mu\) in Eq.~\eqref{eq:K_mu_definition} is built from \(A_i\) and \(F_{ij}\), it follows that \(K^\mu\to 0\) at spatial infinity. Therefore, the second term on the right-hand side of Eq.~\eqref{eq:decomposition_surface_integral} vanishes. 
Substituting the definition of \(K^0\) from Eq.~\eqref{eq:K_mu_definition}, we obtain
\begin{align}
  \label{eq:relation_winding_number_Chern_Simons}
  \frac{g^2}{32\pi^2} \int d^4 x\ F_{\mu\nu}^a \tilde{F}^{a\mu\nu} &= \int d^3 x\ K^0\bigg\vert^{t=+ \infty}_{ t=-\infty} \nonumber \\
  &= \frac{1}{24\pi^2 } \int d^3 x\ \epsilon^{ijk} \mathrm{Tr}\left( U\partial_i U^{-1} U\partial_j U^{-1} U\partial_k U^{-1} \right)\bigg\vert^{t=+ \infty}_{ t=-\infty} \nonumber \\
  &= N_{\rm CS}(t=+\infty) - N_{\rm CS}(t=-\infty) \ .
\end{align}
Thus, we have shown that the winding number \(\nu\) defined in Eq.~\eqref{eq:winding_number_SU(2)} equals the difference of the Chern--Simons number between the two vacua at \(t=+\infty\) and \(t=-\infty\). This is a very important conclusion, because it shows that the tunneling winding number \(\nu\) is determined by the difference in Chern-Simons numbers between the two vacua. 
The winding number is not limited to \(\nu=\pm 1\); higher values can occur and, within the dilute instanton gas approximation, be decomposed into individual instantons and anti-instantons; see relevant discussions in Appendix.~\ref{sec:instanton_QM} and Ref.~\cite{Coleman:1985rnk}.

\subsection{Baryon Number Violation from Sphaleron}
\label{sec:sphaleron_baryon_violation}
\flag A full understanding of the sphaleron mechanism requires a solid grasp of the instanton mechanism. For readers who are not familiar with instantons, we provide a pedagogical discussion in Appendix~\ref{sec:instanton_QM}, where simple quantum-mechanics examples are used to illustrate the key features of instanton tunneling. In addition, Appendix~\ref{sec:instanton_baryon_violation} presents a detailed analysis of instanton-induced baryon-number violation in gauge theories with fermions. Since one of the main focuses of this thesis is sphaleron-induced BNV, we discuss the sphaleron mechanism and its relation to the instanton mechanism in greater detail in the main text, while leaving the instanton mechanism itself to the appendix. 

Sphaleron is a static solution of Yang--Mills--Higgs theory, which is intrinsically different from the imaginary-time instanton solution. However, as mentioned at the beginning of this section, the introduction of the scalar field does not change the vacuum structure of the non-Abelian gauge theory. Therefore, the picture of baryon-number violation from transitions between vacua with different \flag Chern-Simons numbers remains valid. Here, the transition can refer either to quantum tunneling at zero temperature or to thermal transitions at finite temperature. While the instanton describes the quantum-tunneling process at zero temperature, the sphaleron describes the thermal transition process at finite temperature. 

In the following, we first illustrate the key differences and connections between instanton- and sphaleron-induced BNV. We then explain why the sphaleron-induced BNV rate can be much larger than the instanton-induced BNV rate at finite temperature.

\subsubsection{difference between sphaleron and instanton mechanisms}
\label{sec:instanton_sphaleron_difference}

In the previous instanton case, the action is written as \(S_E=-\int d^4 x\,\mathcal{L}\), and only gauge and fermion fields are involved. For the sphaleron case, we consider two modifications. \flag
First, let us consider the thermal Euclidean action. In contrast to zero-temperature field theory, thermal field theory requires a Wick rotation of the time coordinate, $t \to -i\tau$, followed by restricting the Euclidean time $\tau$ to the interval $[0, \beta]$, where $\beta = 1/T$ \cite{Laine:2016hma}. Peoridic (anti-peoridic) conditions are then applied depending on the spin statistics of the fields. This is known as the imaginary-time (Matsubara) formalism of thermal field theory; a detailed discussion of Matsubara frequencies is given in Appendix~\ref{app:matsubara_frequencies}. The Euclidean action then takes the form
\begin{align}
  S_E = -\int_0^\beta d\tau \int d^3x \, \mathcal{L} \sim -\frac{1}{T} \int d^3x \, \mathcal{L} \ .
\end{align}
In the second approximation, we assume the high-temperature limit. In this limit, the fermion fields decouple because their energy levels are proportional to \(\pi T\) \footnote{
  This follows from Eq.~\eqref{eq:matsubara_frequencies}, where the lowest Matsubara frequency for fermions is $\pi T$. In contrast, bosonic fields possess a zero Matsubara mode corresponding to $n=0$.
}. This effective three-dimensional Euclidean action and the associated thermal field theory will be introduced in detail in Sec.~\ref{sec:thermal_effective_potential_DR}. Second, we include scalar fields in the Lagrangian, such as the SM Higgs field. The 3D Euclidean Lagrangian density then becomes
\begin{align}
  \mathcal{L}_{\rm 3D} = -\frac{1}{4} F_{ij}^a F_{ij}^a - (D_i \phi)^\dagger (D_i \phi) - V(\phi) \ .
\end{align}
Here \(i,j=1,2,3\) are spatial indices, and \(V(\phi)\) is the scalar potential. Note that we have assumed temporal gauge, \(A_0^a=0\), and a static field configuration, i.e. \(\partial_\tau A_i^a=0\) and \(\partial_\tau \phi=0\). Therefore, the field-strength components \(F_{0i}\) and the covariant derivative \(D_0\phi\) vanish. As mentioned earlier, under this gauge choice the Hamiltonian density equals minus the Lagrangian density, as seen from Eq.~\eqref{eq:Hamiltonian_density}. The sphaleron solution is the static solution of the nontrivial, spherically symmetric equations of motion derived from the above 3D Euclidean Lagrangian. The sphaleron field ansatz and the numerical solutions were introduced in Sec.~\ref{sec:manton_sphaleron}.

\begin{figure}[htbp]
\centering

\begin{minipage}{0.48\linewidth}
  \centering
  \begin{tikzpicture}
    \begin{axis}[
      width=\linewidth,
      height=0.45\linewidth,
      domain=0:18.85, samples=401,
      axis x line=middle,
      axis y line=middle,
      xlabel={$j$},
      ylabel={$E$},
      xmin=0, xmax=19.85,
      ymin=0, ymax=2.6,
      xtick={0.01,6.283,12.566,18.849},
      xticklabels={$0$,$1$,$2$,$3$},
      ytick=\empty,
      title={(a) Instanton}]
      \addplot[blue, thick] {1 - cos(deg(x))};
      \addplot[only marks, mark=*, mark options={fill=black}] coordinates {(0,0) (6.283,0) (12.566,0) (18.849,0)};
      \draw[->, dashed, thick, purple] (axis cs:6.283,0.25) -- (axis cs:12.566,0.25);
    \end{axis}
  \end{tikzpicture}
\end{minipage}\hfill
\begin{minipage}{0.48\linewidth}
  \centering
  \begin{tikzpicture}
    \begin{axis}[
      width=\linewidth,
      height=0.45\linewidth,
      domain=0:18.85, samples=401,
      axis x line=middle,
      axis y line=middle,
      xlabel={$j$},
      ylabel={$E$},
      xmin=0, xmax=19.85,
      ymin=0, ymax=2.6,
      xtick={0.01,6.283,12.566,18.849},
      xticklabels={$0$,$1$,$2$,$3$},
      ytick=\empty,
      title={(b) Sphaleron}]
      \addplot[blue, thick] {1 - cos(deg(x))};
      \addplot[only marks, mark=*, mark options={fill=black}] coordinates {(0,0) (6.283,0) (12.566,0) (18.849,0)};
      \addplot[only marks, mark=*, mark options={fill=red}] coordinates {(9.4248,2)};
      \draw[->, thick, red] (axis cs:8.0,2.0) to[out=70,in=110] (axis cs:10.8,2.0);
    \end{axis}
  \end{tikzpicture}
\end{minipage}
\caption{Comparison of the instanton and sphaleron mechanisms for baryon-number violation. (a) Instanton mechanism: quantum tunneling from the vacuum with Chern-Simons number \(j=1\) to the vacuum with Chern-Simons number \(j=2\). (b) Sphaleron mechanism: thermal transition over the energy barrier from the vacuum with Chern-Simons number \(j=1\) to the vacuum with Chern-Simons number \(j=2\). The static sphaleron solution is located at the top of the energy barrier (red point).}
\label{fig:instanton_sphaleron_mechanisms_comparison}
\end{figure}
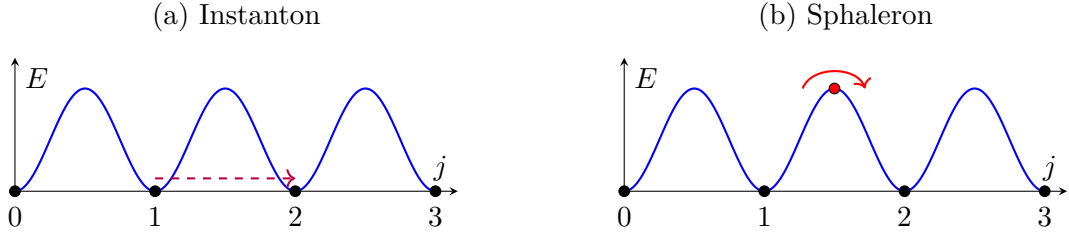

Fig.~\ref{fig:instanton_sphaleron_mechanisms_comparison} illustrates the key difference between the instanton and sphaleron mechanisms for baryon-number violation. In the instanton case, BNV is often pictured as quantum tunneling between vacua with different Chern-Simons numbers, as shown in Fig.~\ref{fig:instanton_sphaleron_mechanisms_comparison}(a). In the sphaleron case, BNV occurs via thermal transitions over the energy barrier between vacua with different Chern-Simons numbers, as shown in Fig.~\ref{fig:instanton_sphaleron_mechanisms_comparison}(b). However, as discussed in Appendix~\ref{sec:instanton_baryon_violation}, the physically relevant tunneling amplitude is $\langle \theta| e^{-H\tau}| \theta \rangle$, where $|\theta\rangle$ is the true vacuum state defined in Eq.~\eqref{eq:theta_vacuum_definition}. By contrast, the amplitude $\langle j^\prime | e^{-H\tau}| j \rangle$ violates the cluster decomposition theorem for chirality-violating operators in the presence of massless quarks \cite{Callan:1976je}. Therefore, Fig.~\ref{fig:instanton_sphaleron_mechanisms_comparison} should be understood only as a schematic comparison of the instanton and sphaleron configurations, rather than as a literal representation of the physical transition amplitude.

In the previous subsection, we introduced the general winding-number functional \(\mathcal{P}(t_a,t_b)\) in Eq.~\eqref{eq:definition_P_ta_tb}. For the sphaleron, we take the interval \(t\in(-\infty,t_0]\), with \(t_0\) the time at which the fields sit on the sphaleron configuration. The corresponding sphaleron charge is defined as \cite{Klinkhamer:1984di}
\begin{equation}
\begin{aligned} \label{eq:Q_B sphaleron definition new}
Q_{B}&=\int d^{3} x\, K^{0}\bigg\vert^{t=t_{0}}_{t=-\infty}+\int^{t_{0}}_{-\infty}dt \int d^3 x\, \nabla\cdot \vec{K} \ .
\end{aligned}
\end{equation}
At \(t=-\infty\), the gauge fields are in the vacuum, while at \(t=t_0\), they are located at the top of the energy barrier between adjacent vacua. Here \(K^\mu\) is defined previously in Eq.~\eqref{eq:K_mu_definition}.
The sphaleron charge definition $Q_B$ is analogous to the definition of the winding number, except that the limit $t=+\infty$ is replaced by $t=t_0$. Since the sphaleron refers to the static solution at $t=t_0$, we can further define
\begin{align}
  Q_B\equiv N_{\rm CS}({\rm sph}, t_0) - N_{\rm CS}({\rm vac},t=-\infty) \ .
\end{align}
If we choose $N_{\rm CS}({\rm vac},t=-\infty)=0$, then $Q_B=N_{\rm CS}({\rm sph}, t_0)$. 
We will show explicitly that the sphaleron has a half-integer Chern-Simons number, i.e. \(j_{\rm sph}=1/2\) (note that we denote the Chern-Simons number of the sphaleron as \(j_{\rm sph}\)).
For the sphaleron configuration in Eq.~\eqref{eq:sphaleron_multiplet_confg_manton}, one finds \(K^0=0\), so the first term in Eq.~\eqref{eq:Q_B sphaleron definition new} vanishes. The remaining term reduces to
\begin{align} \label{eq:Q_B sphaleron definition simplified}
j_{\rm sph}&=\int^{t_{0}}_{-\infty}dt \int d^3 x\, \nabla\cdot \vec{K} \nonumber \\
&=-\int _{-\infty}^{t_{0}}dt \,\frac{2}{\pi} f^2(\xi)\times \sin^{2}(\mu(t))\frac{d\mu(t)}{dt}\bigg\vert^{r=\infty}_{r=0} \ ,
\end{align}
where we (i) expand \(\nabla\cdot \vec{K}\) in spherical coordinates and (ii) insert the sphaleron ansatz, Eq.~\eqref{eq:sphaleron_multiplet_confg_manton}, into \(K^i\). Using the boundary conditions \(f(0)=0\) and \(f(\infty)=1\), we obtain
\begin{equation}
\begin{aligned}
j_{\rm sph}(\text{MK}) =-\frac{1}{\pi}\left(\mu-\frac{1}{2}\sin(2\mu)\right)\bigg\vert^{\mu=\frac{\pi}{2}}_{\mu=0}=-\frac{1}{2} \ .
\end{aligned}
\end{equation}

We now compute the transition amplitude at finite temperature. Since we considered the pure quantum-tunneling effect in Appendix.~\ref{sec:instanton_QM}, it is pedagogically natural to use the same quantum-mechanics picture to illustrate the crossover from tunneling to thermal barrier crossing. However, instead of considering all three potential scenarios, we focus only on the double-well potential \flag \footnote{
One may wonder why we do not consider a periodic potential, which more closely resembles the vacuum structure labeled by different Chern–Simons numbers. The reason is that our purpose here is not to reproduce the full vacuum structure, but rather to illustrate the effect of finite temperature on barrier crossing between neighboring vacua. For this purpose, a double-well potential already captures the essential physics of thermally activated transitions and the role of the sphaleron configuration at the top of the barrier. Although the double-well and periodic potentials differ in their global structures, this difference is not crucial for the present discussion, where we focus only on transitions between adjacent vacua.
}, as illustrated in Fig.~\ref{fig:finite-temperature-tunneling}.

\begin{figure}[htbp]
\centering
\begin{minipage}{0.4\linewidth}
  \centering
  \begin{tikzpicture}
    \begin{axis}[
      width=\linewidth,
      domain=-2.5:2.5, samples=301,
      axis lines=middle,
      xlabel={$x$}, ylabel={$V(x)$},
      xmin=-2, xmax=2,
      ymin=0, ymax=2.5,
      xtick={-1.225,0.005,1.225},
      xticklabels={$-a$, $0$, $a$},
      extra x ticks={-0.636,0.636},
      extra x tick labels={$-x(E)$, $x(E)$},
      extra x tick style={tick label style={font=\scriptsize, xshift=3pt}},
      ytick=\empty,
      title={}]
      \addplot[red, thick] {((x^2-1.5)^2/1.5)};
      \addplot[only marks, mark=*, mark options={fill=black}] coordinates {(-1.225,0) (1.225,0)};
      \addplot[only marks, mark=*, mark options={fill=black}] coordinates {(0,1.5)};
      \node[anchor=west] at (axis cs:0.02,1.52) {$V_0$};
      \draw[dashed] (axis cs:-0.636,0.8) -- (axis cs:0.636,0.8);
      \draw[dashed] (axis cs:-0.636,0.8) -- (axis cs:-0.636,0);
      \draw[dashed] (axis cs:0.636,0.8) -- (axis cs:0.636,0);
      \addplot[only marks, mark=*, mark options={fill=red}] coordinates {(0,0.8)};
      \node[anchor=west, text=red] at (axis cs:0.02,0.92) {$E$};
    \end{axis}
  \end{tikzpicture}
\end{minipage}
\caption{Finite-temperature barrier crossing in one-dimensional quantum mechanics with a double-well potential.}
\label{fig:finite-temperature-tunneling}
\end{figure}
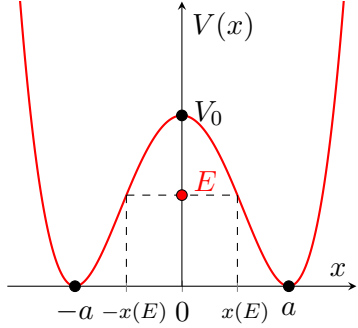

According to the WKB approximation in quantum mechanics, the tunneling amplitude is dominated by the exponential of the Euclidean action. Therefore, the leading-order tunneling rate is \cite{Landau1981Quantum}
\begin{align}
  \Gamma_{\rm tunneling} \sim e^{-2 S_0} \ ,
\end{align}
where $S_0$ is the Euclidean action we will discuss below.
Recalling that the particle has vanishing kinetic energy, the equation of motion for the particle is given in Eq.~\eqref{eq:classical_path_equation_QM}. The solution of this classical path equation connects \(x=-a\) at \(\tau=-\infty\) and \(x=a\) at \(\tau=+\infty\). Integrating Eq.~\eqref{eq:classical_path_equation_QM} over \(dx\), we obtain \(dx/d\tau=\sqrt{2V}\). The Euclidean action \(S_0\) can thus be computed as
\begin{align} \label{eq:euclidean_action_instanton_QM}
  S_0&=\int_{-\infty}^{\infty} d\tau \left[\frac{1}{2}\left(\frac{dx}{d\tau}\right)^2 + V(x)\right] \nonumber \\
 &= \int_{-\infty}^{\infty}d\tau\ \left(\frac{dx}{d\tau}\right)^2 \nonumber \\
 & = \int_{-a}^{a}dx\ \sqrt{2V(x)} \ .
\end{align}
In the last line, we changed the integration variable from \(\tau\) to \(x\), and used \(d\tau=dx/\sqrt{2V(x)}\) in both the second and third lines. The last line of Eq.~\eqref{eq:euclidean_action_instanton_QM} is the same action obtained from the WKB method in ordinary quantum mechanics. This action is also analogous to the instanton action in four-dimensional non-Abelian gauge theory, where the action value is \(8\pi^2/g^2\).

We now account for finite-temperature effects in the double-well potential. At temperature \(T\), the particle can occupy excited states with energy \(E\), weighted by the Boltzmann factor \(e^{-E/T}\). The transition rate is therefore obtained by summing the tunneling contributions from all excited states \cite{Affleck:1980ac}:
\begin{align}
  \Gamma(T) &\sim \sum_n e^{-E_n/T}\,\Gamma_{\rm tunneling}(E_n) \nonumber \\
 &= \sum_n \exp\!\left[-\frac{E_n}{T}-2S(E_n)\right] \ .
\end{align}
Here the WKB action at fixed energy is
\begin{align}
  S(E)=\int_{-x(E)}^{x(E)} dx\,\sqrt{2\bigl(V(x)-E\bigr)} \ ,
\end{align}
and the turning point \(x(E)\) satisfies \(V\!\bigl[x(E)\bigr]=E\), as shown in Fig.~\ref{fig:finite-temperature-tunneling}. In the high-temperature limit \(T\gg m\), with \(m\) the particle mass, the sum can be approximated by an integral over \(x\) \cite{Rubakov:1996vz}:
\begin{align}
  \Gamma (T) & \sim \int dx \, \exp\left[-\frac{V(x)}{T} - 2 \int_{-x}^{x} dy \sqrt{2(V(y)-V(x))}\right] \nonumber \\
  & \sim \exp\left(-\frac{V_0}{T} \right) \ ,
  \label{eq:finite_temperature_barrier_crossing_rate_QM}
\end{align}
where \(V_0\) is the height of the potential barrier. The second line is obtained from the saddle-point approximation around \(x=0\). This result can be understood as follows: the particle is thermally excited to the top of the potential barrier with energy \(V_0\), after which it can freely roll down to the other side of the potential well. Therefore, in the high-temperature limit, the dominant process is no longer controlled by zero-temperature instanton tunneling, but by thermal barrier crossing with Boltzmann suppression \(e^{-V_0/T}\). This barrier-crossing rate is analogous to the theory of vacuum decay at finite temperature \cite{Linde:1977mm,Linde:1980tt}. The result in Eq.~\eqref{eq:finite_temperature_barrier_crossing_rate_QM} is valid under the conditions \cite{Rubakov:1996vz}
\begin{align}
  \label{eq:high_temperature_conditions}
  \frac{U_0}{T}<2S_0 \ , \ \ \frac{U_0}{T} \gg 1 \ ,
\end{align}
where the first condition ensures that quantum tunneling is much smaller than the classical thermal transition, while the second ensures that the thermal transition is valid within the saddle-point approximation. Motivated by this, in the high-temperature limit one can use a {\it fully classical} calculation to compute the transition rate, which is equal to the probability flux in one direction, e.g. from left to right, at the point \(x=0\) \cite{Hanggi:1990zz}:
\begin{align}
\gamma
= \left\langle \delta(x)\,\theta(\dot{x})\,\dot{x}\right\rangle
= \frac{\displaystyle \int dp\,dx\, e^{-H/T}\,\delta(x)\,\theta(p)\,p}
{\displaystyle \int dp\,dx\, e^{-H/T}}
= \frac{m}{2\pi}\,e^{-U_0/T} \ .
\end{align}

We can now apply the above quantum-mechanics result to sphaleron-induced baryon-number violation at finite temperature \footnote{One may wonder what the connection is between the quantum-mechanics example discussed above and the static sphaleron solution in field theory, which is an infinite-dimensional system. Since the sphaleron is a static field configuration, its energy is given by $E_{\rm sph}=\int d^3 x\ \mathcal{H}$. At zero temperature, the Euclidean action $S_E=\int_{-\infty}^{\infty} d\tau\, E_{\rm sph}$ diverges, so it cannot contribute to the zero-temperature semiclassical approximation \cite{Arnold:1987mh}. This is analogous to the quantum-mechanics example, where the $S_0$ defined in Eq.~\eqref{eq:euclidean_action_instanton_QM} is very large. Therefore, the first condition in Eq.~\eqref{eq:high_temperature_conditions} is automatically satisfied, and the zero-temperature tunneling contribution can be neglected.}.
Consider the partition function $\mathcal{Z}$ evaluated around the sphaleron background:
\begin{align}
  \label{eq:sphaleron_path_integral_finite_T}
  \mathcal{Z}_{\rm sph}&\sim \int [d\varphi]e^{-\int_0^\beta d\tau \int d^3 x \, \mathcal{L}_E(\phi_{\rm sph}+\varphi)} \nonumber \\
  & \sim \int [d\varphi] e^{-\frac{1}{T} \int d^3 x \, \mathcal{L}_E(\phi_{\rm sph}+\varphi)} \ .
\end{align}
Here \(\phi_{\rm sph}\) is the static sphaleron solution, and \(\varphi\) represents fluctuations around the sphaleron solution. For simplicity, we do not distinguish between gauge and scalar fields in the notation. The symbol \(\sim\) means that we are not keeping track of the overall normalization factor in the path integral. \flag
In the second line above, we have taken the high-temperature limit, with the condition that the Higgs vacuum expectation value should be replaced by its temperature-dependent value \cite{Carson:1990jm}. However, this replacement is not actually implemented in this work.\footnote{What we actually do is the following: one can define a three-dimensional effective theory, as in the second line above, whose coefficients and field normalizations are determined by matching to the parent four-dimensional theory (the first line). This matching procedure will be introduced in detail in the next section.} Here, our aim is simply to provide some physical intuition.
  Since the sphaleron is located at the top of the energy barrier between vacua with different Chern-Simons numbers, the sphaleron energy \(E_{\rm sph}\) plays the role of the barrier height \(V_0\) in the quantum-mechanics example. If we define the sphaleron energy as
\begin{align}
  E_{\rm sph} &= -\int d^3 x \, \mathcal{L}_{\rm 3D} \nonumber \\
  &= \int d^3 x \, \left[\frac{1}{4} F_{ij}^a F_{ij}^a + (D_i \phi)^\dagger (D_i \phi) + V(\phi)  \right] \ ,
\end{align}
and, correspondingly, the sphaleron-induced baryon-number-violation rate at finite temperature can be estimated as \footnote{Formally, this rate can be expressed in terms of the imaginary part of the free energy evaluated at the sphaleron configuration; although we do not discuss this property here, see Refs.~\cite{Affleck:1980ac,Arnold:1987mh}.}
\begin{align}
  \Gamma_{\rm sphaleron} (T) \sim T^4 \exp\left(-\frac{E_{\rm sph}}{T} \right) \ .
\end{align}
The factor \(T^4\) is introduced to give the rate the correct mass dimension. As one can see, this rate can be much larger than the instanton-induced baryon-number-violation rate at finite temperature, because \(E_{\rm sph}/T \ll 8\pi^2/g^2\) for \(T\) around the electroweak scale. The value of \(E_{\rm sph}\) was calculated in the previous section for the zero-temperature case, especially in Eq.~\eqref{eq:sphaleron_energy_final_definition}. At finite temperature, the broken-phase sphaleron energy can be approximated as \(E_{\rm sph}= 4\pi B_{\rm sph} v(T)/g(T)\), where \(v(T)\) is the temperature-dependent scalar-field expectation value, \(g(T)\) is the temperature-dependent gauge coupling, and \(B_{\rm sph}\) is a numerical factor of order one. There are important theoretical issues involving gauge dependence and higher-order corrections, which will be discussed in later sections. Thus, after the electroweak phase transition (EWPT), we can write the sphaleron-induced baryon-number-violation rate as
\begin{align}
  {\rm After\ EWPT:\ } \Gamma_{\rm sph} (T) \sim T^4 \exp\left(-\frac{4\pi B_{\rm sph} }{g(T)} \frac{v(T)}{T}\right) \ .
\end{align}
From this expression, we see that at high temperature, when \(v(T)\rightarrow 0\), the sphaleron energy \(E_{\rm sph}\rightarrow 0\), and therefore the baryon-number-violation rate \(\Gamma_{\rm sphaleron}(T)\) is no longer exponentially suppressed. The sphaleron rate above the EWPT cannot be calculated within the semiclassical approximation; instead, it must be computed using real-time lattice studies. On the lattice, one measures the so-called Chern--Simons-number diffusion rate, which is twice the sphaleron rate. This rate is defined as \cite{Bodeker:1999gx,Moore:2000ara}
\begin{align}
  \Gamma_{\rm CS} = \lim_{V,t \rightarrow \infty} \frac{\langle (N_{\rm CS}(t) - N_{\rm CS}(0))^2 \rangle}{V t} \ ,
\end{align}
where \(N_{\rm CS}\) is the Chern--Simons number defined in Eq.~\eqref{eq:Chern_Simons_number_definition}, \(V\) is the system volume, and \(t\) is real time. The sphaleron rate is then given by \(\Gamma_{\rm sphaleron}=\Gamma_{\rm CS}/2\). Therefore, above the EWPT, the lattice result for the sphaleron rate is \cite{Bodeker:1999gx,Moore:2000ara}
\begin{align}\label{eq:sphaleron_rate_above_EWPT}
  {\rm Before\ EWPT:\ } \Gamma_{\rm sph} (T) \sim N \alpha_W^5 T^4 \ ,
\end{align}
where \(N = (25.4 \pm 2.0)/2\) is a numerical factor from lattice calculations, and \(\alpha_W = g^2/(4\pi)\). There have been significant developments in lattice computations of the sphaleron rate in recent years; see Refs.~\cite{Burnier:2005hp,DOnofrio:2014rug,BarrosoMancha:2022mbj,Annala:2025aci} for more details. As will be seen in later sections, sphaleron-mediated baryon-number violation plays an important role in electroweak baryogenesis \cite{Kuzmin:1985mm}.

\subsubsection{large baryon-number-violation rate from the sphaleron}
\label{sec:BNV_rate_sphaleron}

What we analyzed in the previous subsection is the transition rate between vacua with different Chern-Simons numbers. The transition rates for increasing and decreasing the Chern-Simons number by one unit are the same. However, the physically relevant quantity is the baryon-number-violation rate, which we denote by \(\dot{B}\), where \(B\) is the baryon number. Therefore, it is crucial to relate \(\dot{B}\) to the sphaleron rate, \(\Gamma_{\rm sph}\). Establishing this relation requires non-equilibrium methods, as analyzed in Refs.~\cite{Khlebnikov:1988sr,Khlebnikov:1996vj}. The basis of that non-equilibrium analysis is the so-called reduced description of statistical systems. A full discussion of this method is lengthy and lies beyond the main scope of this thesis. In the following, we only present the results and comment on their numerical consequences.

The non-equilibrium analysis in Ref.~\cite{Khlebnikov:1988sr} gives the baryon-number-violation rate as
\begin{align}
  \label{eq:BNV_rate_sphaleron}
 \dot{B}=-n_f^2 \frac{\Gamma(B)}{T} \frac{\partial F(B)}{\partial B} \ ,
\end{align}
where \(F(B)\) is the free energy of the system with baryon number \(B\), and \(\Gamma(B)\) is the sphaleron rate, which is one half of the Chern--Simons-number diffusion rate\footnote{In some references, such as Ref.~\cite{Burnier:2005hp}, \(\Gamma(B)\) is defined as the Chern--Simons-number diffusion rate.}. This expression itself follows from non-equilibrium statistical mechanics, namely the Fokker--Planck equation, while the free energy \(F(B)\) can be computed using equilibrium statistical mechanics. Ref.~\cite{Khlebnikov:1996vj} discussed the computation of \(F(B)\) in detail and found
\begin{align}
  \label{eq:F_B_expression}
F(B)=\kappa\left(\frac{v}{T}\right) \frac{(B-B_0)^2}{V T^2} \ ,
\end{align}
where \(v\) is the scalar-field expectation value, \(B_0\) is the equilibrium baryon number, \(V\) is the system volume, and \(\kappa(v/T)\) is a dimensionless function of \(v/T\). The equilibrium value \(B_0\) can be written as \cite{Khlebnikov:1996vj}
\begin{align}
  B_0=\chi\left(\frac{v}{T}\right) (B-L)_{\rm eq} \ ,
\end{align}
where \((B-L)_{\rm eq}\) is the equilibrium value of \(B-L\), and \(\chi(v/T)\) is another dimensionless function of \(v/T\). Substituting the expression for \(F(B)\) into Eq.~\eqref{eq:BNV_rate_sphaleron}, we obtain
\begin{align}
  \dot{B}=-2 n_f^2 \kappa\left(\frac{v}{T}\right) \frac{\Gamma(B)}{VT^3} (B-B_0) \ .
\end{align}
Defining the sphaleron rate per unit volume as \(\Gamma_{\rm sph} = \Gamma(B)/V\), and the baryon number density as \(n_B = B/V\), we can rewrite the above equation as
\begin{align}
  \label{eq:BNV_rate_sphaleron_final}
  \dot{n}_B=-2 n_f^2 \kappa\left(\frac{v}{T}\right) \frac{\Gamma_{\rm sph}}{T^3} (n_B-n_{B_0}) \ .
\end{align}
The above equation shows that the baryon number density \(n_B\) relaxes toward the equilibrium value \(n_{B_0}\) with a relaxation rate given by \(2 n_f^2 \kappa(v/T)\Gamma_{\rm sph}/T^3\). For electroweak baryogenesis, one has \(n_{B_0}=0\).

The expressions for \(\kappa(x)\) and \(\chi(x)\) are \cite{Khlebnikov:1996vj}
\begin{align}
 \kappa\left(x\right) &= \frac{3(4 + 2 n_f + n_s) (22 n_f + 13 n_s) + 9 (24 n_f + 13 (2 + n_s)) x^2}{2 n_f [2 (4 + 2 n_f + n_s) (5 n_f + 3 n_s) + 3(11 n_f + 6 (2 + n_s)) x^2]} \ , \\
 \chi(x) &= \frac{4(2 n_f + n_s) (4 + 2 n_f + n_s)  + 12(2 + 2 n_f + n_s) x^2}{ (4 + 2 n_f + n_s) (22 n_f + 13 n_s) + 3(24 n_f + 13 (2 + n_s)) x^2} \ ,
\end{align}
where \(n_f\) is the number of fermion generations and \(n_s\) is the number of scalar doublets. For the pure Standard Model, \(n_f=3\) and \(n_s=1\). In many beyond-the-Standard-Model theories, multiple scalar doublets are predicted. However, it is easy to verify that both \(\kappa(x)\) and \(B_0(x)\) change negligibly when \(n_s\) is increased. Another feature is that both \(\kappa(x)\) and \(B_0(x)\) are rather insensitive to \(x=v/T\). Therefore, for a large class of theories, it is numerically a good approximation to take
\begin{align}
  \label{eq:kappa_chi_approximate_constants}
  \kappa \approx \frac{13}{12} \ ,
\end{align}
and one obtains the approximate constant values for \(\chi\) as
\begin{align}
  &{\rm After\ EWPT:\ } \frac{v}{T} \gg 1: \quad \chi \approx \frac{12}{37} \ , \\
  &{\rm Before\ EWPT:\ } \frac{v}{T} \ll 1: \quad \chi \approx \frac{28}{79} \ .
\end{align}
In practical computations, one may use these approximate constant values. As noted above, the introduction of additional scalar multiplets in many BSM models has only negligible effects on these quantities.

\subsection{Baryon Number Violation from Monopole}
\label{sec:monopole_induced_bnv}
Eq.~\eqref{eq:topology_for_monopole} in the previous section shows that monopole solutions can arise in many BSM theories. We now discuss how such monopoles can induce BNV. Dirac first incorporated monopoles into field theory \cite{Dirac:1931kp,Dirac:1948um}, thereby providing an explanation for charge quantization: all charged particles must carry charges that are integer multiples of a fundamental unit $e$. In this thesis, however, our main interest is the 't Hooft-Polyakov monopole \cite{tHooft:1974kcl,Polyakov:1974ek}, which can also give rise to baryon-number-violating effects. Magnetic monopoles have been the subject of numerous collider and cosmic searches; for a recent review, see Ref.~\cite{Mavromatos:2020gwk}.

In contrast to the sphaleron, the monopole is topologically stable, as explained in Sec.~\ref{sec:homotopy_and_TFCs}. In the presence of a monopole background, one can have the scattering process\footnote{To simplify the presentation, we suppress the chirality labels and any possible flavor quantum numbers of the quarks and leptons.},
\begin{equation}
  \label{eq:monopole_induced_bnv}
    u+u\rightarrow \bar{d}+e^+ \ ,
\end{equation}
where $u$, $\bar{d}$, and $e^+$ represent the up quark, anti-down quark, and positron, respectively. The process violates baryon plus lepton number by $\Delta (B+L)=-2$. According to the analyses of Rubakov and Callan \cite{Rubakov:1981rg,Rubakov:1982fp,Callan:1982ac}, this $B+L$-violating scattering is generated by time-dependent fluctuations about the static monopole solution. We write these fluctuations as $a_\mu(r,t)$, namely
\begin{equation}
\begin{aligned}
A_0&=T^a \hat{x}^a a_0(r,t)/i \ ,\\
A_i&=T^a \hat{x}^a  \hat{x}_i a_1(r,t)/i+A_i^{\text{cl}} \ ,\\
\Phi &= \Phi^{\text{cl}} \ ,
\end{aligned}
\end{equation}
where $T^a$ are the SU(2) generators, $\hat{\vec{x}}=\vec{x}/r$, and $A_i^{\text{cl}}$ and $\Phi^{\text{cl}}$ denote the classical monopole background fields, using the conventions of Ref.~\cite{Rubakov:1982fp}. The fluctuation fields $a_0(r,t)$ and $a_1(r,t)$ obey
\begin{align}
    a_0(r,\pm \infty)=a_1(r,\pm \infty)=0 \ .
\end{align}
An important point is that some fluctuation configurations satisfying these boundary conditions (1) carry an integer winding number and (2) render the action functional arbitrarily small. The latter property implies that the fermion-number-violating matrix element is not suppressed by a factor of $\exp(-\text{const}/\text{coupling}^2)$ \cite{Rubakov:1982fp}. 

The relevant monopole-induced BNV process comes from $s$-wave fermion--monopole scattering, and we denote its cross section by $\sigma_{\Delta B\neq 0}$ \footnote{The BNV process in Eq.~\eqref{eq:monopole_induced_bnv} is tied to the ``selection rule'' associated with fermion zero modes in the monopole background. The same underlying idea appears in instanton-induced BNV, where the fermion zero modes in the instanton background are likewise essential, as discussed in Appendix.~\ref{sec:instanton_baryon_violation}. A very clear explanation of this phenomenon is given in Ref.~\cite{Rubakov:2002fi}.}. 
Before seeing the explicit result, one might expect this cross section to depend on the monopole size, on the scattering kinematics ({\it e.g.}, the center-of-mass energy), and perhaps on a non-perturbative exponential factor involving the gauge coupling, $g$. However, Refs.~\cite{Rubakov:1981rg,Rubakov:1982fp,Callan:1982ac} show that $\sigma_{\Delta B\neq 0}$ is suppressed neither by the geometric size of the monopole nor by a factor of the form $\exp(-f(v)/g^2)$, where $f(v)$ is some function of the multiplet vev, $v$. If the only remaining relevant scale is the ``fermion energy,'' $E_f$, then one may write \cite{Ellis:1982bz}
\begin{align}
\sigma_{\Delta B\neq 0} \simeq v^{-1}\frac{c}{E_f^2} \ ,
\end{align}
where $v$ is the relative velocity between the monopole and the fermion, and $c$ is an appropriate constant. In the monopole case, $\sigma_{\Delta B\neq 0}$ plays a role analogous to that of $\Gamma_\text{sph}$ in the sphaleron case through the right-hand side of Eq.~\eqref{eq:BNV_rate_sphaleron_final}, although the physical origin of the two quantities is different. 

To formulate a Boltzmann-equation description of monopole-catalyzed BNV, in analogy with Eq.~\eqref{eq:BNV_rate_sphaleron_final}, one needs not only the cross section but also the monopole and fermion number densities in the early-Universe plasma. For the monopole density, monopoles may be generated during EWSB either by bubble collisions \cite{Preskill:1984gd}, if the transition is first order, or by thermal monopole--antimonopole pair production \cite{Patel:2012pi}. In what follows, we assume a first-order EWPT. Over a broad range of parameters, the thermal monopole--antimonopole pair production rate is estimated to be larger than the bubble-collision production rate \cite{Patel:2012pi}. The corresponding equilibrium monopole density is
\begin{equation} \label{eq:monopole_thermal_pair_production_thermal_equilibrium}
	\frac{n_M}{T^3}\bigg\vert_{\text{eq.}}=\left(\frac{m_M(T)/T_{\text{nuc}}}{2\pi} \right)^{3/2}e^{-m_M(T)/T} \ ,
\end{equation}
where $n_M$ is the monopole density, $m_M(T)$ is the monopole mass, and $T_{\text{nuc}}$ is the nucleation temperature. 

The resulting Boltzmann equation for monopole-catalyzed BNV takes the form\footnote{A detailed derivation of this equation is given in the appendix of Ref.~\cite{Wu:2023mjb}, so we do not reproduce it here.}
\begin{align} \label{eq:monopole_boltzmann_eq}
    \frac{\partial \bar{n}_B}{\partial t}+3H\bar{n}_B = -\langle\sigma_{\Delta B\neq 0} v\rangle g_f \bar{n}_f n_M \ ,
\end{align}
where $\langle \ldots \rangle$ denotes the thermal average of the enclosed quantity, $g_f$ is the number of fermion degrees of freedom, $n_f$ and $n_{\bar{f}}$ are the fermion and antifermion number densities, $\bar{n}_f\equiv n_f - n_{\bar{f}}$ is the net fermion number density, and $n_M$ is the monopole density. In this study, we take the monopole density to be in thermal equilibrium, so that it is given by Eq.~\eqref{eq:monopole_thermal_pair_production_thermal_equilibrium}, which contains the exponential suppression factor $e^{-m_M/T}$.

To summarize, in the leading-order computation of the BNV rate, we focus only on the exponential factor $e^{-E/T}$, where $E$ denotes either the sphaleron energy or the monopole mass. It is important to emphasize that the origin of this factor is different in the two cases. For the sphaleron, $e^{-E/T}$ is associated with the transition rate between neighboring vacua, whereas for the monopole, it comes from the equilibrium monopole density.

\newpage 
\section{Electroweak Baryogenesis: Roles of Thermal Effective Field Theory}
\label{sec:EWBG_thermal_effective_potential}
In this section, we discuss the role of thermal field theory, especially the thermal effective potential, in electroweak baryogenesis (EWBG). The material in this section is important not only for the properties of the electroweak phase transition (EWPT), but also for the computation of the sphaleron rate in the next section. As mentioned in the Introduction, EWBG is an attractive mechanism for generating the observed baryon asymmetry of the Universe (BAU), because it connects BAU generation to physics at the electroweak scale, which is testable at current and near-future collider experiments. In this section, we first give a brief introduction to EWBG, then discuss thermal field theory with the technique of dimensional reduction. We also comment on the gauge-dependence issue of the thermal effective potential.

\subsection{Four Processes in Electroweak Baryogenesis}
\label{sec:EWBG_four_processes}
The pure SM cannot realize successful EWBG, because it fails to satisfy two of the three Sakharov conditions \cite{Sakharov:1967dj}: (i) the amount of CP violation from the Cabibbo--Kobayashi--Maskawa (CKM) matrix is too small to generate sufficient baryon asymmetry \cite{Gavela:1993ts,Gavela:1994dt,Huet:1994jb}; (ii) the electroweak phase transition (EWPT) in the SM is a smooth crossover \cite{Kajantie:1995kf,Kajantie:1996mn,Kajantie:1996qd,Csikor:1998ge,Aoki:1999fi}, rather than a strong first-order phase transition (SFOEWPT), which is required to drive the system out of equilibrium and to suppress the sphaleron process inside the broken-phase bubbles. Therefore, successful EWBG requires physics BSM to provide additional sources of CP violation and to catalyze a SFOEWPT. Many review papers on electroweak baryogenesis can be found in the literature, e.g.~\cite{Cohen:1993nk,Rubakov:1996vz,Morrissey:2012db,Bodeker:2020ghk}.

A first-order electroweak phase transition triggered by BSM physics enables baryogenesis through the sequence of processes below \cite{Morrissey:2012db}; see Fig.~\ref{fig:EWBG_big_picture}. We label these processes as (i)--(iv), in accordance with the subfigures in Fig.~\ref{fig:EWBG_big_picture}. We will see that the thermal effective potential plays an important role in processes (i) and (iv).

\begin{figure}[ht!]
  \centering
  \includegraphics[width=0.7\textwidth]{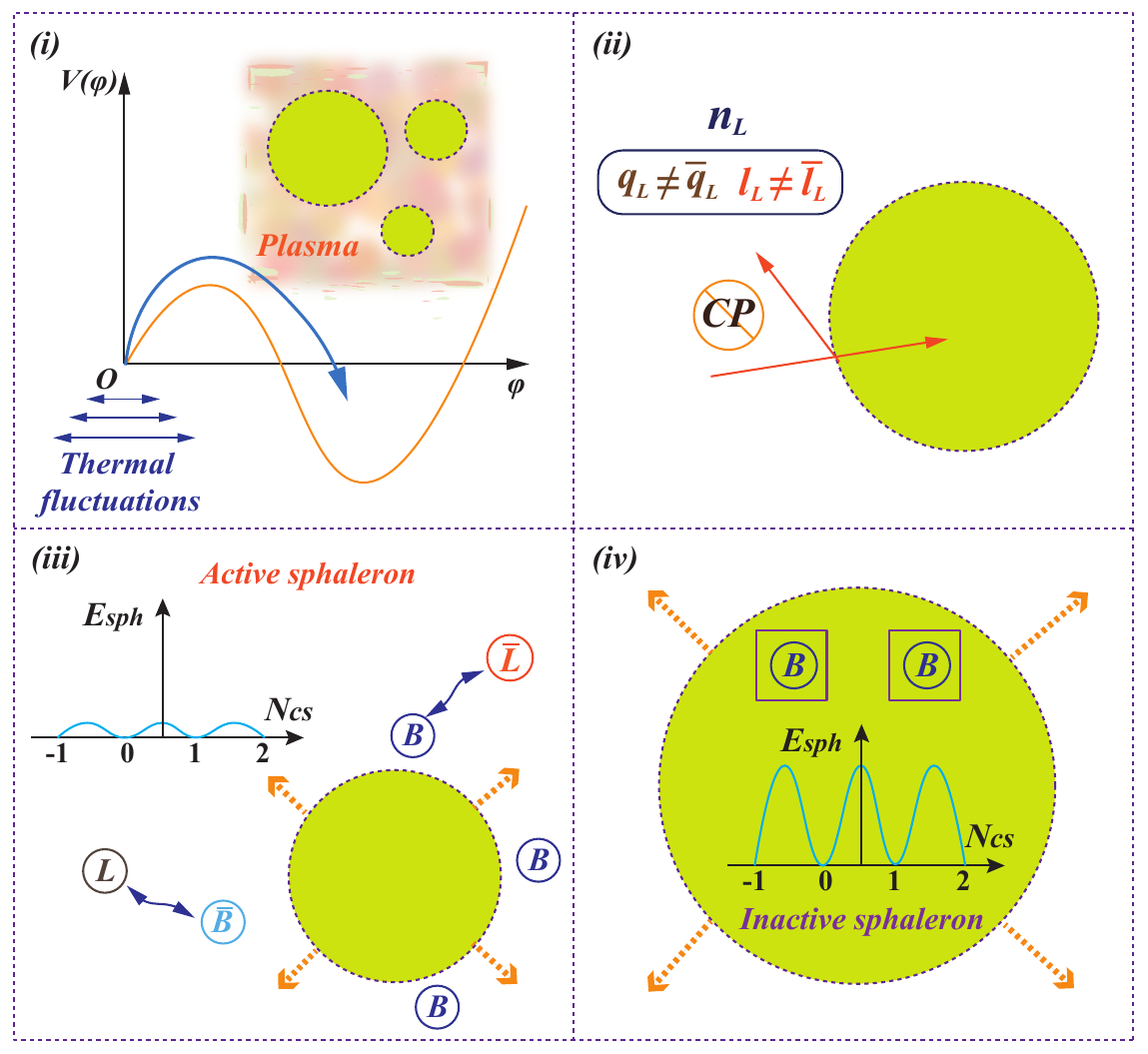}%
  \caption{
    Schematic stages of electroweak baryogenesis:
    (i) a first-order phase transition proceeds through bubble nucleation from the false vacuum to the true vacuum;
    (ii) CP-violating interactions with the expanding wall generate a nonzero left-handed fermion density;
    (iii) in the symmetric phase, rapid sphalerons convert the left-handed bias into a baryon excess that is swept into the bubbles;
    (iv) inside the bubbles, the sphaleron rate is suppressed, so the asymmetry is preserved, partially washed out, or erased depending on the washout exponent.\reproducedfromref{Li:2025kyo}
  }
  \label{fig:EWBG_big_picture}
\end{figure}

{\bf (i)} In the expanding Universe, once the temperature drops below the critical value, thermal fluctuations can overcome the barrier between the high-temperature symmetric phase and the low-temperature Higgs phase. This thermal transition process is analogous to the thermal transition discussed in the quantum-mechanics example of Fig.~\ref{fig:finite-temperature-tunneling}. From a perturbative perspective, such a phase transition is analyzed through the thermal effective potential. A first-order phase transition is characterized by the presence of a potential barrier between the two phases at the critical temperature \(T_c\), and by a sizable jump in the scalar-field expectation value across the bubble wall \cite{Dolan:1973qd,Arnold:1992rz,Dine:1992wr}. By contrast, a second-order phase transition or a crossover is characterized by the absence of such a potential barrier. However, perturbative computations of the thermal effective potential suffer from gauge-dependence issues \cite{Jackiw:1974cv,Patel:2011th} and infrared divergences \cite{Linde:1978px,Linde:1980ts}, which will be discussed in later subsections. The robust method for determining the order of the phase transition is non-perturbative lattice simulation \cite{Kajantie:1995kf,Kajantie:1996mn,Kajantie:1996qd,Csikor:1998ge,Aoki:1999fi}. Nevertheless, in the perturbative picture, the transition proceeds through bubble nucleation and growth, and is described on macroscopic scales by relativistic hydrodynamics. At microscopic scales, the released latent heat drives the plasma out of equilibrium. Additional sources of nonequilibrium behavior include wall interactions, turbulence, and shock fronts.

The critical bubble is described by a classical \textit{bounce} solution \(\varphi_\text{b}\) \cite{Coleman:1977py,Callan:1977pt}. The equilibrium nucleation rate is governed by a Boltzmann factor \(e^{-S_{\text{eff}}(\varphi_\text{b})}\), with quantum corrections encoded in the fluctuation determinant. This Boltzmann factor is analogous to that appearing in the finite-temperature sphaleron rate discussed in the previous section. 

{\bf (ii)} C- and CP-violating scatterings off the expanding wall produce unequal diffusion of particles and antiparticles, generating a nonzero left-handed density \(n_L\) in front of the wall. The left-handed density \(n_L\) is obtained by solving a set of transport equations that account for particle diffusion and CP-violating sources in the plasma outside the wall \cite{Huet:1995sh,Chung:2008aya,Chung:2009cb,Chung:2009qs}. Such a computation involves the closed-time-path (CTP) formalism of non-equilibrium quantum field theory, which we will not discuss in detail here. For more details, see the review \cite{Morrissey:2012db} and references therein. A compensating right-handed asymmetry \(n_R\) is also produced, but it does not bias baryogenesis because sphalerons act only on \(\mathrm{SU}(2)_L\) fermions. The generation time for \(n_L\) is typically shorter than the sphaleron time scale, so step (iii) can be treated as decoupled from step (ii).

{\bf (iii)} Outside the bubble, rapid sphaleron transitions violate \(B+L\) while conserving \(B-L\). As shown by lattice studies, this transition rate is not Boltzmann suppressed in the symmetric phase, and its expression was given previously in Eq.~\eqref{eq:sphaleron_rate_above_EWPT}. The left-handed asymmetry biases these transitions, producing a net baryon excess that is swept into the broken phase. The generated baryon asymmetry outside the wall is given by \cite{Huet:1995sh,Chung:2009qs}
\begin{align}
  n_B=-3 \frac{\Gamma_{\rm ws}}{v_w} \int_{-\infty}^{0} d z\, n_{L}(z) e^{\frac{15}{4} \frac{\Gamma_{\rm ws}}{v_w} z} \ ,
\end{align}
where \(\Gamma_{\rm ws}\) is the weak sphaleron rate in the symmetric phase, \(v_w\) is the wall velocity, and \(z\) is the coordinate normal to the wall in the wall frame.

{\bf (iv)} Inside the bubble, the sphaleron rate is Boltzmann suppressed because the sphaleron energy is large. One may naively expect that the baryon asymmetry generated outside the bubble is then fully preserved inside. However, the sphaleron rate may remain faster than the Hubble rate for some time after nucleation, leading to partial or complete washout of the asymmetry. A detailed treatment of this process is one of the main topics of this thesis, and it will be studied in detail in the next section. In terms of the baryon-number preservation condition (BNPC), three possibilities follow \cite{Li:2025kyo}:
\begin{itemize}
  \item[(a)] Immediately after nucleation, the baryon-number-violation rate falls below the Hubble rate, so washout is negligible (strong BNPC).
   
  \item[(b)] The violation rate exceeds the Hubble rate for a finite interval after the nucleation temperature \(T_n\), erasing part of the asymmetry (weak BNPC):
  \begin{align} \label{eq:washout_equation_sec2}
    n_B(T_*) = e^{-\mathcal{W}} n_B(T_n) \ ,
  \end{align}
  where \(T_*<T_n\) is the decoupling temperature at which the rates coincide, and \(\mathcal{W}\) is the washout exponent, see Sec.~\ref{sec:gauge_invariant_BPC}. In this case, CP-violating sources must overproduce the asymmetry.
  
  \item[(c)] The washout exponent is so large that essentially all baryon asymmetry is erased.
\end{itemize}
For the third possibility, successful EWBG cannot be achieved, whereas for the first two possibilities, successful EWBG remains possible for appropriate choices of model parameters. Although the second possibility requires a larger CP-violating source to overproduce the asymmetry, it is still viable.

\subsection{Thermal Effective Potential: Dimensional Reduction}
\label{sec:thermal_effective_potential_DR}
In this subsection, we give a brief introduction to thermal field theory and the technique of dimensional reduction (DR), which is widely used in computations of the thermal effective potential and in electroweak baryogenesis. For more details on thermal field theory and DR, see Refs.~\cite{Kajantie:1995dw,Ekstedt:2022bff,Laine:2016hma} and references therein \footnote{For recent and related developments in dimensional reduction and thermal field theory, see Refs.~\cite{Chapman:1994vk,Brauner:2016fla,Andersen:2017ika,Niemi:2018asa,Gorda:2018hvi,Kainulainen:2019kyp,Gould:2019qek,Niemi:2020hto,Gould:2021dzl,Schicho:2021gca,Niemi:2021qvp,Camargo-Molina:2021zgz,Gould:2021oba,Gould:2021ccf,Lofgren:2021ogg,Schicho:2022wty,Ekstedt:2022ceo,Gould:2022ran,Ekstedt:2022zro,Ekstedt:2022bff,Biondini:2022ggt,Niemi:2022bjg,Aarts:2023vsf,Gould:2023ovu,Gould:2023jbz,Kierkla:2023von,Gould:2024chm,Niemi:2024axp,Ekstedt:2024etx,Chala:2024xll,Qin:2024idc,Niemi:2024vzw,Camargo-Molina:2024sde,Gould:2024jjt,Chakrabortty:2024wto,Hallfors:2025key,Kierkla:2025qyz,Bernardo:2025vkz,Chala:2025aiz,Chala:2025oul,Li:2025kyo,Annala:2025aci,Bhatnagar:2025jhh,Chala:2025xlk,Chala:2025cya,Keus:2025ova,Biekotter:2025npc,Jahedi:2025yjz,Liu:2025ipj,Liu:2026ask,Fuentes-Martin:2026bhr,Gil:2026cqz}.}.  We adopt the convention \(\hbar=1\) and \(k_B=1\) in this subsection.

The difference between thermal field theory and zero-temperature field theory is that, in thermal field theory, the Euclidean time dimension is compactified on a circle with circumference \(\beta=1/T\), where \(T\) is the temperature of the system. Thus, when we perform the Fourier transform of a field \(\phi(\tau,\vec{x})\) in thermal field theory, the time component of the four-momentum becomes discrete:
\begin{align}
  \phi(\tau,\vec{x}) &= \frac{1}{\beta} \sum_{n=-\infty}^{\infty} \int \frac{d^3 k}{(2\pi)^3} e^{i(\omega_n \tau + \vec{k}\cdot \vec{x})} \phi(\omega_n, \vec{k}) \ , \\
  \omega_n &= \begin{cases}
    2 n \pi T & \text{for bosons} \ , \\
    (2 n + 1) \pi T & \text{for fermions} \ ,
  \end{cases} \quad n\in \mathbb{Z} \ .
\end{align}
The discrete frequencies \(\omega_n\) are called Matsubara frequencies. The difference between bosonic and fermionic Matsubara frequencies arises from the different commutation relations of bosonic and fermionic fields. These Matsubara frequencies are derived in detail in Appendix.~\ref{app:matsubara_frequencies} from the KMS relations of the Green's functions.

For a scalar field with zero-temperature mass \(m_i^2\), the one-loop thermal correction to the mass term is given by
\begin{align}
  m_i^2(T) &= \gamma_i T^2 + m_i^2 \ ,
\end{align}
where \(\gamma_i\sim g^2\) is a numerical factor. Different field components have different energy scales at finite temperature. For the spatial components of gauge fields, \(A_i\), we have \(\gamma_i=m_i^2=0\). For the temporal gauge components, \(A_0\), we have \(\gamma_i\sim g^2\) and \(m_i^2=0\). For scalar fields, \(\phi\), we have \(\gamma_i\sim g^2\) and \(m_i^2\neq 0\). Two scenarios may occur for scalar fields \cite{Kajantie:1995dw,Lofgren:2021ogg}: (i) if the zero-temperature mass \(m_i^2\) differs from \(-\gamma_i T^2\), the scalar field has mass of order \(gT\); (ii) if the zero-temperature mass \(m_i^2\) is close to \(-\gamma_i T^2\), the scalar mass squared is smaller than \(g^2T^2\), which we may parameterize as \(g^3T^2\) or \(g^4T^2\). In the following discussion, we assume that the scalar field is in the second scenario and has mass squared of order \(g^4T^2\). From this analysis, we see that there are three energy scales for different fields: (i) \(\pi T\) for the non-zero Matsubara modes; this scale includes all fermionic modes, and we call it the ``superheavy'' scale; (ii) \(gT\) for the temporal gauge fields \(A_0\), which have mass of order \(gT\); we call it the ``heavy'' scale; (iii) \(g^2T\) for the scalar fields; we call it the ``light'' scale. Note that the spatial gauge fields \(A_i\) acquire a magnetic mass of order \(g^2T\) from non-perturbative effects \cite{Linde:1978px} even before symmetry breaking.

The above scale analysis is important for thermal gauge field theory, because there exists the so-called infrared (IR) problem in thermal gauge field theory \cite{Linde:1980ts}. Consider the Green's function of the gauge fields, \(G_{ij}^{ab}=\delta^{ab}(\delta_{ij}-k_ik_j/k^2) G(k)\), where \(a,b\) are the gauge-group indices and \(i,j\) are the spatial indices. The Linde IR problem states that at small momentum \(k\ll g^2T\), the Green's function \(G(k)\) has the following expansion \cite{Linde:1980ts}:
\begin{align}
  G(k)^{-1} \sim \sum_n a_n k^2 \left(\frac{g^2 T}{k}\right)^n \ ,
\end{align}
where \(a_n\) are numerical coefficients of order one. From this expression, we see that when \(k\sim g^2T\), the perturbative expansion becomes divergent, which signals the breakdown of perturbation theory at this scale. However, when \(k\gg g^2T\), the series is convergent and perturbation theory is valid. The idea of dimensional reduction (DR) is to integrate out the ``superheavy'' and ``heavy'' scales, so that we obtain a three-dimensional (3D) effective field theory (EFT) for the ``light''-scale physics, where perturbation theory remains.

\begin{table}[h!]
\centering
\caption{Dimensional reduction for the SM \(SU(2)\)+Higgs theory. The \(U(1)\) gauge field is omitted because its contribution is beyond the order considered here. \(A_\mu^a\) denotes the \(SU(2)\) gauge field, \(\Phi\) the Higgs doublet, and \(\psi\) the fermion fields (with only the top quark retained). \(\mu^2\) and \(\lambda\) are the Higgs mass parameter and quartic coupling, while \(g\) and \(g_Y\) are the \(SU(2)\) gauge and top Yukawa couplings, respectively. Parameters with subscript 3 are those of the 3D EFT after integrating out the superheavy scale, whereas parameters with a bar are those of the 3D EFT after integrating out both the superheavy and heavy scales.}
\label{tab:dr_hierarchy}
\renewcommand{\arraystretch}{1.2}
\begin{tabular}{c c c c c}
\hline
\textbf{Scale}  & \textbf{Dimension} & \textbf{Lagrangian} & \textbf{Fields} & \textbf{Parameters} \\
\hline
\textit{Superheavy} \((\pi T)\) & \(3+1\) & \(\mathcal{L}_{4\mathrm{d}}\) \eqref{eq:4d_lagrangian} & \(A_\mu^a,\Phi,\psi\) & \(\mu^2,\lambda,g,g_Y\) \\
\multicolumn{5}{c}{\(\downarrow\) \textit{Step 1: Integrate out \(n\neq 0\) Matsubara modes}} \\
\textit{Heavy} \((gT)\)  & \(3\) & \(\mathcal{L}_{3\mathrm{d}}\) \eqref{eq:3d_lagrangian} & \(A_i^a,A_0^a,\Phi_3\) & 
\(\mu_3^2,\lambda_3,g_3,m_D,h_3,\lambda_A\) \\
\multicolumn{5}{c}{\(\downarrow\) \textit{Step 2: Integrate out temporal scalars \(A_0^a\)}} \\
\textit{Light} \((g^2T)\) & \(3\) & \(\bar{\mathcal{L}}_{3\mathrm{d}}\) \eqref{eq:3d_lagrangian_light} & \(\bar{A}_{3,i},\bar{\Phi}_3\) & \(\bar{\mu}_3^2,\bar{\lambda}_3,\bar{g}_3\) \\
\hline
\end{tabular}
\end{table}

In this thesis, we discuss the DR technique in the context of the SM \(SU(2)\)+Higgs theory, which is the setup used later to study the sphaleron rate. The DR procedure for the SM \(SU(2)\)+Higgs theory is shown in Table.~\ref{tab:dr_hierarchy}. The first step of DR is to integrate out the non-zero Matsubara modes, which gives a 3D EFT containing all zero modes. The second step of DR is to integrate out the temporal gauge fields \(A_0^a\), which gives a 3D EFT containing only spatial gauge fields and scalar fields. First, in the parent 4D theory, the Lagrangian is given by
\begin{align} \label{eq:4d_lagrangian}
\mathcal{L}_{4\mathrm{d}} &= \frac{1}{4} F_{\mu\nu}^a F_{\mu\nu}^a + (D_\mu \Phi)^\dagger (D_\mu \Phi) - \mu^2 \Phi^\dagger \Phi + \lambda (\Phi^\dagger \Phi)^2 + \mathcal{L}_{\rm fermion}(\psi) + \mathcal{L}_{\rm Yukawa}(g_Y,\psi) \ ,
\end{align}
where \(F_{\mu\nu}^a\) is the field-strength tensor of the \(SU(2)\) gauge field, \(D_\mu\) is the covariant derivative, \(\Phi\) is the Higgs doublet, \(\psi\) represents fermion fields (with only the top quark retained), \(\mu^2\) and \(\lambda\) are the mass term and quartic coupling of the Higgs field, and \(g\) and \(g_Y\) are the \(SU(2)\) gauge and top Yukawa couplings, respectively.

In the following, we focus on the first step of DR, namely integrating out the non-zero Matsubara modes. In particular, we discuss how \(\lambda_3\) is matched to the parameters in the parent 4D theory. We will see in the next section that \(\lambda_3\) is an important parameter in the computation of the sphaleron energy and sphaleron rate. We leave the remaining details of DR to the following subsections, the appendix, and the references therein. The logic of DR is as follows: (i) one first writes down the most general Lagrangian at the Heavy or Light scale, consistent with 3D gauge symmetry and the other relevant symmetries; (ii) one then performs matching between the parent 4D theory and the 3D EFT by computing the same Green's functions in both theories and requiring them to agree at low momentum. At the Heavy scale, the 3D theory need to reproduce the Heavy- and Light-scale Green's functions of the parent 4D theory for $k \lesssim gT$ up to order $\mathcal{O}(g^4)$ \cite{Kajantie:1995dw}. At the Light scale, the 3D theory need to reproduce the Light-scale Green's functions of the parent 4D theory for $k \lesssim g^2T$ up to order $\mathcal{O}(g^4)$. For example, for the \(\lambda_3\) matching, one may compute the four-point Green's function of the scalar field in both theories and require them to coincide for momenta \(k\lesssim gT\).

The Heavy-scale Lagrangian after integrating out the non-zero Matsubara modes is given by
\begin{align} \label{eq:3d_lagrangian}
\mathcal{L}_{3\mathrm{d}} & = \frac{1}{4} F_{ij}^a F_{ij}^a  + (D_i \Phi_3)^\dagger (D_i \Phi_3) + \mu_3^2 \Phi_3^\dagger \Phi_3 + \lambda_3 (\Phi_3^\dagger \Phi_3)^2 \nonumber \\
&\quad + \frac{1}{2} (D_i A_0)^a (D_i A_0)^a + \frac{1}{2}m_D^2 A_0^a A_0^a + h_3 \Phi_3^\dagger \Phi_3 A_0^a A_0^a + \frac{1}{4}\lambda_A (A_0^a A_0^a)^2 \ ,
\end{align}
where \(F_{ij}^a\) is the field-strength tensor of the spatial gauge fields, \(\mu_3^2\) and \(\lambda_3\) are the mass term and quartic coupling of the scalar field in the 3D EFT, \(m_D\) is the Debye mass for the temporal gauge fields, \(h_3\) is the coupling between the scalar field and temporal gauge fields, and \(\lambda_A\) is the quartic coupling of the temporal gauge fields. The above Lagrangian maintains the 3D gauge symmetry. The temporal gauge fields \(A_0^a\) can be further integrated out to obtain the Light-scale Lagrangian, which is given by
\begin{align} \label{eq:3d_lagrangian_light}
\bar{\mathcal{L}}_{3\mathrm{d}} & = \frac{1}{4} \bar{F}_{ij}^a \bar{F}_{ij}^a  + (D_i \bar{\Phi}_3)^\dagger (D_i \bar{\Phi}_3) + \bar{\mu}_3^2 \bar{\Phi}_3^\dagger \bar{\Phi}_3 + \bar{\lambda}_3 (\bar{\Phi}_3^\dagger \bar{\Phi}_3)^2 \ ,
\end{align}
A detailed discussion of the second step of DR is beyond the scope of this thesis, and we refer to Ref.~\cite{Kajantie:1995dw} and recent papers \cite{Niemi:2018asa,Schicho:2021gca} for more details.

Below we discuss two aspects of the first step of DR: (i) matching of the two-point and four-point Green's functions of the scalar and gauge fields, which gives the relation between \(\Phi_3\) and \(\Phi\), and determines the other 3D gauge couplings \((g_3,m_D^2,\lambda_A)\) and the portal coupling \(h_3\); (ii) matching of the two- and four-point Green's functions of the scalar field through the effective-potential approach, which determines the mass term \(\mu_3^2\) and quartic coupling \(\lambda_3\) of the scalar field in the 3D EFT.

\subsubsection{matching of two-point Green's functions}

The two-point Green's function of the scalar field in the parent 4D theory is given by \cite{Kajantie:1995dw}
\begin{align}
  G_{2,4\mathrm{d}}(k)^{-1}
  & =  k^2 - \mu^2 + \Pi(k^2) \nonumber \\
  & = k^2 - \mu^2 + \Pi_3(k^2) + \bar{\Pi}(k^2)\ ,
\end{align}
where \(\Pi(k^2)\) is the scalar self-energy. It can be decomposed into two parts: \(\Pi_3(k^2)\), which contains the contributions from the Heavy and Light modes only, and \(\bar{\Pi}(k^2)\), which contains the contribution from the non-zero Matsubara modes. The two-point Green's function of the scalar field at the Heavy scale is
\begin{align}
  G_{2,3\mathrm{d}}(k)^{-1}
  & = k^2 + \mu_3^2 + \Pi_3(k^2)\ .
\end{align}
Note that there is no IR problem in \(\bar{\Pi}(k^2)\), so it can be expanded analytically in powers of \(k^2\) as \cite{Kajantie:1995dw}
\begin{align}
  \bar{\Pi}(k^2) = \bar{\Pi}(0) + k^2 \bar{\Pi}'(0) + \mathcal{O}\!\left(g^2 k^4/T^2 \right)\ .
\end{align}
Assuming \(\bar{\Pi}(0)\sim T^2(g^2+g^4)\) with two-loop accuracy, and \(\bar{\Pi}'(0)\sim g^2\) with one-loop accuracy, the 4D Green's function can be written as
\begin{align}
  G_{2,4\mathrm{d}}(k)^{-1}
  & = [1+\bar{\Pi}'(0)] \Bigl(k^2 + [\mu^2+\bar{\Pi}(0)][1-\bar{\Pi}'(0)] + \Pi_3(k^2)\Bigr)\ .
\end{align}
From the matching condition \(G_{2,4\mathrm{d}}(k)^{-1} = G_{2,3\mathrm{d}}(k)^{-1}\) for \(k\lesssim gT\), we obtain the matching relations for the fields and the mass term,
\begin{align}
  \Phi_3^2 & = \frac{1}{T}[1+\bar{\Pi}'(0)] \Phi^2 \ , \\
  \mu_3^2 & = [\mu^2+\bar{\Pi}(0)][1-\bar{\Pi}'(0)] \ .
\end{align}
Here the factor of \(1/T\) in the field matching arises from dimensional reduction. Therefore, one needs to compute the 4D scalar and gauge-field self-energies from the non-zero Matsubara modes in order to determine the matching of the mass term and the field normalization. The details of this computation can be found in \cite{Kajantie:1995dw}, and we quote the results here. For the mass term, we will present its computation from the effective-potential approach in the next subsubsection, which is more intuitive and straightforward. The Heavy-scale fields are matched through
\begin{align}
  \label{eq:field_matching_scalar}
\Phi_3^{2}
&= \frac{1}{T}\Phi^{2}\left\{1+\frac{1}{16\pi^{2}}\left[-\frac{9}{4}g^{2}L_b+3g_Y^{2}L_f\right]\right\}\ ,\\
\left(A_{0}^{3d}\right)^{2}
&= \frac{1}{T}\left(A_{0}\right)^{2}\left\{1+\frac{g^{2}}{16\pi^{2}}\left[\frac{4n_F}{3}(L_f-1)+\frac{N_s}{6}(L_b+2)-\frac{13}{3}L_b+\frac{8}{3}\right]\right\}\ ,\\
\left(A_{i}^{3d}\right)^{2}
&= \frac{1}{T}\left(A_{i}\right)^{2}\left\{1+\frac{g^{2}}{16\pi^{2}}\left[\frac{4n_F}{3}L_f+\frac{N_s}{6}L_b-\frac{13}{3}L_b-\frac{2}{3}\right]\right\}\ ,
\end{align}
where \(N_s\) and \(n_F\) are the numbers of scalar and fermion fields, respectively, and \(L_b\) and \(L_f\) are the logarithmic terms defined by
\begin{align}
  \label{eq:log_terms_Lb_Lf}
L_b & = \log \frac{\bar{\mu}^2}{T^2}-2c_B \ , \\
L_f & = \log \frac{\bar{\mu}^2}{T^2} -2 c_F \ ,
\end{align}
where \(\bar{\mu}\) is the 4D renormalization scale in the \(\overline{\text{MS}}\) scheme, and \(c_B=\log 4\pi-\gamma_E\approx 1.953808\) and \(c_F=c_B-2\log 2\approx 0.567514\), with \(\gamma_E\) the Euler--Mascheroni constant. It can be shown that \(\Phi_3\), \(A_{0}^{3d}\), and \(A_{i}^{3d}\) are independent of the renormalization scale \(\bar{\mu}\). For example, for the scalar field, \(\bar{\Pi}'(0)\) can be computed from the one-loop diagrams shown in the first two panels of Fig.~\ref{fig:scalar_self_energy}, with the third panel representing the counterterm contribution. Note that only non-zero Matsubara modes contribute to the loop diagrams. We may therefore treat the gauge and fermion propagators in the loop as massless, since their thermal masses are of order \(gT \ll \pi T\) and can be neglected at the order of interest. However, this massless treatment is more subtle in the second step of dimensional reduction; see the recent discussion in Ref.~\cite{Bernardo:2025vkz} for more details. These diagrams give
\begin{align}
  \bar{\Pi}'(0) = \frac{1}{16\pi^2}\left(-\frac{9}{4}g^2 L_b + 3 g_Y^2 L_f\right)\ ,
\end{align}
and this result is used in Eq.~\eqref{eq:field_matching_scalar}.

\begin{figure}[t]
\centering
\begin{minipage}{0.32\linewidth}
\centering
\begin{tikzpicture}
  \begin{feynman}
    \vertex (i) at (-1.2,0);
    \vertex (a) at (-0.2,0);
    \vertex (b) at (0.8,0);
    \vertex (f) at (1.8,0);
    \vertex (t) at (0.3,0.7);
    \vertex (u) at (0.3,-0.7);
    \diagram*{
      (i) -- [scalar] (a),
      (b) -- [scalar] (f),
      (a) -- [boson, half left] (b),
      (b) -- [scalar, half left] (a),
    };
  \end{feynman}
\end{tikzpicture}
\end{minipage}\hfill
\begin{minipage}{0.32\linewidth}
\centering
\begin{tikzpicture}
  \begin{feynman}
    \vertex (i) at (-1.2,0);
    \vertex (a) at (-0.2,0);
    \vertex (b) at (0.8,0);
    \vertex (f) at (1.8,0);
    \diagram*{
      (i) -- [scalar] (a),
      (b) -- [scalar] (f),
      (a) -- [fermion, half left] (b),
      (b) -- [fermion, half left] (a),
    };
  \end{feynman}
\end{tikzpicture}
\end{minipage}\hfill
\begin{minipage}{0.32\linewidth}
\centering
\begin{tikzpicture}
  \begin{feynman}
    \vertex (i) at (-1.2,0);
    \vertex (a) at (0,0);
    \vertex (f) at (1.2,0);
    \diagram*{
      (i) -- [scalar] (a) -- [scalar] (f),
    };
    \node[fill=black, circle, inner sep=2pt] at (a) {};
  \end{feynman}
\end{tikzpicture}
\end{minipage}
\caption{One-loop diagrams for the two-point Green's function of the scalar field that contribute to scalar wave-function renormalization. The dashed, wavy, and solid lines denote the scalar, gauge, and fermion fields, respectively. The first two diagrams arise from the non-zero Matsubara modes, while the last diagram is the counterterm contribution.}
\label{fig:scalar_self_energy}
\end{figure}
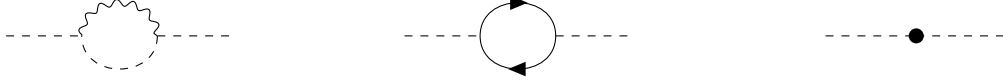

The couplings \(g_3^2\) and \(h_3\) can be obtained from the non-zero Matsubara-mode contributions to the \((\Phi\Phi A_i A_j)\) and \((\Phi\Phi A_0 A_0)\) four-point Green's functions, respectively. The coupling \(\lambda_A\) can be obtained from the non-zero Matsubara-mode contribution to the \((A_0 A_0 A_0 A_0)\) four-point Green's function. The Debye mass \(m_D^2\) can be obtained from the non-zero Matsubara-mode contribution to the two-point Green's function of the temporal gauge field, \((A_0 A_0)\). The details of these computations can be found in \cite{Kajantie:1995dw}, and we quote the results here:
\begin{align}
g_3^2
&= g^2(\bar{\mu})\,T\left[1+\frac{g^2}{16\pi^2}\left(\frac{44-N_s}{6}L_b-\frac{4n_F}{3}L_f+\frac{2}{3}\right)\right]\ ,\\
h_3
&= \frac{1}{4}g^2(\bar{\mu})\,T\left[1+\frac{g^2}{16\pi^2}\left(\frac{44-N_s}{6}L_b-\frac{4n_F}{3}L_f+\frac{53}{6}-\frac{N_s}{3}+\frac{4n_F}{3}+\frac{3}{2}h^2-3t^2\right)\right]\ , \\
m_D^2
&= \left(\frac{2}{3}+\frac{N_s}{6}+\frac{n_F}{3}\right) g^2 T^2 \ ,\\
\lambda_A
&= T\,\frac{g^4}{16\pi^2}\,\frac{16+N_s-4n_F}{3}\ ,
\end{align}
where the variables \(h\) and \(t\) are defined by \(h^2=8\lambda/g^2\) and \(t^2=2g_Y^2/g^2\).

\subsubsection{Green's functions from the effective potential}

The effective potential contains the information of the one-particle-irreducible Green's functions of the scalar field at zero external momentum:
\begin{align} \label{eq:effective_potential_and_green_function}
  V(\phi) = \sum_{n} \frac{1}{n!} \Gamma_{n}(0) \phi^n \ ,
\end{align}
where \(\phi\) is the background field of the scalar field. Thus, the matching for the mass term and quartic coupling of the scalar field in the 3D EFT can also be obtained by computing the effective potential. Note that this method for obtaining the matching coefficients is equivalent to the direct computation of Feynman diagrams discussed in the previous subsubsection. An explicit demonstration of this equivalence can be found in \cite{Schicho:2021gca}. Considering only the non-zero Matsubara-mode contribution to the effective potential, the coefficients of the \(\phi^2/2\) and \(\phi^4/4\) terms give the matching of the mass term and quartic coupling of the scalar field in the 3D EFT, respectively. The effective potential is gauge dependent. However, we will show in the next subsubsection that there is a gauge cancellation in the dimensional-reduction matching. In the present setup, it is convenient to perform the computation in Landau gauge, where the vector bosons are transverse. We will demonstrate the gauge cancellation in the dimensional-reduction matching, together with other gauge-dependence issues, in the next subsubsection.

The effective potential can be computed by standard techniques; see \cite{Kajantie:1995dw} and references therein. The one-loop effective potential in Landau gauge is
\begin{align}
  V_{\rm eff}(\phi)
  & = J_B(m_h^2) + 3 J_B(m_G^2) + 9 J_B(m_A^2) + 12 J_F(m_t^2) \ ,
\end{align}
where we derived this expression in Eq.~\eqref{eq:effective_potential_landau_gauge} of Appendix.~\ref{app:effective_potential_computation}. The functions \(J_B\) and \(J_F\) are the bosonic and fermionic one-loop integrals, defined in Eq.~\eqref{eq:J_B_integral} and Eq.~\eqref{eq:J_F_integral} of Appendix.~\ref{app:effective_potential_computation}. The field-dependent masses \(m_h^2\), \(m_G^2\), \(m_A^2\), and \(m_t^2\) are
\begin{align}
  m_h^2 & = -\mu^2 + 3 \lambda \phi^2 \ , \\
  m_G^2 & = -\mu^2 + \lambda \phi^2 \ , \\
  m_A^2 & = \frac{1}{4} g^2 \phi^2 \ , \\
  m_t^2 & = \frac{1}{2} g_Y^2 \phi^2 \ .
\end{align}
Although the effective potential above is written in Landau gauge, a more general expression in \(R_\xi\) gauge was derived in Eq.~\eqref{eq:effective_potential_R_xi_gauge}. We will discuss the gauge dependence of the effective potential and the gauge cancellation in dimensional-reduction matching in the next subsection.

Now we explain how to obtain the matching relation for the quartic coupling \(\lambda_3\) from the effective potential. The strategy is to equate
\begin{align} \label{eq:lambda3_matching_effective_potential}
    \frac{1}{4}\lambda_3 T \Phi_3^4
    = \frac{1}{4}\lambda \Phi^4 + {\rm coefficient\ of\ }\frac{\Phi^4}{4}{\rm\ in\ } V_{\rm eff}(\Phi)\ ,
\end{align}
where the factor of \(T\) on the left-hand side comes from dimensional reduction. Note that we already established the matching relation between \(\Phi_3\) and \(\Phi\) in Eq.~\eqref{eq:field_matching_scalar}, from which \(\Phi\) can be expressed in terms of \(\Phi_3\) as
\begin{align} \label{eq:phi_in_terms_of_phi3}
  \Phi^4
  & \approx T^2 \Phi_3^4 \left(1 - \frac{2}{16\pi^2}\left[-\frac{9}{4}g^2 L_b + 3 g_Y^2 L_f \right]  \right)\ ,
\end{align}
where we expanded \(\frac{1}{1+x}\sim 1-x+\cdots\) for small \(x\). Substituting Eq.~\eqref{eq:phi_in_terms_of_phi3} into Eq.~\eqref{eq:lambda3_matching_effective_potential}, we obtain the matching for \(\lambda_3\) as
\begin{align}
  \label{eq:lambda3_matching_effective_potential_final}
  \lambda_3
  & = T \lambda(\bar{\mu})\left\{ 1 - \frac{g^2}{16\pi^2} \left[ \left(\frac{9}{16}\frac{g^2}{\lambda} - \frac{9}{2} + 12 \frac{\lambda}{g^2} \right)L_b + \left(6 \frac{g_Y^2}{g^2} - 3\frac{g_Y^4}{\lambda g^2}\right)L_f-\frac{3}{8}\frac{g^2}{\lambda} \right] \right\}\ .
\end{align}
The three-dimensional mass term \(\mu_3^2\) can be obtained in a similar way, except that one must include the two-loop contribution to the effective potential in order to reach \(\mathcal{O}(g^4)\) accuracy for \(\mu_3^2\). The details can be found in \cite{Kajantie:1995dw}.

\subsection{Gauge Dependence of the Effective Potential}
\label{sec:gauge_dependence_effective_potential}

In Table.~\ref{tab:dr_hierarchy}, we showed the two steps of dimensional reduction for the SM \(SU(2)\)+Higgs theory. We explicitly presented the matching for the scalar quartic coupling in the first step, obtained from the effective potential. After these two steps, we arrive at the Light-scale 3D EFT given by Eq.~\eqref{eq:3d_lagrangian_light}. However, the scalar potential still does not contain the cubic barrier term required for a first-order phase transition. Therefore, after the two steps of dimensional reduction, one usually integrates out the Light-scale gauge fields as well, in order to obtain a 3D EFT containing only scalar fields, for which a cubic barrier term appears in the scalar potential. This step corresponds to the computation of the effective potential within the 3D EFT, and the perturbative result can be checked against 3D lattice simulations. As we will see in the next section, such a barrier term is crucial for the sphaleron rate in a first-order phase transition. We therefore postpone the necessary scale analysis, including the sphaleron scale, to the next section, where it will justify all of these dimensional-reduction and integration steps. In the present subsection, we restrict our discussion to the gauge dependence of the effective potential and the gauge cancellation in dimensional-reduction matching.

From the discussion above, gauge dependence may enter in two ways:
\begin{itemize}
  \item Fermi-gauge dependence in the two steps of dimensional reduction.
  \item \(R_\xi\)-gauge dependence in the effective-potential computation within the 3D EFT after the two steps of dimensional reduction.
\end{itemize}
Note that we use two different gauge-fixing conditions for these two cases. For the gauge-fixing Lagrangian \(F^2/\xi\), the Fermi and \(R_\xi\) gauge-fixing conditions are
\begin{align}
F_{\rm Fermi}^a & = \partial_\mu A_\mu^a \ , \\
F_{R_\xi}^a & = \partial_\mu A_\mu^a -  \xi g \phi \Phi_2 \ ,
\end{align}
where \(\Phi_2\) is the Goldstone field, \(\phi\) is the scalar background field, and \(\xi\) is the gauge-fixing parameter. The \(R_\xi\) gauge is discussed in more detail in Appendix.~\ref{sec:gauge_field_effective_potential_R_xi_gauge}. The Fermi gauge corresponds to the zero-background-field limit of the \(R_\xi\) gauge. The advantage of the \(R_\xi\) gauge is that the kinetic mixing between gauge fields and Goldstone bosons is cancelled, which makes the effective-potential computation more straightforward. Therefore, the \(R_\xi\) gauge is usually used for the effective-potential computation within the 3D EFT after the two steps of dimensional reduction \footnote{See also the discussions in Refs.~\cite{Laine:1994bf,Arnold:1992fb}, where it is emphasized that the use of the \(R_\xi\) gauge is not \textit{a priori} guaranteed to be valid in this context.}. By contrast, dimensional-reduction matching is usually performed in Fermi gauge (also called general covariant gauge). Since the matching is carried out at zero background field and the Fermi gauge is therefore more convenient \cite{Croon:2020cgk}. We emphasize that the dimensional-reduction matching and the effective-potential computation within the 3D EFT after the two steps of dimensional reduction are two separate procedures, so there is no conflict in using different gauge-fixing conditions for them.

We will show that, if the power counting is performed consistently, the gauge dependence in the dimensional-reduction matching cancels up to \(\mathcal{O}(g^4)\), and the final 3D EFT after the two steps of dimensional reduction is gauge invariant up to \(\mathcal{O}(g^4)\). We first discuss the gauge cancellation in dimensional-reduction matching, focusing again on the first step and using the matching of the quartic coupling \(\lambda_3\) as an example. We then discuss the gauge dependence in the effective-potential computation within the 3D EFT after the two steps of dimensional reduction, which is relevant for first-order phase-transition calculations.

\subsubsection{gauge cancellation in dimensional reduction}
\label{sec:gauge_cancellation_dimensional_reduction}

In Eq.~\eqref{eq:lambda3_matching_effective_potential_final}, we presented the matching of the quartic coupling \(\lambda_3\) after integrating out the non-zero Matsubara modes, obtained by computing the effective potential in Landau gauge. We now derive the same \(\lambda_3\) by direct computation of correlators in a general Fermi gauge. Starting from Eq.~\eqref{eq:lambda3_matching_effective_potential}, we rewrite the matching relation for \(\lambda_3\) as \cite{Croon:2020cgk}
\begin{align}
  \lambda_3 T \Phi_3^4
  & = \left(\lambda - \frac{1}{2}\hat{\Gamma}_{(\Phi^\dagger \Phi)^2}\right) \Phi^4 \nonumber \\
  & \approx \left(\lambda - \frac{1}{2}\hat{\Gamma}_{(\Phi^\dagger \Phi)^2}\right) T^2 \Phi_3^4 \left(1 - 2 \hat{\Pi}^\prime_{\Phi^\dagger \Phi} \right) \nonumber \\
  & \approx \left(\lambda - \frac{1}{2}\hat{\Gamma}_{(\Phi^\dagger \Phi)^2} - 2 \lambda \hat{\Pi}^\prime_{\Phi^\dagger \Phi} \right) T^2 \Phi_3^4\ ,
\end{align}
where \(\hat{\Gamma}_{(\Phi^\dagger \Phi)^2}\) is the contribution from the non-zero Matsubara modes to the four-point Green's function of the scalar field at zero external momentum. In the second line, we used Eq.~\eqref{eq:phi_in_terms_of_phi3} to express \(\Phi\) in terms of \(\Phi_3\), and we replaced the previous Landau-gauge result \(\bar{\Pi}'(0)\) by the more general Fermi-gauge quantity \(\hat{\Pi}^\prime_{\Phi^\dagger \Phi}\). Therefore, the matching for \(\lambda_3\) becomes
\begin{align}
  \lambda_3
  & = T \left(\lambda - \frac{1}{2}\hat{\Gamma}_{(\Phi^\dagger \Phi)^2} - 2 \lambda \hat{\Pi}^\prime_{\Phi^\dagger \Phi} \right)\ .
\end{align}

The gauge dependence of \(\hat{\Gamma}_{(\Phi^\dagger \Phi)^2}\) and \(\hat{\Pi}^\prime_{\Phi^\dagger \Phi}\) can be obtained by computing the relevant Feynman diagrams in a general Fermi gauge. The details can be found in \cite{Croon:2020cgk}; here we quote the results up to \(\mathcal{O}(g^4)\)\footnote{Note that we correct an overall-sign error in Eq.~\eqref{eq:gamma_phi4_and_pi_prime_phi2_fermi_gauge} relative to the original result in \cite{Croon:2020cgk}.}:
\begin{align}
  \label{eq:gamma_phi4_and_pi_prime_phi2_fermi_gauge}
\hat{\Gamma}_{(\phi^\dagger\phi)^2}
&=\frac{1}{(4\pi)^2}\Bigg(
-\frac{3}{4}g^4
-6g_Y^4 L_f
+\Bigg[
\frac{9}{8}g^4
+24\lambda^2
-3\lambda g^2\xi
\Bigg]L_b
\Bigg)\ , \\
\hat{\Pi}'_{\phi^\dagger\phi}
&= \frac{1}{(4\pi)^2}
\left[
-\frac{3L_b}{4}(3-\xi)g^2 
+ 3L_f g_Y^2
\right]\ ,
\end{align}
where \(\xi\) is the gauge-fixing parameter in Fermi gauge, and \(L_b\) and \(L_f\) are the logarithmic terms defined in Eq.~\eqref{eq:log_terms_Lb_Lf}. We see that both \(\hat{\Gamma}_{(\Phi^\dagger \Phi)^2}\) and \(\hat{\Pi}^\prime_{\Phi^\dagger \Phi}\) are gauge dependent. However, this gauge dependence cancels in the final expression for \(\lambda_3\), which is identical to the previous Landau-gauge result in Eq.~\eqref{eq:lambda3_matching_effective_potential_final}. This demonstrates the gauge cancellation in dimensional-reduction matching. The same cancellation can be shown for the mass term \(\mu_3^2\) and for the other couplings in the 3D EFT. We therefore conclude that the final 3D EFT after the two steps of dimensional reduction is gauge invariant up to \(\mathcal{O}(g^4)\).

\subsubsection{gauge independence of the effective potential within the 3D EFT}
\label{sec:gauge_independence_effective_potential_3d_eft}

After the two steps of dimensional reduction, we obtain the 3D EFT in Eq.~\eqref{eq:3d_lagrangian_light}, which is gauge invariant up to \(\mathcal{O}(g^4)\). If we then integrate out the Light-scale gauge fields to obtain a 3D EFT containing only scalar fields, we must introduce a gauge-fixing condition in order to compute the effective potential within the 3D EFT. We now use the \(R_\xi\) gauge for this effective-potential computation, and show that the final effective potential is gauge independent up to \(\mathcal{O}(g^4)\), which is sufficient for first-order phase-transition analyses. In Eq.~\eqref{eq:effective_potential_R_xi_gauge} of Appendix.~\ref{sec:gauge_field_effective_potential_R_xi_gauge}, we derived the 4D effective potential in a general \(R_\xi\) gauge. We now adapt that analysis to the 3D EFT. Note, however, that the parameters of the 3D EFT differ from those of the 4D theory, and that the top-quark contribution enters through matching in the 3D EFT. Note that the gauge-fixing term is now added to the 3D Lagrangian in Eq.~\eqref{eq:3d_lagrangian_light}, rather than to the parent 4D theory. We denote the gauge-fixing parameter in the 3D EFT by \(\xi_3\), to distinguish it from the parameter \(\xi\) in the parent 4D theory. The effective potential in a general \(R_\xi\) gauge within the 3D EFT is
\begin{align}
  V_{\rm eff}(\phi)
  & = J_3(m_{h,3}^2) + 3 J_3(m_{G,3}^2) + 6 J_3(m_{A,3}^2) + 3 J_3(m_{c,3}^2) - 6 J_3(m_{c,3}^2)\nonumber \\
  & = J_3(m_{h,3}^2)  + 6 J_3(m_{A,3}^2) + \underbrace{3 J_3(m_{G,3}^2) - 3 J_3(m_{c,3}^2)}_{\rm beyond\ \mathcal{O}(g^4)} \ ,
\end{align}
where \(J_3\) is the three-dimensional one-loop integral defined in Eq.~\eqref{eq:J3_and_J4_integrals} of Appendix.~\ref{app:effective_potential_computation}. The numerical coefficients in front of the \(J_3\) functions count the corresponding degrees of freedom, as explained below Eq.~\eqref{eq:effective_potential_R_xi_gauge}. The masses of the Higgs boson, Goldstone bosons, gauge bosons, and ghost fields in the 3D EFT, denoted by \(m_{h,3}^2\), \(m_{G,3}^2\), \(m_{A,3}^2\), and \(m_{c,3}^2\), respectively, are
\begin{align}
  m_{h,3}^2 & = \bar{\mu}_3^2 + 3 \bar{\lambda}_3 \bar{\phi}_3^2 \ , \\
  m_{G,3}^2 & = \bar{\mu}_3^2 + \bar{\lambda}_3 \bar{\phi}_3^2 + m_{c,3}^2 \ , \\
  m_{A,3}^2 & = \frac{1}{4} \bar{g}_3^2 \bar{\phi}_3^2 \ , \\
  m_{c,3}^2 & = \frac{1}{4} \bar{g}_3^2 \xi_3 \bar{\phi}_3^2 \ , 
\end{align}
where all barred parameters are those of the 3D EFT after the two steps of dimensional reduction, and \(\bar{\phi}_3\) is the scalar background field in that EFT. We see that the gauge dependence enters only through the Goldstone and ghost sectors. Now note that
\begin{align}
  J_3(m_{G,3}^2) - J_3(m_{c,3}^2)
  &= J_3(\bar{\mu}_3^2 + \bar{\lambda}_3 \bar{\phi}_3^2 + m_{c,3}^2) - J_3(m_{c,3}^2) \nonumber \\
  &= \underbrace{(\bar{\mu}_3^2 + \bar{\lambda}_3 \bar{\phi}_3^2)\frac{d J_3(m_{c,3}^2)}{d m_{c,3}^2}}_{\mathcal{O}(g^5)}
  + \cdots \ .
\end{align}
If \(\bar{\phi}_3\) were the tree-level minimum, the mass \(\bar{\mu}_3^2 + \bar{\lambda}_3 \bar{\phi}_3^2\) would vanish, and the gauge dependence would cancel exactly. However, at one loop the minimum of the effective potential is shifted away from the tree-level minimum, so \(\bar{\mu}_3^2 + \bar{\lambda}_3 \bar{\phi}_3^2\neq 0\), leaving a residual gauge dependence in the effective potential. Nevertheless, this residual gauge dependence is of higher order than \(\mathcal{O}(g^4)\) \cite{Hirvonen:2021zej}, and can therefore be neglected in the present analysis. If we count \(\bar{\lambda}_3\sim g^4\), then tunneling requires \(\bar{\mu}_3^2 \sim g^4 \sigma^2\) \cite{Metaxas:1995ab}, where \(\sigma\) is a characteristic scale associated with the initial value of the tunneling solution. It then follows that \(m_{h,3}^2,(m_{G,3}^2- m_{c,3}^2) \sim g^4\sigma^2\), while \(m_{A,3}^2,m_{c,3}^2\sim g^2\sigma^2\). Using the expression for \(J_3\) in Eq.~\eqref{eq:J3_and_J4_integrals}, one finds \(d J_3(m_{c,3}^2)/d m_{c,3}^2 \sim g \sigma\). Therefore, the gauge dependence in the effective potential is of order \(\mathcal{O}(g^5)\), which is beyond the \(\mathcal{O}(g^4)\) accuracy relevant here. 

\newpage
\section{Electroweak Baryogenesis: Roles of Sphaleron}
\label{sec:electroweak_baryogenesis_sphaleron}
In this section, we discuss the perturbative computation of the sphaleron rate for either the SM crossover or a general first-order phase transition, as well as the application of this computation to the baryon-preservation condition in electroweak baryogenesis. We adopt the dimensional-reduction technique of thermal field theory, introduced in the previous section, to compute the sphaleron rate. Specifically, for first-order phase transitions triggered by BSM fields, we discuss the following two state-of-the-art techniques for computing the sphaleron rate:
\begin{itemize}
  \item A gauge-invariant computation of the sphaleron rate within the 3D EFT, based on the power-counting analysis of the sphaleron action and the sphaleron scale. Previous work on this topic has focused mainly on crossover phase transitions \cite{Carson:1990jm,Carson:1989rf,Baacke:1993aj,Baacke:1994ix,Burnier:2005hp} or on lattice studies \cite{Moore:1998swa,Moore:1999fs,DOnofrio:2014rug}. We extend these earlier studies to the first-order phase-transition case.
  \item An accurate computation of baryon washout due to sphaleron processes inside the broken phase. This computation is related to the so-called condition for a ``strongly first-order phase transition''. However, previous studies of baryon washout used approximate washout integrals or simply imposed the condition that the baryon-number-violation rate be smaller than the Hubble rate \cite{Patel:2011th,Ahriche:2014jna}. Here we compute the washout integrals accurately by solving the Boltzmann equation with the sphaleron rate obtained from the first principle.
\end{itemize}

\subsection{Sphaleron Power Counting in 3D EFT}
\label{sec:sphaleron_power_counting}

We decompose the sphaleron rate as
\begin{align}
  \Gamma_{\rm sph} = A_{\rm dyn}\times A_{\rm static} \ ,
\end{align}
where $A_{\rm dyn}$ and $A_{\rm static}$ denote the dynamical (non-equilibrium) and static (equilibrium) contributions to the sphaleron rate, respectively. We note that the dynamical part requires real-time formalism computations \cite{Bodeker:1998hm,Bodeker:1999ey}, which cannot be described solely by the 3D EFT and are therefore beyond the scope of this thesis. In the following analysis, we assume $A_{\rm dyn}\sim T$ on dimensional grounds, and focus on the static part $A_{\rm static}$, which can be computed in the high-temperature limit using the 3D EFT \cite{Arnold:1987mh,Carson:1989rf,Carson:1990jm,Baacke:1993aj,Baacke:1994ix}. This high-temperature limit means that the classical transition rate dominates over the quantum-tunneling rate, as discussed previously around Eq.~(\ref{eq:sphaleron_path_integral_finite_T}).

We note that this EFT corresponds to the Lagrangian in the last line of Table~\ref{tab:dr_hierarchy}, where only the spatial gauge fields and scalar fields are retained, while the temporal gauge fields and non-zero Matsubara modes are integrated out. To simplify the notation, in this section we denote these fields and couplings without bars. The 3D action can then be written as
\begin{equation}
\label{eq:S3d}
S_{\text{3D}} = \int \mathrm{d}^{3} x \biggl[ \frac{1}{4}F_{ij,3}^{a}F^{a}_{ij,3}+(D_{i}\phi)_{3}^{\dagger}(D_{i}\phi)_{3}
+\mu^{2}_{3}\phi^{\dagger}_{3}\phi_{3}+\lambda_{3} (\phi_{3}^{\dagger}\phi_{3})^{2} 
\biggr] \ .
\end{equation}
Here \(a\) labels the \(SU(2)\) adjoint index, and \(i,j\) are spatial indices. In three dimensions, the fields \(\phi_3\) and \(A_{i,3}^a\) have mass dimension \(T^{1/2}\), while \(\mu_3\), \(\lambda_3\), and \(g_3^2\) have dimension \(T\). The field-strength tensor is
\begin{align}
F_{ij,3}^{a} \;=\; \partial_{i} A_{j,3}^{a} - \partial_{j} A_{i,3}^{a} + g_3 f^{abc}A_{i,3}^{b}A_{j,3}^{c} \ ,
\end{align}
with \(f^{abc}\) the \(SU(2)\) structure constants. The scalar covariant derivative is
\begin{align}
(D_{i}\phi)_3 \;=\; \partial_{i} \phi_3 - i g_3 A_{i,3}^{a} T^{a}\phi_3 \ ,
\end{align}
where \(T^a=\sigma_a/2\) are the generators in the fundamental representation. For simplicity, we neglect the subleading effects of the \(U(1)\) sector in the 3D EFT \cite{Kajantie:1996qd}.

The sphaleron is a static, non-trivial solution of the equations of motion derived from the above 3D action. The properties of the sphaleron solution, such as the field configuration and sphaleron energy, were discussed in Sec.~\ref{sec:manton_sphaleron}.

The static part of the sphaleron rate, $A_{\rm static}$, can be further decomposed into an exponential factor and a pre-exponential factor,
\begin{equation}
\label{eq:static}
    A_{\text{static}} \simeq \text{[det]}_{\text{sph}} e^{-S_{\text{3D}}(\phi_{3,\text{sph}},\vec{A}_{3,\text{sph}})} \ ,
\end{equation}
where $S_{\text{3D}}(\phi_{3,\text{sph}},\vec{A}_{3,\text{sph}})$ is the sphaleron action, i.e.\ the 3D action evaluated at the sphaleron solution. The determinant factor $\text{[det]}_{\text{sph}}$ arises from the one-loop fluctuations around the sphaleron solution. The LO sphaleron rate is determined by the exponential factor, while the NLO sphaleron rate includes the determinant contribution.

We now perform the power-counting analysis for the sphaleron scale, which is crucial for constructing the sphaleron EFT. This analysis also justifies the use of the 3D EFT for the sphaleron-rate computation, as well as the integration steps in the dimensional-reduction procedure discussed in the previous section. To make the scale hierarchy explicit, we perform power counting in the soft-scale 3D action of Eq.~\eqref{eq:S3d}, using \(g\) as the expansion parameter. At leading order, all terms in the action must scale parametrically in the same way. Suppressing indices, we formally have
$
F_3^2 \sim (\partial A_3)^2 \sim m_{W,3}^2 A_3^2
$,
where the last relation follows by matching the Yang--Mills term to the 3D \(W\)-mass term from \((D_i\phi_3)^\dagger(D_i\phi_3)\). With \(\phi_3\sim T^{1/2}\), the characteristic sphaleron momentum satisfies
\begin{align}
  k \sim \partial \sim m_{W,3} \sim g_3\phi_3 \sim gT \ .
\end{align}
Hence the sphaleron probes the soft scale \cite{Li:2025kyo}, with size
$
R_{\text{sph}} \sim (gT)^{-1}
$.
Balancing the interaction terms in the gauge sector,
$
g_3 A_3^2\,\partial A_3 \sim g_3^2 A_3^4
$,
gives \(A_3\sim \phi_3\sim T^{1/2}\). This counting implies that the kinetic terms in Eq.~\eqref{eq:S3d} scale as \(\sim g^2T^3\). Importantly, the sphaleron scale in the Higgs phase is soft and does not depend on the detailed scalar potential \cite{Li:2025kyo}. Therefore, although the semiclassical sphaleron-rate calculation is technically analogous to bounce-rate calculations, it does not require constructing an additional EFT below the soft scale.

\begin{figure}[t]
\centering
\begin{tikzpicture}[x=1cm,y=1cm,>=Latex,font=\large]
  \draw[->,line width=0.9pt] (0,0.4) -- (0,6.6);

  \fill (0,5.7) circle (1.8pt);
  \fill (0,3.6) circle (1.8pt);
  \fill (0,1.2) circle (1.8pt);

  \node[left=3pt] at (0,5.7) {$\pi T$};
  \node[left=3pt] at (0,3.6) {$gT$};
  \node[left=3pt] at (0,1.2) {$\dfrac{g^{3/2}T}{\sqrt{x}}$};

  \node[anchor=west] at (0.45,5.7) {$n\neq 0,\ \psi_{n}$};

  \node[anchor=west] at (0.45,4.15) {$A_{0}$};
  \node[anchor=west] at (0.45,3.45) {$A_{i,3},\ \phi_{3}$};
  \node[anchor=west] at (2.45,3.45) {Sphaleron};

  \node[anchor=west] at (0.45,1.2) {$\phi_{3}$};
  \node[anchor=west] at (2.45,1.2) {Bubble nucleation};
\end{tikzpicture}
\caption{Hierarchy of thermal scales together with the corresponding sphaleron and bubble-nucleation processes.}
\label{fig:scale_hierarchy_sphaleron_bubble}
\end{figure}
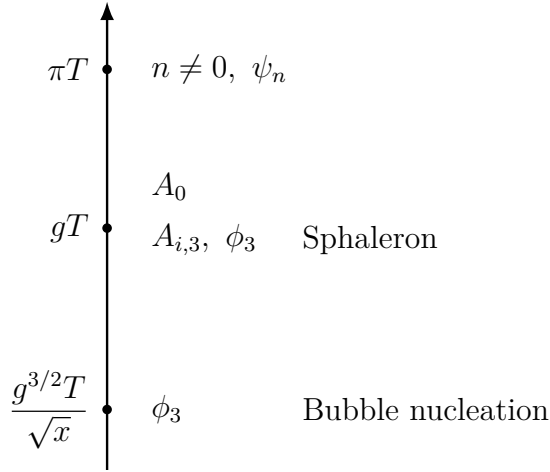

To illustrate the sphaleron scale hierarchy more clearly, Fig.~\ref{fig:scale_hierarchy_sphaleron_bubble} shows the thermal scale hierarchy together with the corresponding modes and processes. The non-zero Matsubara modes and fermions lie at the hard scale \(\pi T\) and are integrated out in the first step of dimensional reduction. The temporal gauge fields lie at the intermediate scale \(gT\)\footnote{Here we comment on a subtle point regarding the temporal gauge components. In standard treatments, these soft modes are often integrated out in a second step of dimensional reduction to construct an EFT below the soft scale \cite{Kajantie:1995dw}. In our setup, this may appear questionable because the sphaleron itself is also a soft-scale configuration. Nevertheless, in the Higgs phase the temporal adjoint mode \(A_0^a\) is parametrically heavier than the scales relevant for the sphaleron, so integrating it out remains justified \cite{Moore:2000mx}.}, and are integrated out in the second step of dimensional reduction. The sphaleron process probes the soft scale \(gT\), whereas the bubble-nucleation process probes the ultrasoft scale \(g^{3/2}T/\sqrt{x}\) \cite{Gould:2021ccf,Gould:2023ovu}, where \(x=\lambda_3/g_3^2\). This ultrasoft scale is also the characteristic scale of a first-order phase transition. Note that the static spatial gauge fields are integrated out at the bubble-nucleation scale, leaving only the scalar field. As mentioned in the previous section, integrating out the static spatial gauge field \(A_{i,3}\) can generate a cubic barrier term in the scalar potential, which is crucial for a first-order phase transition. Thus, the sphaleron and bubble-nucleation processes are essentially decoupled, since they probe different scales and different EFTs. From a physical point of view, the ``sphaleron size'' is much smaller than the ``bubble size''.

We now return to the sphaleron 3D action in Eq.~(\ref{eq:S3d}). The thermal dynamics of this action is governed by two dimensionless parameters,
\begin{align}
  \label{eq:x_y_parameters}
x  = \frac{\lambda_3}{g_3^2} \ , \ \quad
y  = \frac{\mu_3^2}{g_3^4} \ .
\end{align}
The parameter \(x\) is the ratio of the scalar quartic coupling to the gauge coupling, and controls the strength of the phase transition, while the parameter \(y\) is the ratio of the scalar mass term to the gauge coupling, and controls the temperature evolution of the phase transition. As the temperature decreases, \(y\) decreases and changes from positive to negative, while \(x\) remains approximately constant during the phase transition \cite{Li:2025kyo}. Different values of \(x\) therefore determine the nature of the phase transition: a first-order phase transition can occur only when \(x\) is smaller than a critical value, \(x_c\simeq 0.1\) \cite{Kajantie:1996mn}.

\begin{figure}[t!]
\center
\includegraphics[width=8cm]{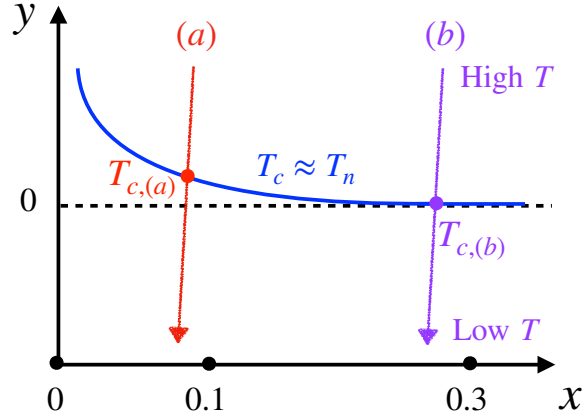}
\caption{Phase diagram in the \(x\)-\(y\) plane, with \(x\) and \(y\) defined in Eq.~(\ref{eq:x_y_parameters}). The blue line denotes the critical-temperature line, which is very close to the nucleation-temperature line when the phase transition does not involve significant supercooling. As the temperature decreases, the system evolves from the upper part of the phase diagram to the lower part, as indicated by the red and purple arrows in cases (a) and (b), respectively. The intersections of trajectories (a) and (b) with the blue line represent the critical temperatures of the corresponding phase transitions. Case (a), with \(x<x_c\approx 0.1\), corresponds to a first-order phase transition, whereas case (b), with \(x>x_c\), corresponds to a crossover.}
\label{fig:x_y_phase_diagram}
\end{figure}

The phase diagram in the \(x\)-\(y\) plane is shown in Fig.~\ref{fig:x_y_phase_diagram}. The blue line represents the critical-temperature line. We show two representative trajectories, indicated by the red and purple arrows. The red line, with \(x<x_c\approx 0.1\), corresponds to a first-order phase transition, while the purple line, with \(x>x_c\), corresponds to a crossover. 
Since the perturbative sphaleron calculation requires a non-zero vev of the scalar field, the condition for the existence of this non-zero vev differs between crossover and first-order phase transitions. If one directly uses Eq.~(\ref{eq:S3d}) to describe both crossover and first-order phase transitions, a problem arises in the first-order case. Here we demonstrate the origin of this problem.
For the crossover case in Eq.~(\ref{eq:S3d}), we always have \(\mu_3^2<0\) in the broken phase, so a non-zero vev can be obtained from Eq.~(\ref{eq:S3d}). By contrast, for a first-order phase transition, immediately in the broken phase one has \(\mu_3^2>0\), as shown by case (a) in Fig.~\ref{fig:x_y_phase_diagram}, and a perturbative non-zero vev cannot be obtained from Eq.~(\ref{eq:S3d}).
Previously, the sphaleron rate for this first-order phase-transition case was studied using non-perturbative lattice simulations \cite{Moore:1998swa,Annala:2025aci,DOnofrio:2014rug}. Since the lattice calculation does not require a non-zero perturbative vev, there is no problem in using Eq.~(\ref{eq:S3d}) to calculate the sphaleron rate for FOPT.
To carry out a perturbative sphaleron-rate computation for this case, we adopt the method developed in \cite{Li:2025kyo}, in which the spatial quantum fluctuations of the gauge fields, \(A_{i,3}\), are integrated out to generate a term proportional to \(-|\phi_3|^3\) in the effective potential. This term can trigger a non-zero vev for the scalar field and hence allow a sphaleron solution. In other words, we consider the following two forms of the potential term in the 3D action of Eq.~(\ref{eq:S3d}):
\begin{align}
V(\phi_3) & = \mu_3^2  \phi_3^\dagger \phi_3 + \lambda_3 (\phi_3^\dagger \phi_3)^2 \quad &&{\rm (Crossover\ phase\ transition)} \ , \\
V(\phi_3) & =  \mu_3^2 \phi_3^\dagger \phi_3 - c (\phi_3^\dagger \phi_3)^{3/2} + \lambda_3 (\phi_3^\dagger \phi_3)^2  \quad &&{\rm (First\ order\ phase\ transition)} \ ,
\label{eq:potential_cubic_term}
\end{align}
where \(c\) is the coefficient of the cubic term generated by integrating out the spatial gauge fields, whose value will be specified later.

One may wonder whether it is legitimate to integrate out the spatial gauge fields, given that the sphaleron solution itself is a non-trivial configuration of the spatial gauge fields. As pointed out in Ref.~\cite{Li:2025kyo}, the justification for this step can be seen from the numerical analysis in the later subsection: the detailed form of the potential, i.e.\ whether it contains a cubic term or not, affects mainly the boundary conditions of the sphaleron configuration, while the dominant contribution to the sphaleron action comes from the kinetic terms. Although the sphaleron is a soft-scale configuration, the boundary condition for the scalar profile is set by supersoft physics. In practice, this supersoft sector determines the Higgs minimum \(v_3\)—or, beyond the leading-order semiclassical approximation, the Higgs condensate—across the \((x,y)\) plane \cite{Ekstedt:2024etx}. At leading order in perturbation theory, this information is encoded in the 3D effective scalar potential, whose minimum defines \(v_3\).

\subsection{Sphaleron Scaling Properties and Numerical Fits}
\label{sec:sphaleron_scaling_properties_and_numerical_fittings}
For numerical convenience, we scale all fields and coordinates appearing in the 3D action in Eq.~\eqref{eq:S3d} to dimensionless quantities.

A commonly used normalization in earlier work is to express the fields in units of the (temperature-dependent) Higgs vev \(v\) in the full theory using the high-\(T\) expansion \cite{Carson:1990jm}.  
That choice is known to require care when discussing gauge dependence \cite{Patel:2011th}.  
Another possible normalization is through \(\bar{\mu}\equiv\sqrt{-\mu_3^2}\), but this is meaningful only for \(\mu_3^2<0\).  
It is therefore not suitable for a unified treatment of first-order transitions, for which \(\mu_3^2\) in the Higgs phase can be either positive or negative (cf.\ case (a) in Fig.~\ref{fig:x_y_phase_diagram}).

We therefore follow Ref.~\cite{Li:2025kyo} and normalize with \(g_3^2\), which is positive definite and gauge invariant at all temperatures \cite{Croon:2020cgk}.  
Conceptually, this parallels the gauge-cancellation discussion for quartic matching in Sec.~\ref{sec:gauge_cancellation_dimensional_reduction}.  
We define
\begin{align}
A_3=g_3\hat{A}_3,\quad \phi_3=g_3 \hat{\phi}_3,\quad x_i=\frac{\xi_i}{g_3^2}\,,
\end{align}
so that Eq.~\eqref{eq:S3d} becomes
\begin{align} \label{eq:S3D_dimensionless_version}
  S_\text{3D}\rightarrow \hat{S}_{\text{3D}} = \int \mathrm{d}^{3}\xi \left[\frac{1}{4}\hat{F}_{ij,3}^{a}\hat{F}^{a}_{ij,3}+(\hat{D}_{i}\hat{\phi})_{3}^{\dagger}(\hat{D}_{i}\hat{\phi})_{3}+V_{3}(\hat{\phi}_3,x,y)\right] \,,
\end{align}
with \(\hat{F}_{ij,3}^{a}\) and \((\hat{D}_{i}\hat{\phi})_{3}\) free of explicit gauge-coupling factors.

Although the tree-level scalar part corresponding to Eq.~\eqref{eq:S3d} contains only quadratic and quartic operators, motivated by the discussion around Eq.~\eqref{eq:potential_cubic_term}, we consider a more general cubic potential of the form
\begin{equation}
\label{eq:general_cubic_potential}
V_{3}(\hat{\phi}_3,x,y)=y\hat{\phi}_3^\dagger\hat{\phi}_3-q (\hat{\phi}_3^\dagger\hat{\phi}_3)^{3/2}+x (\hat{\phi}_3^\dagger\hat{\phi}_3)^2 \,,
\end{equation}
where \(x\) and \(y\) are defined in Eq.~\eqref{eq:x_y_parameters}.

At this stage, we keep \(q\) symbolic in the formulas, but it is fixed by consistency with the leading-order effective potential in each physical case.  
Concretely, \(q=0\) for a second-order transition (used below for the SM crossover study), and
\(q=\frac{1}{4\sqrt{2}\pi}\) for first-order transitions.  
The latter value follows from the consistent treatment of gauge modes and yields a gauge-invariant setup, as discussed in Sec.~\ref{sec:gauge_independence_effective_potential_3d_eft}.

For the sphaleron in this 3D EFT, we use the ansatz%
\footnote{
The gauge-field ansatz adopted here is not unique, and the \(U(1)\) gauge field is omitted for simplicity. Alternative sphaleron conventions, including treatments of the \(U(1)\) sector, are discussed in Ref.~\cite{Wu:2023mjb} and references therein.
}
\begin{align} \label{eq_app:scalar_gauge_sphaleron_confg_in_3DEFT}
\hat{\phi}_{3}&=h(\xi)\left(\begin{array}{c} 
 0  \\ 
 v_3(x,y)
  \end{array}\right) \,, \\
\label{eq_app:gauge_sphaleron_asatz_in_3DEFT}
\hat{A}_{i,3}^{a} T^{a} \mathrm{d}x^{i} &= [1-f(\xi)] \sum_{a=1}^{3} F_{a} T^{a} \,,
\end{align}
where \(h(\xi)\) and \(f(\xi)\) are radial profiles, and \(F_a\) are the usual one-forms depending on \(\theta\), \(\varphi\), and the loop parameter \(\mu\) \cite{Manton:1983nd}.  
Here \(v_3(x,y)\) is the minimum of \(V_3(\hat{\phi}_3,x,y)\).  
From this point onward, hatted fields denote the variables appearing in the sphaleron ansatz.  
The consistency of this 3D gauge ansatz with Ref.~\cite{Klinkhamer:1990fi} is shown in the appendix of Ref.~\cite{Li:2025kyo}, where additional technical details are also given.

From Eq.~\eqref{eq:general_cubic_potential}, the minimum is%
\footnote{
Here \(v_3(x,y)\) is defined as the vacuum expectation value, i.e.\ the minimum, of the magnitude of the complex scalar field \(|\hat{\phi}_3|\).
}
\begin{align}\label{eq:definition_vev}
v_3(x,y)=\frac{3q+\sqrt{-32xy+9q^2}}{8x} \,.
\end{align}

Even though Eq.~\eqref{eq:S3D_dimensionless_version} is already dimensionless, it is numerically useful to perform one further rescaling with \(v_3(x,y)\):
\begin{align}
\label{eq:rescaling}
   \hat{\phi}_3 = v_3(x,y)\hat{\phi}_3^\prime \,,\ \ \hat{A}_3 = v_3(x,y)\hat{A}_3^\prime \,,\ \ \xi=\frac{1}{v_3(x,y)}\xi^\prime \,,
\end{align}
As shown below, this improves the numerical extraction of the action.  
This \(v_3\)-rescaling is not the same as the \(v\)-scaling of Ref.~\cite{Carson:1990jm}: here \(v_3\) is dimensionless and manifestly gauge invariant, being the minimum of the leading-order effective potential.  
By contrast, \(v\) in Ref.~\cite{Carson:1990jm} is dimensional; while the tree-level \(v\) with a thermal mass is gauge invariant, later loop-improved implementations of that framework can become gauge dependent.  
Operationally, the present rescaling is chosen because, in first-order transitions, the action is approximately of the form \(v_3(x,y)\times \text{constant}\), which simplifies the washout analysis in the later subsection on baryon washout.

With this transformation, the action becomes
\begin{align}
\label{eq:Shat}
 \hat{S}_{\text{3D}} = v_3(x,y) \mathcal{C}_{\text{sph}}(x,y) \,,
\end{align}
where
\begin{align} 
\label{eq:general_C_sph-main}
\mathcal{C}_{\text{sph}}(x,y) = \int \mathrm{d}^{3}\xi \left[\frac{1}{4}\hat{F}_{ij,3}^{a}\hat{F}^{a}_{ij,3} + (\hat{D}_{i}\hat{\phi})_{3}^{\dagger}(\hat{D}^{i}\hat{\phi})_{3} + \overline{V}_{3}(\hat{\phi}_3,x,y)\right] \,,
\end{align}
and
\begin{align} \label{eq:vev_transformed_cubic_potential}
   \overline{V}_{3}(\hat{\phi}_3,x,y) = \frac{y}{v_3^2(x,y)} \hat{\phi}_3^\dagger \hat{\phi}_3 - \frac{q}{v_3(x,y)} (\hat{\phi}_3^\dagger \hat{\phi}_3)^{3/2} + x (\hat{\phi}_3^\dagger \hat{\phi}_3)^2 \,,
\end{align}
after dropping primes for notational simplicity.  
Henceforth, \(\hat A_3\) and \(\hat\phi_3\) denote the \(v_3\)-rescaled fields, and \(\overline{V}_3\) denotes the correspondingly rescaled scalar potential.  
The Yang--Mills and covariant-derivative terms do not contain explicit \(x\) or \(y\) dependence, while \(\overline{V}_3\) does.

After this step, the gauge ansatz is unchanged, while the scalar ansatz reduces to
\begin{align}
\hat{\phi}_{3}&=h(\xi)\left(\begin{array}{c} 
 0  \\ 1
 \end{array}\right) \,.
\end{align}
The sphaleron equations of motion obtained from \(\mathcal{C}_{\text{sph}}\) are
\begin{align}
f^{\prime \prime} - \frac{2f(-1+f)(-1+2f)}{\xi^2}-\frac{2}{3}(-1+f)J(1+J)h^2 = 0 \,, \\
h^{\prime \prime}+\frac{2h^\prime}{\xi}-\frac{8}{3\xi^2}J(1+J)(-1+f)^2h-\frac{1}{2} \frac{\partial \overline{V}_3}{\partial h}=0 \,,
\end{align}
with prime denoting the radial derivative.  
The equation for \(f\) has no explicit \(x\) or \(y\) dependence, whereas the equation for \(h\) depends on \(x\) and \(y\) only through \(\overline{V}_3\).  
Therefore, all \(x\)- and \(y\)-dependence of the action and equations of motion enters through \(\overline{V}_{3}(\hat{\phi}_3,x,y)\).  
We now discuss the second- and first-order cases separately.

\subsubsection{second-order phase transitions}
\label{sec:2nd-order}

\begin{figure}[t]
\centering
\includegraphics[width=0.5\textwidth]{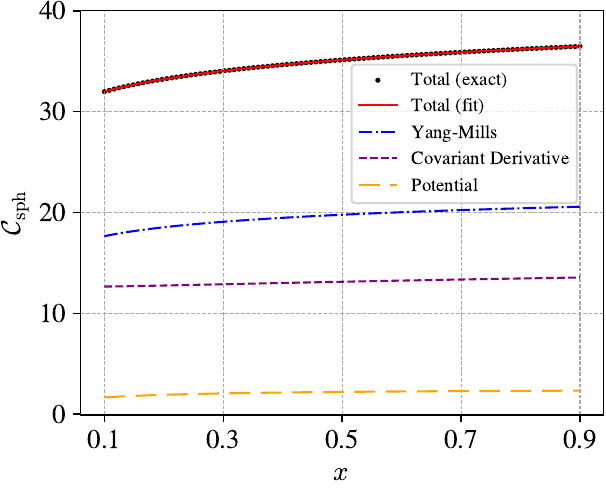}
\caption{
Rescaled sphaleron action \(\mathcal{C}_{\text{sph}}\) as a function of \(x\). The black points show the exact numerical values of \(\mathcal{C}_{\text{sph}}\), while the red curve shows the fit in Eq.~\eqref{eq:fit-crossover}. The agreement is so good that the two are nearly indistinguishable. The dashed curves indicate the separate contributions to \(\mathcal{C}_{\text{sph}}\): blue for the Yang--Mills term, purple for the covariant-derivative term, and orange for the scalar-potential term in Eq.~\eqref{eq:general_C_sph-main}.\reproducedfromref{Li:2025kyo}
}
\label{fig:Csph_SM_scaling_fit}
\end{figure}

For the tree-level case \(q=0\), the transition is second order; the second derivative of the pressure is discontinuous at the critical point, which occurs at \(y=0\).  
In the Higgs phase (\(y<0\)), sphalerons can be computed with
\begin{align}
\label{eq:minimum-2nd-order}
v_3(x,y)=\sqrt{\frac{-y}{2x}} \,,
\end{align}
and the scaled potential becomes
\begin{align}
\overline{V}_3(\hat{\phi}_3,x)=-2x\hat{\phi}_3^\dagger\hat{\phi}_3+ x (\hat{\phi}_3^\dagger\hat{\phi}_3)^2 \,.
\end{align}
Both terms are linear in \(x\), with the first term becoming linear because of the scaling transformation.

Solving the sphaleron equations of motion numerically then yields \(\mathcal{C}_{\text{sph}}(x)\), which is accurately fit by
\begin{align}
\label{eq:fit-crossover}
\mathcal{C}_{\text{sph}}(x)=A+B x^C \log(Dx) \,,
\end{align}
with
\begin{align}
A=26.12,\ B=-2.145,\ C=0.4237,\ D=0.00717 \,.
\end{align}
Figure~\ref{fig:Csph_SM_scaling_fit} compares the fit with the exact numerical result and separates the contributions from the various terms in Eq.~\eqref{eq:general_C_sph-main}.

A key outcome is that, after the scaling, the scalar-potential contribution to \(\mathcal{C}_{\text{sph}}\) is numerically much smaller than the kinetic contributions.  
Formally, all terms scale as \(\int d^3x\,(g^2T)^{-3}\sim 1/g\), but numerically the potential mainly determines the location of the Higgs minimum through \(v_3\) in Eq.~\eqref{eq:minimum-2nd-order}, whereas \(\mathcal{C}_{\text{sph}}\) is dominated by the kinetic terms.  
This observation is central to the first-order case discussed below.

Equation~\eqref{eq:fit-crossover} is directly useful: once a model is mapped to this SU(2)+Higgs 3D EFT and its leading Higgs-phase minimum is given by Eq.~\eqref{eq:minimum-2nd-order} for \(y<0\), the sphaleron action follows from this fit.  
The temperature dependence then enters through the dimensional-reduction matching in \(x\) and \(y\), which can be computed, for example, with {\tt DRalgo} \cite{Ekstedt:2022bff,Fonseca:2020vke}.

\subsubsection{first-order phase transitions}
\label{sec:1st-order}

\begin{figure}[!t]
\centering
\includegraphics[width=0.4\textwidth]{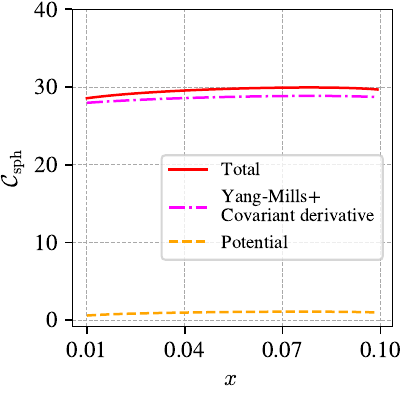}
\includegraphics[width=0.4\textwidth]{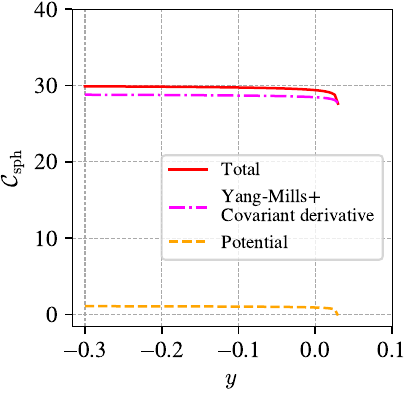} \\
\centering \includegraphics[width=0.5\textwidth]{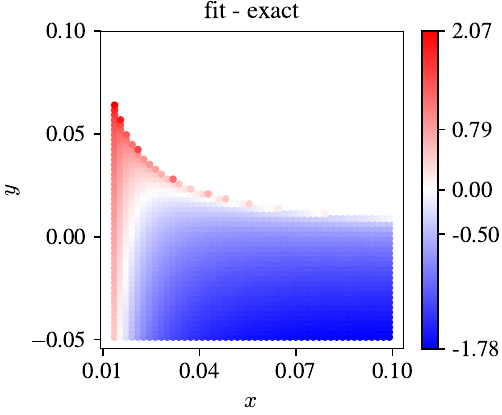}
\caption{ 
Top row: \(v_3\)-rescaled sphaleron action for the cubic potential.  
Left: \(\mathcal{C}_{\text{sph}}\) as a function of \(x\) at fixed \(y=0.008\). The red curve shows the total action, the magenta curve shows the combined Yang--Mills and covariant-derivative contributions, and the orange curve shows the scalar-potential contribution.  
Right: \(\mathcal{C}_{\text{sph}}\) as a function of \(y\) at fixed \(x=0.03\), with the same color coding.  
Bottom row: Difference between the approximate sphaleron action in Eq.~\eqref{eq:vev_scaled_sphaleron_action_fitting} and the full numerical action in the \((x,y)\) plane. The small residual confirms that the approximation reproduces the exact numerical sphaleron action well over a broad parameter range.\reproducedfromref{Li:2025kyo}
}
\label{fig:sph_cubic_scan_under_x_or_y}
\end{figure}

For a first-order transition, the leading-order perturbative potential must contain a barrier between the symmetric and Higgs phases, and this barrier determines the Higgs-phase minimum.  
In what follows, we set
\(
q=\frac{1}{4\sqrt{2}\pi}
\),
so that both \(x\) and \(y\) enter \(\overline{V}_3\).  
This value corresponds to the leading-order scalar potential of the \(SU(2)\) gauge--Higgs theory \cite{Ekstedt:2022zro}.

In this setup, obtaining a simple closed-form fit for \(\mathcal{C}_{\text{sph}}(x,y)\) as a function of two variables is difficult.  
Numerically, however, the potential contribution is small, as in the second-order case, and \(\mathcal{C}_{\text{sph}}\) is approximately constant.  
This is illustrated in the top row of Fig.~\ref{fig:sph_cubic_scan_under_x_or_y}, where we scan \(\mathcal{C}_{\text{sph}}\) at fixed \(x\) or fixed \(y\).  
The potential contribution is subdominant, and the \(v_3\)-rescaled sphaleron action remains nearly constant throughout the relevant parameter region.

Motivated by this behavior, we use the approximation
\begin{align}
\label{eq:vev_scaled_sphaleron_action_fitting}
\hat{S}_{\text{3D}} \simeq 29\, v_3(x,y) \,,
\end{align}
with \(v_3(x,y)\) from Eq.~\eqref{eq:definition_vev} and \(q=\frac{1}{4\sqrt{2}\pi}\).

We further validate Eq.~\eqref{eq:vev_scaled_sphaleron_action_fitting} by comparing it with the full numerical result across the \((x,y)\)-plane, as shown in the bottom row of Fig.~\ref{fig:sph_cubic_scan_under_x_or_y}.  
The agreement indicates that the dependence of the equations of motion and of \(\mathcal{C}_{\text{sph}}\) on \(x\) and \(y\) is weak, especially in the phenomenologically relevant region \(y>0\).

\subsection{Sphaleron Rate Beyond Leading Order}
\label{sec:sphaleron_rate_beyond_leading_order}
The sphaleron scaling and numerical analysis above concern only the leading-order exponential factor in the sphaleron rate. In this subsection, we discuss the one-loop fluctuation corrections around the sphaleron solution, which are formally encoded in the determinant factor \([{\rm det}]_{\rm sph}\) in Eq.~\eqref{eq:static}. This determinant can be written schematically as \cite{Carson:1990jm,Carson:1989rf,Baacke:1993aj,Baacke:1994ix}
\begin{align}
  [{\rm det}]_{\rm sph} = \mathcal{N} \times |{\rm negative\ modes}| \times |{\rm zero\ modes}| \times |{\rm positive\ modes}| \,,
\end{align}
where \(\mathcal{N}\) is an overall normalization factor, and the negative, zero, and positive modes correspond to the eigenvalues of the fluctuation operator around the sphaleron solution, which can be formally defined by
\begin{align}
  \hat{\mathcal{O}}_{\rm sph} \psi_n = \omega_n^2 \psi_n \,,
\end{align}
where \(\hat{\mathcal{O}}_{\rm sph}\) is the second variation of the action around the sphaleron solution, and \(\psi_n\) and \(\omega_n^2\) are the corresponding eigenfunctions and eigenvalues. The negative, zero, and positive modes are then defined by \(\omega_n^2<0\), \(\omega_n^2=0\), and \(\omega_n^2>0\), respectively. We already encountered these modes in the earlier discussion of instantons in quantum mechanics and in the Standard Model, for example in Eq.~\eqref{eq:transition_amplitudes_QM_semi_classical} of Appendix.~\ref{sec:instanton_QM} and Eq.~\eqref{eq:instanton_eigenvalue_equations} of Appendix.~\ref{sec:instanton_baryon_violation}. Since the sphaleron transition is dominated by the classical transition rate in the high-temperature limit, we may expand Eq.~\eqref{eq:sphaleron_path_integral_finite_T} as
\begin{align}
  \mathcal{Z}_{\rm sph} & \sim \int [d\varphi] e^{-\frac{1}{T}\int d^3 x \mathcal{L}_E(\phi_{\rm sph}+\varphi)} \,, \nonumber \\
& \sim \int [d\varphi] e^{-S_{\rm 3D}(\phi_{\rm sph}+\varphi)} \,, \nonumber \\
\label{eq:sphaleron_path_integral_finite_T_fluctuation_expansion}
& \sim e^{-S_{\rm 3D}(\phi_{\rm sph})} \int [d\varphi_0] [d\varphi_+]  e^{-\frac{1}{2} \varphi_+ \frac{\delta^2 S_{\rm 3D}}{\delta \varphi^2} \varphi_+} \int [d\varphi_-] e^{+\frac{1}{2} \varphi_- \frac{\delta^2 S_{\rm 3D}}{\delta \varphi^2} \varphi_-} \,,
\end{align}
where in the second line we define the effective 3D action, which is the same as that in Eq.~\eqref{eq:S3D_dimensionless_version}, although for simplicity we do not distinguish gauge and scalar fields in the notation \(\varphi\). This is also the approach adopted in Ref.~\cite{Carson:1989rf}. In the third line, we separate the fluctuation modes into negative, zero, and positive modes, denoted by \(\varphi_-\), \(\varphi_0\), and \(\varphi_+\), respectively.

It has been shown in Refs.~\cite{Carson:1990jm,Carson:1989rf} that (i) the magnitude of the negative mode is small compared with the zero- and positive-mode contributions; (ii) the zero-mode contribution is of order \(\mathcal{O}(e^{8})\) and is insensitive to the value of \(\lambda/g^2\) for SU(2)+Higgs theory; and (iii) the positive-mode contribution is very sensitive to \(\lambda/g^2\), reaching \(\mathcal{O}(e^{-40}-e^{-10})\) for \(\lambda/g^2\lesssim 0.1\), while becoming much larger for \(\lambda/g^2\gtrsim 0.1\). Moreover, an exact computation of the positive modes is technically difficult, as it involves lengthy algebra. However, for smaller values of \(\lambda/g^2\), the one-loop effective potential provides a good approximation to the exact determinant computation \cite{Baacke:1993aj,Baacke:1994ix}. In this subsection, we introduce the zero modes for SU(2)+Higgs theory, which provide the main correction for the SM crossover phase transition, where $\lambda/g^2\sim 0.3$. In the next subsection, we present the full result and compare it with lattice simulations.

We now consider the computation of the zero modes, \([d\varphi_0]\), in Eq.~\eqref{eq:sphaleron_path_integral_finite_T_fluctuation_expansion}. First, we need to convert the integration variable into collective coordinates \cite{Carson:1989rf},
\begin{align}
  \label{eq:zero_mode_collective_coordinate_relation}
  d\varphi_0 = \sum_\alpha \psi_\alpha dq_\alpha \,,
\end{align}
where \(\psi_\alpha\) are the zero-mode eigenfunctions satisfying the normalization condition
\begin{align}
  \label{eq:zero_mode_normalization_condition}
  \int d^3 \xi \psi_\alpha \psi_\beta = \delta_{\alpha\beta} \,,
\end{align}
where \(\xi\) is the dimensionless spatial coordinate, and \(q_\alpha\) are the corresponding collective coordinates.

The zero modes are related to the broken symmetries of the classical sphaleron solution \(\phi_{\rm sph}\). This can be understood by analogy with Goldstone's theorem. Consider a classical potential \(V(\phi_i)\) with a minimum at \(\phi=\phi_i\), and suppose it is invariant under the transformation \(\phi_i \rightarrow \phi_i + i \alpha \phi_i\), where \(\alpha\) is an infinitesimal parameter. Expanding the potential gives
\begin{align}
  V(\phi_i) & = V(\phi_i + i \alpha \phi_i) \,, \nonumber \\
& = V(\phi_i) + \frac{1}{2} i\alpha \phi_i \frac{\partial^2 V}{\partial \phi_i \partial \phi_j} i \alpha \phi_j + \cdots \,.
\end{align}
If \(\phi_i\neq 0\) and \(\alpha\neq 0\), we must have
\begin{align}
  \frac{\partial^2 V}{\partial \phi_i \partial \phi_j} \phi_j = 0 \,,
\end{align}
which means that there is a zero mode in the spectrum of the fluctuation operator \(\frac{\partial^2 V}{\partial \phi_i \partial \phi_j}\). This is analogous to the zero modes in the sphaleron case. In the simple example above, the classical solution \(\phi_i\) is analogous to the sphaleron solution \(\phi_{\rm sph}\), and the symmetry transformation is analogous to a gauge transformation. Therefore, the zero modes should satisfy the following two properties:
\begin{itemize}
  \item The classical sphaleron solution \(\phi_{\rm sph}\) is \textbf{not invariant} under the zero-mode transformation.
  \item The classical action is \textbf{invariant} under the zero-mode transformation.
\end{itemize}

Since the sphaleron solutions for the Higgs and gauge fields read, from Eq.~\eqref{eq:sphaleron_multiplet_confg_manton},
\begin{align}
H &= \frac{v}{\sqrt{2}}h(\xi)U_{\text{sph}}(\mu,\theta,\phi)\begin{pmatrix}
    0\\
    1 
\end{pmatrix} \,, \\
A_i^a T^a dx^i &= -\frac{i}{g}f(\xi) (\partial_i U_{\text{sph}} )U_{\text{sph}}^{-1}dx^i \,,
\end{align}
where \(\xi\), \(\theta\), and \(\phi\) are the spherical coordinates, and \(U_{\text{sph}}\) is the SU(2) matrix defined in Eq.~\eqref{eq::Manton_Umatrix_fundamental}, we see that (i) the sphaleron solution is not invariant under spatial translations \((\xi_i\rightarrow \xi_i + \delta \xi_i)\) or spatial rotations \((\theta \rightarrow \theta + \delta \theta,\ \phi \rightarrow \phi + \delta \phi)\); and (ii) the sphaleron action is invariant under both spatial translations and rotations. We therefore conclude that the sphaleron solution has six zero modes, corresponding to three spatial translations and three spatial rotations.

We now discuss the zero modes associated with infinitesimal spatial translations, \(\xi_i \rightarrow \xi_i + \epsilon_i\). We must relate the coordinates to the collective coordinates \(q_\alpha\) defined in Eq.~\eqref{eq:zero_mode_collective_coordinate_relation}. Since all field transformations are accounted for by the path-integral measure, we need to determine which transformations of the fields in the measure correspond to spatial translations. Equivalently, we ask how a spatial translation acts on the fluctuation fields. We adopt this latter viewpoint for the gauge and scalar fluctuations in the measure, denoted by \(a_k^a\) and \(\eta\), respectively. The spatial transformation \(\xi\rightarrow \xi + \epsilon_i\) induces the following transformations:
\begin{align}
  \label{eq:spatial_translation_zero_mode_transformation}
a_k^a & \rightarrow a_k^a + \epsilon_j F_{jk}^a \,, \\
\eta & \rightarrow \eta + \epsilon_j D_j \Phi \,,
\end{align}
where \(F_{ij}^a\) is the field strength of the classical sphaleron solution, and \(D_j \Phi\) is the covariant derivative of the scalar field in the classical sphaleron background. We now specialize Eqs.~\eqref{eq:zero_mode_collective_coordinate_relation} and \eqref{eq:zero_mode_normalization_condition} to SU(2)+Higgs theory:
\begin{align}
  \label{eq:zero_mode_collective_coordinate_relation_SU2_Higgs}
a_i^\alpha &= \psi_{i \alpha}^a q_\alpha + \cdots \,, \\
\eta &= \tilde{\psi}_\alpha q_\alpha + \cdots \,,
\end{align}
where the ellipses denote the contributions from the non-zero modes, which are not considered here, and \(\psi_{i \alpha}^a\) and \(\tilde{\psi}_\alpha\) are the zero-mode eigenfunctions for the gauge and scalar fluctuations, respectively. The index \(\alpha\) runs from 1 to 3, corresponding to the three spatial-translation zero modes. The normalization of \(\psi_{i \alpha}^a\) and \(\tilde{\psi}_\alpha\) must satisfy the combined normalization condition for gauge and scalar fluctuations \cite{Carson:1989rf},
\begin{align}
  \label{eq:normalization_condition_zero_modes_SU2_Higgs}
  2\pi \delta_{\alpha\beta} & = \int d^3 \xi \left( \psi_{i \alpha}^a \psi_{i \beta}^a + \tilde{\psi}_\alpha^\dagger \tilde{\psi}_\beta + \tilde{\psi}_\beta^\dagger \tilde{\psi}_\alpha \right) \,.
\end{align}
To connect Eqs.~\eqref{eq:spatial_translation_zero_mode_transformation} and \eqref{eq:zero_mode_collective_coordinate_relation_SU2_Higgs}, we assume the relation
\begin{align}
  \label{eq:collective_coordinate_spatial_translation_relation}
q_j= N_{\rm tr} \epsilon_j \,, \qquad j=1,2,3 \,,
\end{align}
where all parameters in this relation are real.

We now consider the following integral:
\begin{align}
  &\int d^3 \xi \left( a_i^\alpha a_i^\alpha + \eta^\dagger \eta + \eta \eta^\dagger \right) \nonumber \\
   & \overset{\rm Eq.~(\ref{eq:zero_mode_collective_coordinate_relation_SU2_Higgs})}{=} \int d^3 \xi \left[ (\psi_{i \alpha}^a q_\alpha)(\psi_{i \beta}^a q_\beta) + (\tilde{\psi}_\alpha q_\alpha)^\dagger (\tilde{\psi}_\beta q_\beta) + (\tilde{\psi}_\alpha q_\alpha )(\tilde{\psi}_\beta q_\beta )^\dagger \right] \nonumber \\
   & \overset{\rm Eq.~(\ref{eq:collective_coordinate_spatial_translation_relation})}{=} N_{\rm tr}^2 \epsilon_\alpha \epsilon_\beta \int d^3 \xi \left(\psi_{i \alpha}^a \psi_{i \beta}^a + \tilde{\psi}_\alpha^\dagger \tilde{\psi}_\beta + \tilde{\psi}_\beta^\dagger \tilde{\psi}_\alpha \right) \nonumber \\
   & \overset{\rm Eq.~(\ref{eq:normalization_condition_zero_modes_SU2_Higgs})}{=} 2\pi N_{\rm tr}^2 \epsilon_\alpha \epsilon_\beta \delta_{\alpha\beta} \nonumber \\
   & \overset{\rm Eq.~(\ref{eq:spatial_translation_zero_mode_transformation})}{=} \epsilon_\alpha \epsilon_\beta\int d^3 \xi \left(  F_{\alpha i}^a  F_{\beta i}^a +  (D_\alpha \Phi)^\dagger  (D_\beta \Phi) + (D_\alpha \Phi) (D_\beta \Phi)^\dagger \right) \,,
\end{align}
where each step is labeled by the corresponding equation number for clarity. From this relation, we extract the normalization factor for the spatial-translation zero modes:
\begin{align}
  \label{eq:normalization_factor_spatial_translation_zero_modes}
  N_{\rm tr}^2 = \frac{1}{6\pi} \int d^3 \xi \left[  F_{ij}^a  F_{ij}^a +  2(D_i \Phi)^\dagger  (D_i \Phi) \right] \,,
\end{align}
where the extra factor of 3 in the denominator comes from the sum over the spatial indices \(\alpha\) and \(\beta\). Since the right-hand side is just the kinetic part of the sphaleron action, with the covariant-derivative term carrying an extra factor of 2, it can be computed numerically once the sphaleron solution is known.

Finally, the contribution from the translation zero modes is
\begin{align}
  \label{eq:zero_mode_contribution_spatial_translation}
  \int [d a]_{\rm tr} [d\eta]_{\rm tr} & = \int d^3 q  = \int d^3 \epsilon\, N_{\rm tr}^3  = N_{\rm tr}^3 V  \equiv (\mathcal{N})_{\rm tr} V \,,
\end{align}
where the subscript ``tr'' denotes the spatial-translation zero modes, and \(V\) is the system volume. The Jacobian factor in the first equality is unity because of the normalization condition in Eq.~\eqref{eq:normalization_condition_zero_modes_SU2_Higgs}.

The contribution from the rotation zero modes can be computed in a similar way. In this case, one analyzes how the fields transform under the spatial rotation \(\vec{\xi}\rightarrow \vec{\epsilon}\times \vec{\xi}\), which yields
\begin{align}
  \int [d a]_{\rm rot} [d\eta]_{\rm rot} & = \int d^3 q  = \int d^3 \epsilon\, N_{\rm rot}^3  = N_{\rm rot}^3 V_{\rm rot}  \equiv (\mathcal{NV})_{\rm rot} \,,
\end{align} 
where the subscript ``rot'' denotes the spatial-rotation zero modes, and \(V_{\rm rot}=8\pi^2\) is the volume of the rotation group. The normalization factor for the rotation zero modes can be obtained by the same method as in the translation case, and is given by \cite{Carson:1989rf}
\begin{align}
\label{eq:normalization_factor_rotation_zero_modes}
N_{\mathrm{rot}}^{2}
=\frac{1}{6\pi}\int d^{3}\xi\,
\Big\{
(\xi^{2}\delta_{ik}-\xi_{i}\xi_{k})
\big[
F_{ij}^{a}F_{kj}^{a}
+2\,(D_{i}\Phi)^{\dagger}(D_{k}\Phi)
\big]
+\epsilon_{jkl}\,\tilde{\Lambda}_{j}^{a}F_{kl}^{a}
\Big\} \,,
\end{align}
where \(\tilde{\Lambda}_{j}^{a}\) is
\begin{equation}
\tilde{\Lambda}_i^{\,a}
=4\left(\delta_{ia}-\hat{\xi}_i\hat{\xi}_a\right)P(\xi)
+4\,\hat{\xi}_i\hat{\xi}_a\,Q(\xi) \,,
\end{equation}
and \(P(\xi)\) and \(Q(\xi)\) satisfy
\begin{align}
\frac{1}{\xi}\frac{df}{d\xi}
&=
\left[
-\frac{d^2}{d\xi^2}
-\frac{2}{\xi}\frac{d}{d\xi}
+\frac{2+4f(f-1)}{\xi^2}
+\frac{h^2}{4}
\right]P
+\frac{2(2f-1)}{\xi^2}\,Q \,, \\
\frac{2f(1-f)}{\xi^2}
&=
\left[
-\frac{d^2}{d\xi^2}
-\frac{2}{\xi}\frac{d}{d\xi}
+\frac{4+8f(f-1)}{\xi^2}
+\frac{h^2}{4}
\right]Q
+\frac{4(2f-1)}{\xi^2}\,P \,,
\end{align}
where \(f\) and \(h\) are the numerical solutions of the sphaleron equations of motion. 

In summary, the full zero-mode contribution can be written as
\begin{align}
  \label{eq:full_contribution_zero_modes_main_text}
  \int [d a]_{\rm zero} [d\eta]_{\rm zero} &= V\times v_3^3\times g_3^6 v_3^3\times (\mathcal{N})_{\rm tr} \times (\mathcal{NV})_{\rm rot} \,, \nonumber \\
& \approx V\times g_3^6 v_3^6 \times \exp(8.8) \,,
\end{align}
where the numerical value in the second line is insensitive to \(\lambda/g^2\) for SU(2)+Higgs theory. We have manually introduced the extra factor \(g_3^6 v_3^6\), which arises from the scaling transformation used here and the mismatch between the numerator and denominator in the determinant. A full derivation of this factor is given in Eq.~\eqref{eq:full_contribution_zero_modes} of Appendix.~\ref{app:scaling_constants_sphaleron_zero_modes}. The volume factor \(V\) is usually placed in the denominator of the sphaleron rate, so that the latter is expressed as a rate per unit volume.

\subsection{Sphaleron Rate for the Standard Model Crossover}
\label{sec:sphaleron_rate_crossover}
We now utilize the above results to compute the sphaleron rate for the Standard Model crossover. The leading-order (LO) sphaleron rate can be written as
\begin{align}
  \label{eq:sphaleron_rate_LO}
  \Gamma_{\rm sph, LO} & =  T^4 \exp\left(-v_3(x,y)\mathcal{C}_{\rm sph}(x,y) \right) \,,
\end{align}
where \(v_3\) and \(\mathcal{C}_{\rm sph}\) are defined in Eqs.~\eqref{eq:minimum-2nd-order} and \eqref{eq:fit-crossover}, respectively. The prefactor \(T^4\) follows from dimensional analysis. For the next-to-leading-order (NLO) sphaleron rate, we include the zero-mode contribution, which gives
\begin{align}
  \label{eq:sphaleron_rate_NLO}
  \Gamma_{\rm sph, NLO} & =  T g_3^6(x,y) v_3^6(x,y) \mathcal{N}_{\rm tr} \mathcal{(NV)}_{\rm rot} \kappa(x,y) \exp\left(-v_3(x,y)\mathcal{C}_{\rm sph}(x,y)  \right) \,,
\end{align}
where the zero-mode factor is taken from Eq.~\eqref{eq:full_contribution_zero_modes_main_text}. The positive-mode contribution is encoded in \(\kappa(x,y)\), which we set to unity in the present discussion. Since the analysis is carried out within the 3D EFT, the connection to the parent 4D Standard Model parameters is provided by the dimensional-reduction matching, which can be computed with {\tt DRalgo} \cite{Ekstedt:2022bff}.

\begin{figure}[t]
\centering
\includegraphics[width=0.5\textwidth]{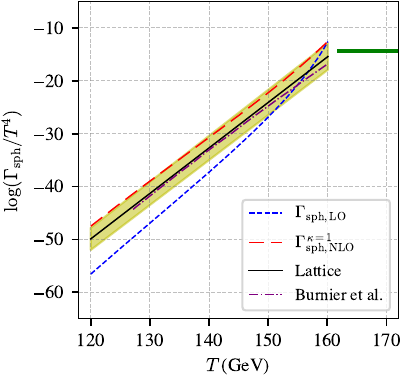}
\caption{
Sphaleron rate for the Standard Model crossover. The blue curve shows the LO sphaleron rate in Eq.~\eqref{eq:sphaleron_rate_LO}, while the red curve shows the NLO sphaleron rate in Eq.~\eqref{eq:sphaleron_rate_NLO}, including the zero-mode contribution. The black curve is the lattice result from Ref.~\cite{Annala:2023jvr}, and the yellow band indicates its uncertainty. The purple dash-dotted curve shows the earlier perturbative result of Ref.~\cite{Burnier:2005hp}.\reproducedfromref{Li:2025kyo} Note that the NLO sphaleron-rate curve shown here has been updated relative to the one in Ref.~\cite{Li:2025kyo}.
}
\label{fig:sph_SM_vev_scaling_fit}
\end{figure}

We present the sphaleron rate for the Standard Model crossover in Fig.~\ref{fig:sph_SM_vev_scaling_fit}, where we compare our LO and NLO results with the lattice result from Ref.~\cite{Annala:2023jvr} and the earlier perturbative result from Ref.~\cite{Burnier:2005hp}. We see that our LO result is qualitatively consistent with the lattice result, while the NLO result including the zero modes is quantitatively consistent with it. We emphasize that, although the result of Ref.~\cite{Burnier:2005hp} is also quantitatively consistent with the lattice result, it was obtained in Landau gauge, whereas our setup is gauge invariant from the outset. This is a significant improvement over the previous approach. In addition, Ref.~\cite{Burnier:2005hp} argued that using the two-loop effective potential effectively incorporates the positive-mode contribution, which is not included in our current NLO result. We leave the explicit inclusion of the positive modes to future work. Another advantage of our approach is that it can also be applied to the sphaleron rate for a first-order phase transition, which is relevant for baryogenesis and will be discussed in detail in Sec.~\ref{sec:application_to_RSM}.

\subsection{Sphaleron Decoupling and Baryon Preservation Condition}
\label{sec:sphaleron_decoupling_and_BPC}
In this subsection, we compute the baryon washout integral and the baryon preservation condition in the broken phase of electroweak baryogenesis. We will see that our formalism leads to a gauge-invariant baryon preservation condition, in contrast to the previously used gauge-dependent condition,
\begin{align}
  \frac{v(T)}{T}\bigg\vert_{T_c} \gtrsim 1 \,,
\end{align}
where \(v(T)\) is the minimum of the finite-temperature effective potential and is therefore gauge dependent if one does not compute it in a gauge-invariant way, and \(T_c\) is the critical temperature. We show how to derive a gauge-invariant baryon preservation condition within our formalism. In this subsection, we assume a general first-order phase transition; a specific BSM example will be discussed in the next subsection.

\subsubsection{sphaleron decoupling temperature}
The first step is to define the sphaleron decoupling temperature, \(T_{*}\), as the temperature at which the baryon-number-violation rate, \(\Gamma_{\rm B}(T)\), equals the Hubble expansion rate, i.e.\ \(\Gamma_{\rm B}(T_{*}) = H(T_{*})\). The rate \(\Gamma_{\rm B}(T)\) is defined through the standard Boltzmann equation for the baryon-number density \(n_B\),%
\footnote{We omit the cosmological-expansion term \(3 H n_B\) on the left-hand side of the Boltzmann equation, since we assume \(\Gamma_{\rm B}(T) \gg H(T)\). Although this is not an excellent approximation if the sphaleron decoupling temperature is extremely close to the nucleation temperature, we will see numerically that such a region is very narrow, so this approximation is adequate for the baryon-washout integral over most of parameter space.}
\begin{align}
  \label{eq:boltzmann_equation_baryon_number_density}
  \frac{dn_B}{dt} = -\Gamma_{\rm B}(T) n_B \,.
\end{align}
Since we previously derived this structure in Eq.~\eqref{eq:BNV_rate_sphaleron_final}, we can read off
\begin{align}
  \Gamma_{\rm B}(T) &= 2 n_f^2 \kappa \frac{\Gamma_{\rm sph}}{T^3}  \approx \frac{39}{2} \frac{\Gamma_{\rm sph}}{T^3} \,,
\end{align}
where in the second step we set the number of fermion generations to \(n_f=3\), and use the approximate value \(\kappa \approx 13/12\) from Eq.~\eqref{eq:kappa_chi_approximate_constants}. Assuming a radiation-dominated Universe, the Hubble rate is
\begin{align}
  H(T) = \sqrt{\frac{4\pi^3 g_*}{45}} \frac{T^2}{M_{\rm pl}} \,,
\end{align}
where \(g_*\) is the effective number of relativistic degrees of freedom, and \(M_{\rm pl}=1.22 \times 10^{19}\,\mathrm{GeV}\) is the Planck mass.

For the SU(2)+Higgs theory, it is convenient to express the condition \(\Gamma_{\rm B}(T_{*}) = H(T_{*})\) in the \((x,y)\) plane, where \(x\) and \(y\) are the dimensionless Higgs self-coupling and mass parameter defined in Eq.~\eqref{eq:x_y_parameters}. Since the properties of a first-order phase transition are well understood in the \((x,y)\) plane within the 3D EFT \cite{Ekstedt:2022zro,Ekstedt:2024etx}, it is natural to analyze sphaleron decoupling there as well. As introduced at the beginning of this section, we write the sphaleron rate in the broken phase as
\begin{align}
  \Gamma_{\rm sph} = A_{\rm dyn}\times [{\rm det}]_{\rm sph} e^{-\hat{S}_{\rm 3D}(x,y)} \,.
\end{align}
For a first-order phase transition, we assign \(A_{\rm dyn}\times [{\rm det}]_{\rm sph}=T^4\) by dimensional analysis. The sphaleron action \(\hat{S}_{\rm 3D}(x,y)\) was computed for a first-order transition in the previous subsection; see Eq.~\eqref{eq:vev_scaled_sphaleron_action_fitting}. The decoupling condition \(\Gamma_{\rm B}(T_{*}) = H(T_{*})\) can then be rewritten as \cite{Li:2025kyo}
\begin{align}
  \hat{S}_{\rm 3D}(x,y) & = \log \left(\frac{39}{2} \frac{T_*}{ H(T_*)} \right) = \mathcal{S}_0 \approx 39 \,,
\end{align}
where the logarithm is approximately constant, which we denote by \(\mathcal{S}_0\). Solving this equation therefore gives a relation between \(x\) and \(y\) at the sphaleron decoupling temperature, which we denote by
\begin{align}
  y = y_f(x) \,.
\end{align}
This curve is independent of the specific BSM model, as long as the final theory is the effective 3D SU(2)+Higgs theory. We discuss this line in the \((x,y)\) plane in more detail below, after deriving the gauge-invariant baryon preservation condition.

\subsubsection{gauge-invariant baryon preservation condition}
\label{sec:gauge_invariant_BPC}
For electroweak baryogenesis, we assume that a net baryon-number density \(n_B\) is generated outside the bubble wall and diffuses into the broken phase. The details of this process were briefly introduced in Sec.~\ref{sec:EWBG_four_processes}. The question we now address is how much of this generated baryon-number density can be preserved in the broken phase. To answer this, we solve the Boltzmann equation in Eq.~\eqref{eq:boltzmann_equation_baryon_number_density} with the initial condition \(n_B(t_{0})\), where \(t_{0}\) is the nucleation time, corresponding to \(T=T_n\). If the sphaleron rate in the broken phase were zero, then \(\Gamma_B=0\), and all of the generated baryon asymmetry would be preserved. In reality, however, the sphaleron rate in the broken phase can still be sizable, depending on the value of \(x=\lambda_3/g_3^2\). Integrating over time, we find that the baryon density at a later time \(t\) is
\begin{align}
  \frac{n_B(t)}{n_B(0)}=\exp\left(-\frac{39}{2}\int_{t=0}^{t} \frac{\Gamma_{\rm sph}(T(t^\prime))}{T^3(t^\prime)}dt^\prime \right) \,,
\end{align}
where \(t=0\) denotes the nucleation time, and we used the expression for \(\Gamma_B\) in terms of \(\Gamma_{\rm sph}\). We can further convert the integration variable from time to temperature using \(dt=-dT/(HT)\) in a radiation-dominated Universe. Note that the upper limit in time corresponds to the sphaleron decoupling temperature \(T_*\), since below that point the sphaleron rate is smaller than the Hubble rate and the baryon-number density effectively freezes out. Thus, we obtain \cite{Li:2025kyo}
\begin{align}
  \frac{n_B(T_*)}{n_B(T_n)}=\exp\left(-\frac{39}{2}\int_{T^*}^{T_n} \frac{\Gamma_{\rm sph}(T)}{H(T) T^4}dT \right) \,.
\end{align}
Since the sphaleron rate in the broken phase can be written as
\begin{align}
  \Gamma_{\rm sph} = T^4 \exp\left(-v_3(x,y)\mathcal{C}_{\rm sph}(x,y) \right) \,,
\end{align}
and for a first-order phase transition we have \(\mathcal{C}_{\rm sph}(x,y)\approx 29\), see Eq.~\eqref{eq:vev_scaled_sphaleron_action_fitting}, the above integral can be evaluated analytically. We note that this exact integral is much more accurate than the approximations used previously in Refs.~\cite{Patel:2011th,Ahriche:2014jna,Funakubo:2009eg,Fuyuto:2014yia}. Furthermore, the formalism developed here is gauge invariant, unlike the ones used in Refs.~\cite{Ahriche:2014jna,Funakubo:2009eg,Fuyuto:2014yia}. We define the baryon-washout exponent
\begin{align}
  \mathcal{W}=\frac{39}{2}\int_{T^*}^{T_n} \frac{\Gamma_{\rm sph}(T)}{H(T) T^4}dT \,,
\end{align}
so that the baryon-number density at the sphaleron decoupling temperature becomes
\begin{align}
  n_B(T_*)=\exp(-\mathcal{W})n_B(T_n) \,.
\end{align}
In other words, a fraction \(\exp(-\mathcal{W})\) of the generated baryon asymmetry is preserved in the broken phase. Ref.~\cite{Li:2025kyo} showed that \(\mathcal{W}\) is highly sensitive to the value of \(x=\lambda_3/g_3^2\): it can change by several orders of magnitude when \(x\) changes by \(\mathcal{O}(0.1)\). This means that the baryon preservation condition is correspondingly very sensitive to the value of \(x\).

\begin{figure}[t]
\centering
\includegraphics[width=0.6\textwidth]{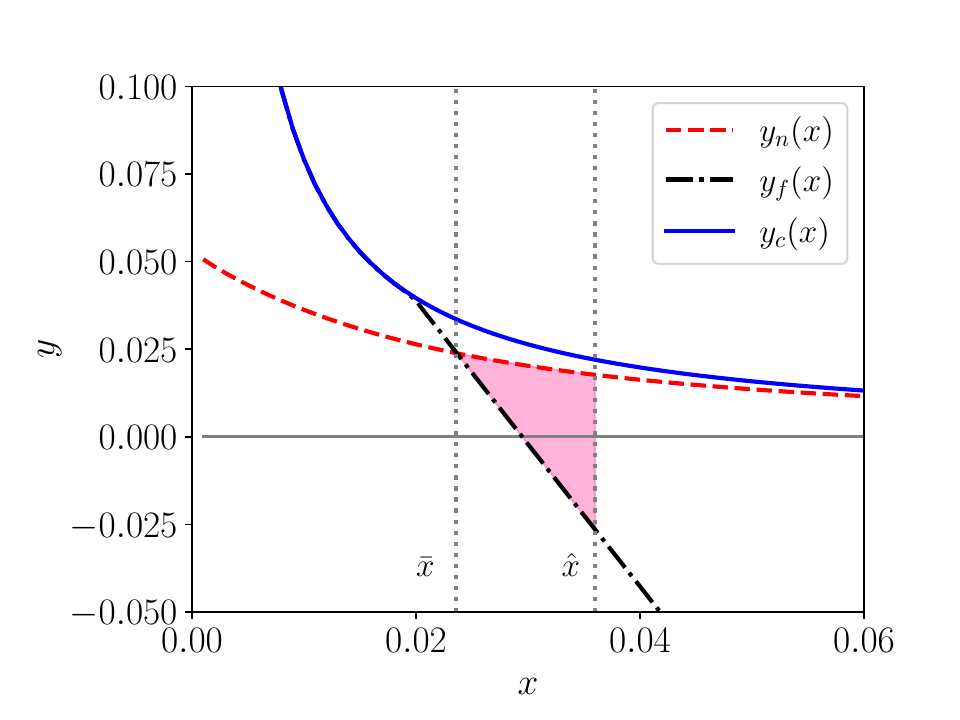}
\caption{Phase-transition diagram and sphaleron-decoupling line in the \((x,y)\) plane. The solid blue and dashed red curves denote the critical- and nucleation-temperature lines, respectively. The black dash-dotted curve marks sphaleron decoupling, defined by \(\Gamma_{\rm B}=H\). As the Universe cools, \(y\) decreases while \(x\) remains approximately constant, so the system moves downward in the diagram. Crossing the nucleation line triggers the transition from the symmetric to the broken phase. Above the decoupling line, \(\Gamma_{\rm B}>H\), while below it \(\Gamma_{\rm B}<H\). Two benchmark values, \(\bar{x}=0.025\) and \(\hat{x}=0.035\), are highlighted.\reproducedfromref{Li:2025kyo}}
\label{fig:yplot}
\end{figure}

To illustrate this point more clearly, we show in Fig.~\ref{fig:yplot} the phase-transition diagram and sphaleron-decoupling line in the \((x,y)\) plane. The solid blue and dashed red curves denote the critical- and nucleation-temperature lines, respectively. The black dash-dotted curve marks sphaleron decoupling, defined by \(\Gamma_{\rm B}=H\). As the Universe cools, \(y\) decreases while \(x\) remains approximately constant, so the system moves downward in the diagram. We highlight two benchmark values, \(\bar{x}=0.025\) and \(\hat{x}=0.035\). For any BSM model that can be mapped to the 3D SU(2)+Higgs theory, there are three possible scenarios for sphaleron decoupling and baryon preservation:
\begin{itemize}
  \item If \(x < \bar{x}\), the sphaleron-decoupling line lies above the nucleation line. This means that there is no baryon-number washout in the broken phase after the phase transition occurs, and therefore all of the generated baryon asymmetry can be preserved. This is the most favorable scenario for successful electroweak baryogenesis, since there is no need to compensate for any washout by generating a larger initial baryon asymmetry.
  \item If \(\bar{x} < x < \hat{x}\), there is some baryon-number washout in the broken phase after the phase transition, as indicated by the pink region in Fig.~\ref{fig:yplot}. In this case, only a fraction of the generated baryon asymmetry is preserved. The value of \(\mathcal{W}\) increases from 0 to \(\mathcal{O}(100)\) as \(x\) increases from \(\bar{x}\) to \(\hat{x}\), which means that the preserved baryon asymmetry decreases by several orders of magnitude across this interval. In this situation, one may compensate by choosing CP-violating parameters in the BSM model that generate a larger initial baryon asymmetry. Thus, successful electroweak baryogenesis is still possible, although it is more difficult than in the previous case.
  \item If \(x > \hat{x}\), then \(\mathcal{W}\gtrsim \mathcal{O}(100)\), so the preserved baryon asymmetry is exponentially suppressed. In this case, successful electroweak baryogenesis becomes very difficult. For generic BSM models, any physically reasonable CP-violating parameters will be unable to produce a sufficiently large initial baryon asymmetry to compensate for such a large washout factor, e.g.\ \(e^{-100}\). Thus, the final preserved asymmetry will typically be far below the observed value. Of course, this conclusion can be model dependent, and in principle there may exist special BSM scenarios that generate an exceptionally large initial baryon asymmetry. Nevertheless, in general it is difficult to realize successful electroweak baryogenesis in this regime.
\end{itemize}
Remarkably, our perturbative value of \(\bar{x}\) has been confirmed by the recent lattice simulation of Ref.~\cite{Annala:2025aci}, which studied the non-perturbative version of the \(y_f\) curve in Fig.~\ref{fig:yplot} using lattice determinations of the sphaleron rate.

\subsection{Application to the Real Triplet Extension of the SM}
\label{sec:application_to_RSM}
Among the various BSM theories, the real triplet extension of the Standard Model is one of the simplest extensions. It can lead to a strong first-order phase transition and can be searched for at collider experiments \cite{FileviezPerez:2008bj,Du:2018eaw,Chiang:2020rcv,Niemi:2018asa}. \flag The real triplet has also been extensively studied in the context of dark matter~\cite{Cirelli:2005uq,Cirelli:2008id,AbdusSalam:2013eya,Fischer:2013hwa}. Applying the 3D EFT technique to this model can therefore provide valuable insights and constraints for phase-transition dynamics, dark matter phenomenology, and collider analyses. Another advantage of the triplet model is that it serves as a useful benchmark, as it has been investigated in non-perturbative lattice simulations~\cite{Niemi:2020hto}, allowing for direct comparison between perturbative and lattice results. Moreover, triplet scalars appear in many BSM scenarios, such as the Georgi--Machacek model~\cite{Georgi:1985nv,Chanowitz:1985ug,Zhou:2018zli,Bian:2019bsn,Chen:2022zsh} and extended supersymmetric models~\cite{Bandyopadhyay:2015oga}. For these reasons, the real triplet model provides a well-motivated starting point for our analysis of thermal phase transitions.

\subsubsection{dimensional reduction and the relation to \texorpdfstring{$\overline{\rm MS}$}{MSbar} parameters in \texorpdfstring{$\Sigma$}{Sigma}SM}
In the real triplet extension of the Standard Model, the scalar sector is extended by an additional real scalar field, \(\Sigma\), which transforms as a triplet under the SU(2) gauge group. It can be written as
\begin{equation}
  \Sigma^a = 
  \left(\begin{array}{c} 
    \sigma^1 \\ 
    \sigma^2 \\ 
    \sigma^3 
  \end{array}\right) \ .
\end{equation}
Here \(\sigma^a\) are the components of the real triplet scalar field. The Lagrangian for the real triplet extension of the Standard Model is
\begin{align}
  \label{eq:lagrangian_RSM}
  \mathcal{L} & = \mathcal{L}_{\rm SM} + \frac{1}{2} (D_\mu \Sigma)^a (D^\mu \Sigma)^a + \frac{1}{2} \mu_\Sigma^2 \Sigma^a \Sigma^a + \frac{b_4}{4} (\Sigma^a \Sigma^a)^2 + \frac{1}{2} a_2 (H^\dagger H)(\Sigma^a \Sigma^a) \ ,
\end{align}
where the SM Lagrangian and parameters are given in Eq.~\eqref{eq:4d_lagrangian}. Following Ref.~\cite{Niemi:2018asa}, we assume \(\mu_\Sigma^2>0\)\footnote{Note that our definition in Eq.~\eqref{eq:lagrangian_RSM} differs by a sign in \(\mu_\Sigma^2\) from that in Ref.~\cite{Niemi:2018asa}.}, so that the zero modes of the triplet field are classified as heavy scales and can therefore be integrated out during the second step of dimensional reduction, recalling Table~\ref{tab:dr_hierarchy}. Under this setup, the final 3D EFT remains the SU(2)+Higgs theory, but with different values of the parameters \(x\) and \(y\) compared with the pure SM case. The non-zero Matsubara modes and zero Matsubara modes of the triplet field affect the Higgs mass parameter and self-coupling during the first and second steps of dimensional reduction, respectively, and thus modify the values of \(x\) and \(y\) in the final 3D EFT.

In the following, we use the Higgs self-coupling to illustrate how the triplet field can reduce its value. For clarity, we focus mainly on the one-loop corrections. After integrating out all non-zero Matsubara modes in the first step of dimensional reduction, the Higgs self-coupling becomes
\begin{align}
  \label{eq:lambda3_matching_effective_potential_final_RSM} 
\lambda_{3,{\rm (hard)}}=\lambda_{3,{\rm (hard)}}^{\rm SM} -\frac{3}{4(4\pi)^2} T a_2^2 L_b \ ,
\end{align}
where the subscript ``hard'' denotes the contribution from the non-zero Matsubara modes. The quantity \(\lambda_{3,{\rm (hard)}}^{\rm SM}\) is the pure-SM result derived in Eq.~\eqref{eq:lambda3_matching_effective_potential_final}\footnote{Note that the notation in this section differs from that in Table~\ref{tab:dr_hierarchy}: (i) after the first step of dimensional reduction, the notation changes from \(\lambda_3\) to \(\lambda_{3,{\rm (hard)}}\); (ii) after the second step of dimensional reduction, the notation changes from \(\bar{\lambda}_3\) to \(\lambda_3\). The reason for this change is that the sphaleron is discussed mainly after two steps of dimensional reduction, so we remove the bar in the final 3D EFT to simplify the notation. For example, the definitions of \(x\) and \(y\) in Eq.~\eqref{eq:x_y_parameters} are after two steps of dimensional reduction.}. The expression for \(L_b\) was defined previously in Eq.~\eqref{eq:log_terms_Lb_Lf}. We see that the non-zero Matsubara modes of the triplet field reduce the Higgs self-coupling, which is favorable for a strong first-order phase transition. After the second step of dimensional reduction, the zero modes of the triplet field further reduce the Higgs self-coupling:
\begin{align}
  \label{eq:lambda3_matching_effective_potential_final_RSM_after_second_step_DR}
\lambda_{3}=\lambda_{3,{\rm (hard)}} - \frac{3}{8\pi} \frac{h_3^2}{m_D} - \frac{1}{8\pi} \frac{3}{4} \frac{a_{2,3}^2}{\mu_{\Sigma,3}} \ ,
\end{align}
where \(m_D\) is the SU(2) Debye mass, and \(h_3\) is the Higgs--temporal portal coupling. The quantities \(a_{2,3}\) and \(\mu_{\Sigma,3}\) are the triplet--Higgs portal coupling and triplet mass parameter after the first step of dimensional reduction, respectively. The expressions for \(m_D\), \(h_3\), \(a_{2,3}\), and \(\mu_{\Sigma,3}\) can be found in Ref.~\cite{Niemi:2018asa}. Below, we give only the expressions for \(m_D\), \(h_3\), and \(a_{2,3}\):
\begin{align}
  \label{eq:parameters_after_first_step_DR_RSM}
  m_D^2 & = g^2 T^2 \left(\frac{4+N_d+2N_t}{6}+ \frac{N_f}{3} \right) \ , \\
  h_3 & = \frac{g^2 T}{4} \ , \\
  a_{2,3}
&= T\Bigg\{
a_2
+\frac{1}{(4\pi)^2}\Bigg[
2g^4
-3a_2 y_t^2 L_f \nonumber \\
& \qquad
-L_b\Bigg(
2a_2^2+5a_2 b_4+3g^4+6a_2\lambda-\frac{33}{4}a_2 g^2
\Bigg)
\Bigg]
\Bigg\}\ ,
\label{eq:a2_after_first_step_DR_RSM}
\end{align}
while the expression for \(\mu_{\Sigma,3}\) is lengthy and will not be shown here. Here \(N_d=1\) is the number of Higgs doublets, \(N_t=1\) is the number of triplet fields, and \(N_f=3\) is the number of fermion generations. We see that the zero modes of the triplet field further reduce the Higgs self-coupling, which is again favorable for a strong first-order phase transition. The larger the triplet--Higgs portal coupling \(a_2\), the smaller the Higgs self-coupling \(\lambda_3\). However, since the parameter that determines the characteristics of the phase transition is \(x=\lambda_3/g_3^2\), we must also examine how the triplet field affects the value of \(g_3^2\) in the final 3D EFT. The expression for \(g_3^2\) is
\begin{align}
  \label{eq:g3_matching_effective_potential_final_RSM_after_second_step_DR}
  g_3^2 = g_{3,{\rm (hard)}}^2 \left[1 - \frac{g_{3,{\rm (hard)}}^2}{24\pi}\left(\frac{1}{\mu_{\Sigma,3}} + \frac{1}{m_D} \right)\right] \ ,
\end{align}
where \(g_{3,{\rm (hard)}}^2\) is the contribution from the non-zero Matsubara modes:
\begin{align}
g_{3,{\rm (hard)}}^2
= g^2\,T
\left[
1+\frac{g^2}{(4\pi)^2}
\left(
\frac{44-N_d-2N_t}{6}\,L_b
+\frac{2}{3}
-\frac{4N_f}{3}\,L_f
\right)
\right] \ .
\end{align}
From Eq.~\eqref{eq:g3_matching_effective_potential_final_RSM_after_second_step_DR}, we see that the zero modes of the triplet field also reduce the SU(2) gauge coupling in the final 3D EFT, but this effect is suppressed by the small factor \(g_{3,{\rm (hard)}}^2/(24\pi)\). Thus, the effect of the triplet field on \(g_3^2\) is much smaller than its effect on \(\lambda_3\). As a result, the triplet field reduces the value of \(x=\lambda_3/g_3^2\) in the final 3D EFT, and can therefore lead to a strong first-order phase transition.

Before presenting the numerical results, we discuss how the physical input parameters are mapped to the parameters \(x=\lambda_3/g_3^2\) and \(y=\mu_3^2/g_3^4\) in the final 3D EFT. We use \(\lambda_3\) as an illustration. As can be seen from Eqs.~\eqref{eq:parameters_after_first_step_DR_RSM} and \eqref{eq:a2_after_first_step_DR_RSM}, the final expanded expression for \(\lambda_3\) in Eq.~\eqref{eq:lambda3_matching_effective_potential_final_RSM_after_second_step_DR} is a function of the parent 4D parameters, such as the Higgs mass parameter \(\mu^2\), the Higgs self-coupling \(\lambda\), the SU(2) gauge coupling \(g\), the top Yukawa coupling \(y_t\), the Higgs--triplet portal coupling \(a_2\), and other parameters. The question is how these parameters are determined by the physical input parameters. For the pure-SM parameters, such as \(\mu^2\), \(\lambda\), \(y_t\), and \(g\), they can be determined from physical input quantities such as the Higgs mass \(M_H\), the \(W\)- and \(Z\)-boson masses \(M_W\) and \(M_Z\), the top-quark mass \(M_t\), and the Fermi constant \(G_F\). These 4D parameters can be related to the physical input parameters in the \(\overline{\rm MS}\) renormalization scheme, as shown in detail in Appendix~\ref{app:MSbar_parameters_and_physical_input_quantities}. As demonstrated there, these parent 4D parameters depend on the renormalization scale \(\bar{\mu}\). For example, the \(g^2(\bar{\mu})\) appearing in Eq.~\eqref{eq:parameters_after_first_step_DR_RSM} is related to the physical input quantities and depends on the renormalization scale, as shown in Eq.~\eqref{eq:one_loop_correction_gauge_coupling_a}. Similar expressions for \(\lambda(\bar{\mu})\), \(y_t(\bar{\mu})\), and \(\mu^2(\bar{\mu})\) can be found in the same appendix. For the portal coupling \(a_2\) and the parameters associated with the triplet field, since we have no direct measurements of them, we treat them as free parameters and vary them to study how they affect the phase transition and sphaleron rate. We note that the results presented in Appendix~\ref{app:MSbar_parameters_and_physical_input_quantities} apply only to the pure SM. The triplet field modifies these expressions. Such effects appear at least in two places: (i) the self-energy expressions in Eq.~\eqref{eq:Higgs_self_energy_one_loop} are modified by the triplet field, although the resulting expressions are lengthy and will not be shown here; (ii) the renormalization-group equations for the running of these parameters in Eq.~\eqref{eq:RGEs_SM_parameters_mu2} are also modified by the triplet field, as shown in Ref.~\cite{Niemi:2018asa}. For illustration, we list the modifications to the beta functions of the Higgs mass parameter and quartic coupling, the SU(2) gauge coupling, and the relevant \(a_2\) and \(\mu_\Sigma^2\) parameters:
\begin{align}
  \bar{\mu} \frac{d }{d \bar{\mu}}\mu^2(\bar{\mu}) & = \beta_{\mu^2}^{\rm SM} + \frac{1}{16\pi^2} 3 a_2 \mu_\Sigma^2 \ , \\
  \bar{\mu} \frac{d }{d \bar{\mu}}g^2(\bar{\mu}) & = \beta_{g^2}^{\rm SM} + \frac{1}{8\pi^2} \frac{1}{3} N_t g^4 \ , \\
  \bar{\mu} \frac{d }{d \bar{\mu}}\lambda(\bar{\mu}) & = \beta_{\lambda}^{\rm SM} + \frac{3}{32\pi^2} a_2^2 \ , \\
  \bar{\mu} \frac{d}{d\bar{\mu}}\mu_\Sigma^{2}(\bar{\mu})
&= \frac{1}{16\pi^{2}}\,2\left(2a_{2}\mu^{2}-6g^{2}\mu_\Sigma^{2}+5b_{4}\mu_\Sigma^{2}\right) \ , \\
\bar{\mu} \frac{d }{d\bar{\mu}}a_2(\bar{\mu})
&= \frac{1}{16\pi^2}\,2\left[
a_2\left(-\frac{33}{4}g^2+3y_t^2+2a_2+5b_4+6\lambda\right)+3g^4
\right] \ ,
\end{align}
where \(\beta_{\mu^2}^{\rm SM}\), $\beta_{g^2}^{\rm SM}$ and \(\beta_{\lambda}^{\rm SM}\) are the beta functions for the Higgs mass parameter, SU(2) gauge coupling, and quartic coupling in the pure SM, which can be read from Eqs.~\eqref{eq:RGEs_SM_parameters_mu2} to \eqref{eq:RGEs_SM_parameters_lambda}. We see that the triplet field modifies the running of these parameters. When these parameters are evolved from \(\bar{\mu}=M_Z\) to the matching scale \(\bar{\mu}=4\pi T/e^{\gamma_E}\), the final values of \(x\) and \(y\) in the 3D EFT are affected by the triplet field. We also note that including two-loop corrections in dimensional reduction largely cancels the renormalization-scale dependence of these parameters \cite{Gould:2021oba}.

\subsubsection{sphaleron rate and baryon preservation condition in \texorpdfstring{$\Sigma$}{Sigma}SM}
\label{sec:sphaleron_rate_and_BPC_in_RSM}
Using the method above, we can map the parameters in the \(\Sigma\)SM to the parameters \(x\) and \(y\) in the final 3D EFT. Since we are interested in the baryon preservation condition, the value of \(x\) is the most important quantity, according to the discussion around Fig.~\ref{fig:yplot}. We can vary the portal coupling \(a_2\) and the triplet mass \(M_\Sigma\), for a chosen value of \(b_4\), to study the value of \(x\) in the \((a_2,M_\Sigma)\) plane. Before presenting the numerical results, we point out that the \(\Sigma\)SM setup used here has no additional CP-violation source, since all BSM parameters are assumed to be real. However, we use this model only as a benchmark to illustrate how the sphaleron decoupling condition and baryon preservation condition can be applied to a specific BSM model, and we assume that a net baryon-number density is generated outside the bubble wall by some CP-violation source, which is not the focus of this work. Since the final 3D EFT remains the SU(2)+Higgs theory, the new baryon preservation condition developed here, based on the value of \(x\), is not restricted to this specific model. Any BSM model that can be mapped to the 3D SU(2)+Higgs theory obeys the same baryon preservation condition in terms of \(x\). Therefore, the results presented here for the \(\Sigma\)SM can also be applied to other BSM models, provided that they can be mapped to the 3D SU(2)+Higgs theory.

\begin{figure}[t]
\centering
\includegraphics[width=0.7\textwidth]{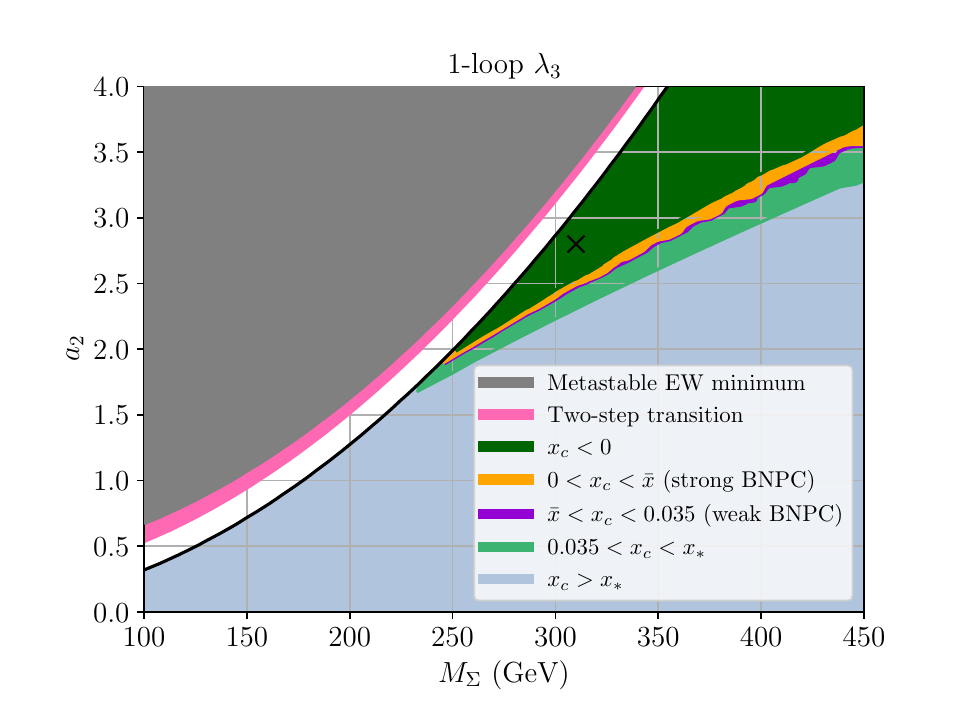}
\caption{Phase structure and baryon-number preservation condition in the \((a_2,M_\Sigma)\) plane from the one-loop result for \(\lambda_3\) in the \(\Sigma\)SM. The solid black line denotes \(\mu_\Sigma^2=0\). Below this line, \(\mu_\Sigma^2>0\), where our EFT is valid; the white region corresponds to \(\mu_\Sigma^2<0\), where our EFT is not valid. The pink region denotes a two-step phase transition, which is not the focus of this work. The grey region indicates that the electroweak minimum is unstable. Below the solid black line, the different colors represent different values of \(x\) for the one-step phase transition from the symmetric phase to the electroweak minimum. This is the region of primary interest here.\reproducedfromref{Li:2025kyo}}
\label{fig:sigmaSM_x_a2_Msigma_one_loop}
\end{figure}

We first choose \(b_4=0.25\), and present the one-loop numerical results for the value of \(x\) in the \((a_2,M_\Sigma)\) plane in Fig.~\ref{fig:sigmaSM_x_a2_Msigma_one_loop}. The region of interest is below the solid black line, where \(\mu_\Sigma^2>0\), so our EFT is valid. Below this line, the phase transition is also the one-step transition from the symmetric phase to the electroweak minimum, which is our main focus. The orange region in the figure corresponds to \(x<\bar{x}=0.025\), which implies no baryon-number washout in the broken phase after the phase transition, so all generated baryon asymmetry can be preserved. The purple region corresponds to \(\bar{x}=0.025<x<\hat{x}=0.035\), which implies partial baryon-number washout in the broken phase after the phase transition, so only a fraction of the generated baryon asymmetry can be preserved. The light-green region corresponds to \(\hat{x}=0.035<x<x_*\approx0.1\), i.e. it is still in the first-order phase-transition region, but the preserved baryon asymmetry is exponentially suppressed, making successful electroweak baryogenesis very difficult. The light-blue region corresponds to \(x>x_*\approx0.1\), where there is no first-order phase transition.

We note that there is a large parameter region with \(x<0\), meaning that the one-loop result for \(\lambda_3\) is negative and the perturbative result is therefore not reliable there. We mark one such point with a cross (\(\textsf{x}\)), at \(M_\Sigma=310\) GeV and \(a_2=2.8\), which has been verified by lattice simulation \cite{Niemi:2020hto} to exhibit a first-order phase transition. This indicates that our one-loop result for \(\lambda_3\) in Eq.~\eqref{eq:lambda3_matching_effective_potential_final_RSM_after_second_step_DR} is not reliable in that region, and that two-loop corrections are needed for better accuracy. The origin of this failure can be inferred from Eq.~\eqref{eq:lambda3_matching_effective_potential_final_RSM_after_second_step_DR}, where the triplet expansion term is proportional to \(a_2^2/\mu_{\Sigma,3}\). Since a large value of \(a_2\) is required to trigger a first-order phase transition, the convergence of the triplet contribution is much slower than that of the other terms, so two-loop corrections are necessary \cite{Li:2025kyo}.

The two-loop correction from the triplet field to the Higgs self-coupling is \cite{Li:2025kyo,Niemi:2018asa}
\begin{align}
\lambda_3
&= \lambda_{3,\mathrm{one-loop}}
+\frac{1}{(4\pi)^2\mu_{\Sigma,3}^2}
\left(
\frac{3}{4}a_{2,3}^3
-\frac{3}{2}a_{2,3}^2 g_3^2
+\frac{3}{8}a_{2,3}g_3^4
-\frac{1}{128}\left(3g_3^6+g_3^4 g_3'^2\right)
\right) \ ,
\end{align}
where \(\lambda_{3,\mathrm{one-loop}}\) is the one-loop result for \(\lambda_3\) in Eq.~\eqref{eq:lambda3_matching_effective_potential_final_RSM_after_second_step_DR}. The two-loop triplet correction can increase \(\lambda_3\), and therefore can shift the negative-\(x\) region in Fig.~\ref{fig:sigmaSM_x_a2_Msigma_one_loop} to positive values, improving the reliability of perturbation theory there.

\begin{figure}[t]
\centering
\includegraphics[width=0.5\textwidth]{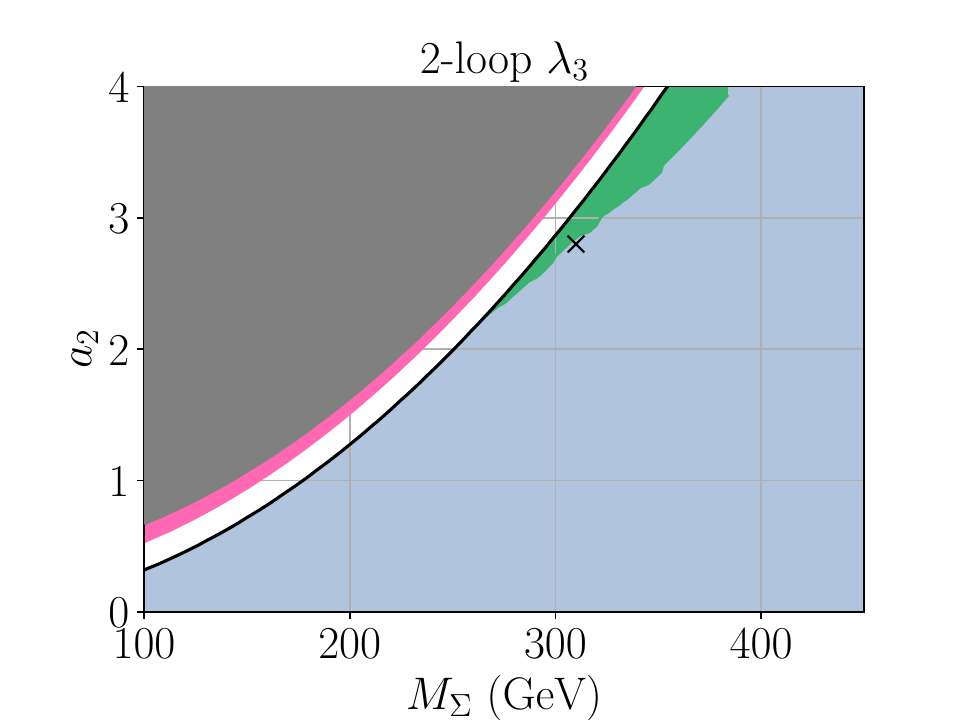}
\caption{Phase structure and baryon-number preservation condition in the \((a_2,M_\Sigma)\) plane from the two-loop result for \(\lambda_3\) in the \(\Sigma\)SM. The meaning of the different color regions is the same as in the one-loop result shown in Fig.~\ref{fig:sigmaSM_x_a2_Msigma_one_loop}.\reproducedfromref{Li:2025kyo}}
\label{fig:sigmaSM_x_a2_Msigma_two_loop}
\end{figure}

We now present the two-loop numerical results for the value of \(x\) in the \((a_2,M_\Sigma)\) plane in Fig.~\ref{fig:sigmaSM_x_a2_Msigma_two_loop}. Below the solid black line, the different colors represent different values of \(x\) for the one-step phase transition from the symmetric phase to the electroweak minimum. We see that the two-loop correction increases \(x\), and therefore shifts the negative-\(x\) region in Fig.~\ref{fig:sigmaSM_x_a2_Msigma_one_loop} to positive values. We also see that the entire first-order phase-transition region has \(0.035<x<0.1\), which implies strong washout of the generated baryon asymmetry in the broken phase after the phase transition, making successful electroweak baryogenesis very difficult in this case. Therefore, the baryon-preservation conclusion from the two-loop result is qualitatively different from that from the one-loop result, indicating that two-loop corrections are crucial for studying both the phase transition and baryon preservation.

\newpage
\section{Electroweak Baryogenesis: Impact of the First-Order Phase Transition}
\label{sec:DM-from-TFC-AND_PT}

Since a first-order phase transition is a key ingredient for successful electroweak baryogenesis, it can have important cosmological implications for the Universe. In this section, we discuss some of these implications, including the production of primordial black holes and the indirect probing of a first-order phase transition through collider searches for BSM particles.

\subsection{Primordial Black Hole Production from a Delayed First-Order Phase Transition}
\label{sec:PBH_from_delayed_FOPT}

Since PBH are potential dark-matter candidates \cite{Zeldovich:1967lct,Hawking:1971ei,Carr:1974nx,Carr:2016drx,Carr:2020gox}, it is interesting to study PBH production from a first-order phase transition. Since the total dark-matter relic density of our Universe is known, if a first-order phase transition (FOPT) produces PBH with a relic density exceeding the observed dark-matter relic density, then the corresponding FOPT parameter space can be excluded. We note that many existing studies of PBH production are carried out in simplified scalar models, although the underlying ideas and methods can also be extended to more realistic scenarios. In this thesis, we employ a simple scalar model to illustrate the essential phase-transition dynamics and their implications for PBH dark matter. At the same time, this framework can be generalized to more realistic models, such as the \(\Sigma\)SM discussed in the previous section.

PBH production from a delayed (or supercooled) first-order phase transition has been studied in the literature. The basic idea is that, during a first-order phase transition, bubbles of the new phase nucleate and expand. If some patch of the Universe remains in the false vacuum for a sufficiently long time, its energy density can become large enough, compared with that of the surrounding regions that have already transitioned to the true vacuum. If the density contrast between the true-vacuum and false-vacuum ``island'' exceeds a certain threshold, the false-vacuum region can collapse to form a primordial black hole \cite{Liu:2021svg}. This criterion is analogous to the curvature-perturbation criterion for PBH formation from the inflationary epoch. Ref.~\cite{Flores:2024lng} cast doubt on this density-contrast criterion, and it was studied recently in Ref.~\cite{Ning:2026nfs} that the above condition remains valid in the thin-wall approximation. We will comment further on this point when we discuss it in more detail below.

Following Refs.~\cite{Kanemura:2024pae,Dine:1992wr,Wu:2024lrp}, we consider the following simple model with a real scalar field \(\phi\):
\begin{equation}
 \begin{aligned}
V(\phi,T)&=\frac{1}{2}\left( \frac{\mu_3 \omega - m_\phi^2}{2}+c T^2\right)\phi^2 -\frac{\mu_3}{3}\phi^3 + \frac{m_\phi^2}{8\omega^2} \left( 1+\frac{\mu_3 \omega}{m_\phi^2}\right)\phi^4 \ .
\end{aligned}
\end{equation}
Here \(m_\phi\), \(\omega\), \(c\), and \(\mu_3\) are free parameters of the model. The parameter \(\mu_3\) measures the height of the barrier of this potential, while the parameter \(c\) controls the thermal-mass correction to the scalar field. We can vary these parameters to study how they affect the phase-transition dynamics and the resulting PBH production. One benefit of this model is that it admits an analytic expression for the bubble-nucleation rate. The bubble-nucleation rate per unit volume can be expressed as
\begin{align}
  \Gamma(T)\simeq T^4 \left(\frac{S_3(T)}{2\pi T}\right)^{3/2}e^{-S_3(T)/T}\ .
\end{align}
Here \(S_3(T)\) is the three-dimensional Euclidean action. The factor \(T^4\) provides the correct mass dimension, and the factor \(\left(\frac{S_3(T)}{2\pi T}\right)^{3/2}\) comes from the zero-mode contribution. One can find a similar zero-mode contribution for the SM \(SU(2)\)+Higgs theory in Sec.~\ref{sec:sphaleron_rate_beyond_leading_order}, from which the exponent \(3/2\) can be understood. The expression for \(S_3(T)\) in this model is
\begin{align}
\frac{S_3(T)}{T}\simeq \frac{123.48 (\mu^2+c T^2)^{3/2}}{2^{3/2}T\mu_3^2}f\left(\frac{9(\mu^2+c T^2)\lambda}{2\mu_3^2} \right)\ ,
\end{align}
where
\begin{align}
\mu^2=\frac{\mu_3 \omega - m_\phi^2}{2},\quad \lambda=\frac{\mu_3 \omega + m_\phi^2}{2 \omega^2}\ ,
\end{align}
and
\begin{align}
  f(\mu)=1+\frac{\mu}{4}\left(1+\frac{2.4}{1-\mu}+\frac{0.26}{(1-\mu)^2}\right)\ .
\end{align}
Note that the above expressions work well for \(\mu\in[0,1]\).

We now consider the process of a delayed first-order phase transition and the resulting PBH production. First, we consider the fraction of the Universe that remains in the false vacuum at time \(t\) per unit volume, which can be expressed as \cite{Turner:1992tz}
\begin{align}
  \label{eq:false_vacuum_fraction}
P(t)=e^{-F(t)}\ .
\end{align}
Here
\begin{align}  \label{eq:F_t_expression}
F(t)=\int_{t_c}^t dt^\prime \Gamma(t^\prime) V_{\rm phys}(t,t^\prime)\ ,
\end{align}
where \(t_c=0\) represents the time at which the Universe is at the critical temperature \(T_c\), and \(F(t)\) measures how many times a point in the Universe is swept by the bubble wall. In other words, \(F(t)\) denotes the average number of bubbles that have passed through a given point in space by time \(t\). Suppose bubble nucleation follows a Poisson distribution. Then \(P(t)\) in Eq.~\eqref{eq:false_vacuum_fraction} can be understood as the probability that a given point in space has not been swept by any bubble wall by time \(t\). It therefore represents the probability that the point is still in the false vacuum at time \(t\), and can also be understood as the fraction of false vacuum at time \(t\) per unit volume. We now consider an infinitesimal time interval \([t^\prime,t^\prime+dt^\prime]\), and compute how many times a point in the Universe at a later time \(t\) is swept by the bubble wall during this interval. For a bubble produced at time \(t^{\prime\prime}\in [t^\prime,t^\prime+dt^\prime]\), the bubble grows with velocity \(v_w\). It is benefit to revert to the viewpoint of a specific point \(P\), for example. At time \(t\), the bubble wall passes through the point \(P\) if the bubble is produced within a distance \(v_w(t-t^{\prime\prime})\) from \(P\). Furthermore, it is convenient to work in comoving coordinates, where the volume around \(P\) can be written as \(V_{\rm com}=4\pi\gamma(t,t^\prime)^3/3\), with \(\gamma(t,t^\prime)\) the comoving distance:
\begin{align}
\gamma(t,t^\prime)=\int_{t^\prime}^t \frac{v_w}{a(t^{\prime\prime})} dt^{\prime\prime}\ ,
\end{align}
where \(a(t)\) is the scale factor and \(v_w\) is the bubble-wall velocity. In this study, we simply set the wall velocity equal to the speed of light. The computation of the wall velocity is also important for electroweak baryogenesis and for predicting the gravitational-wave signal from a first-order phase transition. For dedicated studies of the wall velocity, see Refs.~\cite{Steinhardt:1981ct,Ignatius:1993qn,Arnold:1993wc,Moore:1995ua,Moore:1995si,Espinosa:2010hh,Konstandin:2014zta,Bodeker:2017cim,Azatov:2021ifm,Gouttenoire:2021kjv,Laurent:2022jrs,DeCurtis:2022hlx,Azatov:2023xem,Ai:2023suz,Li:2023xto,Wang:2023lam,Ai:2023see,Ramsey-Musolf:2025jyk,Ai:2025bjw,Si:2025vdt,vandeVis:2025plm,Branchina:2025adj,Munzenberg:2025nwa}. Thus, for a bubble produced during \([t^\prime,t^\prime+dt^\prime]\) with an initial zero radius, the physical volume, from the viewpoint of \(P\), that can be swept by the bubble wall at time \(t\) can be expressed as
\begin{align}
  V_{\rm phys}(t,t^\prime) & = a(t^\prime)^3 V_{\rm com}(t,t^\prime) \nonumber \\
  & = \frac{4\pi}{3} a(t^\prime)^3 \gamma(t,t^\prime)^3 \ .
\end{align}
Thus, the expression for \(F(t)\) becomes
\begin{align}
F(t) & = \int_{t_c}^t dt^\prime \Gamma(t^\prime) V_{\rm phys}(t,t^\prime) \nonumber \\
& = \frac{4\pi}{3} \int_{t_c}^t dt^\prime \Gamma(t^\prime)  a(t^\prime)^3 \gamma(t,t^\prime)^3 \ .
\end{align}
Here \(\Gamma(t^\prime)\) denotes the nucleation rate at time \(t^\prime\). One may insist on replacing \(t_c\) here by the time when the first bubble is nucleated, \(t_i\) \cite{Liu:2021svg}. We note that, at the critical time, the bubble-nucleation rate is zero. Therefore, the interval error from replacing \(t_c\) with \(t_i\) may be negligible, which can be verified numerically. Before the critical time, \(F=0\); after the critical time, \(F\) increases with time.

We now discuss the PBH production mechanism from a delayed first-order phase transition proposed in Ref.~\cite{Liu:2021svg}; this mechanism is also called the accumulating mechanism. The key point is that bubble nucleation is a stochastic process, which allows us to divide the Universe into normal patches and delayed patches. A delayed patch is defined as a patch that has not been swept by any bubble wall until a certain time \(t_d\). In other words, for a delayed patch, we can replace the lower limit of the integral in \(F(t)\) by \(t_d\), where \(t_d=t_c+\Delta t\), and \(\Delta t\) is the delay time induced by stochasticity. Since the vacuum-energy density and the radiation-energy density scale differently with the expansion of the Universe, after some time the vacuum-energy density in the delayed patch can become dominant compared with that in the normal patch, and the delayed patch can collapse to form a PBH \cite{Liu:2021svg,Cai:2024nln,Kawana:2022olo}. This idea is illustrated in Fig.~\ref{fig:PBH_production_from_delayed_FOPT}. In that figure, the blue (green) background represents the false (true) vacuum. In panels (a) and (b), for illustration, we divide the Universe into nine patches, where the middle patch is the delayed patch that has not been swept by any bubble wall until a certain time \(t_d\). Panels (a) and (b) represent the time interval between the critical time and \(t_d\), and the time after \(t_d\), respectively. After some time, the energy density of the middle delayed patch is higher than that of its surrounding normal patches, and there is a probability that the delayed patch collapses to form a PBH.

\begin{figure}[t]
\centering
\includegraphics[width=1.0\textwidth]{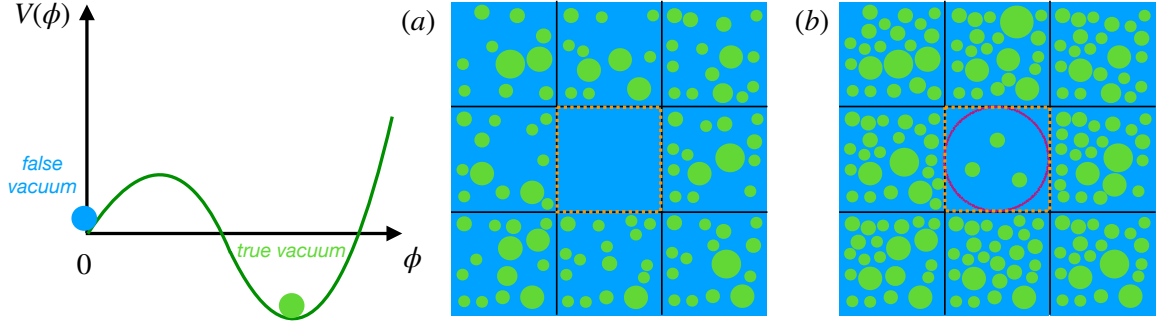}
\caption{Illustration of PBH production from a delayed first-order phase transition. The blue (green) background represents the false (true) vacuum, as shown in the left panel of the effective potential. The true-vacuum regions arise from bubble nucleation, as indicated by the green circles. In the right panels, (a) and (b), for illustration, the Universe is divided into nine patches, where the middle patch is the delayed patch with delay time \(t_d\). Panel (a) represents the time interval between the critical time and \(t_d\), while panel (b) represents the time after \(t_d\). After some time, the energy density of the middle patch is higher than that of the surrounding normal patches, and there is a probability that the delayed patch collapses to form a PBH, as shown by the red circle in panel (b).}
\label{fig:PBH_production_from_delayed_FOPT}
\end{figure}

We now discuss the evolution of the energy density of the Universe. The total energy density is composed of two parts, the vacuum-energy density and the radiation-energy density:
\begin{align}
\rho_{\rm tot}=\rho_{\rm vac}+\rho_{\rm rad}\ .
\end{align}
Assuming a superfast wall velocity, the energy of the bubble wall is included in the radiation energy \cite{Liu:2021svg}. Thus, the radiation-energy density can be expressed as \cite{Kawana:2022olo}
\begin{align}
\rho_{\rm rad}= \rho_{\rm SM} + \rho_{\rm heat} + \rho_{\rm wall}\ ,
\end{align}
where \(\rho_{\rm SM}\) is the original radiation-energy density from SM particles, \(\rho_{\rm heat}\) is the energy density associated with the latent heat released during the phase transition, and \(\rho_{\rm wall}\) is the energy density stored in the bubble walls. The vacuum-energy density can be expressed as\footnote{Strictly speaking, the vacuum-energy density discussed here is the ``average'' vacuum-energy density, since \(P(t)\) is defined as the average fraction of the Universe that remains in the false vacuum at time \(t\) per unit volume. One should in principle place a bar over these quantities \cite{Kawana:2022olo}. For notational clarity, we do not do so; one should simply keep in mind that these are average quantities.}
\begin{align}
\rho_{\rm vac}= P(t) \Delta V\ ,
\end{align}
where \(\Delta V\) is the vacuum-energy difference between the false and true vacua. The quantity \(P(t)\) is the average fraction of the Universe that remains in the false vacuum at time \(t\) per unit volume, as given in Eq.~\eqref{eq:false_vacuum_fraction}. The evolution of the radiation- and vacuum-energy densities reads
\begin{align}
  \frac{d \rho_{\rm rad}}{d t} + 4 H \rho_{\rm rad} & = -\frac{d \rho_{\rm vac}}{d t}\ .
\end{align}
Here the Hubble parameter \(H\) is determined from the Friedmann equation:
\begin{align}
H^2=\frac{8\pi }{3 M_{\rm Pl}^2} \rho_{\rm tot}\ ,
\end{align}
where \(M_{\rm Pl} = 1.22 \times 10^{19} \, \text{GeV}\) is the Planck mass. Since the decrease of the vacuum energy in the delayed patch is slower than that in the normal patch, the energy-density contrast between the delayed and normal patches can increase with time. There is then a probability that the delayed patch collapses to form a PBH if the contrast exceeds the critical limit:
\begin{align}
  \label{eq:PBH_formation_condition}
  \frac{\rho_{\rm del}}{\rho_{\rm nor}} - 1 > \delta_c\ ,
\end{align}
where \(\rho_{\rm del}\) and \(\rho_{\rm nor}\) are the energy densities in the delayed and normal patches, respectively, and \(\delta_c \sim 0.45\) is the critical density contrast computed from numerical simulations of curvature perturbations during radiation domination \cite{Niemeyer:1999ak,Shibata:1999zs,Musco:2004ak,Harada:2013epa}. We note that there have been recent criticisms of using the critical density contrast \(\delta_c\) for PBH formation from a first-order phase transition \cite{Flores:2024lng}. Recent full numerical simulations \cite{Ning:2026nfs} indicate that the above critical condition is still valid if the thin-wall approximation holds. However, if the thin-wall approximation is violated, the critical condition can have large uncertainties. In the present study, we still assume the thin-wall approximation, and defer a careful analysis of the more general condition to future work.

The mass of the PBH can be estimated as
\begin{align}
M_{\rm pbh} \simeq \gamma \rho_{\rm del}(t_{\rm pbh}) V_{\rm del}(t_{\rm pbh})\ ,
\end{align}
where \(\gamma \sim 0.2\) is a numerical factor defined as the ratio between the PBH mass and the Hubble mass \cite{Carr:1975qj}. For a recent accurate numerical simulation of this factor, see Ref.~\cite{Musco:2020jjb}. Here \(t_{\rm pbh}\) is the time at which the PBH forms, or equivalently, the time at which the energy-density contrast exceeds the critical limit given in Eq.~\eqref{eq:PBH_formation_condition}. The quantity \(\rho_{\rm del}(t_{\rm pbh})\) is the energy density in the delayed patch at time \(t_{\rm pbh}\), and \(V_{\rm del}(t_{\rm pbh})=4\pi H_{\rm del}^{-3}(t_{\rm pbh})/3\) is the physical volume of the delayed patch at that time. The PBH energy density reads
\begin{align}
  \rho_{\rm PBH} = P(t_d) \frac{M_{\rm pbh}}{V_{\rm nor}}\ ,
\end{align}
where \(V_{\rm nor}\) is the physical volume of the normal patch at time \(t_{\rm pbh}\). The quantity \(P(t_d)\) is defined as the probability that there is no bubble nucleation in the delayed patch from the critical time to the delayed time \(t_d\), which can be expressed as
\begin{align}
  P(t_d)= \exp\left(-\int_{t_0}^{t_d} \Gamma_{\rm del}(t) a_{\rm del}^3(t) V_{\rm com}(t_{\rm pbh}) \, dt\right)\ ,
\end{align}
where \(\Gamma_{\rm del}(t)\) is the bubble-nucleation rate in the delayed patch at time \(t\), \(a_{\rm del}(t)\) is the scale factor in the delayed patch at time \(t\), and \(V_{\rm com}(t_{\rm pbh})=a_{\rm del}^{-3}(t_{\rm pbh}) H_{\rm del}^{-3}(t_{\rm pbh})\) is the comoving volume of the delayed patch at time \(t_{\rm pbh}\). The physical meaning of \(P(t_d)\) can be understood from panel (a) of Fig.~\ref{fig:PBH_production_from_delayed_FOPT}, where it accounts for the probability that there is no bubble nucleation in the delayed patch from the critical time to the delayed time \(t_d\). In principle, the probability shown in panel (a) of Fig.~\ref{fig:PBH_production_from_delayed_FOPT} should read \(P(t_d)\times (1-P(t_d))^n\), where \(n\) is the number of normal patches surrounding the delayed patch. However, since \(P(t_d)\) is typically very small \cite{Wu:2024lrp}, we can approximate this probability simply by \(P(t_d)\).

Finally, we consider the relic density of the PBH. In our simple model, the formed PBH mass is of order the solar mass \cite{Wu:2024lrp,Kanemura:2024pae}, so the Hawking-radiation effect can be neglected, and its relic density scales as cold matter with entropy conservation. We define the PBH relic abundance as
\begin{align}
  \Omega_{\rm pbh} = \frac{\rho_{\rm pbh}^0}{\rho_0}\ ,
\end{align}
where \(\rho_{\rm pbh}^0=\rho_{\rm pbh} s_0/s\) is the PBH energy density today, \(s\) is the entropy density at the time of PBH formation, and \(s_0\) and \(\rho_0\) are the entropy density and total energy density today, respectively. These cosmological values can be found in Ref.~\cite{ParticleDataGroup:2024cfk}. It is also useful to define \(f_{\rm pbh}\) as the fraction of PBH in the total dark-matter relic density:
\begin{align}
  f_{\rm pbh} = \frac{\Omega_{\rm pbh}}{\Omega_{\rm DM}}\ ,
\end{align}
where \(\Omega_{\rm DM}\) is the total dark-matter relic density.

We now present the numerical results for the PBH relic-density ratio \(f_{\rm pbh}\) in the simple model considered above. We first choose five benchmark points (BMs) in the parameter space of this model, as shown in Table~\ref{tab:BM_values_in_fpbh_u3_u3star}. For all benchmarks, we fix \(m_\phi=300\) MeV and choose \(\mu_3^*\) such that \(f_{\rm pbh}=1\). We then present the numerical results for \(f_{\rm pbh}\) as a function of \(\mu_3\) for these benchmark points in Fig.~\ref{fig:fpbh_mu3_mu3star}. The upper panel shows \(f_{\rm pbh}\) versus \(\mu_3\) for BM1. The lower panel displays \(f_{\rm pbh}\) versus \(\mu_3/\mu_3^*\) for all five benchmark points. We see that the value of \(\mu_3\) has a significant impact on the resulting PBH relic density. Any small variation in \(\mu_3\) can lead to a large change in the PBH relic density, placing a strong constraint on the parameter space of this model.

\begin{table}[t]
  \centering
  \caption{Summary of the five benchmark models. For all benchmarks, we fix \(m_\phi = 300\) MeV and choose \(\mu_3^*\) such that \(f_{\text{pbh}}=1\). The quantities labeled with the subscript 1 denote the BM1 input values, while for BM2--BM5 the listed parameters are given as shifts relative to BM1. The last column reports the phase-transition strength, \(v_c/T_c\), where \(v_c\) is the vacuum expectation value at the critical temperature \(T_c\). This table is reproduced from Ref.~\cite{Wu:2024lrp}.}
  \begin{tabular}{lcccc}
    \toprule
     & $\boldsymbol{\omega}$ (MeV) & $\boldsymbol{c}$ & $\boldsymbol{\mu_3^*}$ (MeV) & $\boldsymbol{v_c/T_c}$ \\ \midrule
    \textbf{BM1:} & $\omega_1=$860 & $c_1=$0.140 & 160.203 & 2.26 \\ \hline
    \textbf{BM2:} & $\omega_1+34.29$ & $c_1-0.017$ & $153.659$ & 2.20 \\ \hline
    \textbf{BM3:} & $\omega_1+34.29$ & $c_1-0.06$ & $160.066$ & 1.85 \\ \hline
    \textbf{BM4:} & $\omega_1+11.43$ & $c_1-0.026$ & $160.419$  & 2.10 \\ \hline
    \textbf{BM5:} & $\omega_1+80$ & $c_1-0.009$ & $142.557$ & 2.33 \\ \bottomrule
  \end{tabular}
\label{tab:BM_values_in_fpbh_u3_u3star}
\end{table}

\begin{figure}[t]
  \centering
\includegraphics[width=0.6\textwidth]{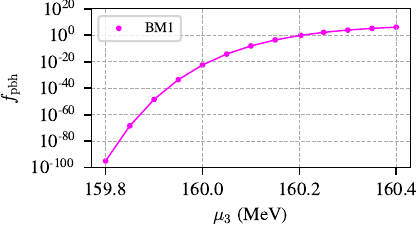}\\%
\includegraphics[width=0.6\textwidth]{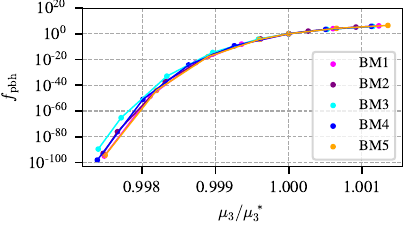}%
  \caption{\(f_{\text{pbh}}\) as a function of \(\mu_3\) for the benchmark points listed in Table~\ref{tab:BM_values_in_fpbh_u3_u3star}. The upper panel shows \(f_{\text{pbh}}\) versus \(\mu_3\) for BM1. The lower panel displays \(f_{\text{pbh}}\) versus \(\mu_3/\mu_3^*\) for all five benchmark points, where \(\mu_3^*\) is defined in each benchmark by the condition \(f_{\text{pbh}}=1\).\reproducedfromref{Wu:2024lrp}}
\label{fig:fpbh_mu3_mu3star}
\end{figure}

In Ref.~\cite{Wu:2024lrp}, this strong dependence of the PBH relic density on the model parameters was analyzed using a saddle-point approximation for the PBH relic density. The key point is that the PBH relic density is exponentially sensitive to the model parameters, especially to \(S_3/T\). For this simple model, the value of \(f_{\rm PBH}\) can be approximated as
\begin{align}
  \label{eq:fpbh_approx}
  f_{\text{pbh}}\simeq \mathcal{M} \exp\left(-\mathcal{Q}\exp\left(-S_3(T_p)/T_p\right)\right)\ ,
\end{align}
where \(\mathcal{M}\approx 4\times 10^7\) and \(\mathcal{Q}\approx 1\times 10^{77}\) are two large numerical factors that are less sensitive to the model parameters. The quantity \(S_3(T_p)/T_p\) is the three-dimensional Euclidean action at the time of PBH formation. The temperature \(T_p\) is defined as the temperature at which \(S_3(T)/T\) reaches its minimum value. Thus, \(f_{\rm PBH}\) is super-exponentially sensitive to the value of \(S_3(T_p)/T_p\), and any small variation in the model parameters that induces a small change in \(S_3(T_p)/T_p\) can lead to a large change in \(f_{\rm PBH}\). This explains the strong dependence of the PBH relic density on the model parameters.

\subsection{Indirect Probe of the First-Order Phase Transition by Collider Searches for BSM Particles}
\label{sec:collider_probe_FOPT_light_singlet}

As mentioned earlier, electroweak baryogenesis requires a first-order phase transition, which can be achieved by extending the scalar sector of the Standard Model. Typically, such new scalar particles have portal couplings to the SM Higgs boson, which can modify the Higgs potential and render the electroweak phase transition first order. If such new scalar particles exist, they can be produced at colliders, and collider searches for these particles can therefore provide an indirect probe of a first-order phase transition. In this subsection, we consider a simple example of such a scenario, in which the SM is extended by adding a real scalar singlet field with a portal coupling to the SM Higgs boson. In addition, we consider the case in which the scalar singlet is very light \cite{Kozaczuk:2019pet,Wang:2022dkz,Wang:2023zys}.

The scalar potential of the singlet extension of the SM can be written as
\begin{align}
V=-\mu^{2}|H|^{2}+\lambda|H|^{4}+\frac{1}{2} a_{1}|H|^{2} S+\frac{1}{2} a_{2}|H|^{2} S^{2}+b_{1} S+\frac{1}{2} b_{2} S^{2}+\frac{1}{3} b_{3} S^{3}+\frac{1}{4} b_{4} S^{4}\ ,
\end{align}
where \(H\) is the SM Higgs doublet, and \(S\) is the real scalar singlet. The parameters \(a_1\) and \(a_2\) are the portal couplings between the scalar singlet and the SM Higgs boson. After electroweak symmetry breaking, we express the Higgs doublet and the scalar singlet as
\begin{align}
H=\frac{1}{\sqrt{2}}\begin{pmatrix}0\\ v+h
\end{pmatrix},\quad S=s\ ,
\end{align}
where \(v\approx 246\) GeV is the vacuum expectation value (vev) of the SM Higgs field at zero temperature, and \(h\) and \(s\) are the fluctuation fields of the Higgs boson and scalar singlet, respectively. Since we have introduced the tadpole term for the singlet field in the potential, we can choose the vev of the singlet field to be zero \cite{Profumo:2007wc}. Since the Higgs field and singlet field mix, we define the mass eigenstates as
\begin{align}
   h_{1} & =h \cos \theta+s \sin \theta\ , \\
 h_{2} & =-h \sin \theta+s \cos \theta\ ,
\end{align}
where \(h_1\) and \(h_2\) are the mass eigenstates with masses \(m_{1}\) and \(m_{2}\), and \(\theta\) is the mixing angle between the Higgs boson and the scalar singlet. We assign \(h_1\) and \(h_2\) to be the singlet-like and SM-Higgs-like mass eigenstates, respectively. This implies that we take \(m_2\approx 125\) GeV. We also take \(\cos\theta=0.01\) as our benchmark choice \cite{Kozaczuk:2019pet}.

Since our goal is to investigate collider searches for the light scalar singlet, we mainly focus on the sensitivity of future lepton colliders, such as CEPC \cite{CEPCStudyGroup:2018ghi}, FCC-ee \cite{FCC:2018evy}, and ILC \cite{ILC:2013jhg}. In this thesis, specifically, we focus on the CEPC sensitivity. For a summary of new-physics opportunities at the CEPC, see Ref.~\cite{Ai:2025cpj}. Specifically, we consider the process \(e^+ e^- \to Z h_2\), where the \(Z\) boson further decays into a pair of leptons, and the SM-Higgs-like mass eigenstate \(h_2\) decays into a pair of singlet-like mass eigenstates \(h_1\), each of which subsequently decays into a pair of \(b\) quarks. The final state is therefore \(\ell^+ \ell^- b \bar{b} b \bar{b}\), where \(\ell\) can be either an electron or a muon. The reason for choosing the four-\(b\)-quark final state is that this process has a large branching ratio and is therefore easy to probe at CEPC \cite{Wang:2022dkz}. The Feynman diagram for this process is shown in Fig.~\ref{fig:collider_search_for_light_scalar_singlet}.

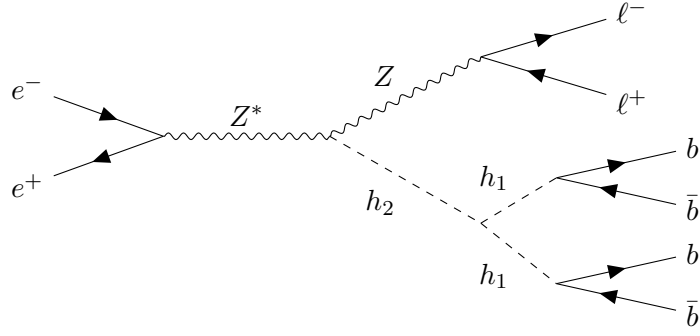
\begin{figure}[t]
\centering
\begin{tikzpicture}
  \begin{feynman}
  \vertex (em) at (0,1.4) {\(e^{-}\)};
  \vertex (ep) at (0,0.1) {\(e^{+}\)};
  \vertex (v1) at (1.8,0.75);

  \vertex (v2) at (4.0,0.75);

  \vertex (zv) at (6.0,1.8);
  \vertex (lp) at (8.0,2.4) {\(\ell^{-}\)};
  \vertex (lm) at (8.0,1.2) {\(\ell^{+}\)};

  \vertex (hv) at (6.0,-0.4);

  \vertex (h1a) at (7.0,0.2);
  \vertex (h1b) at (7.0,-1.2);

  \vertex (bb1) at (8.8,0.6) {\(b\)};
  \vertex (bb2) at (8.8,-0.2) {\(\bar b\)};

  \vertex (bb3) at (8.8,-0.8) {\(b\)};
  \vertex (bb4) at (8.8,-1.6) {\(\bar b\)};

  \diagram*{
    (em) -- [fermion] (v1) -- [boson, edge label=\(Z^\ast\)] (v2),
    (ep) -- [anti fermion] (v1),

    (v2) -- [boson, edge label=\(Z\)] (zv),
    (zv) -- [fermion] (lp),
    (zv) -- [anti fermion] (lm),

    (v2) -- [scalar, edge label'=\(h_2\)] (hv),
    (hv) -- [scalar, edge label=\(h_1\)] (h1a),
    (hv) -- [scalar, edge label'=\(h_1\)] (h1b),

    (h1a) -- [fermion] (bb1),
    (h1a) -- [anti fermion] (bb2),

    (h1b) -- [fermion] (bb3),
    (h1b) -- [anti fermion] (bb4),
  };
  \end{feynman}
\end{tikzpicture}
\caption{Feynman diagram for \(e^+e^- \to Z h_2\), followed by \(Z\to \ell^+\ell^-\) (\(\ell = e, \mu\)), \(h_2 \to h_1 h_1\), and \(h_1 \to b\bar b\), giving the final state \(\ell^+\ell^- b\bar b b\bar b\)\ .}
\label{fig:collider_search_for_light_scalar_singlet}
\end{figure}

The decay widths of the singlet-like and SM-like scalars read
\begin{align}
  &\Gamma(h_1)=\left.\cos^2{\theta}\Gamma^{\mathrm{SM}}\right|_{m_{1}}\ , \\
&\Gamma\left(h_{2}\right)=\left.\sin ^{2} \theta \Gamma^{\mathrm{SM}}\right|_{m_{2}}+\Gamma\left(h_{2} \rightarrow h_{1} h_{1}\right)\ .
\end{align}
Here \(\Gamma^{\mathrm{SM}}\) is the decay width of the SM Higgs boson, and \(\Gamma\left(h_{2} \rightarrow h_{1} h_{1}\right)\) is the decay width of \(h_2\) into a pair of \(h_1\), which we refer to as an exotic Higgs decay:
\begin{align}
  \Gamma\left(h_{2} \rightarrow h_{1} h_{1}\right)=\frac{1}{32 \pi m_{2}} \lambda_{211}^{2} \sqrt{1-\frac{4 m_{1}^{2}}{m_{2}^{2}}}\ .
\end{align}
Here \(\lambda_{211}\) is the trilinear coupling between \(h_2\) and \(h_1\), which can be expressed as
\begin{align}
  \lambda_{211}=2 s^{2} c b_{3}+\frac{a_{1}}{2} c\left(c^{2}-2 s^{2}\right)+\left(2 c^{2}-s^{2}\right) s v a_{2}-6 \lambda s c^{2} v\ ,
\end{align}
with \(s=\sin\theta\) and \(c=\cos\theta\). This effective coupling \(\lambda_{211}\) appears in the following effective interaction term:
\begin{align}
V \supset \frac{1}{6}\lambda_{111} h_1^3 + \frac{1}{2}\lambda_{211} h_2 h_1^2 + \frac{1}{2}\lambda_{221} h_2^2 h_1 + \frac{1}{6}\lambda_{222} h_2^3\ .
\end{align}

We now discuss the input and free parameters of this model. We choose the SM-like Higgs mass \(m_2 = 125\) GeV and the vev of the SM Higgs field, \(v=246\) GeV, as input parameters. The other parameters in the model can be expressed as
\begin{align*}
\mu^2 &= \lambda v^2, \qquad
\lambda = \frac{1}{2v^2}\left(m_1^2\cos^2\theta + m_2^2\sin^2\theta\right), \\
b_1 &= -\frac{1}{4}a_1 v^2, \qquad
b_2 = -\frac{1}{2}a_2 v^2 + m_2^2\cos^2\theta + m_1^2\sin^2\theta, \qquad
a_1 = \frac{1}{v}\left(m_2^2 - m_1^2\right)\sin 2\theta \ .
\end{align*}
Note that these are tree-level relations between the model parameters and the physical input parameters. In principle, one could include the one-loop vacuum renormalization to relate the model parameters to the physical input parameters in the \(\overline{\rm MS}\) scheme, as discussed in detail in Appendix~\ref{app:MSbar_parameters_and_physical_input_quantities}. However, in the present study, we use only these tree-level relations. The free parameters of the model are \(m_1\), \(\theta\), \(a_2\), \(b_3\), and \(b_4\). We fix \(\cos\theta=0.01\) in this study, while varying all other free parameters to study the implications for the first-order phase transition and collider searches for the light scalar singlet. In the small-\(|\cos\theta|\) limit, the coupling \(\lambda_{211}\) can be approximated as
\begin{align}
  \lambda_{211} \approx -a_2 v + \mathcal{O}\left[\left(\frac{\pi}{2} - \theta \right)^2 \right]\ .
\end{align}
As we will show below, a first-order phase transition requires a larger value of the portal coupling \(a_2\), which also controls the branching ratio of the process \(h_2\to h_1 h_1\). Thus, there is a strong connection between the first-order phase transition and collider searches for the light scalar singlet in this model.

From the theory side, we now consider two physical aspects of the real-singlet model introduced above. First, we study the parameter space of the model that can lead to a strong first-order phase transition. Second, we compute the branching ratio of the process \(h_2\to h_1 h_1\).

For the study of the FOPT in this model, the one-loop finite-temperature effective potential for the SM is discussed in Appendix~\ref{app:effective_potential_computation}, where the scalar one-loop thermal master integral is given in Eq.~\eqref{eq:J_B_integral}. As mentioned in Sec.~\ref{sec:EWBG_thermal_effective_potential}, the state-of-the-art computation of the effective potential for a FOPT uses the dimensional-reduction technique. For recent applications of this technique to the real-singlet model, see Refs.~\cite{Brauner:2016fla,Niemi:2021qvp,Niemi:2024axp,Ekstedt:2024etx}. However, in this singlet model, we still use the one-loop finite-temperature effective potential to study the FOPT, since the main purpose of this subsection is to illustrate the connection between the FOPT and collider searches for the light scalar singlet. The numerical package \texttt{CosmoTransitions} \cite{Wainwright:2011kj} is used to compute the effective potential and the phase-transition properties. As shown in Ref.~\cite{Kozaczuk:2019pet}, the parameter space that yields a strong first-order phase transition typically imposes a lower bound on the portal coupling \(a_2\). According to the semi-analytic argument given there, the lower bound on \(a_2\) can be approximated as
\begin{align}
  a_2 \gtrsim \frac{m_1^2}{4v^2} \frac{\Delta}{1-\Delta}\ ,
\end{align}
where \(\Delta\) is a numerical parameter between 0.1 and 1 that can be fixed to reproduce the numerical results. For example, the value \(\Delta=0.7\) is used to fit the numerical results in Ref.~\cite{Kozaczuk:2019pet}. This semi-analytic analysis is consistent with the numerical results shown in Refs.~\cite{Kozaczuk:2019pet,Wang:2022dkz,Wang:2023zys}, where the other three parameters are scanned over the ranges \(m_1\in[5,60]\) GeV, \(b_3/v\in[10^{-4},1]\) GeV, and \(b_4\in[10^{-5},1]\).

Then, for the set of input parameters that can lead to a strong first-order phase transition, we compute the branching ratio of the process \(h_2\to h_1 h_1\). Strictly speaking, we should compute the following scaled branching ratio:
\begin{align}
  \overline{\rm Br}(h_2\to h_1 h_1)=\frac{\sigma(e^+e^-\to Z h_2) \times {\rm Br}(h_2\to h_1 h_1)}{\sigma_{\rm SM}(e^+e^-\to Z h_{\rm SM})}\ .
\end{align}
Here \(\sigma(e^+e^-\to Z h_2)\) is the production cross section of \(e^+e^-\to Z h_2\) in this model, \({\rm Br}(h_2\to h_1 h_1)\) is the branching ratio of \(h_2\to h_1 h_1\), and \(\sigma_{\rm SM}(e^+e^-\to Z h_{\rm SM})\) is the production cross section of \(e^+e^-\to Z h_{\rm SM}\) in the SM. These processes can be computed using the \texttt{MadGraph5\_aMC@NLO} package \cite{Alwall:2014hca}. The reason for dividing by the SM cross section is that we want to compare directly with the experimental limits on the exotic Higgs-decay branching ratio. For example, if the experimental limit on the exotic Higgs-decay branching ratio is 0.1, then a value \({\rm Br}(h_2\to h_1 h_1)=0.2\) is not necessarily excluded, because the production cross section of \(h_2\) can be smaller than that of the SM Higgs boson. If the ratio \(\sigma(e^+ e^- \to Z h_2)/\sigma_{\rm SM}(e^+ e^- \to Z h_{\rm SM})\) is smaller than 0.5, then the overall scaled branching ratio \(\overline{\rm Br}(h_2\to h_1 h_1)\) can be smaller than 0.1, and is therefore not excluded by the experimental limit. Another reason for using the scaled branching ratio is that the experimental limits do not necessarily come from lepton colliders. The LHC can also probe exotic Higgs decays; for example, the process \(pp\to W(Z)h_2\), followed by \(h_2 \to h_1 h_1 \to b\bar b b\bar b\), can be probed at the LHC \cite{ATLAS:2018pvw}. The constraint on the branching ratio of \(h_2\to h_1 h_1\) reported in Ref.~\cite{ATLAS:2018pvw} should therefore be interpreted as a constraint on our scaled branching ratio \(\overline{\rm Br}(h_2\to h_1 h_1)\), since the Higgs production cross section used there is the SM one.

\begin{figure}[t]
\centering
\includegraphics[width=0.6\textwidth]{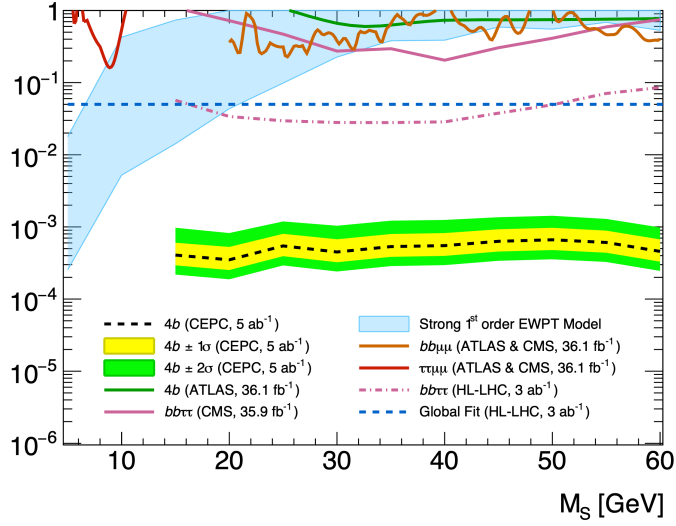}
\caption{The vertical axis shows the scaled branching ratio \(\overline{\rm Br}(h_2\to h_1 h_1)\), and the horizontal axis shows the mass of the singlet-like scalar eigenstate \(h_1\). The black dotted curve gives the expected BDT sensitivity, with the \(1\sigma\) and \(2\sigma\) uncertainty bands shown in yellow and green, respectively. The blue shaded region indicates parameter points that yield a strong first-order electroweak phase transition with successful tunneling. Also shown are the current (solid) and projected (dashed) LHC sensitivities in the \(\tau\tau\mu\mu\) (red) \cite{ATLAS:2015unc,CMS:2019spf}, \(bb\mu\mu\) (orange) \cite{ATLAS:2018emt,CMS:2018nsh}, \(bb\tau\tau\) (pink) \cite{CMS-PAS-FTR-18-035,CMS:2018zvv}, and \(4b\) (green) channels \cite{ATLAS:2018pvw}. The blue dashed curve denotes the projected HL-LHC indirect bound on the total exotic Higgs branching fraction \cite{Cepeda:2019klc}.\reproducedfromref{Wang:2023zys}}
\label{fig:experimental_limits_on_scaled_branching_ratio}
\end{figure}

We now present the final result and the experimental sensitivity, especially at CEPC. The details of the experimental analysis are beyond the scope of this thesis, and we refer to Refs.~\cite{Wang:2022dkz,Wang:2023zys} for more details. The expected sensitivity at CEPC is shown in Fig.~\ref{fig:experimental_limits_on_scaled_branching_ratio}, where the blue shaded region indicates parameter points that yield a strong first-order electroweak phase transition with successful tunneling. The black dotted curve gives the expected BDT sensitivity, with the \(1\sigma\) and \(2\sigma\) uncertainty bands shown in yellow and green, respectively. Also shown are the current (solid) and projected (dashed) LHC sensitivities. We now discuss the physical implications of this figure. We see that the CEPC sensitivity is much better than the (HL-)LHC sensitivity, and can probe a large portion of the parameter space that yields a strong first-order phase transition. If such a light scalar singlet exists in our Universe, it would correspond to a point in this figure. Suppose, for illustration, that its mass is 25 GeV. Then three possible scenarios can occur:
\begin{itemize}
  \item If the point lies in the blue shaded region, this indicates that the electroweak phase transition is strongly first order, and therefore can provide the necessary condition for electroweak baryogenesis. It also automatically lies above the CEPC sensitivity curve, which means that CEPC can probe this benchmark point.
  \item If the point lies outside the blue shaded region but still above the CEPC sensitivity curve, this indicates that the electroweak phase transition is not strongly first order. However, such a point can still be probed by CEPC, thereby providing a strong constraint on the parameter space of this model.
  \item If the point lies below the CEPC sensitivity curve, this indicates that such a light scalar singlet cannot be probed by CEPC, and therefore CEPC cannot provide information about this model.
\end{itemize}
Therefore, combining the theoretical study of the first-order phase transition with the experimental search for the light scalar singlet provides a powerful way to probe new scalars and their connection to electroweak baryogenesis. We note that the method presented here is not unique to this singlet model, but can also be applied to other models that can lead to a first-order phase transition. On the theory side, future work could improve the accuracy of the effective-potential computation for the first-order phase transition, for example by using the dimensional-reduction technique and including the one-loop vacuum renormalization to relate the model parameters to the physical input parameters in the \(\overline{\rm MS}\) scheme.

\newpage
\section{Conclusion and Outlook}
\label{sec:conclusion}

\subsection{Summary of My Works}
The main works of this thesis can be summarized as follows.

First, this thesis develops a self-contained framework for understanding baryon-number violation in electroweak theory from the viewpoints of topology, anomaly, and thermal field theory. We reviewed the roles of homotopy theory in determining the existence of finite-energy topological field configurations, and discussed representative examples, including cosmic string, monopole, sphaleron, and instanton. We also presented the detailed discussion of the vacuum structure of non-Abelian gauge theories, winding number and Chern-Simons number, and sphaleron- and monopole-induced baryon-number violation mechanisms. A detailed discussion of the Adler--Bell--Jackiw anomaly, and instantons in quantum mechanics and instanton-induced baryon number violation are also discussed in the appendices. In this way, the thesis is written so that readers without prior background in these topics can still follow the later developments on electroweak baryogenesis and sphaleron-rate calculations.

Second, we generalized the construction of electroweak topological field configurations to a general \(SU(2)\) scalar multiplet. For a single multiplet with isospin \(J\) and hypercharge \(Y\), we showed that the topology of the resulting configuration is determined by the hypercharge: multiplets with \(Y\neq 0\) lead to sphalerons, while multiplets with \(Y=0\) lead to monopoles. We constructed the sphaleron matrix for a general-dimensional \(SU(2)\) representation, demonstrated the representation independence of the relevant one-form \(F_a\), and established the connection between the general sphaleron and monopole constructions. We then derived the corresponding masses and equations of motion, and discussed how these configurations appear in multi-step phase transitions. In particular, we showed that an intermediate phase with a zero-hypercharge multiplet can support a monopole configuration whose mass may be much larger than the Standard Model sphaleron energy, with important implications for baryon-number preservation.

Third, the main technical advance of this thesis is the development of a gauge-invariant perturbative formalism for the sphaleron rate in the three-dimensional thermal effective field theory. Using dimensional reduction and careful power counting, we showed that the sphaleron is a soft-scale configuration and that its action can be expressed in terms of the gauge-invariant 3D-EFT parameters \(x=\lambda_3/g_3^2\) and \(y=\mu_3^2/g_3^4\). For the crossover case, we derived an accurate fit to the sphaleron action as a function of \(x\). For the first-order case, we showed numerically that after an appropriate \(v_3\)-rescaling, the sphaleron action is well approximated by a nearly constant coefficient times the Higgs-phase minimum \(v_3(x,y)\). We also discussed the one-loop fluctuation corrections around the sphaleron, especially the zero-mode contribution. Applying this formalism to the Standard Model crossover, we obtained perturbative sphaleron rates that agree well with recent lattice results, while maintaining gauge invariance in the setup.

Fourth, we used this gauge-invariant sphaleron-rate formalism to derive a new baryon-preservation condition for electroweak baryogenesis. Instead of the commonly used approximate criterion \(v_c/T_c \gtrsim 1\), we showed that the physically relevant quantity is the 3D-EFT parameter \(x\). We computed the sphaleron-decoupling line in the \((x,y)\) plane and evaluated the baryon-washout integral explicitly, obtaining a gauge-invariant washout exponent. This leads to three qualitatively different regimes: no washout, partial washout, and strong washout. We then applied this framework to the real triplet extension of the Standard Model. This analysis showed that, although the triplet can strengthen the phase transition, the baryon washout in the broken phase can still be severe in much of the parameter space. We also found that two-loop thermal corrections are necessary for a reliable treatment of both the phase transition and the sphaleron rate in this model.

Finally, we explored broader cosmological and phenomenological implications of first-order phase transitions. On the cosmology side, we studied primordial-black-hole production from delayed first-order phase transitions in a simple scalar model, and showed that the relic abundance can depend super-exponentially on the phase-transition parameters. On the phenomenology side, we discussed how collider searches for BSM scalar particles can indirectly probe the first-order phase transition required by electroweak baryogenesis, using a light-singlet example to illustrate the connection between exotic Higgs decays and phase-transition dynamics.

\subsection{Future Directions}
The works presented in this thesis open up several interesting directions for future research. Here we list a few of them.

First, we showed that the sphaleron rate and the baryon-preservation condition are very sensitive to the 3D-EFT parameter \(x=\lambda_3/g_3^2\). Specifically, as we demonstrated in the real triplet extension of the Standard Model, BSM effects generally have a larger impact on the Higgs self-coupling \(\lambda_3\) than on the gauge coupling \(g_3\). Thus, the value of the Higgs self-coupling is crucial. It can also be constrained by collider searches that probe the Higgs self-coupling. Although the current sensitivity to the Higgs self-coupling at the LHC is still weak, future colliders such as FCC-hh can probe it with much better precision \cite{Contino:2016spe,Banerjee:2018yxy}. Many studies of the Higgs self-coupling, including both zero-temperature loop corrections and its implications for BSM electroweak baryogenesis, can be found in the literature; see, for example, Refs.~\cite{Grojean:2004xa,Kanemura:2004ch,Profumo:2014opa,Hashino:2016rvx,Basler:2017uxn,Biekotter:2022kgf,Bahl:2023eau,Bahl:2025wzj,Bittar:2025lcr,Braathen:2025qxf,Zhang:2025zkn}.

Below, we discuss a novel constraint scenario based on the 3D EFT. Since we formulate the baryon-preservation condition in the \((x,y)\) plane, it would be interesting to map future experimental bounds onto this same \((x,y)\) plane. This method can be applied to any BSM model whose heavy degrees of freedom can be integrated out. Since we identified two representative values of \(x\), namely \(\bar{x}=0.025\) and \(\hat{x}=0.035\), future experimental bounds on \(x\) can be interpreted directly in terms of the viability of electroweak baryogenesis. If future experimental bounds exclude the region \(x<\hat{x}\), then electroweak baryogenesis may no longer be a viable mechanism for explaining the baryon asymmetry of the Universe. In that case, mechanisms with an initial non-vanishing baryon-minus-lepton charge, \(B-L\), may be favored. On the other hand, if future experimental bounds still allow the region \(x<\bar{x}\), then electroweak baryogenesis remains a viable mechanism for explaining the baryon asymmetry of the Universe. However, the analysis above is subject to an important restriction: the BSM states must be heavy enough to be integrated out, so that the SM-like 3D EFT provides an effective description of our Universe.

Second, for the sphaleron-rate NLO computation, we have primarily focused on the zero-mode contribution to the one-loop fluctuation determinant. Moreover, this effect has so far been included only for the Standard Model crossover case. It would be interesting to include the zero-mode contribution also in the first-order case, and to include the positive modes in both cases. This would lead to a more complete and accurate treatment of the sphaleron rate, and therefore to a more reliable baryon-preservation condition. In addition, our current first-order sphaleron-rate computation is based on the one-loop effective potential. It would be interesting to extend this analysis by using the two-loop effective potential. However, the justification for using either the one-loop or two-loop effective potential in the sphaleron-rate computation is still lacking. In principle, for example, the one-loop effective potential may have some overlap with the one-loop fluctuation determinant. While the one-loop fluctuation determinant corresponds to the quadratic fluctuation contribution around a non-homogeneous sphaleron background, including both gauge and scalar background fields, the one-loop effective potential corresponds to the quadratic fluctuation contribution around a homogeneous background, involving only the scalar background field. It would therefore be interesting to clarify the relation between these two one-loop contributions and to justify the use of the one-loop or two-loop effective potential in the sphaleron-rate computation.

\iffullversion{
\newpage
\section*{Acknowledgments}
\addcontentsline{toc}{section}{Acknowledgments}

First and foremost, I would like to express my heartfelt gratitude to my supervisor, Prof.~Michael Ramsey-Musolf. Throughout my PhD, he taught me to approach research with rigor and care, to ground my work in a thorough study of the literature and in deep reflection, and to develop a clear understanding of the originality, value, and significance of my work, while presenting my ideas and writing with clarity and logic. These lessons gradually shaped the way I think about physics. They taught me not only to handle technical details carefully, but also to step back and see the broader structure and meaning of the problems I work on. During my PhD, I had many one-on-one discussions with him, ranging from long-term research planning and academic career development to the details of a single point in a paper. Through these discussions, I gradually came to understand the importance of clear physical thinking, careful research planning, and precise scientific expression. His guidance, patience, and support played a crucial role in helping me grow from a PhD student into a more independent researcher. I am also deeply grateful to him for supporting my participation in many conferences and workshops during my PhD, including international visits and conferences in the United States, Hong Kong, China, and Japan, as well as several domestic conferences in China. These experiences greatly broadened my horizons, strengthened my communication skills, and gave me the opportunity to meet many outstanding collaborators in China and abroad. During my visits to the United States, I am especially grateful to Lorenzo Sorbo at the University of Massachusetts Amherst and Stefano Profumo at the University of California, Santa Cruz, for their warm hospitality and stimulating discussions. Their academic styles were quite different from that of my advisor, and through these interactions I was able to learn different ways of doing research and thinking about physics, which was extremely valuable for my development.

At the same time, I would like to thank the postdoctoral researchers and students in our group for their discussions, support, and friendship. I am especially grateful for my collaboration with Tuomas Tenkanen, who helped me greatly in understanding finite-temperature field theory. He also gave me much valuable guidance on how to move projects forward and how to write papers effectively. I also benefited a great deal from discussions with other members of the group, both during group meetings and in private conversations. These exchanges often helped me recognize the limitations of my own thinking in a timely way, which was very important for improving my understanding of physics and developing deeper insight. I would like to thank the postdoctoral researchers Leon Friedrich, Wenxing Zhang, Van Que Tran, Lauri Niemi, Yong Du, Jiang Zhu, Mohamed Younes Sassi, Cristian Sierra, and Anjan Barik, as well as the students Guotao Xia, Yihong Zhong, Xiyuan Jin, Chenzi Liao, Ziyang Huang, Qin Yu, Xinchen Li, and Haifeng Xu. Of course, my interactions with the postdocs and students in the group went far beyond research. Over time, genuine friendships grew among us, and we supported one another in many ways. I would also like to thank other postdocs and students with whom I had the pleasure to collaborate, including Xuxiang Li, Philipp Schicho, Xuliang Zhu and Zhen Wang. I shared many enjoyable and fruitful discussions with them, and together we accomplished a great deal.

I am grateful to the Tsung-Dao Lee Institute (TDLI) for providing an outstanding research environment and strong institutional support. Prof.~Shu Li offered me much help when I first joined the institute. I also had the opportunity to collaborate with Prof.~Qingdong Jiang from the condensed-matter division, and I thank him for the discussions and support during our collaboration.
I also thank Prof.~Tobias Diez and Prof.~Tudor Stefan Ratiu from the School of Mathematical Sciences for discussions on homotopy theory and its applications in field theory. I am also grateful to Prof.~Nicholas Manton from University of Cambridge, whose insights greatly helped our understanding of the properties of the sphaleron. I thank Prof. Wenyuan Ai for providing many helpful comments and suggestions on this thesis. I am also grateful to many professors and doctors outside TDLI for carefully reading this thesis, including Björn Garbrecht, Shinya Kanemura, Philipp Schicho, and Carlos E.M. Wagner. I am thankful for the support of the TDLI ``Tsung-Dao Lee PhD Student'' program, which allowed me to focus on my academic work more fully and smoothly. I also had many stimulating academic exchanges with other students at TDLI, including my office mate Mingwei Li and my roommate Junxiao Hui. I am grateful to the other faculty members and students at TDLI as well. We often exchanged ideas and held discussions during the institute's academic activities, and these interactions contributed greatly to my growth. I also thank the administrative staff at TDLI, especially our group secretary Wanfen Zeng, for their many forms of help and support during my PhD, which allowed me to devote myself more fully to research. Finally, I am grateful for the rich and diverse cultural, recreational, and campus activities at TDLI and Shanghai Jiao Tong University, which made my doctoral years much more colorful and memorable.

Last but not least, I would like to thank my family. I am deeply grateful to my father, Shanzong Wu, and my mother, Jiqin Xie, for their selfless love and unwavering support throughout my PhD. Since my undergraduate years, they have firmly supported my decision to pursue an academic path in theoretical physics, allowing me to focus on research without worrying about practical matters in life. Whether I faced difficulties in research or challenges in life, they were always there to encourage and support me. Their understanding and support gave me the strength to continue along the path of physics and to keep moving forward with determination even in difficult times. Every time I returned home during a holiday, it felt like a renewal of strength and peace. Home has always been my safe harbor, giving me warmth, comfort, and courage. I would like to thank my girlfriend, Shanshan Hu, for her companionship and support throughout my PhD. Being with her has brought me great happiness, warmth, and a deep sense of peace. We shared joyful moments, traveled together to many places, and created many beautiful memories. We also faced difficulties and challenges together, helping each other and walking side by side through the ups and downs of these doctoral years. The beauty and strength of love are among the most precious and unique memories of my PhD. We understand each other deeply, and our hearts remain closely connected across the universe. No matter what the future may bring, we will continue to support and accompany one another. At the same time, I would also like to thank her family, including her father Yangcheng Hu, her mother Manmei Teng, and her younger brother Shuaishuai Hu. We shared many warm and happy moments together, and their kindness and friendliness made me feel truly welcome. Their support and encouragement have brought additional warmth and strength to both my academic journey and my life.

}{}

\newpage
\phantomsection
\addcontentsline{toc}{section}{Publication List}
\thispagestyle{plain}
\begin{minipage}{0.96\textwidth}
  \centering
  {\Large\bfseries Publication List\par}
  \vspace{0.3cm}
  {\small
  \begin{enumerate}
    \setlength{\itemsep}{0.65em}
    \setlength{\parskip}{0.15em}

    \publicationentry
      {An Effective Sphaleron Awakens}
      {Xu-Xiang Li, Michael J.~Ramsey-Musolf, Tuomas V.~I.~Tenkanen, \textbf{Yanda Wu}.}
      {Fourth author.}
      {arXiv:2506.01585}

    \publicationentry
      {Phase transitions, anomalous baryon number violation and electroweak multiplet dark matter}
      {\textbf{Yanda Wu}, Wenxing Zhang, Michael J.~Ramsey-Musolf.}
      {First author.}
      {Phys.\ Rev.\ D 112 (2025) 053003; arXiv:2307.02187}

    \publicationentry
      {Super-exponential Primordial Black Hole Production via Delayed Vacuum Decay}
      {\textbf{Yanda Wu}, Stefano Profumo.}
      {First author.}
      {Phys.\ Rev.\ D 111 (2025) 103524; arXiv:2412.10666}

    \publicationentry
      {Study of Electroweak Phase Transition in Exotic Higgs Decays at the CEPC}
      {Zhen Wang, Xuliang Zhu, Elham E Khoda, Shih-Chieh Hsu, Nikolaos Konstandinidis, Ke Li, Shu Li, Michael J.~Ramsey-Musolf, \textbf{Yanda Wu}, Yuwen E.~Zhang.}
      {Ninth author.}
      {Contribution to Snowmass 2021; arXiv:2203.10184}

    \publicationentry
      {Probing Electroweak Phase Transition at CEPC via Exotic Higgs Decays with 4b Final States}
      {Zhen Wang, Xuliang Zhu, Elham E Khoda, Shih-Chieh Hsu, Nikolaos Konstandinidis, Ke Li, Shu Li, Michael J.~Ramsey-Musolf, \textbf{Yanda Wu}, Yuwen E.~Zhang.}
      {Ninth author.}
      {Published in LHEP 2023 (2023) 436}
  \end{enumerate}}
\end{minipage}

\newpage
\appendix
\counterwithin{equation}{section}

\section{Anomalous Axial Ward Identity and ABJ Anomaly}
\label{sec:ABJ_anomaly}
The ABJ anomaly \cite{Adler:1969gk,Bell:1969ts} describes the anomalous breaking of axial-current conservation in quantum theory. It plays a central role in understanding baryon-number violation in gauge theories. In this subsection, we introduce the ABJ anomaly from the viewpoint of the anomalous axial Ward identity. Historically, the ABJ anomaly was first discovered in the study of neutral-pion decay, $\pi^0 \to \gamma\gamma$, where the anomalous axial Ward identity arises from triangle Feynman diagrams. For pedagogical reasons, it is useful to start from a sigma model describing this pion-decay process \cite{Bell:1969ts,Treiman:1986ep}. Consider the $\sigma$ model with Lagrangian
\begin{align} \label{eq:sigma_model_lagrangian}
  \mathcal{L} &= \bar{\psi} i \gamma^\mu (\partial_\mu - i e A_\mu) \psi + \frac{1}{2} \partial_\mu \phi \partial^\mu \phi - \frac{1}{2}\mu^2 \phi^2 + \frac{1}{2} \partial_\mu \sigma \partial^\mu \sigma \\
  &\quad -\frac{1}{2}(\mu^2+2\lambda F^2)\sigma^2 + g\bar{\psi}(\sigma+\phi \gamma_5)\psi -\lambda \left[ (\sigma^2 + \phi^2)^2 - 2 F \sigma (\sigma^2 + \phi^2) \right] \ ,
\end{align}
where $\psi$, $\phi$, $\sigma$, and $A_\mu$ denote the proton, pion, $\sigma$-particle, and photon fields, respectively; $m$, $e$, $g$, $\mu$, $\lambda$, and $F$ are constants, with $F=2m g^{-1}$. This model has an axial current
\begin{align} \label{eq:axial_current_sigma_model}
  J_5^\mu = i \bar{\psi} \gamma^\mu \gamma_5 \psi + 2(\sigma \partial^\mu \phi - \phi \partial^\mu \sigma) - F \partial^\mu \phi \ ,
\end{align}
whose divergence is proportional to the pion field,
\begin{align} \label{eq:axial_Ward_identity_sigma_model}
  \partial_\mu J_5^\mu = \mu^2 F \phi \ .
\end{align}
At the same time, the electromagnetic current remains conserved, $\partial_\mu J^\mu = 0$, with $J^\mu = \bar{\psi} \gamma^\mu \psi$.

For the process $\pi^0 \to \gamma\gamma$, the matrix element determining the decay amplitude is $M(p,q)=\langle \gamma,p; \gamma^\prime,q | \pi, k\rangle$, where $p$ and $q$ are the photon momenta and $k$ is the pion momentum. We decompose the matrix element as
\begin{align}
  M(p,q) = \epsilon_\mu(p) \epsilon_\nu^\prime(q) T^{\mu\nu}(p,q) \ ,
\end{align}
where $\epsilon_\mu(p)$ and $\epsilon_\nu^\prime(q)$ are the polarization vectors of the two photons. Following Sutherland and Veltman \cite{Sutherland:1967vf,Veltman:1967ceb}, we write the tensor $T^{\mu\nu}(p,q)$ in terms of the off-shell pion amplitude as
\begin{align} \label{eq:T_munu_definition}
  T^{\mu\nu}(p,q) = e^2 (\mu^2 - k^2) \int d^4x\, d^4y\, e^{-ip\cdot x }e^{-iq \cdot y} \langle 0 | T(J^\mu(x) J^\nu(y) \phi(0)) | 0 \rangle \ ,
\end{align}
where $J^\mu$ is the electromagnetic current defined above. This tensor has several important properties: (i) $p_\mu T^{\mu\nu}=q_\nu T^{\mu \nu}=0$, which is the vector Ward identity guaranteed by gauge invariance, or equivalently by vector-current conservation; (ii) $T^{\mu\nu}(p,q)=T^{\nu\mu}(q,p)$, which expresses the Bose symmetry of the two photons.

Now consider the axial Ward identity. According to Eq.~\eqref{eq:axial_Ward_identity_sigma_model}, we have
\begin{align} \label{eq:pion_field_from_axial_current_divergence}
  \phi(0)=\frac{\partial_\mu J^\mu_5 (0)}{F \mu^2} \ ,
\end{align}
and therefore
\begin{align} \label{eq:axial_Ward_identity_T_munu}
  T^{\mu\nu}(p,q) &= \frac{e^2(\mu^2 - k^2)}{F \mu^2} \int d^4x\, d^4y\ e^{-ip\cdot x }e^{-iq \cdot y} \langle 0 | T(J^\mu(x) J^\nu(y) \partial_\rho J_5^\rho(0)) | 0 \rangle \nonumber \\
  &=  \frac{e^2(\mu^2 - k^2)}{F \mu^2} \int d^4x\, d^4y\ e^{-ip\cdot x }e^{-iq \cdot y} \partial_\rho \langle 0 | T(J^\mu(x) J^\nu(y)  J_5^\rho(0)) | 0 \rangle \nonumber \\
  &= \frac{(\mu^2 - k^2)}{F \mu^2} k_\rho T^{\rho \mu \nu}(p,q) \ ,
\end{align}
where we have defined
\begin{align} \label{eq:T_rho_munu_definition}
  T^{\rho \mu \nu}(p,q) = -i e^2 \int d^4x\, d^4y\ e^{-ip\cdot x }e^{-iq \cdot y} \langle 0 | T(J^\mu(x) J^\nu(y)  J_5^\rho(0)) | 0 \rangle \ .
\end{align}
In the last step of Eq.~\eqref{eq:axial_Ward_identity_T_munu}, a factor of $-ik_\rho$ is extracted from the integral, with $k=p+q$ the pion momentum. This is easily seen if one keeps the factor $\int d^4z\, e^{ikz}J_5^\rho(z)\,\delta^4(k-(p+q))$ in the integral. We therefore summarize the vector and axial Ward identities for $T^{\rho \mu \nu}(p,q)$ as
\begin{align} \label{eq:vector_axial_Ward_identity_T_rho_munu}
 {\rm Vector\ Ward\ Identity:\ } & p_\mu T^{\rho \mu \nu}(p,q) = q_\nu T^{\rho \mu \nu}(p,q) = 0 \ , \\
 \label{eq:axial_Ward_identity_T_rho_munu}
 {\rm Axial\ Ward\ Identity:\ } & k_\rho T^{\rho \mu \nu}(p,q) = \frac{F \mu^2}{\mu^2 - k^2} T^{\mu\nu}(p,q) \ .
\end{align}
The first line follows from $p_\mu T^{\mu\nu}=q_\nu T^{\mu \nu}=0$. Below, we show within the $\sigma$ model that
\begin{tcolorbox}[colback=yellow!10!white, colframe=yellow!50!black]
The vector and axial Ward identities in Eq.~\eqref{eq:vector_axial_Ward_identity_T_rho_munu} and Eq.~\eqref{eq:axial_Ward_identity_T_rho_munu} cannot be satisfied simultaneously. Since the vector Ward identity originates from gauge invariance, it is preferable to preserve it and modify the axial Ward identity. This is the origin of the anomalous axial Ward identity, and hence of the ABJ anomaly.
\end{tcolorbox}

We now compute the tensors $T^{\mu \nu}$ and $T^{\rho\mu\nu}$ in the $\sigma$ model of Eq.~\eqref{eq:sigma_model_lagrangian}. The leading diagrams contributing to $T^{\mu\nu}$ are shown in Fig.~\ref{fig: Triangle diagram for pion decay} (a1) and (a2), while the leading diagrams contributing to $T^{\rho\mu\nu}$ are shown in Fig.~\ref{fig: Triangle diagram for pion decay} (b1)--(b4). The crossed vertex in the figure represents the insertion of the axial current: (b1) and (b2) correspond to the first term of Eq.~\eqref{eq:axial_current_sigma_model}, while (b3) and (b4) correspond to the third term of Eq.~\eqref{eq:axial_current_sigma_model}. We do not reproduce the full diagrammatic calculation here; it can be found in \cite{Bell:1969ts,Treiman:1986ep}. The final results are
\begin{align} \label{eq:T_munu_final_result}
  T^{\mu\nu}(p,q) &= \Gamma^{\mu\nu}(p,q) + \Gamma^{\nu\mu}(q,p) \ , \\
  \Gamma^{\mu\nu}(p,q) &= ige^2\int \frac{d^4 r}{(2\pi)^4} {\rm Tr}[\gamma_5 (\slashed{r}+\slashed{p}-m)^{-1}\gamma^\mu (\slashed{r}-m)^{-1}\gamma^\nu (\slashed{r}-\slashed{q}-m)^{-1}] \ ,
\end{align}
where $r^\mu$ is the loop momentum, and
\begin{align} \label{eq:T_rho_munu_final_result}
  T^{\rho \mu \nu}(p,q) &= T_1^{\rho \mu \nu}(p,q) + T_2^{\rho \mu \nu}(p,q) \ , \\
  \label{eq:T1_rho_munu_final_result}
  T_1^{\rho \mu \nu}(p,q) &= \Gamma^{\rho \mu \nu}(p,q) + \Gamma^{\rho \nu \mu}(q,p) \ , \\
  \Gamma^{\rho \mu \nu}(p,q) &= ie^2 \int \frac{d^4 r}{(2\pi)^4} {\rm Tr}[\gamma_5 \gamma^\rho (\slashed{r}+\slashed{p}-m)^{-1}\gamma^\mu (\slashed{r}-m)^{-1}\gamma^\nu (\slashed{r}-\slashed{q}-m)^{-1}] \ , \\
  T_2^{\rho \mu \nu}(p,q) &= -\frac{F}{k^2-\mu^2} k^\rho T^{\mu\nu}(p,q) \ .
\end{align}
Here $T_1^{\rho \mu \nu}(p,q)$ comes from diagrams (b1) and (b2), while $T_2^{\rho \mu \nu}(p,q)$ comes from diagrams (b3) and (b4).

\begin{figure}[t]
\centering

\begin{minipage}{0.48\linewidth}
\centering
\begin{tikzpicture}
  \begin{feynman}
    \vertex (pion) at (-2,0);
    \vertex (v1) at (-0.6,0);
    \vertex (v2) at (0.8,0.8);
    \vertex (v3) at (0.8,-0.8);
    \vertex (g1) at (2,0.8);
    \vertex (g2) at (2,-0.8);

    \diagram*{
      (pion) -- [scalar, edge label=$\pi^0$] (v1),
      (v1) -- [fermion, edge label=$P$] (v2) -- [fermion] (v3) -- [fermion] (v1),
      (v2) -- [photon, edge label=$\gamma$] (g1),
      (v3) -- [photon, edge label=$\gamma$] (g2),
    };
  \end{feynman}
\end{tikzpicture}
\par\textbf{(a1)}
\end{minipage}\hfill
\begin{minipage}{0.48\linewidth}
\centering
\begin{tikzpicture}
  \begin{feynman}
    \vertex (pion) at (-2,0);
    \vertex (v1) at (-0.6,0);
    \vertex (v2) at (0.8,0.8);
    \vertex (v3) at (0.8,-0.8);
    \vertex (g1) at (2,0.8);
    \vertex (g2) at (2,-0.8);

    \diagram*{
      (pion) -- [scalar] (v1),
      (v1) -- [fermion] (v2) -- [fermion] (v3) -- [fermion] (v1),
      (v2) -- [photon] (g2),
      (v3) -- [photon] (g1),
    };
  \end{feynman}
\end{tikzpicture}
\par\textbf{(a2)}
\end{minipage}

\par\medskip

\begin{minipage}{0.48\linewidth}
\centering
\begin{tikzpicture}
  \begin{feynman}
    \vertex (v1) at (-0.6,0);
    \vertex (v2) at (0.8,0.8);
    \vertex (v3) at (0.8,-0.8);
    \vertex (g1) at (2,0.8);
    \vertex (g2) at (2,-0.8);

    \diagram*{
      (v1) -- [fermion] (v2) -- [fermion] (v3) -- [fermion] (v1),
      (v2) -- [photon] (g1),
      (v3) -- [photon] (g2),
    };
  \end{feynman}
  \node[cross out, draw] at (-0.6,0) {};
    \node[black] at (2.2,-0.8) {\small $q$};
    \node[black] at (2.2,0.8) {\small $p$};
    \node[black] at (1.2,0.0) {\small $r$};
    \node[black] at (0.0,-0.7) {\small $r-q$};
    \node[black] at (0.0,0.7) {\small $r+p$};
\end{tikzpicture}
\par\textbf{(b1)}
\end{minipage}\hfill
\begin{minipage}{0.48\linewidth}
\centering
\begin{tikzpicture}
  \begin{feynman}
    \vertex (v1) at (-0.6,0);
    \vertex (v2) at (0.8,0.8);
    \vertex (v3) at (0.8,-0.8);
    \vertex (g1) at (2,0.8);
    \vertex (g2) at (2,-0.8);

    \diagram*{
      (v1) -- [fermion] (v2) -- [fermion] (v3) -- [fermion] (v1),
      (v2) -- [photon] (g2),
      (v3) -- [photon] (g1),
    };
  \end{feynman}
  \node[cross out, draw] at (-0.6,0) {};
\end{tikzpicture}
\par\textbf{(b2)}
\end{minipage}

\par\medskip

\begin{minipage}{0.48\linewidth}
\centering
\begin{tikzpicture}
  \begin{feynman}
    \vertex (pion) at (-2,0);
    \vertex (v1) at (-0.6,0);
    \vertex (v2) at (0.8,0.8);
    \vertex (v3) at (0.8,-0.8);
    \vertex (g1) at (2,0.8);
    \vertex (g2) at (2,-0.8);

    \diagram*{
      (pion) -- [scalar] (v1),
      (v1) -- [fermion] (v2) -- [fermion] (v3) -- [fermion] (v1),
      (v2) -- [photon] (g1),
      (v3) -- [photon] (g2),
    };
  \end{feynman}
  \node[cross out, draw] at (-2,0) {};
   \node[black] at (-1.3,0.3) {\small $k=p+q$};
    \node[black] at (2.2,-0.8) {\small $q$};
    \node[black] at (2.2,0.8) {\small $p$};
    \node[black] at (1.2,0.0) {\small $r$};
    \node[black] at (0.0,-0.7) {\small $r-q$};
    \node[black] at (0.0,0.7) {\small $r+p$};
\end{tikzpicture}
\par\textbf{(b3)}
\end{minipage}\hfill
\begin{minipage}{0.48\linewidth}
\centering
\begin{tikzpicture}
  \begin{feynman}
    \vertex (pion) at (-2,0);
    \vertex (v1) at (-0.6,0);
    \vertex (v2) at (0.8,0.8);
    \vertex (v3) at (0.8,-0.8);
    \vertex (g1) at (2,0.8);
    \vertex (g2) at (2,-0.8);

    \diagram*{
      (pion) -- [scalar] (v1),
      (v1) -- [fermion] (v2) -- [fermion] (v3) -- [fermion] (v1),
      (v2) -- [photon] (g2),
      (v3) -- [photon] (g1),
    };
  \end{feynman}
  \node[cross out, draw] at (-2,0) {};
\end{tikzpicture}
\par\textbf{(b4)}
\end{minipage}

\caption{Triangle diagrams for $\pi^0 \to \gamma\gamma$. The dashed, solid, and wavy lines denote the pion, proton, and photon fields, respectively. Panels (a1) and (a2) contribute to $T^{\mu\nu}$ in Eq.~\eqref{eq:T_munu_definition}, while panels (b1)--(b4) contribute to $T^{\rho\mu\nu}$ in Eq.~\eqref{eq:T_rho_munu_definition}. The crossed vertex denotes the insertion of the axial current: (b1) and (b2) correspond to the first term of Eq.~\eqref{eq:axial_current_sigma_model}, while (b3) and (b4) correspond to the third term of Eq.~\eqref{eq:axial_current_sigma_model}. The momentum labels shown in (b1) also apply to (a1), (a2), and (b2), while the momentum labels shown in (b3) also apply to (b4).}
\label{fig: Triangle diagram for pion decay}
\end{figure}
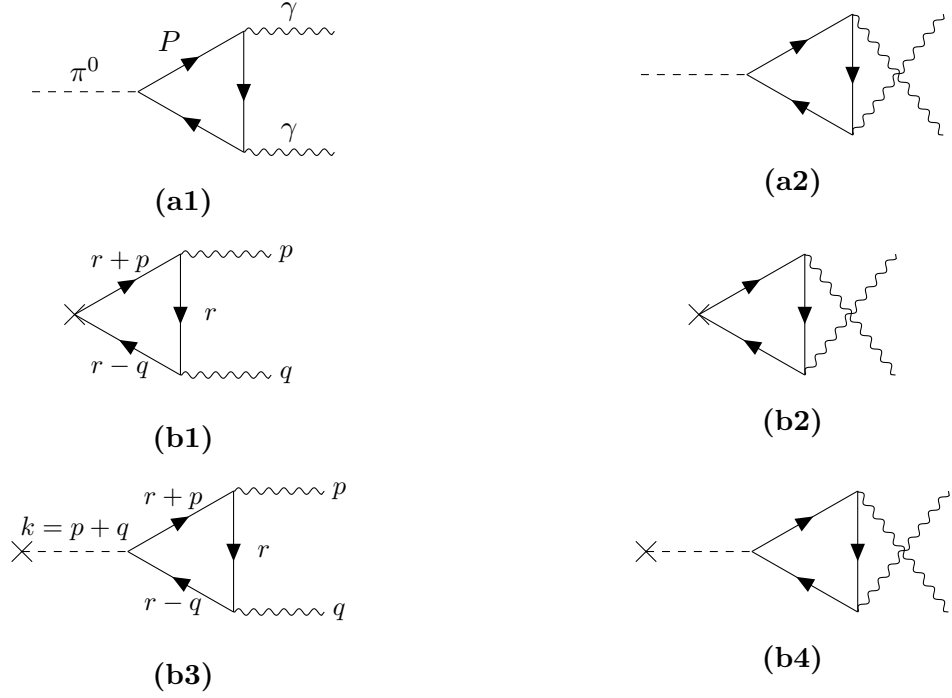

\begin{figure}[t]
\centering

\begin{minipage}{0.48\linewidth}
\centering
\begin{tikzpicture}
  \begin{feynman}
    \vertex (v1) at (-0.6,0);
    \vertex (v2) at (0.8,0.8);
    \vertex (v3) at (0.8,-0.8);
    \vertex (g1) at (2,0.8);
    \vertex (g2) at (2,-0.8);

    \diagram*{
      (v1) -- [fermion] (v2) -- [fermion] (v3) -- [fermion] (v1),
      (v2) -- [photon] (g1),
      (v3) -- [photon] (g2),
    };
  \end{feynman}
  \node[cross out, draw] at (-0.6,0) {};
    \node[black] at (2.2,-0.8) {\small $q$};
    \node[black] at (2.2,0.8) {\small $p$};
    \node[black] at (1.4,0.0) {\small $r+a$};
    \node[black] at (-0.4,-0.7) {\small $r-q+a$};
    \node[black] at (-0.4,0.7) {\small $r+p+a$};
\end{tikzpicture}
\par\textbf{(b1)}
\end{minipage}\hfill

\caption{Triangle diagram (b1) of Fig.~\ref{fig: Triangle diagram for pion decay} after shifting the loop momentum.}
\label{fig: Triangle diagram with shifted momentum}
\end{figure}

We now check the vector and axial Ward identities in Eq.~\eqref{eq:vector_axial_Ward_identity_T_rho_munu}. First, the axial Ward identity is satisfied only if
\begin{align} \label{eq:axial_Ward_identity_T1_rho_munu}
  k_\rho T_1^{\rho \mu \nu}(p,q) = F T^{\mu\nu}(p,q) \ .
\end{align}
On the other hand, the vector Ward identity requires
\begin{align} \label{eq:vector_Ward_identity_T1_rho_munu}
  p_\mu T_1^{\rho \mu \nu}(p,q) = q_\nu T_1^{\rho \mu \nu}(p,q) = 0 \ .
\end{align}
However, it can be shown that Eq.~\eqref{eq:axial_Ward_identity_T1_rho_munu} and Eq.~\eqref{eq:vector_Ward_identity_T1_rho_munu} cannot be satisfied simultaneously \cite{Treiman:1986ep}. The reason lies in the treatment of diagrams (b1) and (b2). When computing these diagrams, we initially choose the momentum flowing between the two photon vertices to be $r^\mu$. This choice is not unique: one may always shift the loop momentum as $r^\mu \to r^\mu + a^\mu$, where $a^\mu$ is an arbitrary constant four-vector, as illustrated in Fig.~\ref{fig: Triangle diagram with shifted momentum}. In ordinary Feynman-diagram calculations, such a shift does not change the final result, and one is free to choose any $a^\mu$. In the present case, however, this shift changes the value of $T_1^{\rho \mu \nu}(p,q)$ by a finite amount. The reason is that the momentum integral has a nontrivial superficial degree of divergence, even though the final answer is finite. Below, we present the result for $T_1^{\rho \mu \nu}(p,q)$ with a general shift $a^\mu$. The detailed calculation can be found in \cite{Treiman:1986ep}. According to the definition of $T_1^{\rho \mu \nu}(p,q)$ in Eq.~\eqref{eq:T1_rho_munu_final_result}, we first define a regulated version of $\Gamma^{\rho \mu \nu}(p,q|a)$ by
\begin{align}
  \Gamma^{\rho \mu \nu}(p,q|a) &= \Gamma^{\rho \mu \nu}(p,q) + \Delta^{\rho \mu \nu}(p,q|a) \ , \\
  \Delta^{\rho \mu \nu}(p,q|a) &= ie^2 \int \frac{d^4 r}{(2\pi)^4} {\rm Tr}\ \gamma_5 \gamma^\rho\times \\
  &\quad \left[ (\slashed{r}+\slashed{a}+\slashed{p}-m)^{-1}\gamma^\mu (\slashed{r}+\slashed{a}-m)^{-1}\gamma^\nu (\slashed{r}+\slashed{a}-\slashed{q}-m)^{-1} \right. \\
  &\qquad \left. - (\slashed{r}+\slashed{p}-m)^{-1}\gamma^\mu (\slashed{r}-m)^{-1}\gamma^\nu (\slashed{r}-\slashed{q}-m)^{-1}\right] \ ,
\end{align}
where $\Delta^{\rho \mu \nu}(p,q|a)$ is the surface term induced by the loop-momentum shift. After some algebra, one finds
\begin{align}
  \Delta^{\rho \mu \nu}(p,q|a) = -\frac{e^2}{8\pi^2} \epsilon^{\rho \mu \nu\lambda} a_\lambda \ ,
\end{align}
where the arbitrary vector is further parameterized as $a_\lambda = (a+b)p_\lambda + b q_\lambda$. Since the photon momenta $p$ and $q$ are the only available vectors, the shift vector $a_\lambda$ must be a linear combination of them. The parameterization with two coefficients, $a$ and $b$, is chosen only for later convenience. Substituting this result back into $T_1^{\rho \mu \nu}(p,q|a)$ gives
\begin{align}
  T_1^{\rho \mu \nu}(p,q|a) = T_1^{\rho \mu \nu}(p,q) - \frac{e^2}{8\pi^2} a\, \epsilon^{\rho \mu \nu \lambda} (p_\lambda-q_\lambda) \ .
\end{align}
We now have one free parameter, $a$. Next, we check the axial and vector Ward identities again. Since the Ward identities require $T_1$ to satisfy Eq.~\eqref{eq:axial_Ward_identity_T1_rho_munu} and Eq.~\eqref{eq:vector_Ward_identity_T1_rho_munu}, we test these relations with the modified $T_1^{\rho \mu \nu}(p,q|a)$. After some algebra, we find
\begin{align}
  k_\rho T_1^{\rho \mu \nu}(p,q|a) &= F T^{\mu\nu}(p,q) + \frac{e^2}{4\pi^2} a \epsilon^{\mu \nu \lambda \beta} p_\lambda q_\beta \ , \\
  p_\mu T_1^{\rho \mu \nu}(p,q|a) &= \frac{e^2}{4\pi^2} \epsilon^{\rho \mu \nu \beta} p_\mu q_\beta \left[ 1+ \frac{a}{2} \right] \ , \\
  q_\nu T_1^{\rho \mu \nu}(p,q|a) &= -\frac{e^2}{4\pi^2} \epsilon^{\rho \mu \nu \beta} p_\mu q_\beta \left[ 1+ \frac{a}{2} \right] \ .
\end{align}
From these expressions, two conclusions follow:
\begin{itemize}
  \item[(i)] The axial Ward identity is satisfied if we choose $a=0$. This corresponds to the original choice of loop momentum in diagrams (b1) and (b2). However, with this choice the vector Ward identity is violated.
  \item[(ii)] The vector Ward identity is satisfied if we choose $a=-2$. However, with this choice the axial Ward identity is violated.
\end{itemize}
Thus, the two Ward identities cannot be satisfied simultaneously for any choice of $a$. Since the vector Ward identity originates from gauge invariance, we preserve it and modify the axial Ward identity. Choosing $a=-2$, the conventional axial Ward identity in Eq.~\eqref{eq:axial_Ward_identity_T1_rho_munu} is replaced by the anomalous axial Ward identity,
\begin{align} \label{eq:anomalous_axial_Ward_identity_sigma_model}
  k_\rho T^{\rho \mu \nu}(p,q) =  \frac{F \mu^2}{\mu^2 - k^2} T^{\mu\nu}(p,q) - \frac{e^2}{2\pi^2} \epsilon^{\mu \nu \lambda \beta} p_\lambda q_\beta \ .
\end{align}

Finally, we must understand how this anomalous axial Ward identity leads to the anomalous divergence of the axial current, i.e. the ABJ anomaly. In the following, we use a slightly different line of reasoning from \cite{Treiman:1986ep}. First, we use the relation \cite{Peskin:1995ev}
\begin{align}
  \frac{e^2}{2\pi^2} \epsilon^{\mu \nu \lambda \beta} p_\lambda q_\beta \epsilon_\mu(p)\epsilon_\nu(q) & = \frac{e^2}{8\pi^2}\langle \gamma,p; \gamma^\prime,q | F_{\mu\nu}(0) \tilde{F}^{\mu\nu}(0) | 0 \rangle \\
  & = \frac{e^2}{8\pi^2}(ie)^2 \int d^4x\  \int d^4y\ \epsilon_\mu(p)\epsilon_\nu(q)  \\
  &\quad \times e^{-ip\cdot x}e^{-iq\cdot y} \langle 0 | T(J^\mu(x)J^\nu(y)F_{\mu\nu}(0) \tilde{F}^{\mu\nu}(0)) | 0 \rangle \ ,
\end{align}
where $\tilde{F}^{\mu\nu}=\frac{1}{2}\epsilon^{\mu\nu\rho\sigma}F_{\rho\sigma}$ is the dual field-strength tensor. In the second line, the LSZ reduction formula for two photons has been applied \cite{Srednicki:2007qs}. We may further factorize this expression according to the right-hand side of Eq.~\eqref{eq:anomalous_axial_Ward_identity_sigma_model} as
\begin{align}
  \frac{e^2}{2\pi^2} \epsilon^{\mu \nu \lambda \beta} p_\lambda q_\beta &= \frac{F \mu^2}{\mu^2-k^2} \nonumber \\
  &\qquad \times \left[e^2(\mu^2-k^2)   \int d^4x\  \int d^4y\ e^{-ip\cdot x}e^{-iq\cdot y} \right. \nonumber \\
  &\qquad \times \left. \langle 0 | T(J^\mu(x)J^\nu(y)\left(-\frac{e^2}{8\pi^2 F \mu^2}\right)F_{\mu\nu}(0) \tilde{F}^{\mu\nu}(0)) | 0 \rangle \right] \ .
\end{align}
Comparing this with the definition of $T^{\mu\nu}(p,q)$ in Eq.~\eqref{eq:T_munu_definition} and Eq.~\eqref{eq:axial_Ward_identity_T_munu}, we see that the expression in square brackets is an additional contribution to $T^{\mu\nu}(p,q)$. This contribution appears if the relation in Eq.~\eqref{eq:pion_field_from_axial_current_divergence} is replaced by
\begin{align} \label{eq:anomalous_axial_divergence_sigma_model}
  \phi(0)=\frac{\partial_\mu J^\mu_5 (0)}{F \mu^2}  + \frac{e^2}{8\pi^2 F\mu^2} F_{\mu\nu}(0) \tilde{F}^{\mu\nu}(0) \ .
\end{align}
In other words, if one modifies Eq.~\eqref{eq:pion_field_from_axial_current_divergence} to Eq.~\eqref{eq:anomalous_axial_divergence_sigma_model}, then the anomalous axial Ward identity in Eq.~\eqref{eq:anomalous_axial_Ward_identity_sigma_model} follows. Since our logic is the reverse---namely that the anomalous axial Ward identity implies a modified divergence equation---the final anomalous divergence of the axial current is
\begin{align}
  \partial_\mu J^\mu_5 = F \mu^2 \phi - \frac{e^2}{8\pi^2} F_{\mu\nu} \tilde{F}^{\mu\nu} \ .
\end{align}
Generalizing this result to Abelian gauge theory gives
\begin{align} \label{eq:ABJ_anomaly_Abelian}
  \partial_\mu J^\mu_5 =  -\frac{e^2}{16\pi^2}\epsilon^{\mu\nu\rho\sigma} F_{\mu\nu} {F}_{\rho\sigma} \ ,
\end{align}
where the mass term has been dropped for massless fermions. This is the familiar ABJ anomaly in Abelian gauge theory.

The same conclusion generalizes straightforwardly to non-Abelian gauge theory. For a field in representation $R$ of a non-Abelian gauge group with gauge coupling $g$, the ABJ anomaly reads
\begin{align}
  \partial_\mu J_5^\mu = -\frac{g^2}{16\pi^2} T(R) F_{\mu\nu}^a \tilde{F}^{a\mu\nu} \ ,
\end{align}
where $T(R)$ is the Dynkin index of the representation $R$, defined by ${\rm Tr}[t^a_R t^b_R] = T(R) \delta^{ab}$, with $t^a_R$ the generators of the gauge group in representation $R$. For example, for the fundamental representation of $SU(N)$ one has $T(R)=\frac{1}{2}$. For chiral $SU(2)_L$ gauge theory, the anomalous variation of left-handed baryons and leptons equals the total anomalous variation of baryon and lepton number, so\footnote{One may note that there can be a minus-sign difference between the left-handed fermion current and the chiral current. This sign difference is not important for the present discussion. Either sign implies baryon-number violation, while the physically relevant violation rate is determined by the baryonic free energy, as discussed in Sec.~\ref{sec:BNV_rate_sphaleron}.}
\begin{align}
  \label{eq:ABJ_anomaly_non_Abelian}
  \partial_\mu j^\mu_{B} = \partial_\mu j^\mu_{L}=\frac{n_f}{2}\partial_\mu J_5^\mu = n_f \frac{g^2}{32\pi^2} F_{\mu\nu}^a \tilde{F}^{a\mu\nu} \ ,
\end{align}
where $n_f$ is the number of fermion generations.

\section{Instanton and Tunneling in Quantum Mechanics}
\label{sec:instanton_QM}
As mentioned at the beginning of this chapter, we first introduce instantons and tunneling in quantum mechanics, which provides a simpler setting than gauge theory. The quantum-mechanics examples already contain the essential physics of instantons, tunneling, and the semiclassical approximation. It is therefore pedagogically useful to introduce these concepts first in quantum mechanics before turning to the more complicated gauge-theory case.

We consider a spinless particle moving in one dimension in a potential \(V(x)\). The Hamiltonian is
\begin{align}
  H = \frac{p^2}{2} + V(x) \ .
\end{align}
We will study three different potentials, shown in Fig.~\ref{fig:three-potentials}. Panel (a) is a simple quadratic potential, panel (b) is a double-well potential with two degenerate minima at \(x=\pm a\), and panel (c) is a periodic cosine potential with minima at \(x=na\) for integer \(n\). In all three cases, we normalize the minima to \(V=0\). The key quantity we want to study is the amplitude for the particle to start at one minimum, \(x_i\), and end at another minimum, \(x_f\), in Euclidean time \(\tau\):
\begin{align} \label{eq:transition_amplitues_QMs}
  \langle x_f | e^{-H\tau} | x_i \rangle = \int \mathcal{D}x \, e^{-S_E[x]} \ .
\end{align}
Here we express the amplitude as a Euclidean path integral over all paths \(x(\tau)\) that start at \(x_i\) and end at \(x_f\). Later we will see that this amplitude is analogous to the vacuum-to-vacuum amplitude in gauge theory in the presence of instantons. We emphasize two important properties of Eq.~\eqref{eq:transition_amplitues_QMs}. First, if we expand the position eigenstates in terms of energy eigenstates, \(|x\rangle = \sum_n |n\rangle \langle n | x \rangle\), with \(H|n\rangle = E_n |n\rangle\), then the left-hand side of Eq.~\eqref{eq:transition_amplitues_QMs} becomes
\begin{align} \label{eq:amplitute_eigen_decomposition_QM}
  \langle x_f | e^{-H\tau} | x_i \rangle = \sum_n e^{-E_n \tau} \langle x_f | n \rangle \langle n | x_i \rangle \ ,
\end{align}
so in the large-\(\tau\) limit the amplitude is dominated by the ground state \(n=0\). Second, the right-hand side of Eq.~\eqref{eq:transition_amplitues_QMs} can be evaluated in the semiclassical approximation by expanding around classical paths that extremize the Euclidean action \(S_E[x]\), which is
\begin{align}
  S_E[x] = \int d\tau \left[ \frac{1}{2} \left( \frac{dx}{d\tau} \right)^2 + V(x) \right] \ .
\end{align}
These classical paths satisfy the Euclidean equation of motion
\begin{align} \label{eq:classical_path_equation_QM}
   \frac{d^2 \bar{x}}{d\tau^2} - V^\prime(\bar{x}) = 0 \ .
\end{align}
This equation can be interpreted as the equation of motion of a particle moving in the inverted potential \(-V(x)\), namely the potentials in Fig.~\ref{fig:three-potentials} flipped upside down.

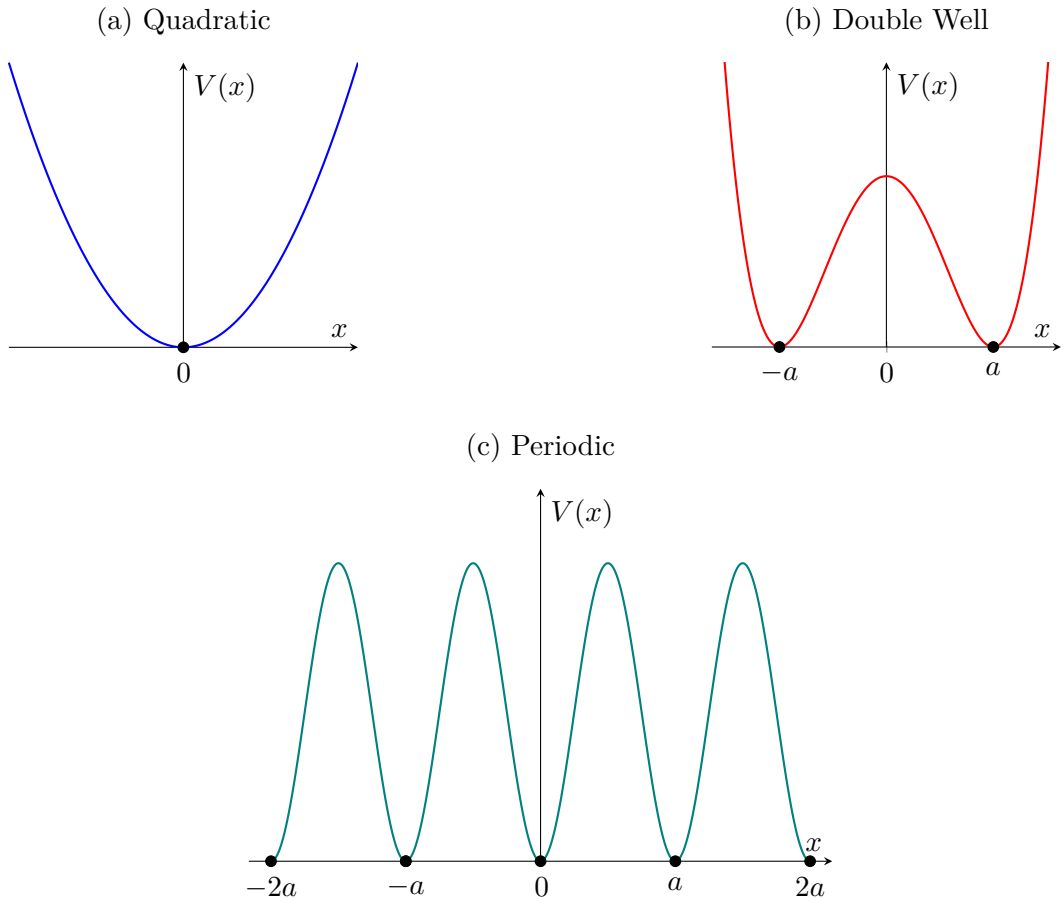
\begin{figure}[htbp]
\centering
\begin{minipage}{0.4\linewidth}
  \centering
  \begin{tikzpicture}
    \begin{axis}[
      width=\linewidth,
      domain=-2:2, samples=201,
      axis lines=middle,
      xlabel={$x$}, ylabel={$V(x)$},
      xmin=-2, xmax=2,
      ymin=0, ymax=4,
      xtick={0.005},
      xticklabels={$0$},
      xticklabel style={anchor=north},
      ytick=\empty,
      title={(a) Quadratic}]
      \addplot[blue, thick] {x^2};
      \addplot[only marks, mark=*, mark options={fill=black}] coordinates { (0,0)};
    \end{axis}
  \end{tikzpicture}
\end{minipage}\hfill
\begin{minipage}{0.4\linewidth}
  \centering
  \begin{tikzpicture}
    \begin{axis}[
      width=\linewidth,
      domain=-2.5:2.5, samples=301,
      axis lines=middle,
      xlabel={$x$}, ylabel={$V(x)$},
      xmin=-2, xmax=2,
      ymin=0, ymax=2.5,
      xtick={-1.225,0.005,1.225},
      xticklabels={$-a$, $0$, $a$},
      ytick=\empty,
      title={(b) Double Well}]
      \addplot[red, thick] {((x^2-1.5)^2/1.5)};
      \addplot[only marks, mark=*, mark options={fill=black}] coordinates {(-1.225,0) (1.225,0)};
    \end{axis}
  \end{tikzpicture}
\end{minipage}

\vspace{1em}

\begin{minipage}{0.6\linewidth}
  \centering
  \begin{tikzpicture}
    \begin{axis}[
      width=\linewidth,
      height=0.7\linewidth,
      domain=-12.566:12.566, samples=401,
      axis lines=middle,
      xlabel={$x$}, ylabel={$V(x)$},
      xmin=-13.6, xmax=13.6,
      ymin=0, ymax=1,
      xtick={-12.566,-6.283,0.05,6.283,12.566},
      xticklabels={$-2a$,$-a$,$0$,$a$,$2a$},
      ytick=\empty,
      title={(c) Periodic}]
      \addplot[teal, thick] {(1-cos(deg(x)))/2.5};
      \addplot[only marks, mark=*, mark options={fill=black}] coordinates {(0,0) (-6.283,0) (-6.283,0) (6.283,0) (12.566,0) (-12.566,0)};
    \end{axis}
  \end{tikzpicture}
\end{minipage}
\caption{Three example potentials, with the minima normalized to \(V=0\). (a) A quadratic potential centered at \(x=0\). (b) A double-well potential with minima at \(x=\pm a\). (c) A periodic potential with multiple minima at \(x=na\) for integer \(n\).}
\label{fig:three-potentials}
\end{figure}

The right-hand side of Eq.~\eqref{eq:transition_amplitues_QMs} can then be evaluated by expanding around the classical path \(\bar{x}(\tau)\):
\begin{align}
  x(\tau) = \bar{x}(\tau) + \sum_n c_n \eta_n(\tau) \ ,
\end{align}
where \(\eta_n(\tau)\) denotes the quantum fluctuation around the classical path and satisfies the boundary conditions \(\eta_n(\tau_i)=\eta_n(\tau_f)=0\). The coefficients \(c_n\) are expansion coefficients. In addition, \(\eta_n(\tau)\) must satisfy a proper normalization condition. The \(\eta_n\) are chosen as eigenfunctions of the operator
\begin{align}
  \left[ - \frac{d^2}{d\tau^2} + V^{\prime\prime}(\bar{x}) \right] \eta_n = \lambda_n \eta_n \ ,
\end{align}
where \(\lambda_n\) are the corresponding eigenvalues. We call \(\eta_n\) positive, zero, or negative modes when \(\lambda_n\) is positive, zero, or negative, respectively. The evaluation at next-to-leading order gives the determinant factor, whose technical details we will not discuss here. We only write the result schematically as
\begin{align}
  \label{eq:transition_amplitudes_QM_semi_classical}
  \int \mathcal{D}x \, e^{-S_E[x]} \approx e^{-S_E[\bar{x}]} K N \tau \ ,
\end{align}
where \(K\) is the determinant factor from the quantum fluctuations containing the positive modes only, and \(N\) is the normalization factor associated with the zero mode. We will see that there is no negative mode in the quantum-mechanics examples. This is also the case for instantons in gauge theory. By contrast, for sphalerons there is one negative mode, which we will discuss later. The approximation above is valid to next-to-leading order in the semiclassical expansion.

Below, we briefly summarize the results for the three potentials in Fig.~\ref{fig:three-potentials}.

(i) For the quadratic potential in Fig.~\ref{fig:three-potentials} (a), there is only one minimum at \(x=0\). Therefore, the only classical path contributing to Eq.~\eqref{eq:transition_amplitues_QMs} is the trivial path \(\bar{x}(\tau)=0\). According to Eq.~\eqref{eq:amplitute_eigen_decomposition_QM}, the amplitude is dominated by the ground-state contribution, \(e^{-E_0 \tau}|\langle x=0 | n=0 \rangle|^2\). The determinant on the right-hand side is evaluated with the result \cite{Coleman:1985rnk}
\begin{align}
  K N \approx \left( \frac{\omega}{\pi} \right)^{1/2} e^{-\omega \tau / 2} \ ,
\end{align}
where \(\omega=\sqrt{V^{\prime\prime}(0)}\). Equating the two sides, we find that the ground-state energy is
\begin{align}
  E_0 = \omega/2 \ ,
\end{align}
which is the well-known result for the harmonic oscillator. Furthermore, we obtain the probability for the particle to be found at \(x=0\) in the ground state:
\begin{align}
  |\langle x=0 | n=0 \rangle|^2 = \left( \frac{\omega}{\pi} \right)^{1/2} \ .
\end{align}

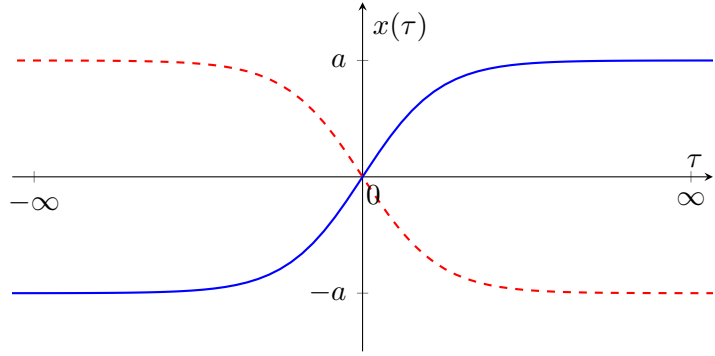
\begin{figure}[htbp]
\centering
\begin{tikzpicture}
  \begin{axis}[
    width=0.7\linewidth,
    height=0.4\linewidth,
    axis lines=middle,
    xlabel={$ \tau $},
    ylabel={$ x(\tau) $},
    xmin=-1.6, xmax=1.6,
    ymin=-1.5, ymax=1.5,
    xtick={-1.5,0,1.5},
    xticklabels={$-\infty$, $0$, $\infty$},
    ytick={-1,0,1},
    yticklabels={$-a$, $0$, $a$},
    samples=200
  ]
    \addplot[thick, blue] {tanh(deg(x/20))};
    \addplot[thick, red, dashed] {-tanh(deg(x/20))};
    \node at (axis cs:0.05,-0.15) {$0$};
  \end{axis}
\end{tikzpicture}
\caption{Instanton-like trajectory interpolating from \(x=-a\) to \(x=a\) in imaginary time (blue solid curve). The red dashed curve shows the anti-instanton trajectory interpolating from \(x=a\) to \(x=-a\).}
\label{fig:instanton_double_well}
\end{figure}

(ii) For the double-well potential in Fig.~\ref{fig:three-potentials} (b), there are two degenerate minima at \(x=\pm a\). Due to parity symmetry, \(\langle -a | e^{-H\tau}| -a \rangle = \langle a | e^{-H\tau}| a \rangle\) and \(\langle a | e^{-H\tau}| -a \rangle = \langle -a | e^{-H\tau}| a \rangle\). We therefore need to consider two types of classical paths, corresponding to a particle that starts at \(x_i=-a\) and ends at either \(x_f=a\) or \(x_f=-a\). In other words, we are interested in the amplitudes \(\langle \pm a | e^{-H\tau}| -a \rangle\). The instanton solution is the classical path that starts at \(x=-a\) at \(\tau=-\infty\) and ends at \(x=a\) at \(\tau=+\infty\). The anti-instanton solution is the classical path that starts at \(x=a\) at \(\tau=-\infty\) and ends at \(x=-a\) at \(\tau=+\infty\). An illustration of the instanton and anti-instanton is shown in Fig.~\ref{fig:instanton_double_well}. Since \(\tau\) is taken to infinity when we evaluate the amplitude, there can be multi-instanton and anti-instanton configurations that contribute. For example, for the amplitude \(\langle a | e^{-H\tau}| -a \rangle\), we can have one instanton, or one instanton followed by one anti-instanton and then another instanton, and so on, i.e. configurations with an odd number of instantons. By contrast, for the amplitude \(\langle -a | e^{-H\tau}| -a \rangle\), we can have zero instanton, or one instanton followed by one anti-instanton, and so on, i.e. configurations with an even number of instantons. For this double-well case, the contribution from each instanton or anti-instanton solution is the same, denoted by \(K \tau e^{-S_0}\), where \(S_0\) is the Euclidean action evaluated on the instanton solution, and \(K\) is the determinant factor from the quantum fluctuations around the instanton solution. The proof is lengthy and will not be repeated here; it can be found in \cite{Coleman:1985rnk}. Furthermore, when we consider multi-instanton and anti-instanton configurations, we must include the combinatorial factor \(1/n!\). Thus, the evaluation of \(\langle \pm a | e^{-H\tau}| -a \rangle\) is obtained by summing over all odd or even numbers of instantons, respectively. The expressions are
\begin{align}
  \langle - a | e^{-H\tau}| -a \rangle & = \left( \frac{\omega}{\pi} \right)^{1/2}  e^{-\omega \tau / 2} \sum_{{\rm even}\ n}^\infty \frac{1}{n!} (K e^{-S_0} \tau)^{n} \ , \\ 
  \label{eq:amplitude_double_well_instanton_odd}
    \langle  a | e^{-H\tau}| -a \rangle & = \left( \frac{\omega}{\pi} \right)^{1/2}  e^{-\omega \tau / 2} \sum_{{\rm odd}\ n}^\infty \frac{1}{n!} (K e^{-S_0} \tau)^{n} \ , 
\end{align}
which can be reorganized as
\begin{align}
  \langle \pm a | e^{-H\tau}| -a \rangle &  = \left( \frac{\omega}{\pi} \right)^{1/2}  e^{-\omega \tau / 2} \frac{1}{2} \left[ \exp(K e^{-S_0} \tau) \mp  \exp(-K e^{-S_0} \tau) \right] \nonumber \\
  & = \frac{1}{2}\left( \frac{\omega}{\pi} \right)^{1/2}   \exp\left[-(\omega/2 - K e^{-S_0}) \tau\right] \mp \frac{1}{2}\left( \frac{\omega}{\pi} \right)^{1/2} \exp\left[-(\omega/2 + K e^{-S_0}) \tau\right] \ .
\end{align}
By comparing with Eq.~\eqref{eq:amplitute_eigen_decomposition_QM}, we can read off the two low-lying energy eigenvalues:
\begin{align} \label{eq:double_well_energy_levels}
  E_\pm = \frac{\omega}{2} \pm K e^{-S_0} \ .
\end{align}
We denote the corresponding energy eigenstates by \(|+\rangle\) and \(|-\rangle\). We also find that
\begin{align}
 |\langle +|\pm a\rangle|^2= |\langle -|\pm a\rangle|^2= \langle a | - \rangle \langle - | -a \rangle = -\langle a | + \rangle \langle + | -a \rangle = \frac{1}{2}\left( \frac{\omega}{\pi} \right)^{1/2} \ ,
\end{align}
which gives the probability for the particle to be found at \(x=\pm a\) in either of the two low-lying energy eigenstates.
Equation~\eqref{eq:double_well_energy_levels} shows that the degeneracy between the two lowest energy levels, namely the oscillator ground states localized around each minimum, is lifted by instanton effects, with an exponentially small splitting proportional to \(e^{-S_0}\).

(iii) For the periodic potential in Fig.~\ref{fig:three-potentials} (c), there are infinitely many degenerate minima at \(x=n\) for integer \(n\)\footnote{For notational clarity, here and in the following discussion we set the spatial separation between two successive minima to be \(1\).}. We now consider the amplitude \(\langle j_f | e^{-H\tau}| j_i \rangle\) for arbitrary integers \(j_i\) and \(j_f\). We assume that the relevant contributions are connected by multi-instanton and anti-instanton solutions. For example, for \(\langle j_f = 2 | e^{-H\tau}| j_i = 0 \rangle\), the possible contributions include two successive instantons, or three instantons followed by one anti-instanton, and so on. Any configuration with \(n\) instantons and \(\bar{n}\) anti-instantons satisfying \(n-\bar{n}=j_f-j_i=2\) contributes to the amplitude. This idea is illustrated in Fig.~\ref{fig:instanton_periodic_potential}. Note that tunneling between minima separated by more than one unit requires multiple instantons and/or anti-instantons. One may wonder whether a classical path can directly connect two minima separated by more than one unit. The answer is negative in two respects: (i) no such classical-path solution is known; (ii) even if such a classical path existed, its Euclidean action would likely be much larger than that of the multi-instanton and anti-instanton configurations, and would therefore be exponentially suppressed. Hence, we only need to consider multi-instanton and anti-instanton configurations. 
We also employ the dilute-instanton-gas approximation, meaning we assume that instantons and anti-instantons are widely separated along the Euclidean-time direction. Ref.~\cite{Coleman:1985rnk} confirms that this approximation is valid: the dominant contributions to the transition amplitude indeed come from instantons and anti-instantons that are well separated.

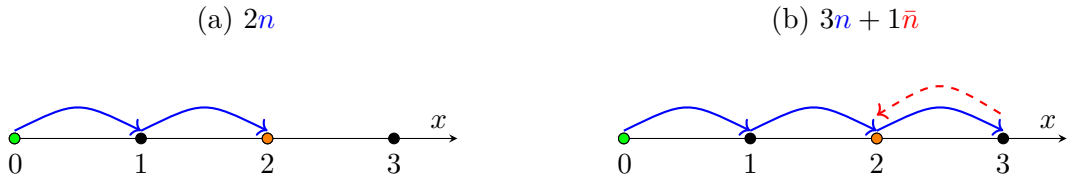
\begin{figure}[htbp]
\centering

\begin{minipage}{0.48\linewidth}
  \centering
  \begin{tikzpicture}
    \begin{axis}[
      width=\linewidth,
      height=0.35\linewidth,
      domain=-10:5, samples=401,
      axis x line=middle,
      axis y line=none,
      xlabel={$x$}, ylabel={$V(x)$},
      xmin=0, xmax=22,
      ymin=0, ymax=1,
      xtick={0.05,6.283,12.566,18.849},
      xticklabels={$0$,$1$,$2$,$3$},
      ytick=\empty,
      title={(a) $2{\color{blue}n}$}]
      \addplot[only marks, mark=*, mark options={fill=black}] coordinates {(0,0) (6.283,0) (12.566,0) (18.849,0)};
      \addplot[only marks, mark=*, mark options={fill=green}] coordinates {(0,0) };
      \addplot[only marks, mark=*, mark options={fill=orange}] coordinates { (12.566,0) };
      \draw[->, thick, blue] (axis cs:0,0.1) .. controls (axis cs:3.1416,0.5) .. (axis cs:6.283,0.1);
      \draw[->, thick, blue] (axis cs:6.283,0.1) .. controls (axis cs:9.4248,0.5) .. (axis cs:12.566,0.1);
    \end{axis}
  \end{tikzpicture}
\end{minipage}\hfill
\begin{minipage}{0.48\linewidth}
  \centering
  \begin{tikzpicture}
    \begin{axis}[
      width=\linewidth,
      height=0.35\linewidth,
      domain=-10:5, samples=401,
      axis x line=middle,
      axis y line=none,
      xlabel={$x$}, ylabel={$V(x)$},
      xmin=0, xmax=22,
      ymin=0, ymax=1,
      xtick={0.05,6.283,12.566,18.849},
      xticklabels={$0$,$1$,$2$,$3$},
      ytick=\empty,
      title={(b) $3{\color{blue}n}+1{\color{red}\bar{n}}$}]
      \addplot[only marks, mark=*, mark options={fill=black}] coordinates {(0,0) (6.283,0) (12.566,0) (18.849,0)};
      \addplot[only marks, mark=*, mark options={fill=green}] coordinates {(0,0) };
      \addplot[only marks, mark=*, mark options={fill=orange}] coordinates { (12.566,0) };
      \draw[->, thick, blue] (axis cs:0,0.1) .. controls (axis cs:3.1416,0.5) .. (axis cs:6.283,0.1);
      \draw[->, thick, blue] (axis cs:6.283,0.1) .. controls (axis cs:9.4248,0.5) .. (axis cs:12.566,0.1);
      \draw[->, thick, blue] (axis cs:12.566,0.1) .. controls (axis cs:15.707,0.5) .. (axis cs:18.849,0.1);
      \draw[<-, thick, red, dashed] (axis cs:12.566,0.3) .. controls (axis cs:15.707,0.8) .. (axis cs:18.849,0.3);
    \end{axis}
  \end{tikzpicture}
\end{minipage}
\caption{Two example instanton/anti-instanton configurations contributing to the amplitude \(\langle j_f = 2 | e^{-H\tau}| j_i = 0 \rangle\) in the periodic potential, where the initial minimum is at \(j_i=0\) (green point) and the final minimum is at \(j_f=2\) (orange point). (a) Two instantons, \(2{\color{blue}n}\). (b) Three instantons and one anti-instanton, \(3{\color{blue}n} + 1{\color{red}\bar{n}}\).}
\label{fig:instanton_periodic_potential}
\end{figure}

We now evaluate the amplitude \(\langle j_f | e^{-H\tau}| j_i \rangle\). We must sum over all possible instanton and anti-instanton configurations satisfying \(n-\bar{n}=j_f-j_i\). As in the double-well case, the contribution from each instanton or anti-instanton solution is the same, namely \(K \tau e^{-S_0}\). Including the combinatorial factors \(1/n!\) and \(1/\bar{n}!\), the evaluation of \(\langle j_f | e^{-H\tau}| j_i \rangle\) gives
\begin{align}
  \langle j_f | e^{-H\tau}| j_i \rangle & = \left( \frac{\omega}{\pi} \right)^{1/2}  e^{-\omega \tau / 2} \sum_{n,\bar{n}=0}^\infty \frac{1}{n!} \frac{1}{\bar{n}!} (K e^{-S_0} \tau)^{n+\bar{n}} \delta_{n-\bar{n},j_f - j_i} \ .
\end{align}
We then use the integral representation of the Kronecker delta,
\begin{align}
  \delta_{n-\bar{n},j_f - j_i} = \int_0^{2\pi} \frac{d\theta}{2\pi} \, e^{i\theta (n-\bar{n} - (j_f - j_i))} \ ,
\end{align}
where \(\theta\) is an auxiliary variable. Later we will see that \(\theta\) represents the true vacuum angle, both in this quantum-mechanics example and in gauge theory. The amplitude therefore becomes
\begin{align} \label{eq:amplitute_periodic_potential}
  \langle j_f | e^{-H\tau}| j_i \rangle & = \left( \frac{\omega}{\pi} \right)^{1/2}  e^{-\omega \tau / 2}  \int_0^{2\pi} \frac{d\theta}{2\pi} \, e^{-i\theta (j_f - j_i)} \sum_{n,\bar{n}=0}^\infty \frac{1}{n!} \frac{1}{\bar{n}!} (K e^{-S_0} \tau)^{n+\bar{n}} e^{i\theta (n-\bar{n})} \nonumber \\
  &= \left( \frac{\omega}{\pi} \right)^{1/2}  e^{-\omega \tau / 2}  \int_0^{2\pi} \frac{d\theta}{2\pi} \, e^{-i\theta (j_f - j_i)} \exp\left[ K e^{-S_0} \tau e^{i\theta} \right]\exp\left[ K e^{-S_0} \tau e^{-i\theta} \right] \nonumber \\
  &= \left( \frac{\omega}{\pi} \right)^{1/2}  e^{-\omega \tau / 2}  \int_0^{2\pi} \frac{d\theta}{2\pi} \, e^{-i\theta (j_f - j_i)} \exp\left[ 2K \cos\theta \, e^{-S_0} \tau \right] \ .
\end{align}
By comparing with Eq.~\eqref{eq:amplitute_eigen_decomposition_QM}, we can read off the energy eigenvalues:
\begin{align}
  E(\theta)  = \frac{\omega}{2} - 2K \cos\theta \, e^{-S_0} \ .
\end{align}
This differs from the double-well case, where there are only two low-lying energy levels. Here we obtain a continuous band of energy eigenvalues parameterized by the vacuum angle \(\theta\), in analogy with the band structure in solid-state physics. In other words, \(\theta\) parameterizes the different energy eigenstates. Rewriting the amplitude in Eq.~\eqref{eq:amplitute_periodic_potential} as
\begin{align} \label{eq:transition_amplitude_periodic_potential}
  \langle j_f | e^{-H\tau}| j_i \rangle & = \int_0^{2\pi}d\theta\ \exp\left[-E(\theta) \tau \right] \frac{1}{2\pi} \, e^{-i\theta (j_f - j_i)} \left( \frac{\omega}{\pi} \right)^{1/2} \nonumber \\
  &=\int_0^{2\pi}d\theta\,  \exp\left[-E(\theta) \tau \right] \, \langle j_f | \theta \rangle \langle \theta | j_i \rangle \ ,
\end{align}
we may identify
\begin{align}
  \langle j | \theta \rangle = \frac{1}{\sqrt{2\pi}} \, \left( \frac{\omega}{\pi} \right)^{1/4} e^{-i\theta j} \ .
\end{align}
From this expression, it is natural to define the \(\theta\)-vacuum state as
\begin{align}
  |\theta \rangle =  \sum_{j^\prime=-\infty}^\infty e^{-i \theta j^\prime} | j^\prime \rangle \ ,
\end{align}
provided we normalize the states as
\begin{align}
  \langle j^\prime | j \rangle =  \frac{1}{\sqrt{2\pi}} \, \left( \frac{\omega}{\pi} \right)^{1/4} \delta_{j^\prime, j} \ .
\end{align}
This reasoning for the definition of the \(\theta\)-vacuum state is a novel approach of this thesis, different from the starting points commonly used in the literature \cite{Coleman:1985rnk}. However, the final result is the same. Another indication that the \(\theta\)-vacuum states are energy eigenstates is that \(\langle \theta^\prime | \mathcal{O} | \theta \rangle \propto \delta(\theta^\prime - \theta)\), where \(\mathcal{O}\) is any operator that commutes with the Hamiltonian. We prove this statement as follows:
\begin{align}
  \langle \theta^\prime | \mathcal{O} | \theta \rangle & = \sum_{j^\prime, j} e^{i \theta^\prime j^\prime} e^{-i \theta j} \langle j^\prime | \mathcal{O} | j \rangle \ .
\end{align}
Defining \(j^\prime = j + n\) \cite{Srednicki:2007qs}, we obtain
\begin{align} \label{eq:theta_vacuum_operator}
  \langle \theta^\prime | \mathcal{O} | \theta \rangle & = \sum_{j, n} e^{i (\theta^\prime - \theta) j} e^{i \theta^\prime n} \langle j + n | \mathcal{O} | j \rangle \nonumber \\
  &= \left(\sum_j e^{i (\theta^\prime - \theta) j} \right) \left( \sum_n e^{i \theta^\prime n} \langle j + n | \mathcal{O} | j \rangle \right) \nonumber \\
  & = \delta(\theta^\prime - \theta) \sum_n e^{i \theta^\prime n} \langle j + n | \mathcal{O} | j \rangle \ .
\end{align}
In the second line, the sums over \(j\) and \(n\) can be separated because \(\langle j + n | \mathcal{O} | j \rangle\) depends only on \(n\), as can be seen from Eq.~\eqref{eq:transition_amplitude_periodic_potential}. This proves the statement.

\newpage
\section{Baryon Number Violation from Non-Abelian Instanton Tunneling}
\label{sec:instanton_baryon_violation}
Now we discuss the vacuum structure of non-Abelian gauge theory, which is similar to the periodic-potential case in quantum mechanics. Since the vacuum manifold of a non-Abelian gauge theory is characterized by the integer Chern-Simons Number \(N_{\rm CS}\), discussed in Eq.~\eqref{eq:Chern_Simons_number_definition} and \eqref{eq:relation_winding_number_Chern_Simons}, this integer is analogous to the integer \(j\) in the periodic-potential case shown in Fig.~\ref{fig:three-potentials} (c). As in that case, we consider only tunneling between vacua that differ by one unit in Chern-Simons number. The instanton solution with unit winding number in non-Abelian gauge theory was first constructed by Belavin, Polyakov, Schwartz, and Tyupkin (BPST) \cite{Belavin:1975fg}, and was introduced in Sec.~\ref{sec:instanton}.

{\bf Note on notation.} In Sec.~\ref{sec:winding_number_gauge_vacuum_structure}, the winding number refers to the map from the first type of \(S^3_{(1)}\) to the gauge group, whereas the Chern--Simons number refers to the map from the second type of \(S^3_{(2)}\) to the gauge group. We will follow the same name convention in this appendix, and denote the winding number as \(\bf \nu\) and the Chern--Simons number as \(j\). The key points are as follows: (i) for the imaginary-time solution with single tunneling process, \(\nu=j(\tau=+\infty)-j(\tau=-\infty)\), and in this thesis we focus on the cases \(\nu=\pm1\), corresponding to instanton and anti-instanton configurations; (ii) the value of \(j\) can be any integer; (iii) if \(|j(\tau=\infty)-j(\tau=-\infty)|>1\), one must consider multi-instanton and anti-instanton configurations, which will be discussed in detail.

As in the periodic-potential case, we can define the \(\theta\)-vacuum state as a superposition of vacuum states with different Chern-Simons numbers:
\begin{align}
  \label{eq:theta_vacuum_definition}
  |\theta \rangle =  \sum_{j=-\infty}^\infty e^{-i \theta j} | j \rangle \ ,
\end{align}
where \(|j \rangle\) is the vacuum state with Chern-Simons number \(j\). For the \(\theta\)-vacuum, we also have the superposition rule in complete analogy with Eq.~\eqref{eq:theta_vacuum_operator}. This \(\theta\)-vacuum plays an important role in discussions of the strong-CP problem and axions in QCD; see Refs.~\cite{Peccei:1977hh,Wilczek:1977pj,Weinberg:1977ma}. In the following, we focus mainly on the physical implications of instanton tunneling between different vacua.

 In the following, we will follow the original computation carried out by 't Hooft \footnote{Note on notations: in the original 't Hooft paper \cite{tHooft:1976rip}, the vacuum is still denoted by $|0\rangle$, which should be interpreted as the physical $|\theta\rangle$ vacuum.}. By analyzing this vacuum-to-vacuum transition amplitude in the background of an instanton, it was first pointed out by 't Hooft \cite{tHooft:1976rip,tHooft:1976snw} that this instanton tunneling induces an effective chiral-violating interaction among fermions coupled to the non-Abelian gauge fields. 't Hooft found that the corresponding chiral-violation rate is proportional to \(|e^{-8\pi^2/g^2}|^2\), where \(g\) is the gauge coupling constant, and \(e^{-8\pi^2/g^2}\) is the instanton factor associated with the action computed in Eq.~\eqref{eq:instanton_action_value}. This computation was later generalized by Ringwald and Espinosa \cite{Ringwald:1989ee,Espinosa:1989qn} to include scalar-field effects. However, the basic principle is the same as in 't Hooft's original computation. In this subsection, we illustrate the physical picture underlying 't Hooft's computation.
For a real process of fermion chiral violation, one should in principle work in Minkowski space. In practice, however, the computation is much easier in Euclidean space. We therefore assume that the quantities computed in Euclidean space can be analytically continued to Minkowski space. Furthermore, as pointed out in Ref.~\cite{Callan:1977gz}, in real Minkowski time a classically forbidden process does not define a stationary path that dominates the functional integral, so there is no simple semiclassical approximation in Minkowski space. Therefore, we perform the computation in Euclidean space. 

In the presence of massless quarks, it has been shown in Ref.~\cite{Callan:1976je} that: (i)$\langle j^\prime | e^{-H\tau}| j \rangle \propto \delta(j^\prime - j)$; (ii) the $|j \rangle$ vacua violate the cluster decomposition principle for operators of non-zero chirality \footnote{The cluster decomposition principle (CDP) requires that the vacuum expectation value of a product of operators factorizes into the product of the vacuum expectation values of the individual operators when the operators are separated by large distances. For example, suppose that $\mathcal{O}_1$ and $\mathcal{O}_2$ are operators. The CDP requires that, for a spacelike interval between $x$ and $y$, we have $\langle \Omega | \mathcal{O}_1(x) \mathcal{O}_2(y) | \Omega \rangle = \langle \Omega | \mathcal{O}_1(x)| \Omega \rangle  \langle \Omega |  \mathcal{O}_2(y) | \Omega \rangle$. For the chiral anomaly, the relevant operator is the chirality-violating operator $D$. It can be shown that, for the $|j\rangle$ state, $\langle j | D^\dagger(x) D(y) | j \rangle\neq 0$, while $\langle j | D^\dagger(x) | j \rangle = 0$, which violates the CDP theorem. This indicates that the true vacuum state should be the $|\theta\rangle$ vacuum. }. These observations are reasonable because the $|j\rangle$ states are not the physical vacua. We also note that the vacuum appearing in the LSZ reduction formula is the physical interacting vacuum.
The relevant quantity is the vacuum-to-vacuum transition amplitude with an insertion of the fermion operators, $
  \left\langle \theta \left| e^{-H\tau} \mathcal{O} \right| \theta \right\rangle \ .
$
Depending on how the insertion $\mathcal{O}$ is implemented, it can play two different roles:
\begin{itemize}
  \item[(i)] If $\mathcal{O}$ is taken to be the identity operator, the chirality-flipping insertion is instead implemented in the path integral through a source term in the Lagrangian, $\bar{\psi} J(x) \psi$ \cite{tHooft:1976rip}, where $\psi$ denotes the fermion field and $J$ is the source. This insertion requires the initial and final vacua to differ by one unit of Chern--Simons number for an instanton configuration.
  \item[(ii)] Alternatively, $\mathcal{O}$ can be taken to be the fermion-number-violating operator relevant to a standard instanton-induced baryon-number-violating process, $\mathcal{O}=\prod_i\psi_i$ \cite{Ringwald:1989ee}. For example, in the three-generation Standard Model, it involves nine (anti-)quarks and three (anti-)leptons.
\end{itemize}
The two approaches are consistent. When the source and instanton are widely separated, the effective amplitude for the chirality-violating process in the first approach has the same spacetime structure as the $N$-point Green's function obtained in the second approach \cite{tHooft:1976snw}. We first examine some general properties of $\left\langle \theta \left| e^{-H\tau} \mathcal{O} \right| \theta \right\rangle$. Although $\mathcal{O}$ plays intrinsically different roles in the two approaches, both describe chirality-violating processes in the instanton background. We therefore use this general notation to discuss their shared features. Using the definition of the $\theta$ vacuum, we find
\begin{align}
  \left\langle \theta \left| e^{-H\tau} \mathcal{O} \right| \theta \right\rangle = \sum_{j,j^\prime} e^{i(j^\prime-j)\theta} \langle j^\prime | e^{-H\tau} \mathcal{O} | j \rangle \ .
\end{align}
We now focus on the instanton background, in which the matrix element is nonvanishing only for a unit increase in the Chern--Simons number, 
$\langle j^\prime | e^{-H\tau} \mathcal{O} | j \rangle \propto \delta_{j^\prime,j+1}$. The above expression then becomes
\begin{align}
  \left\langle \theta \left| e^{-H\tau} \mathcal{O} \right| \theta \right\rangle = \sum_{j} e^{i\theta} \langle j+1 | e^{-H\tau} \mathcal{O} | j \rangle \ .
\end{align}
For the second approach, the corresponding $S$-matrix element is obtained by applying the LSZ reduction formula. In the first approach, the amplitude is obtained from an effective vertex Lagrangian. The overall phase $e^{i\theta}$ cancels upon squaring the amplitude. Moreover, the infinite sum over $j$ requires careful normalization, which can be achieved by dividing by $\langle \theta | \theta\rangle$ \footnote{This is consistent with the requirement that the vacuum-to-vacuum amplitude in the absence of sources must be normalized to 1 \cite{tHooft:1976snw}.}. We therefore focus on the matrix element $\langle j+1 | e^{-H\tau}\mathcal{O}|j\rangle$. In the following, we mainly discuss the details of the first approach. 

We consider a non-Abelian gauge theory with gauge group \(SU(2)\), coupled to \(N_f\) massless fermion doublets. For simplicity, we ignore scalar fields here. The Euclidean Lagrangian is
\begin{align}
  \mathcal{L} = -\frac{1}{4} F_{\mu\nu}^a F^{a\mu\nu} - \sum_{t=1}^{N_f} \bar{\psi}^t \slashed{D} \psi^t \ ,
\end{align}
where \(\psi\) denotes the fermion doublets. The covariant derivative is defined by
\begin{align}
  D_\mu \psi_i^t = \partial_\mu \psi_i^t - \frac{i}{2} g A_\mu^a \tau^a_{ij} \psi_j^t \ ,
\end{align}
where \(\tau^a\) are the Pauli matrices, and \(i,j=1,2\) are the \(SU(2)\) gauge indices (``color'' indices). The index \(t=1,2,\ldots,N_f\) labels the different fermion doublets (``flavor'' indices). The Dirac indices are suppressed.

We first consider the transition amplitude $\langle j+1 | e^{-H\tau}\mathcal{O}|j\rangle$ with $\mathcal{O}=1$. For pedagogical purposes, we initially omit the chirality-flipping source from the Lagrangian. Although this choice does not describe a chirality-violating process, it provides a useful illustration of the role of fermion zero modes.
\begin{align}\label{eq:instanton_tunneling_amplitude}
  &\langle j+1 | e^{-H\tau}\mathcal{O}| j \rangle \nonumber \\
   & = \int \mathcal{D}A_\mu^a \mathcal{D}\bar{\psi} \mathcal{D}\psi \mathcal{D}\phi \,  \exp\left[\!\int\left[\mathcal{L}_{\rm gauge}(A)+\mathcal{L}_{\rm fermion}(\bar{\psi},\psi,A) + \mathcal{L}_{\rm GF}(A) + \mathcal{L}_{\rm FP}(A,\phi)\right]d^4x\right]\mathcal{O} \ ,
\end{align}
where \(A\) denotes the gauge fields, \(\psi\) the fermion fields, and \(\phi\) the Faddeev--Popov ghost fields. \(\mathcal{L}_{\rm GF}\) and \(\mathcal{L}_{\rm FP}\) are the gauge-fixing and Faddeev--Popov ghost Lagrangians, respectively. We do not discuss the gauge-fixing and ghost terms here, since they do not affect the main physical picture that we want to illustrate. We now expand the gauge fields around the instanton solution \(A_\mu^{a,{\rm inst}}\) as
\begin{align}
  A_\mu^a = A_\mu^{a,{\rm inst}} + A_\mu^{a,{\rm qu}} \ ,
\end{align}
where \(A_\mu^{a,{\rm qu}}\) denotes the quantum fluctuation around the instanton solution \(A_\mu^{a,{\rm inst}}\), given in Eq.~\eqref{eq:BPST_instanton_confg}. There is no fermion background field; only fermion quantum fluctuations are present. The Euclidean Lagrangian can then be expanded as
\begin{align}
  \mathcal{L}  = \mathcal{L}(A_\mu^{a,{\rm inst}}) - A^{{\rm qu}} M_1 A^{{\rm qu}} - \bar{\psi} M_2 \psi - \phi^* M_3 \phi + \cdots \ ,
\end{align}
where we have suppressed the gauge and Dirac indices. Here \(M_1\), \(M_2\), and \(M_3\) are the differential operators for the gauge, fermion, and ghost fields, respectively. The dots represent higher-order interaction terms among the quantum fields. In addition, one must replace the functional measure of the fluctuation fields by collective coordinates \((x_0,\lambda)\), where \(x_0\) represents translational collective coordinates and \(\lambda\) represents the scale collective coordinate. The evaluation of $\langle j+1 | e^{-H\tau}| j \rangle$ can be written formally as
\begin{align}
  &\langle j+1 | e^{-H\tau}| j \rangle = \int d^4 x_0 \, d\lambda \, ({\rm det}J)({\rm det}M_1)^{-1/2} ({\rm det}M_2) ({\rm det}M_3) \exp\left[-S_{\rm inst}\right] \ ,
\end{align}
where \(S_{\rm inst}=-\int \mathcal{L}(A^{\rm inst})d^4x= 8\pi^2/g^2\). The factor \(({\rm det}J)\) is the Jacobian associated with the change of variables from the functional measure to the collective coordinates. The factors \(({\rm det}M_1)^{-1/2}\), \(({\rm det}M_2)\), and \(({\rm det}M_3)\) arise from the gauge, fermion, and ghost fluctuations, respectively. These determinants can be determined by solving the eigenvalue equations for the corresponding differential operators,
\begin{align}
  \label{eq:instanton_eigenvalue_equations}
  M_1 A^{\rm qu}_n  = \lambda_{1,n} A^{\rm qu}_n \ , \quad
  M_2 \psi_n  = \lambda_{2,n} \psi_n \ , \quad
  M_3 \phi_n  = \lambda_{3,n} \phi_n \ ,
\end{align}
where \(\lambda_{1,n}\), \(\lambda_{2,n}\), and \(\lambda_{3,n}\) are the corresponding eigenvalues. The operators \(M_1\), \(M_2\), and \(M_3\) all have zero eigenvalues, while the zero eigenvalues of \(M_1\) and \(M_3\) cancel neatly \cite{tHooft:1976rip}. This leaves only the fermion zero modes contributing to the amplitude:
\begin{align}
  M_2 \psi_0  = 0 \ , \qquad  \psi_0=(1+r^2)^{-3/2} u \ ,
\end{align}
where \(u\) is a constant tensor carrying Dirac and gauge indices. It was argued in Ref.~\cite{tHooft:1976rip} that these zero modes make the fermion fluctuation determinant vanish, so that the whole amplitude $\langle j+1 | e^{-H\tau}| j \rangle$ becomes zero.

This viewpoint also has an important implication: in the present calculation, the total fermion chirality in Eq.~\eqref{eq:instanton_tunneling_amplitude} is conserved because we expand only around the instanton solution and take the initial and final fermionic states to be identical (the operator $\mathcal{O}$ is set to unity). By contrast, instanton tunneling itself changes the fermion chirality by \(N_f\) units, as discussed in the ABJ-anomaly subsection. Therefore, to capture the chiral violation induced by instanton tunneling, one must sandwich the amplitude between initial and final states with different chiral quantum numbers. 
To implement such a chirality transition, 
't Hooft suggested \cite{tHooft:1976rip} inserting the following operator into the Lagrangian to encode this chiral difference
\begin{align} \label{eq:source_term_instanton}
  \mathcal{L}\supset \bar{\psi}^s_i J_{(x)}^{st} \psi^t_i \ ,
\end{align}
where \(s,t=1,2,\ldots,N_f\) are flavor indices, and \(i=1,2\) are the \(SU(2)\) gauge indices. 

The external source \(J_{(x)}^{st}\) couples to the fermion bilinear operator \(\bar{\psi}^s_i \psi^t_i\). The addition of this term has two benefits. First, it removes the fermion zero modes, because the eigenvalue equation becomes
\begin{align}
  (M_2 + J_{(x)}) \psi_n  = \lambda_{2,n} \psi_n \ ,
\end{align}
which no longer has a zero eigenvalue for the would-be fermion zero mode \(\psi_0\).

Second, the evaluation of the new determinant goes like
\begin{align}
  \det (M_2+J_{(x)}) \propto \prod_{i=1}^{N_f} (x_i-x_0)^{-6} J_{(x_i)}.
\end{align}
This indicates that, for each generation, there are two fermion propagator connecting the source $J$ and instanton, since each fermion propagator scales as $r^{-3}$. If we apply this to the Standard Model, the picture of instanton induced baryon number violation can be visualized in Fig.~\ref{fig:instanton_baryon_violation}. If we ignore the instanton size, the vacuum-to-vacuum transition leads to an effective Lagrangian \cite{tHooft:1976rip}:
\begin{align}
  \mathcal{L}_{\rm eff} = C g^{-8} \exp(-8\pi^2/g^2) \mathcal{L}_f + {\rm h.c.} \ ,
\end{align}
where \(C\) is a numerical constant, \(g\) is the gauge coupling, and \(\mathcal{L}_f\) denotes the fermionic interaction with the appropriate chiral transformation properties. From this effective Lagrangian, one can obtain the leading-order instanton transition rate as
\begin{align} \label{eq:instanton_one_rate}
  \Gamma_{\rm instanton} = \kappa N e^{-16\pi^2/g^2} \ ,
\end{align}
where the leading-order exponent $e^{-16\pi^2/g^2}$ arises from the square of the instanton action, $e^{-2S_E}$, as can be seen from Eq.~\eqref{eq:instanton_action_value}. The \(\kappa\) and \(N\) are numerical factors arising from the determinant contribution and other normalization constants. This rate is extremely small at zero temperature, since \(e^{-16\pi^2/g^2} \sim 10^{-170}\) for the \(SU(2)\) gauge coupling in the SM. Therefore, instanton-induced baryon-number violation is negligible at zero temperature. At finite temperature, however, baryon-number violation can be greatly enhanced by sphaleron processes, which we discuss in the Sec.~\ref{sec:instanton_sphaleron_difference}.

\begin{figure}[htbp]
\centering
\begingroup
\usetikzlibrary{arrows.meta,decorations.markings}
\definecolor{instInk}{HTML}{20262E}
\definecolor{instSoftInk}{HTML}{65717C}
\definecolor{instFill}{HTML}{EEF1F3}
\definecolor{instQRed}{HTML}{C83F3A}
\definecolor{instQGreen}{HTML}{218657}
\definecolor{instQBlue}{HTML}{2C67B1}
\begin{tikzpicture}[
  x=1cm,y=1cm,
  every node/.style={text=instInk},
  source/.style={circle,fill=instInk,draw=white,line width=0.35pt,
                 minimum size=4.3pt,inner sep=0pt},
  qout/.style={line width=0.78pt,line cap=round,
               postaction={decorate},
               decoration={markings,mark=at position 0.73 with
                 {\arrow{Latex[length=1.85mm,width=1.15mm]}}}},
  qin/.style={line width=0.78pt,line cap=round,
              postaction={decorate},
              decoration={markings,mark=at position 0.39 with
                {\arrowreversed{Latex[length=1.85mm,width=1.15mm]}}}},
  lout/.style={draw=instInk,line width=0.95pt,line cap=round,
               postaction={decorate},
               decoration={markings,mark=at position 0.73 with
                 {\arrow{Latex[length=2.1mm,width=1.35mm]}}}},
  lin/.style={draw=instInk,line width=0.95pt,line cap=round,
              postaction={decorate},
              decoration={markings,mark=at position 0.39 with
                {\arrowreversed{Latex[length=2.1mm,width=1.35mm]}}}},
  generation/.style={font=\sffamily\bfseries\fontsize{7.2}{8.2}\selectfont,
                     text=instSoftInk,align=center},
  fields/.style={font=\fontsize{8.6}{10.6}\selectfont,align=center},
  current/.style={font=\fontsize{9.1}{10.5}\selectfont,align=center},
  note/.style={font=\sffamily\fontsize{6.8}{8.2}\selectfont,
               text=instSoftInk,align=center}
]
  \node[source] (q1) at (-4.65, 3.05) {};
  \node[source] (q2) at ( 0.00, 3.78) {};
  \node[source] (q3) at ( 4.65, 3.05) {};
  \node[source] (l1) at (-4.65,-3.05) {};
  \node[source] (l2) at ( 0.00,-3.78) {};
  \node[source] (l3) at ( 4.65,-3.05) {};

  \node[circle,minimum size=17mm,inner sep=0pt,fill=instFill,
        draw=instInk,line width=0.85pt] (inst) at (0,0) {};
  \node[font=\sffamily\bfseries\fontsize{8.2}{9.2}\selectfont]
    at (0,0) {instanton};

  \draw[qout,draw=instQRed]
    (inst.125) .. controls (-1.52,1.00) and (-3.34,2.18) .. (q1);
  \draw[qout,draw=instQGreen]
    (inst.129) .. controls (-1.56,0.95) and (-3.39,2.13) .. (q1);
  \draw[qout,draw=instQBlue]
    (inst.133) .. controls (-1.60,0.90) and (-3.44,2.08) .. (q1);
  \draw[qin,draw=instQRed]
    (inst.147) .. controls (-1.72,0.67) and (-3.49,1.90) .. (q1);
  \draw[qin,draw=instQGreen]
    (inst.151) .. controls (-1.76,0.62) and (-3.54,1.85) .. (q1);
  \draw[qin,draw=instQBlue]
    (inst.155) .. controls (-1.80,0.57) and (-3.59,1.80) .. (q1);

  \draw[qout,draw=instQRed]
    (inst.103) .. controls (-0.47,1.32) and (-0.49,2.61) .. (q2);
  \draw[qout,draw=instQGreen]
    (inst.100) .. controls (-0.40,1.34) and (-0.42,2.63) .. (q2);
  \draw[qout,draw=instQBlue]
    (inst.97) .. controls (-0.33,1.36) and (-0.35,2.65) .. (q2);
  \draw[qin,draw=instQRed]
    (inst.83) .. controls (0.33,1.36) and (0.35,2.65) .. (q2);
  \draw[qin,draw=instQGreen]
    (inst.80) .. controls (0.40,1.34) and (0.42,2.63) .. (q2);
  \draw[qin,draw=instQBlue]
    (inst.77) .. controls (0.47,1.32) and (0.49,2.61) .. (q2);

  \draw[qout,draw=instQRed]
    (inst.55) .. controls (1.52,1.00) and (3.34,2.18) .. (q3);
  \draw[qout,draw=instQGreen]
    (inst.51) .. controls (1.56,0.95) and (3.39,2.13) .. (q3);
  \draw[qout,draw=instQBlue]
    (inst.47) .. controls (1.60,0.90) and (3.44,2.08) .. (q3);
  \draw[qin,draw=instQRed]
    (inst.33) .. controls (1.72,0.67) and (3.49,1.90) .. (q3);
  \draw[qin,draw=instQGreen]
    (inst.29) .. controls (1.76,0.62) and (3.54,1.85) .. (q3);
  \draw[qin,draw=instQBlue]
    (inst.25) .. controls (1.80,0.57) and (3.59,1.80) .. (q3);

  \draw[lout]
    (inst.230) .. controls (-1.67,-1.02) and (-3.42,-2.18) .. (l1);
  \draw[lin]
    (inst.215) .. controls (-1.83,-0.70) and (-3.60,-1.88) .. (l1);
  \draw[lout]
    (inst.262) .. controls (-0.36,-1.39) and (-0.38,-2.66) .. (l2);
  \draw[lin]
    (inst.278) .. controls (0.36,-1.39) and (0.38,-2.66) .. (l2);
  \draw[lout]
    (inst.310) .. controls (1.67,-1.02) and (3.42,-2.18) .. (l3);
  \draw[lin]
    (inst.325) .. controls (1.83,-0.70) and (3.60,-1.88) .. (l3);

  \node[current,anchor=south] at (-4.65,3.22) {$J_B(x_1)$};
  \node[generation,anchor=south] at (-4.65,3.64) {FIRST GENERATION};
  \node[fields,anchor=south] at (-4.65,3.91)
    {$(u_L^{r},\ d_L^{r})\,/\,(\bar u_R^{r},\ \bar d_R^{r})$};

  \node[current,anchor=south] at (0,3.95) {$J_B(x_2)$};
  \node[generation,anchor=south] at (0,4.37) {SECOND GENERATION};
  \node[fields,anchor=south] at (0,4.64)
    {$(c_L^{r},\ s_L^{r})\,/\,(\bar c_R^{r},\ \bar s_R^{r})$};

  \node[current,anchor=south] at (4.65,3.22) {$J_B(x_3)$};
  \node[generation,anchor=south] at (4.65,3.64) {THIRD GENERATION};
  \node[fields,anchor=south] at (4.65,3.91)
    {$(t_L^{r},\ b_L^{r})\,/\,(\bar t_R^{r},\ \bar b_R^{r})$};

  \node[current,anchor=north] at (-4.65,-3.22) {$J_L(x_4)$};
  \node[generation,anchor=north] at (-4.65,-3.66) {FIRST GENERATION};
  \node[fields,anchor=north] at (-4.65,-3.95)
    {$(\nu_{eL},\ e^-_L)\,/(\,\bar\nu_{eR},\ e^+_R)$};

  \node[current,anchor=north] at (0,-3.95) {$J_L(x_5)$};
  \node[generation,anchor=north] at (0,-4.39) {SECOND GENERATION};
  \node[fields,anchor=north] at (0,-4.68)
    {$(\nu_{\mu L},\ \mu^-_L)\,/(\,\bar\nu_{\mu R},\ \mu^+_R)$};

  \node[current,anchor=north] at (4.65,-3.22) {$J_L(x_6)$};
  \node[generation,anchor=north] at (4.65,-3.66) {THIRD GENERATION};
  \node[fields,anchor=north] at (4.65,-3.95)
    {$(\nu_{\tau L},\ \tau^-_L)\,/(\,\bar\nu_{\tau R},\ \tau^+_R)$};

  \node[note,anchor=east] at (-0.98,0.02)
    {QCD color index $r=1,2,3$\\[-0.2mm]
     \textcolor{instQRed}{\rule{0.25cm}{0.8pt}}\quad
     \textcolor{instQGreen}{\rule{0.25cm}{0.8pt}}\quad
     \textcolor{instQBlue}{\rule{0.25cm}{0.8pt}}};
  \node[note,anchor=west,text width=3.2cm] at (0.98,0.02)
    {for each generation,\\[-0.2mm]$\Delta B=\Delta L=1$};
\end{tikzpicture}
\endgroup
\caption{Instanton-induced baryon number violation in the Standard Model. The $J_B$ and $J_L$ denote the baryon and lepton sources, respectively. For each quark and lepton, the upper index $r$ denotes the QCD color, and the lower index $L$ and $R$ denote the chirality. Each baryonic source is a color singlet and can be written as $J_B(x_i)\propto \epsilon_{rst} Q^r_B(x_i) Q^s_B(x_i) Q^t_B(x_i)$, where $Q^r_B(x_i)$ denotes a source carrying color index $r$. For each generation, $|\Delta B|=|\Delta L|=1$, which can be realized by $u^1_L+u^2_L\rightarrow \bar{d}^3_R+e^+_R$ for the first generation of quark and lepton. For the Standard Model with three generations, we have $|\Delta B|=|\Delta L|=3$ for each instanton transition.}
\label{fig:instanton_baryon_violation}
\end{figure}
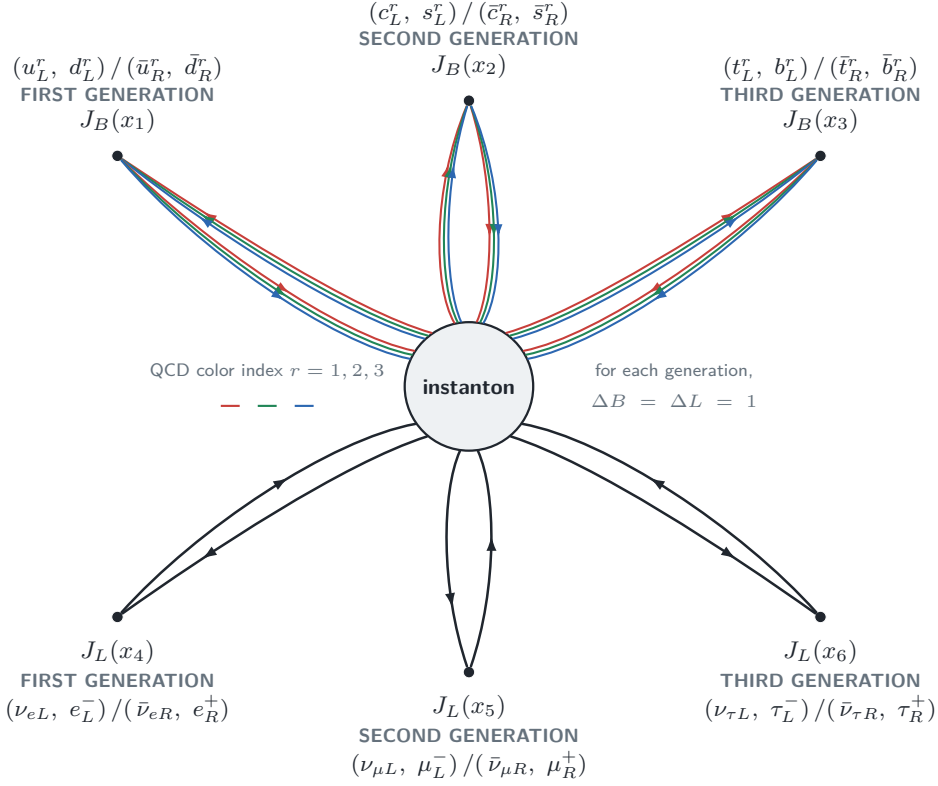

We thus see that the analysis of fermion zero modes is essential for understanding the physics of instanton-induced chiral violation. A related way to understand this point is to analyze the fermion energy levels in the presence of an instanton background, as originally proposed by Callan, Dashen, and Gross \cite{Callan:1976je} (see also Ref.~\cite{Kiskis:1978tb}). They pointed out that when the imaginary time runs from \(-\infty\) to \(+\infty\), a positive-chirality mode has an energy level crossing from negative to positive energy, while a negative-chirality mode has an energy level crossing from positive to negative energy. When instanton tunneling occurs, the positive-chirality mode becomes occupied while the negative-chirality mode becomes unoccupied, leading to a net chirality violation. For a pedagogical introduction to this level-crossing picture, see Ref.~\cite{Rubakov:2002fi}.

\newpage
\section{Matsubara Frequencies in Thermal Field Theory}
\label{app:matsubara_frequencies}

In this appendix, we briefly derive the Matsubara frequencies in thermal field theory. For more details, see the textbook in Ref.~\cite{Das:1997gg}. For a general operator \(\hat{O}\), its thermal expectation value is defined as
\begin{align}
  \langle \hat{O} \rangle = \frac{{\rm Tr} \left(e^{-\beta \hat{H}} \hat{O} \right)}{{\rm Tr} \left(e^{-\beta \hat{H}} \right)} \ ,
\end{align}
where \(\beta = 1/T\) is the inverse temperature, and \(\hat{H}\) is the Hamiltonian operator of the system. The trace is taken over all physical states of the system. In the following discussion, we display only the time coordinates of the operators, since the spatial coordinates are not affected by the thermal formalism. Furthermore, we work in the Heisenberg picture, in which the time evolution of an operator is
\begin{align}
  \hat{O}_H(t) = e^{i \hat{H} t} \hat{O} e^{-i \hat{H} t} \ ,
\end{align}
where \(\hat{O}\) is the operator in the Schr\"odinger picture. Working in the imaginary-time formalism, we perform the Wick rotation \(t \rightarrow -i \tau\), where \(\tau\) is the imaginary time. Thus, the Heisenberg operator in imaginary time reads
\begin{align}
  \hat{O}_H(\tau) = e^{\hat{H} \tau} \hat{O} e^{-\hat{H} \tau} \ .
\end{align}

First, we derive the Kubo--Martin--Schwinger (KMS) relation. Consider the thermal expectation value of the product of two operators \(\hat{A}_H(\tau)\) and \(\hat{B}_H(\tau')\):
\begin{align}
  \langle \hat{A}_H(\tau) \hat{B}_H(\tau') \rangle_\beta &= \frac{{\rm Tr} \left(e^{-\beta \hat{H}} \hat{A}_H(\tau) \hat{B}_H(\tau') \right)}{{\rm Tr} \left(e^{-\beta \hat{H}} \right)} \nonumber \\
  &= Z^{-1} {\rm Tr} \left(\hat{B}_H(\tau') e^{-\beta \hat{H}} \hat{A}_H(\tau)  \right) \nonumber \\
  &= Z^{-1} {\rm Tr} \left(e^{-\beta \hat{H}} e^{\beta \hat{H}}\hat{B}_H(\tau') e^{-\beta \hat{H}} \hat{A}_H(\tau)  \right) \nonumber \\
  &= Z^{-1} {\rm Tr} \left(e^{-\beta \hat{H}} \hat{B}_H(\tau'+\beta) \hat{A}_H(\tau)  \right) \nonumber \\
  &= \langle \hat{B}_H(\tau'+\beta) \hat{A}_H(\tau) \rangle_\beta \ ,
\end{align}
where \(Z={\rm Tr} \left(e^{-\beta \hat{H}} \right)\) is the partition function of the system. In the second line, we used the cyclic property of the trace. This relation holds for both bosonic and fermionic operators. It is known as the Kubo--Martin--Schwinger (KMS) relation. Note that this relation is not unique. We may also write
\begin{align}
  \langle \hat{A}_H(\tau) \hat{B}_H(\tau') \rangle_\beta &= Z^{-1} {\rm Tr} \left( e^{-\beta \hat{H}} \hat{A}_H(\tau) e^{\beta \hat{H}}e^{-\beta \hat{H}}\hat{B}_H(\tau')  \right) \nonumber \\
  &= Z^{-1} {\rm Tr} \left(  \hat{A}_H(\tau-\beta) e^{-\beta \hat{H}}\hat{B}_H(\tau')  \right) \nonumber \\
  &= Z^{-1} {\rm Tr} \left(   e^{-\beta \hat{H}}\hat{B}_H(\tau')\hat{A}_H(\tau-\beta)  \right) \nonumber \\
  &= \langle \hat{B}_H(\tau')\hat{A}_H(\tau-\beta)  \rangle_\beta \ .
\end{align}
Since only the time difference between the two operators matters, the KMS relation implies that, when the order of the two operators is reversed, the time difference changes from \(\tau-\tau'\) to \(\tau'-\tau \pm \beta\).

Next, we define the Green function in the Heisenberg picture as
\begin{align}
  G_{\beta}(\tau,\tau')&=  G_{AB}(\tau-\tau') \nonumber \\ 
  &=\langle P_\tau \left[ \hat{A}_H(\tau) \hat{A}_H^\dagger(\tau') \right] \rangle_\beta \nonumber \\
  &= Z^{-1} {\rm Tr} \left(e^{-\beta \hat{H}} P_\tau \left[ \hat{A}_H(\tau) \hat{A}_H^\dagger(\tau') \right] \right) \ ,
\end{align}
where we assume that \(\hat{A}\) is either a complex bosonic or fermionic operator. The time-ordering operator \(P_\tau\) is defined as
\begin{align}
  P_\tau \left[ \hat{A}_H(\tau) \hat{A}_H^\dagger(\tau') \right] = \begin{cases}
    \hat{A}_H(\tau) \hat{A}_H^\dagger(\tau') \quad & \tau > \tau' \ , \\
    \pm \hat{A}_H^\dagger(\tau') \hat{A}_H(\tau) \quad & \tau < \tau' \ ,
  \end{cases}
\end{align}
where the plus (minus) sign is for bosonic (fermionic) operators. It does not lose generality to set \(\tau = 0\) and \(\tau' = \tau\). For \(\tau > 0\), we have
\begin{align}
  G_{\beta}(0,\tau) &=G_{\beta}(-\tau) \nonumber \\
  &= \pm\langle \hat{A}^\dagger_H(\tau) \hat{A}_H(0) \rangle_\beta \nonumber \\
  &= \pm\langle \hat{A}_H(\beta)\hat{A}^\dagger_H(\tau)  \rangle_\beta \nonumber \\
  &= \pm G_{\beta}(\beta,\tau) \nonumber \\
  &= \pm G_{\beta}(\beta-\tau) \ .
\end{align}
In the third line, we used the KMS relation. Therefore, we see that the bosonic (fermionic) Green function is periodic (anti-periodic) in imaginary time with period \(\beta\). This periodicity (anti-periodicity) condition leads to the Matsubara frequencies in thermal field theory. To see this, we perform the Fourier transformation of the Green function:
\begin{align}
  G_{\beta}(\tau) = \frac{1}{\beta} \sum_{n=-\infty}^{\infty} e^{-i \omega_n \tau} G_{\beta}(\omega_n) \ ,
\end{align}
where
\begin{align}
  G_{\beta}( \omega_n) &= \int_{-\beta/2}^{\beta/2} d\tau \, e^{i \omega_n \tau} G_{\beta}(\tau) \ ,
\end{align}
and \(\omega_n=2n\pi/\beta\), with \(n\) an integer. Changing the integration variable as \(\tau \rightarrow \tau^\prime/2\) in the above integral, we obtain
\begin{align}
  G_{\beta}( \omega_n) &= \frac{1}{2}\int_{-\beta}^{\beta} d\tau \, e^{i n\pi \tau/\beta} G_{\beta}(\tau) \ .
\end{align}
Rewriting the integral gives
\begin{align}
  G_{\beta}( \omega_n) &= \frac{1}{2}\int_{0}^{\beta} d\tau \, e^{i n\pi \tau/\beta} G_{\beta}(\tau) + \frac{1}{2}\int_{-\beta}^{0} d\tau \, e^{i n\pi \tau/\beta} G_{\beta}(\tau) \nonumber \\
  &= \frac{1}{2}\int_{0}^{\beta} d\tau \, e^{i n\pi \tau/\beta} G_{\beta}(\tau) \pm \frac{1}{2}\int_{-\beta}^{0} d\tau \, e^{i n\pi \tau/\beta} G_{\beta}(\beta+\tau) \nonumber \\
  &= \frac{1}{2}\int_{0}^{\beta} d\tau \, e^{i n\pi \tau/\beta} G_{\beta}(\tau) \pm \frac{1}{2}\int_{0}^{\beta} d\tau \, e^{i n\pi (\tau-\beta)/\beta} G_{\beta}(\tau) \nonumber \\
  &= \frac{1}{2}\int_{0}^{\beta} d\tau \, e^{i n\pi \tau/\beta} G_{\beta}(\tau) \left[1\pm e^{-i n\pi} \right] \nonumber \\
  &= \frac{1}{2}\int_{0}^{\beta} d\tau \, e^{i n\pi \tau/\beta} G_{\beta}(\tau) \left[1\pm (-1)^n \right] \ ,
\end{align}
where we used the KMS relation in the second line and shifted the integration variable in the third line. From the last line, we see that for bosonic operators (plus sign), \(G_{\beta}( \omega_n)\) is non-zero for even \(n\), whereas for fermionic operators (minus sign), \(G_{\beta}( \omega_n)\) is non-zero for odd \(n\). Thus, we arrive at the Matsubara frequencies:
\begin{align}
  \label{eq:matsubara_frequencies}
  \omega_n = \begin{cases}
    \frac{2 n \pi}{\beta} = 2 n \pi T \quad & \text{for bosonic operators} \ , \\
    \frac{(2 n  +1) \pi}{\beta} = (2 n +1) \pi T \quad & \text{for fermionic operators} \ ,
  \end{cases}
\end{align}
where \(n\) is an integer. These discrete frequencies are widely used in thermal-field-theory computations.

\newpage 
\section{Computation of the Effective Potential}
\label{app:effective_potential_computation}

\subsection{Useful Integrals}

We first list several useful integrals that are widely used in the computation of the effective potential. We use dimensional regularization to evaluate the integrals, and the spacetime dimension is \(D\). All integrals are evaluated in Euclidean space.

At zero temperature, we have the following one-loop master functional integral:
\begin{align}
  \label{eq:master_integral_JD_zero_temperature}
J_{D}(x) & \equiv \frac{1}{2}\int_p \log(p^2 + x) = -\frac{1}{2} \left( \frac{\bar{\mu}^2 e^\gamma}{2\pi} \right)^\epsilon \frac{x^{\frac{D}{2}}}{(4\pi)^{D/2}} \frac{\Gamma\left(-\frac{D}{2} \right)}{\Gamma(1)} \ ,
\end{align}
where \(D\) denotes the spacetime dimension. The integral symbol \(\int_p\) is defined as
\begin{align} \label{eq:integral_symbol_definition}
  \int_p & \equiv \mu^{2\epsilon} \int \frac{d^D p}{(2\pi)^D} \ ,
\end{align}
where \(\mu\) is the renormalization scale in the MS scheme, which is related to the \(\overline{\rm MS}\) renormalization scale \(\bar{\mu}\) by \cite{Arnold:1992rz}
\begin{align}
  \mu^2=\frac{\bar{\mu}^2 e^\gamma}{4\pi} \ ,
\end{align}
where \(\gamma\) is the Euler--Mascheroni constant. In the following discussion, we use the \(\overline{\rm MS}\) renormalization scheme.

From the master integral above, we obtain two useful cases for \(D=4-2\epsilon\) and \(D=3-2\epsilon\), which we denote by \(J_4\) and \(J_3\), respectively \cite{Hirvonen:2021zej}:
\begin{align}
   J_4(x) &= \frac{1}{16\pi^2}\left[ -\frac{x^2}{4\epsilon} + \frac{x^2}{4}\left( \log\left[ \frac{x}{\bar{\mu}^2} \right] -\frac{3}{2} \right)+\mathcal{O}(\epsilon) \right] \ , \\
   J_3(x) & = -\frac{x^{3/2}}{12\pi} + \mathcal{O}(\epsilon) \ .
   \label{eq:J3_and_J4_integrals}
\end{align}
We see that the \(J_4\) integral contains a UV divergence, which is canceled by the counterterms in the renormalization procedure. By contrast, the \(J_3\) integral is finite and contains no UV divergence.

Now we list some useful finite-temperature integrals. As outlined in the previous appendix, the Matsubara frequencies for bosonic and fermionic operators are different. Accordingly, we have two types of thermal integrals, which we denote by \(J_B\) and \(J_F\), respectively. First, we define the general integral symbols
\begin{align}
\sumint_{P} \equiv T \sum_{\omega_n}\int_{\mathbf p}\,,\quad 
\sumint_{P}^\prime \equiv T \sum_{\omega_n \neq 0}\int_{\mathbf p}\,.
\end{align}
The integral symbol \(\int_{\mathbf p}\) is defined in Eq.~\eqref{eq:integral_symbol_definition}. For the thermal integrals in the \(\overline{\rm MS}\) scheme, we choose the dimension \(D=3-2\epsilon\) in Eq.~\eqref{eq:integral_symbol_definition}. We denote \(P=(\omega_n, \mathbf{p})\), where \(\omega_n\) is the Matsubara frequency. For bosonic operators, \(\omega_n = 2n\pi T\), while for fermionic operators, \(\omega_n = (2n+1)\pi T\), with integer \(n\). For bosonic operators, we have \cite{Arnold:1992rz}
\begin{align}
  \label{eq:J_B_integral}
J_B(x) & \equiv \frac{1}{2} \sumint_{P} \log(P^2 + x)  \nonumber \\
&= -\frac{\pi^2 T^4}{90} + \frac{x T^2}{24} - \frac{x^{3/2} T}{12\pi} -\frac{1}{64\pi^2} x^2 \left[\frac{1}{\epsilon}+\log\left( \frac{\bar{\mu}^2}{T^2}\right)-2c_B \right] \nonumber \\
&\quad + \mathcal{O}(x^3/T^2)+\mathcal{O}(\epsilon) \nonumber \\
& = J_B^\prime(x)+J_B^0(x) \ ,
\end{align}
where \(c_B=\log(4\pi)-\gamma_E\). The divergent part comes from the zero-temperature contribution and is canceled by the counterterms in the renormalization process. Note that the \(\overline{\rm MS}\) scheme does not remove all factors of \(\log(4\pi)\) and \(\gamma_E\), as it does at zero temperature \cite{Arnold:1992rz}, but leaves a finite piece \(-2c_B\) in the thermal contribution. Here \(J_B^\prime(x)\) and \(J_B^0(x)\) are the contributions from the non-zero and zero Matsubara modes, respectively. The zero-mode contribution \(J_B^0(x)\) is
\begin{align}
  J_B^0(x) & = -\frac{x^{3/2}T}{12\pi} + \mathcal{O}(\epsilon) \ ,
\end{align}
so \(J_B^0(x)=J_3(x)T\), where \(J_3(x)\) is defined in Eq.~\eqref{eq:J3_and_J4_integrals}. This term plays an important role in first-order phase transitions because it generates a barrier in the effective potential.

The fermionic integral \(J_F(x)\) is given by \cite{Arnold:1992rz}
\begin{align}
  \label{eq:J_F_integral}
J_F(x) & \equiv -\frac{1}{2} \sumint_{\{P\}} \log(P^2 + x)  \nonumber \\
&= -\frac{7\pi^2 T^4}{720} + \frac{x T^2}{48} + \frac{1}{64\pi^2} x^2 \left[\frac{1}{\epsilon}+\log\left( \frac{\bar{\mu}^2}{T^2}\right)-2c_F \right] + \mathcal{O}(x^3/T^2)+\mathcal{O}(\epsilon) \ ,
\end{align}
where \(c_F=\log(\pi)-\gamma_E=c_B-2\log 2\). The minus sign in the definition of \(J_F\) is due to the Grassmann nature of fermionic operators. The same minus sign also appears in the ghost contribution to the effective potential, as discussed in Sec.~\ref{sec:ghost_and_fermion_contribution_effective_potential}.

\subsection{Introduction to the Effective Potential: A Scalar-Field Example}

At zero temperature, when we expand the field around a constant background field \(\phi\), the effective potential can be computed by standard methods; see Ref.~\cite{Peskin:1995ev} and the references therein. The expanded effective action is
\begin{align}
  \Gamma[\phi] & = S[\phi] + \frac{i}{2}  \log \det \left( -\frac{\delta^2 S}{\delta \Phi^2} \right) + \cdots \ ,
\end{align}
where \(S[\phi]\) is the classical action evaluated at the background field \(\phi\), and the second term is the one-loop contribution to the effective action. We expand the fluctuation field to quadratic order, while higher-order terms are denoted by the ellipsis. For simplicity, we also set \(\hbar = 1\). Since the background field \(\phi\) is constant, the effective potential can be defined through \(\Gamma[\phi]=-\int d^4x \, \mathcal{V}(\phi)\), where \(\mathcal{V}\) is the spacetime volume. Thus, the one-loop effective potential can be written as
\begin{align} \label{eq:one_loop_effective_potential}
  V_{\rm eff}(\phi) & = V_{\rm tree}(\phi) - \frac{i}{2\mathcal{V}} \log \det \left( -\frac{\delta^2 S}{\delta \Phi^2} \right) + \cdots \ ,
\end{align}
where \(V_{\rm tree}(\phi)\) is the tree-level potential. For example, consider a scalar field with action
\begin{align}
S = \int d^4 x \, \left[ \frac{1}{2} (\partial_\mu \Phi)^2 - \frac{1}{2} m^2 \Phi^2 -\frac{1}{4!} \lambda \Phi^4 \right] \ .
\end{align}
We first expand the field around a constant background field \(\phi\) as \(\Phi = \phi + \varphi\), where \(\varphi\) is the fluctuation field. Then we compute the second functional derivative of the action with respect to \(\varphi\) at \(\varphi=0\). For this scalar-field example, the second functional derivative is
\begin{align}
  \frac{\delta^2 S}{\delta \varphi^2}\bigg\vert_{\varphi=0} 
  & = -\partial^2 - m^2 - \frac{\lambda}{2} \phi^2 \nonumber \\
  & \equiv -\partial^2 - m^2(\phi) \ ,
\end{align}
where \(m^2(\phi) = m^2 + \frac{\lambda}{2} \phi^2\) is the shifted mass term that depends on the background field \(\phi\).

We note that the logarithm of the determinant can be converted into the trace of the logarithm, and the latter can be evaluated in momentum space. For the scalar field, we have
\begin{align} \label{eq:log_det_to_trace_log}
 \log \det  \left( -\frac{\delta^2 S}{\delta \varphi^2} \right) & = {\rm Tr} \log \left( \partial^2 + m^2(\phi) \right) \nonumber \\
  &  = \mathcal{V}\int \frac{d^{d+1} p}{(2\pi)^{d+1}} \ln (-p^2 + m^2(\phi)) \ ,
\end{align}
where \(d+1\) is the spacetime dimension. Details of the derivation from the first line to the second line can be found in Ref.~\cite{Peskin:1995ev}. This general integral can be evaluated in dimensional regularization after a Wick rotation to Euclidean space, giving
\begin{align} \label{eq:general_integral_effective_potential}
  \int \frac{d^{d+1} p}{(2\pi)^{d+1}} \ln (-p^2 + m^2(\phi))&=i\int \frac{d^{d+1} p_E}{(2\pi)^{d+1}} \ln (p_E^2 + m^2(\phi)) \nonumber \\
  & \equiv i J_{d+1}(m^2(\phi)) \ ,
\end{align}
where \(J_{d+1}(m^2(\phi))\) is defined by the master integral in Eq.~\eqref{eq:master_integral_JD_zero_temperature}. Thus, the one-loop effective potential becomes
\begin{align}
  V_{\rm eff}(\phi) & = V_{\rm tree}(\phi) + \frac{1}{2} \int \frac{d^{d+1} p_E}{(2\pi)^{d+1}} \ln (p_E^2 + m^2(\phi)) \ .
\end{align}

We now extend the above results to finite temperature. After Wick rotation to imaginary time (\(t=-i\tau\)), the expanded effective action becomes
\begin{align}
  \Gamma[\phi] & = S_E[\phi] + \frac{1}{2} \log \det \left( -\frac{\delta^2 S_E}{\delta \varphi^2} \right) + \cdots \ ,
\end{align}
where \(S_E\) is the Euclidean action, defined by \(e^{iS}\rightarrow e^{-S_E}\). The effective potential is then defined through \(\Gamma[\phi]=\int_0^\beta d\tau \int d^3 x \, V(\phi)\), where \(\beta=1/T\). Therefore, the one-loop effective potential at finite temperature is
\begin{align}
  V_{\rm eff}(\phi) & = V_{\rm tree}(\phi) + \frac{1}{2}T\sum_{n=-\infty}^{\infty} \int \frac{d^d p}{(2\pi)^d} \ln \left( \omega_n^2 + \vec{p}^{\,2} + m^2(\phi) \right) \nonumber \\
  & = V_{\rm tree}(\phi) + J_B(\phi) \ ,
\end{align}
where \(\omega_n\) are the Matsubara frequencies. The second term is the one-loop thermal contribution to the effective potential, and is given by the integral \(J_B(\phi)\) defined in Eq.~\eqref{eq:J_B_integral}.

\subsection{Gauge-Field Effective Potential in a General \texorpdfstring{$R_\xi$}{Rxi} Gauge}
\label{sec:gauge_field_effective_potential_R_xi_gauge}

Many works compute the effective potential in Landau gauge, where \(\xi=0\) and the vector bosons are transverse. However, to illustrate the gauge dependence of the effective potential, we compute it in a general \(R_\xi\) gauge. For pedagogical purposes, we start from the zero-temperature case and then extend the result to finite temperature. Consider the Abelian--Higgs model, namely a \(U(1)\) gauge theory with a complex scalar field. The gauge-fixed Lagrangian is
\begin{align}
  \mathcal{L}_{\rm AH} & = -\frac{1}{4} F_{\mu\nu} F^{\mu\nu} + |D_\mu \Phi|^2 - V(\Phi) - \mathcal{L}_{\rm gf}^{R_\xi} \nonumber \\
  & = -\frac{1}{4} F_{\mu\nu} F^{\mu\nu} + \frac{1}{2}\left(\partial_\mu \Phi_1 - g A_\mu \Phi_2 \right)^2 + \frac{1}{2}\left(\partial_\mu \Phi_2 + g A_\mu \Phi_1 \right)^2 - V(\Phi) - \mathcal{L}_{\rm gf}^{R_\xi}  \ ,
\end{align}
where \(F_{\mu\nu} = \partial_\mu A_\nu - \partial_\nu A_\mu\) is the field-strength tensor of the gauge field \(A_\mu\), and \(D_\mu = \partial_\mu + i g A_\mu\) is the covariant derivative. In the second line, we write the complex scalar field as \(\Phi = (\Phi_1 + i \Phi_2)/\sqrt{2}\). The potential term \(V(\Phi)\) is
\begin{align}
  V(\Phi) & = \frac{\mu^2}{2} |\Phi|^2 + \frac{\lambda}{4!} |\Phi|^4 \ .
\end{align}
We assume that the vacuum expectation value of the scalar field lies along the \(\Phi_1\) direction, and denote the background field by \(\phi\). Thus, \(\langle \Phi_1 \rangle = \phi\) and \(\langle \Phi_2 \rangle = 0\). The last term in \(\mathcal{L}_{\rm AH}\) is the gauge-fixing term in a general \(R_\xi\) gauge, where \(\xi\) is the gauge-fixing parameter. Following Refs.~\cite{Fukuda:1975di,Metaxas:1995ab,Hirvonen:2021zej}, we choose
\begin{align}
  \label{eq:R_xi_gauge_fixing_term}
  \mathcal{L}_{\rm gf}^{R_\xi} & =\frac{F^2}{2\xi}= \frac{1}{2\xi} \left( \partial_\mu A^\mu - \xi g \phi \Phi_2 \right)^2 \ .
\end{align}
According to Ref.~\cite{Fukuda:1975di}, this is a good gauge choice.\footnote{For a comprehensive classification of gauge-fixing conditions, see Ref.~\cite{Martin:2018emo}.}

After shifting the scalar field around the constant background field \(\phi\) as \(\Phi_1 = \phi + \varphi\), the terms quadratic in the gauge field are
\begin{align}
  &\frac{1}{2} \int d^4 x A^\mu(x)\left[\left(\square_x+m^2(\phi)\right) g_{\mu \nu}-\left(1-\frac{1}{\xi}\right) \partial_\mu \partial_\nu\right] A^\nu(x) \nonumber \\
  & \equiv \frac{1}{2} \int d^4 x d^4 y A^\mu(x) \mathcal{D}_{\mu \nu}^{-1}(m^2(\phi) ; x, y) A^\nu(y) \ ,
\end{align}
where \(m^2(\phi) = g^2 \phi^2\) is the gauge-field mass term induced by the background field \(\phi\). The operator \(\mathcal{D}_{\mu \nu}^{-1}(m^2(\phi) ; x, y)\) is
\begin{align}
i \mathcal{D}_{\mu \nu}^{-1}(m^2(\phi) ; x, y)&=\frac{\delta^2 S}{\delta A^\mu \delta A^\nu}(m^2(\phi) ; x, y)\nonumber\\
&=\left\{g_{\mu \nu}\left(\square_x+m^2(\phi)\right)-\left(1-\frac{1}{\xi}\right) \partial_\mu \partial_\nu\right\} \delta^4(x-y) \ ,
\end{align}
which in momentum space becomes
\begin{equation} \label{eq:second_derivative_action_gauge_field}
\frac{\delta^2 S}{\delta A^\mu \delta A^\nu}(m^2(\phi) ; p)=(-p^2+ m^2(\phi)) g_{\mu \nu} + \left(1-\frac{1}{\xi}\right) p_\mu p_\nu \ .
\end{equation}
In analogy with the computation used in Eqs.~\eqref{eq:one_loop_effective_potential} and \eqref{eq:log_det_to_trace_log}, the one-loop contribution of the gauge field to the effective potential is
\begin{align}
  V_{\rm eff}^{\rm gauge}(\phi) & = -\frac{i}{2\mathcal{V}} \log \det \left( -\frac{\delta^2 S}{\delta A^\mu \delta A^\nu} \right) \nonumber \\
  & =-\frac{i}{2} \int \frac{d^{d+1} p}{(2\pi)^{d+1}} \log \det \left[(-p^2+ m^2(\phi)) g_{\mu \nu} + \left(1-\frac{1}{\xi}\right) p_\mu p_\nu \right] \ .
\end{align}
Unlike the scalar-field case, the determinant here is non-trivial because of the Lorentz structure of the operator. It can be evaluated directly in {\tt Mathematica} by representing \(\delta^2 S/\delta A^\mu \delta A^\nu\) as a \(4\times 4\) matrix in momentum space. For Eq.~\eqref{eq:second_derivative_action_gauge_field}, the matrix components are
\begin{align}
&\frac{\delta^2 S}{\delta A^\mu \delta A^\nu}(m^2(\phi) ; p) \nonumber \\
 & = \begin{pmatrix}
\mathcal{W} + \left(1-\frac{1}{\xi}\right) p_0^2 & \left(1-\frac{1}{\xi}\right) p_0 p_1 & \left(1-\frac{1}{\xi}\right) p_0 p_2 & \left(1-\frac{1}{\xi}\right) p_0 p_3 \\
\left(1-\frac{1}{\xi}\right) p_1 p_0 & -\mathcal{W} + \left(1-\frac{1}{\xi}\right) p_1^2 & \left(1-\frac{1}{\xi}\right) p_1 p_2 & \left(1-\frac{1}{\xi}\right) p_1 p_3 \\
\left(1-\frac{1}{\xi}\right) p_2 p_0 & \left(1-\frac{1}{\xi}\right) p_2 p_1 & -\mathcal{W} + \left(1-\frac{1}{\xi}\right) p_2^2 & \left(1-\frac{1}{\xi}\right) p_2 p_3 \\
\left(1-\frac{1}{\xi}\right) p_3 p_0 & \left(1-\frac{1}{\xi}\right) p_3 p_1 & \left(1-\frac{1}{\xi}\right) p_3 p_2 & -\mathcal{W} + \left(1-\frac{1}{\xi}\right) p_3^2
\end{pmatrix} \ ,
\end{align}
where \(\mathcal{W} = -p^2 + m^2(\phi)\). The determinant of this matrix can be computed by {\tt Mathematica}, and the result is
\begin{align}
  \det \left(-\frac{\delta^2 S}{\delta A^\mu \delta A^\nu}(m^2(\phi) ; p)\right)=[-p^2 + m^2(\phi)]^3 \frac{(-p^2+\xi m^2(\phi))}{\xi} \ .
\end{align}
Therefore, the one-loop gauge-field contribution to the effective potential is
\begin{align}
  V_{\rm eff}^{\rm gauge}(\phi) & = -\frac{i}{2} \int \frac{d^{d+1} p}{(2\pi)^{d+1}} \ln \left[(-p^2+ m^2(\phi))^3 \frac{(-p^2+\xi m^2(\phi))}{\xi} \right] \nonumber \\
  & = -\frac{i}{2} \int \frac{d^{d+1} p}{(2\pi)^{d+1}} \left[3 \ln (-p^2+ m^2(\phi)) + \ln (-p^2+\xi m^2(\phi)) - \ln \xi \right] \ .
\end{align}
We see that the final integral structure is similar to the scalar-field case, except that there are two different mass terms, \(m^2(\phi)\) and \(\xi m^2(\phi)\), corresponding to the transverse and longitudinal modes of the gauge field, respectively. The last term is independent of the background field \(\phi\), and can therefore be dropped when computing the effective potential. Using the method in Eq.~\eqref{eq:general_integral_effective_potential}, the final result is
\begin{align}\label{eq:gauge_field_effective_potential_zero_temperature}
  V_{\rm eff}^{\rm gauge}(\phi) & = 3 J_{d+1}(m^2_A)+ J_{d+1}(m_c^2) \ ,
\end{align}
where \(m^2_A = m^2(\phi) = g^2 \phi^2\) is the mass term of the transverse gauge modes, and \(m_c^2 = \xi m^2(\phi) = \xi g^2 \phi^2\) is the mass term of the longitudinal gauge mode, which is also the ghost mass in the \(R_\xi\) gauge. The integral \(J_{d+1}\) is defined in Eq.~\eqref{eq:master_integral_JD_zero_temperature} for spacetime dimension \(D=d+1=4-2\epsilon\).

\subsection{Ghost and Fermion Contributions to the Effective Potential}
\label{sec:ghost_and_fermion_contribution_effective_potential}

At one loop, the effective potential is determined by the quadratic fluctuations, and the Grassmann nature of these fields reverses the sign of the corresponding functional determinant, as seen from \cite{Peskin:1995ev}
\begin{align}
\left(\prod_i \int d\theta_i^{*}\, d\theta_i\right)
e^{-\theta_i^{*} B_{ij}\theta_j}
\propto
\begin{cases}
\det B, & \text{if } \theta_i \text{ are Grassmann numbers} \ ,\\[4pt]
(\det B)^{-1}, & \text{if } \theta_i \text{ are ordinary (commuting) numbers} \ .
\end{cases}
\end{align}
Since the effective potential is proportional to the logarithm of the functional determinant, the contributions from fermion and ghost fields therefore carry an overall minus sign compared with bosonic fields. This explains the minus sign in the definition of the fermionic integral \(J_F\) in Eq.~\eqref{eq:J_F_integral}; the same minus sign appears in the ghost contribution to the effective potential. However, because the fermion field has spinor structure, evaluating its determinant gives an additional factor of 4, which is the dimension of the spinor representation in four dimensions. Thus, the contribution of a Dirac fermion to the effective potential is
\begin{align}
  V_{\rm eff}^{\rm fermion}(\phi) & = 4 J_{F}(m_f^2) \ .
\end{align}

We now discuss the ghost contribution to the effective potential in the \(R_\xi\) gauge defined in Eq.~\eqref{eq:R_xi_gauge_fixing_term}. The ghost Lagrangian arises from \(\det\left(\frac{\delta F(A^\alpha)}{\delta \alpha} \right)\). For the Abelian--Higgs model, the gauge transformation is \(A_\mu \rightarrow A_\mu - \frac{1}{g}\partial_\mu \alpha(x)\) and \(\Phi \rightarrow e^{i \alpha(x)} \Phi\). The \(\Phi_2\) component transforms as \(\Phi_2 \rightarrow \Phi_2 + \alpha(x) \Phi_1=\Phi_2+\alpha(x)(\phi+\varphi)\). Therefore, the gauge-fixing function satisfies
\begin{align}
  \frac{\delta F(A^\alpha)}{\delta \alpha} & = -\frac{1}{g}\partial^2  - \xi g \phi (\phi+\varphi) \ .
\end{align}
After introducing the ghost fields \(c\) and \(\bar{c}\), and rescaling them as \(c \rightarrow  c/\sqrt{g}\) and \(\bar{c} \rightarrow  \bar{c}/\sqrt{g}\), the ghost Lagrangian becomes
\begin{align}
  \mathcal{L}_{\rm ghost} & = \bar{c} \left( -\partial^2 - \xi g^2 \phi^2 -\xi g^2\phi \varphi \right) c \ .
\end{align}
The last term is the interaction term between the ghost field and the fluctuation field \(\varphi\). At one loop, we need only the quadratic term of the ghost field, which gives the ghost mass \(m_c^2\equiv\xi g^2 \phi^2\). Thus, the ghost contribution to the effective potential is
\begin{align}
  V_{\rm eff}^{\rm ghost}(\phi) & = -2\times \frac{1}{2} \int \frac{d^{d+1} p_E}{(2\pi)^{d+1}} \ln (p_E^2 + m_c^2) =- 2 J_{d+1}(m_c^2) \ ,
\end{align}
where the factor of 2 comes from the ghost and anti-ghost fields \(c\) and \(\bar{c}\). The extension to finite temperature is straightforward, and the final result is
\begin{align}
  V_{\rm eff}^{\rm ghost}(\phi) & = - T \sum_{n=-\infty}^{\infty} \int \frac{d^d p}{(2\pi)^d} \ln \left( \omega_n^2 + \vec{p}^{\,2} + m_c^2 \right) = - 2 J_B(m_c^2) \ .
\end{align}

\subsection{Standard Model Effective Potential in the \texorpdfstring{$R_\xi$}{Rxi} Gauge}

The generalization of the above results from the Abelian--Higgs model to the Standard Model is straightforward. The main difference is that the Standard Model has a non-Abelian gauge group, and more fields contribute to the effective potential, such as the \(W\) and \(Z\) bosons. The corresponding ghost fields also carry an \(SU(2)\) color factor of 3. In other words, in the limit of vanishing \(U(1)\) gauge coupling (\(g' \rightarrow 0\)), the gauge- and ghost-field contributions to the Standard Model effective potential are simply the Abelian--Higgs results multiplied by a factor of 3. In addition, the SM Higgs field is a complex \(SU(2)\) doublet, which contains three Goldstone bosons, whereas the Abelian--Higgs model has only one. We write the SM Lagrangian as
\begin{align}
  \mathcal{L}_{\rm SM} & = -\frac{1}{4} F_{\mu\nu}^a F^{a\mu\nu} + |D_\mu \Phi|^2 - V(\Phi) + \mathcal{L}_{\rm Yukawa} - \mathcal{L}_{\rm gf}^{R_\xi} \ ,
\end{align}
where we omit the \(U(1)\) gauge field for simplicity. The Higgs potential is
\begin{align}
  V(\Phi) & = -\mu^2 |\Phi|^2 + \lambda |\Phi|^4 \ ,
\end{align}
where \(\Phi\) is the complex \(SU(2)\) Higgs doublet, which can be written in terms of the background field \(\phi\), the Higgs boson \(h\), and the three Goldstone bosons \(\chi_i\) as
\begin{align}
\Phi =
\begin{pmatrix}
\dfrac{\chi_1 + i\chi_2}{\sqrt{2}} \\
\dfrac{\phi + h + i\chi_3}{\sqrt{2}}
\end{pmatrix} \ .
\end{align}

We first define the field-dependent mass terms as
\begin{align}
  m_h^2(\phi) & = -\mu^2 + 3\lambda \phi^2 \ , \\
  m_G^2(\phi) & = -\mu^2 +  \lambda \phi^2 + m_c^2(\phi) \ , \\
  m_A^2(\phi) & = \frac{g^2}{4} \phi^2 \ , \\
  m_c^2(\phi) & = \xi m_A^2(\phi) = \xi \frac{g^2}{4} \phi^2 \ , \\
  m_t^2(\phi) & = \frac{y_t^2}{2} \phi^2 \ ,
\end{align}
where \(m_h\), \(m_G\), \(m_A\), \(m_c\), and \(m_t\) are the masses of the Higgs boson, Goldstone bosons, gauge bosons, ghost fields, and top quark, respectively. The one-loop effective potential in the \(R_\xi\) gauge at finite temperature is then
\begin{align}
  \label{eq:effective_potential_R_xi_gauge}
  V_{\rm eff}^{\rm 1-loop}(\phi) & = J_B(m_h^2) + 3 J_B(m_G^2) + 9 J_B(m_A^2) + 3 J_B(m_c^2) - 6 J_B(m_c^2) + 12 J_F(m_t^2) \ ,
\end{align}
where \(J_B(x)\) and \(J_F(x)\) are the bosonic and fermionic integrals defined in Eqs.~\eqref{eq:J_B_integral} and \eqref{eq:J_F_integral}, respectively. In this expression, the first four terms are the contributions from the Higgs boson, the three Goldstone bosons, the three gauge bosons, and the longitudinal modes of the gauge bosons, respectively. The numerical factors \(9\) and \(3\) for the gauge bosons and longitudinal modes follow from multiplying the Abelian--Higgs result by the \(SU(2)\) factor of 3, as seen in Eq.~\eqref{eq:gauge_field_effective_potential_zero_temperature}. The fifth term is the anti-ghost contribution, with a minus sign relative to the ghost contribution; here \(6=2\times 3\), where 2 comes from the ghost and anti-ghost fields, and 3 from the \(SU(2)\) factor. The last term is the top-quark contribution, where the factor of 12 comes from the \(SU(3)\) color factor 3 and the spinor factor 4. Note that this factor of 3 is from the \(SU(3)\) color group, unlike the factor of 3 for gauge bosons and ghost fields, which comes from the \(SU(2)\) gauge group.

In Landau gauge, where \(\xi=0\), the effective potential reduces to
\begin{align} \label{eq:effective_potential_landau_gauge}
  V_{\rm eff, Landau}^{\rm 1-loop}(\phi) & = J_B(m_h^2) + 3 J_B(m_G^2) + 9 J_B(m_A^2) + 12 J_F(m_t^2) \ ,
\end{align}
because the contributions from the longitudinal gauge modes and the ghost fields vanish.

\newpage
\section{Scaling Constants of the Sphaleron Zero Modes}
\label{app:scaling_constants_sphaleron_zero_modes}

In this appendix, we analyze the additional scaling constant that contributes to the sphaleron zero modes. As discussed in Sec.~\ref{sec:sphaleron_rate_beyond_leading_order}, the contribution from the sphaleron zero modes is
\begin{align}
  \int [d a]_{\rm zero} [d\eta]_{\rm zero} &= V\times (\mathcal{N})_{\rm tr} \times (\mathcal{NV})_{\rm rot} \ ,
\end{align}
where \(V\) is the spatial volume, \(\mathcal{N}_{\rm tr}\equiv N_{\rm tr}^3\), and \(\mathcal{NV}_{\rm rot}\equiv N_{\rm rot}^3\) are the scaling constants for the translational and rotational zero modes, respectively. Their explicit expressions are given in Eqs.~\eqref{eq:normalization_factor_spatial_translation_zero_modes} and \eqref{eq:normalization_factor_rotation_zero_modes}.

As pointed out in Ref.~\cite{Arnold:1987mh}, there are additional scaling-constant contributions for the zero modes. In particular, such additional constants arise from two sources:
\begin{itemize}
  \item Since the zero modes are separated from the positive-mode contribution, the latter can be written as
    \begin{align}
      \label{eq:scaling_constants_positive_modes}
      \Gamma_{\rm sph,\ positive\ modes}\sim \left(\frac{\det \alpha^{-2} \Omega_0^2}{\det^\prime \alpha^{-2} \Omega_{\rm sph}^2} \right)^{1/2} \ ,
    \end{align}
    where \(\alpha^{-2}\) is the scaling constant appearing in front of the scaled sphaleron action (which is \(g_3^{-2}\) in Ref.~\cite{Arnold:1987mh}). The quantities \(\Omega_0\) and \(\Omega_{\rm sph}\) denote the fluctuation operators evaluated at the vacuum and sphaleron configurations, respectively. The prime in the denominator indicates that the zero-mode contribution has been removed from the determinant. Since the zero modes are removed from the denominator but not from the numerator, the ratio acquires an additional factor of \(\alpha^{-N_0}\), where \(N_0\) is the number of zero modes.

  \item The second source comes from the volume factor \(V\) in the zero-mode contribution. In Ref.~\cite{Arnold:1987mh}, the volume factor is converted into a dimensionless form via
  \begin{align}
    V_\xi = (gv)^3 V \ ,
  \end{align}
  where \(g\) and \(v\) are the gauge coupling and the Higgs vacuum expectation value, respectively. Thus, there is an additional scaling-constant contribution of \((gv)^3\) from the volume factor.
\end{itemize}

Since our sphaleron scaling is completely different from that in Refs.~\cite{Arnold:1987mh,Carson:1989rf,Carson:1990jm,Baacke:1993aj,Baacke:1994ix}, the scaling constants from the two sources above are also different. As introduced in Sec.~\ref{sec:sphaleron_scaling_properties_and_numerical_fittings}, our sphaleron scaling consists of two steps: (i) the dimensional fields and coordinates are scaled by the dimensional quantity \(g_3^2\), with \(x_i=\xi_i/g_3^2\); (ii) the fields and coordinates are further scaled by the dimensionless 3D vev \(v_3\), through \(\xi_i=\xi_i^\prime/v_3\). For the first source mentioned above, the scaled sphaleron action acquires an overall factor of \(v_3^{6/2}=v_3^3\) due to the existance of six zero modes. This follows from the overall factor of \(v_3\) in Eq.~\eqref{eq:Shat} after the two scaling steps.
 We now consider the second source. First, for the translation zero modes, rescaling Eq.~\eqref{eq:zero_mode_contribution_spatial_translation}, we obtain
\begin{align}
\int [da]_{\rm tr}[d\eta]_{\rm tr} = \int d^3 q  \overset{{\color{blue}*}}{=} N_{\rm tr}^3 \int d^3 \xi^\prime  = N_{\rm tr}^3  v_3^3 \int d^3 \xi &= N_{\rm tr}^3 g_3^6 v_3^3 \int d^3 x \nonumber \\
& = N_{\rm tr}^3 g_3^6 v_3^3 V \ ,
\end{align}
where it is important to note that the {\bf initial} volume factor, \(\int d^3 \xi^\prime\), should be the one after the two scaling steps (as marked in the second step), since we assume that the fluctuation fields \(a_{\rm tr}\) and \(\eta_{\rm tr}\) are fields after these two scalings. The final volume factor \(V\) is the one before any scaling, namely the physical volume of the system. Thus, there is an additional scaling-constant contribution \(g_3^6 v_3^3\) for the translation zero modes. Note that the fields and coordinates appearing in the definitions of \(N_{\rm tr}\) and \(N_{\rm rot}\) in Eqs.~\eqref{eq:normalization_factor_spatial_translation_zero_modes} and \eqref{eq:normalization_factor_rotation_zero_modes} should also be understood as those after the two scaling steps. In summary, the full contribution from the zero modes should be
\begin{align}
  \label{eq:full_contribution_zero_modes}
  \int [d a]_{\rm zero} [d\eta]_{\rm zero} &= V\times v_3^3\times g_3^6 v_3^3\times (\mathcal{N})_{\rm tr} \times (\mathcal{NV})_{\rm rot} \nonumber \\
& \approx V\times g_3^6 v_3^6 \times \exp(8.8) \ ,
\end{align}
where we used the numerical fitting results for \(N_{\rm tr}\) and \(N_{\rm rot}\) given in Refs.~\cite{Carson:1989rf,Carson:1990jm}. This result is insensitive to the value of \(\lambda_3/g_3^2\) for the \(SU(2)\)+Higgs theory.

\newpage
\section{Relations Between \texorpdfstring{$\overline{\rm MS}$}{MSbar} Parameters and Physical Input Quantities}
\label{app:MSbar_parameters_and_physical_input_quantities}

In this appendix, we summarize the relations between the \(\overline{\rm MS}\) parameters and the physical input quantities. The physical input quantities are taken from electroweak precision measurements, such as the Fermi constant \(G_F\), the pole masses of the \(W\) and \(Z\) bosons (\(M_W\) and \(M_Z\)), and the top-quark and Higgs-boson masses (\(M_t\) and \(M_H\)); their values can be found in Ref.~\cite{ParticleDataGroup:2024cfk}. Considering the Standard Model Lagrangian
\begin{align} \label{eq:4d_lagrangian_SM}
\mathcal{L}_{4\mathrm{d}} &= \frac{1}{4} F_{\mu\nu}^a F_{\mu\nu}^a + (D_\mu \Phi)^\dagger (D_\mu \Phi) - \mu^2 \Phi^\dagger \Phi + \lambda (\Phi^\dagger \Phi)^2 \nonumber \\
&\quad + \mathcal{L}_{\rm fermion}(\psi) + \mathcal{L}_{\rm Yukawa}(g_Y,\psi) \ ,
\end{align}
the tree-level relations between the model parameters and the physical input quantities are
\begin{align}
  \label{eq:tree_level_relations_parameters_and_physical_quantities}
\mu^2 &= \frac{1}{2} M_H^2 \ , \\
\lambda &= \frac{1}{\sqrt{2}}\,G_F M_H^2
= \frac{g_0^{\,2} M_H^2}{8\,M_W^2} \ , \\
g_Y^2 &= 2\sqrt{2}\,G_F M_t^2
= \frac{g_0^{\,2} M_t^2}{2\,M_W^2} \ ,
\end{align}
where \(g_Y\) is the top-quark Yukawa coupling, and \(g_0\) is the tree-level \(SU(2)\) gauge coupling, defined by
\begin{align}
  \label{eq:tree_level_gauge_coupling}
  g_0^2 & = 8 M_W^2 \frac{G_F}{\sqrt{2}} \ .
\end{align}
We also have the \(U(1)\) gauge coupling \(g'\), which is determined by
\begin{align}
  g'^2 & = g_0^2 \frac{M_Z^2 - M_W^2}{M_W^2} \ .
\end{align}

In the main text, we derive the effective potential in the \(\overline{\rm MS}\) scheme up to \(\mathcal{O}(g^4)\). Correspondingly, to reach the same order of accuracy, we need the one-loop vacuum corrections to the parameters appearing in the tree-level potential \cite{Kajantie:1995dw}. We first consider the \(SU(2)\) gauge coupling. This coupling can be determined from the Fermi constant \(G_F\), which is measured from muon decay in the low-energy effective theory. Usually, the relation between the electroweak theory and the Fermi constant is written in the on-shell scheme:
\begin{align}
  \label{eq:relation_GF_and_g_os}
\frac{G_F}{\sqrt{2}} & = \frac{g^2_{\rm os}}{8 M_W^2} \frac{1}{1-\Delta r} \ ,
\end{align}
where \(\Delta r\) is the radiative correction from gauge bosons and scalars. We now relate this on-shell gauge coupling to the \(\overline{\rm MS}\) gauge coupling. Since all physical quantities are independent of the renormalization scheme, the bare gauge coupling \(g_B\) is the same in both schemes:
\begin{align}
g_B^2 & = g_{\rm os}^2 + \delta \bar{g}_{\rm os}^2 = g^2(\mu) + \delta \bar{g}^2(\mu) \ ,
\end{align}
where \(\delta \bar{g}_{\rm os}^2\) and \(\delta \bar{g}^2(\mu)\) are the one-loop counterterms in the on-shell and \(\overline{\rm MS}\) schemes, respectively. We place a bar on the counterterms to indicate that they still contain the divergent \(1/\epsilon\) part. The \(\overline{\rm MS}\) gauge coupling can then be related to the on-shell gauge coupling by
\begin{align}
g^2(\mu) & = g_{\rm os}^2\left( 1 + \frac{\delta \bar{g}_{\rm os}^2 - \delta \bar{g}^2(\mu)}{g_{\rm os}^2} \right) \nonumber \\
& \equiv g_{\rm os}^2\left( 1 + \frac{\delta g_{\rm os}^2}{g_{\rm os}^2} \right) \ .
\label{eq:relation_g_MSbar_and_g_os}
\end{align}
Since \(\delta \bar{g}^2(\mu)\) cancels the \(1/\epsilon\) piece of \(\delta \bar{g}_{\rm os}^2\), the quantity \(\delta g_{\rm os}^2\) defined in the second line is finite. The one-loop value of \(\Delta r\) in the Feynman--'t Hooft gauge is \cite{Sirlin:1980nh}
\begin{align}
\Delta r
&= \frac{\Pi_W(0)-\operatorname{Re}\Pi_W(-M_W^2)}{M_W^2}
+ \frac{\delta g_{\mathrm{os}}^2}{g_{\mathrm{os}}^2} \nonumber \\
& \quad 
+ \frac{g_{\mathrm{os}}^2}{16\pi^2}
\left[
4\ln\!\left(\frac{\bar{\mu}^2}{M_W^2}\right)
+ \frac{3 M_Z^2 + 4 M_W^2}{2(M_Z^2-M_W^2)}
\ln\!\left(\frac{M_W^2}{M_Z^2}\right)
+ 6
\right] \ ,
\end{align}
where \(\Pi_W(q^2)\) is the \(W\)-boson self-energy, which will be introduced later in Eq.~\eqref{eq:W_boson_self_energy_one_loop}, and \(\bar{\mu}\) is the renormalization scale in the \(\overline{\rm MS}\) scheme. Recalling our definition of the tree-level gauge coupling in Eq.~\eqref{eq:tree_level_gauge_coupling} and the on-shell relation between the Fermi constant and the gauge coupling in Eq.~\eqref{eq:relation_GF_and_g_os}, we can rewrite the above relation as \cite{Kajantie:1995dw,Niemi:2021qvp}
\begin{align}
g_0^2 = \frac{g_{\rm os}^2}{1-\Delta r} \ .
\end{align}
Substituting this relation and the expression for \(\Delta r\) into Eq.~\eqref{eq:relation_g_MSbar_and_g_os}, and replacing \(g_{\rm os}\rightarrow g_0\) inside the one-loop relation (since their difference is higher order), we obtain the relation between the \(\overline{\rm MS}\) gauge coupling and the physical input quantities:
\begin{align}
  \label{eq:one_loop_correction_gauge_coupling_a}
g^2(\bar{\mu}) &= g_0^2\left(1+\frac{\delta g^2}{g_0^2}\right) \ ,\\[4pt]
\frac{\delta g^2}{g_0^2}
&=
\frac{\operatorname{Re}\Pi_W(M_W^2)-\Pi_W(0)}{M_W^2}
-\frac{g_0^2}{16\pi^2}
\left[
4\ln\!\left(\frac{\bar{\mu}^2}{M_W^2}\right)
+\frac{3 M_Z^2 + 4 M_W^2}{2(M_Z^2-M_W^2)}\ln\!\left(\frac{M_W^2}{M_Z^2}\right)
+6
\right] \ .
\label{eq:one_loop_correction_gauge_coupling_b}
\end{align}

We next consider the one-loop corrections to \(\mu^2(\bar{\mu})\), \(\lambda(\bar{\mu})\), and \(g_Y^2(\bar{\mu})\). The tree-level relations between these parameters and the physical input quantities are given in Eqs.~\eqref{eq:tree_level_relations_parameters_and_physical_quantities}. To obtain the one-loop relations, we need the one-loop corrections to the propagators of the Higgs boson, the \(W\) boson, and the top quark. To distinguish the tree-level masses from the physical pole masses, we define the tree-level masses as
\begin{align}
  \label{eq:tree_level_masses}
  m_{H,0}^2  = 2 \mu^2, \
  m_{W,0}^2  = \frac{g^2 \mu^2}{4 \lambda}, \
  m_{t,0}^2  = \frac{g_Y^2 \mu^2}{2 \lambda} \ ,
\end{align}
where the tree-level vev is \(v^2 = \mu^2/\lambda\). The one-loop-corrected propagators of the Higgs field \((\phi)\), \(W\) field \((A_\mu^a)\), and top-quark field \((\Psi)\) are
\begin{align}
\left\langle \phi(-p)\,\phi(p)\right\rangle
&= \frac{1}{p^2 + m_{H,0}^2 - \Pi_H(p^2)} \ , \\
\left\langle A_\mu^a(-p)\,A_\nu^b(p)\right\rangle
&= \delta^{ab}\,
\frac{\delta_{\mu\nu}-p_\mu p_\nu/p^2}{p^2 + m_{W,0}^2 - \Pi_W(p^2)}
\;+\;\text{longitudinal part} \ , \\
\left\langle \Psi_\alpha(p)\,\overline{\Psi}_\beta(p)\right\rangle
&=
\left[
\frac{1}{
i\slashed{p}+m_{t,0}+i\slashed{p}\,\Sigma_v(p^2)
+i\slashed{p}\gamma_5\,\Sigma_a(p^2)
+m_{t,0}\,\Sigma_s(p^2)
}
\right]_{\alpha\beta} \ .
\label{eq:fermion_propagator_one_loop_correction}
\end{align}
The physical masses are defined by the poles of these propagators. This is straightforward for the Higgs boson and the \(W\) boson. For example, for the Higgs boson we have
\begin{align}
  M_H^2 & = m_{H,0}^2 - {\rm Re\ }\Pi_H(-M_H^2) \ ,
\end{align}
and using Eq.~\eqref{eq:tree_level_masses}, this gives
\begin{align}
  \mu^2(\bar{\mu}) & = \frac{1}{2} M_H^2\left[1  +  \frac{{\rm Re\ }\Pi_H(-M_H^2)}{M_H^2} \right] \ .
\end{align}
For the \(W\) boson, \(m_{W,0}^2 = g^2 \mu^2/(4\lambda)=M_W^2+\Pi_W(-M_W^2)\), which gives the relation for \(\lambda(\bar{\mu})\):
\begin{align}
  \lambda(\bar{\mu}) & = \frac{g^2(\mu)\mu^2(\mu)}{4} \frac{1}{M_W^2+\Pi_W(-M_W^2)} \nonumber \\
  & = \frac{1}{4}g_0^2 \left(1 + \frac{\delta g^2}{g_0^2} \right)\frac{M_H^2}{2}\left( 1+ \frac{{\rm Re} \Pi_H(-M_H^2)}{M_H^2} \right)\frac{1}{M_W^2} \left(1 -  \frac{{\rm Re} \Pi_W(-M_W^2)}{M_W^2} \right) + \cdots \nonumber \\
  & = \frac{g_0^2}{8} \frac{M_H^2}{M_W^2} {\rm Re} \left[1 + \frac{\delta g^2}{g_0^2} +  \frac{\Pi_H(-M_H^2)}{M_H^2} - \frac{\Pi_W(-M_W^2)}{M_W^2} \right]+ \cdots \ ,
\end{align}
where the ellipsis denotes higher-order terms beyond the intended accuracy. The expression for \(\delta g^2/g_0^2\) is given in Eq.~\eqref{eq:one_loop_correction_gauge_coupling_b}. Finally, the fermion mass is defined by the pole of the fermion propagator. Note that the \(\gamma_5\) part in Eq.~\eqref{eq:fermion_propagator_one_loop_correction} does not contribute to the physical mass, since \(\bar{u}(p)\gamma_5 u(p)=0\) \cite{Kajantie:1995dw}. The top Yukawa coupling can be determined from \[m_{t,0}^2 = g_Y^2 \mu^2/(2\lambda) = M_t^2 / \left[1 + \Sigma_s(-M_t^2)-\Sigma_v(-M_t^2)\right]^2\]. Substituting the expressions for \(\mu^2(\bar{\mu})\) and \(\lambda(\bar{\mu})\), we obtain the one-loop relation for \(g_Y^2(\bar{\mu})\):
\begin{align}
  g_Y^2(\bar{\mu}) = \frac{g_0^2}{2}\frac{M_t^2}{M_W^2} {\rm Re}\left[1 + \frac{\delta g^2}{g_0^2} -  \frac{\Pi_W(-M_W^2)}{M_W^2} + 2\Sigma_v(-M_t^2) - 2\Sigma_s(-M_t^2) \right] + \cdots \ .
\end{align}

We note that these parameters depend on the renormalization scale \(\bar{\mu}\), which can be fixed to \(\bar{\mu}=M_Z\) when inserting the numerical values of the physical input quantities. However, when these parameters are used to compute the effective potential at finite temperature, they must be run to the matching scale \(\bar{\mu}=4\pi T/e^\gamma\) using the renormalization-group equations (RGEs). The RGEs for the SM parameters are \cite{Kajantie:1995dw}
\begin{align}
  \label{eq:RGEs_SM_parameters_mu2}
\bar{\mu}\,\frac{d}{d\bar{\mu}}\,\mu^2(\bar{\mu})
&=
\frac{1}{8\pi^2}
\left(
-\frac{9}{4}g^2 + 6\lambda + 3g_Y^2
\right)\mu^2 \ , \\
\bar{\mu}\,\frac{d}{d\bar{\mu}}\,g^2(\bar{\mu})
&=
\frac{1}{8\pi^2}
\left(
\frac{8n_F + N_s - 44}{6}
\right)g^4 \ , \\
\label{eq:RGEs_SM_parameters_lambda}
\bar{\mu}\,\frac{d}{d\bar{\mu}}\,\lambda(\bar{\mu})
&=
\frac{1}{8\pi^2}
\left(
\frac{9}{16}g^4
-\frac{9}{2}\lambda g^2
+12\lambda^2
-3g_Y^4
+6\lambda g_Y^2
\right) \ , \\
\bar{\mu}\,\frac{d}{d\bar{\mu}}\,g_Y^2(\bar{\mu})
&=
\frac{1}{8\pi^2}
\left(
\frac{9}{2}g_Y^4
-\frac{9}{4}g^2 g_Y^2
-8g_s^2 g_Y^2
\right) \ ,
\end{align}
where \(n_F=3\) is the number of fermion generations, \(N_s=1\) is the number of scalar doublets, and \(g_s\) is the strong coupling. We do not consider the running of the strong coupling here, since it is not needed for the present problem. We also do not include the running of the \(U(1)\) gauge coupling \(g'\), since according to our power-counting scheme it is of higher order, \(g^\prime \sim g^3\).

Finally, one needs the one-loop self-energies \(\Pi_H(p^2)\), \(\Pi_W(p^2)\), and \(\Sigma_v(p^2), \Sigma_s(p^2)\). These calculations were performed in Ref.~\cite{Bohm:1986rj}; here we summarize the final results. The one-loop Higgs self-energy is
\begin{align}
  \label{eq:Higgs_self_energy_one_loop}
 \Pi_H(-M_H^2) 
&= \frac{3}{8}\frac{g^2}{16\pi^2}M_H^2
\Bigg[
2\!\left(\frac{M_H^2}{M_W^2}+2\frac{M_t^2}{M_W^2}-3\right)\ln\frac{\bar{\mu}^2}{M_W^2}
+3\frac{M_H^2}{M_W^2}\,F(M_H;M_H,M_H) \nonumber\\
&\qquad
+\left(\frac{M_H^2}{M_W^2}-4+12\frac{M_W^2}{M_H^2}\right)F(M_H;M_W,M_W) \nonumber\\
&\qquad -4\,\frac{M_t^2\,(4M_t^2-M_H^2)}{M_W^2M_H^2}\,F(M_H;M_t,M_t) \nonumber\\
&\qquad
-2\frac{M_H^2}{M_W^2}\ln\frac{M_H}{M_W}
-8\frac{M_t^2}{M_W^2}\ln\frac{M_t}{M_W}
-2\frac{M_H^2}{M_W^2}
-2
-12\frac{M_W^2}{M_H^2}
+16\frac{M_t^4}{M_W^2M_H^2}
\Bigg] \ ,
\end{align}
the one-loop \(W\)-boson self-energy is
\begin{align}
  \label{eq:W_boson_self_energy_one_loop}
 \Pi_W(-M_W^2)
&= \frac{3}{8}\frac{g^2}{16\pi^2}M_W^2
\Bigg[
2\!\left(
\frac{16n_F-59}{9}
-\frac{M_H^2}{M_W^2}
-2\frac{M_t^2}{M_W^2}
-6\frac{M_W^2}{M_H^2}
+8\frac{M_t^4}{M_W^2M_H^2}
\right)\ln\frac{\bar{\mu}^2}{M_W^2} \nonumber\\
&\qquad
-22\,F(M_W;M_W,M_W)
+\frac{2}{9}\!\left(\frac{M_H^4}{M_W^4}-4\frac{M_H^2}{M_W^2}+12\right)F(M_W;M_W,M_H) \nonumber\\
&\qquad-\frac{4}{3}\!\left(\frac{M_t^4}{M_W^4}+\frac{M_t^2}{M_W^2}-2\right)F(M_W;M_t,0) \nonumber\\
&\qquad
+4\frac{M_H^2}{M_W^2}\,
\frac{M_H^2-2M_W^2}{M_H^2-M_W^2}\ln\frac{M_H}{M_W}
+\frac{8}{3}\!\left(3\frac{M_t^2}{M_W^2}-2-12\frac{M_t^4}{M_W^2M_H^2}\right)\ln\frac{M_t}{M_W} \nonumber\\
&\qquad
-\frac{22}{9}\frac{M_H^2}{M_W^2}
-4\frac{M_W^2}{M_H^2}
-\frac{4}{3}\frac{M_t^2}{M_W^2}
+16\frac{M_t^4}{M_W^2M_H^2} \nonumber\\
&\qquad +\frac{4}{27}(40n_F-17)
+\frac{8}{3}\!\left(1-\frac{4}{3}n_F\right)\ln(-1-i\epsilon)
\Bigg] \ ,
\end{align}
and the relevant one-loop top-quark self-energy combination is
\begin{align}
  \label{eq:top_quark_self_energy_one_loop}
& \Sigma_v(-M_t^2)-\Sigma_s(-M_t^2) \nonumber \\
&= \frac{3}{16}\frac{g^2}{16\pi^2}
\Bigg[
2\!\left(
\frac{M_t^2}{M_W^2}
-\frac{M_H^2}{M_W^2}
-6\frac{M_W^2}{M_H^2}
+8\frac{M_t^4}{M_W^2M_H^2}
-\frac{32}{3}\frac{g_s^2}{g^2}
\right)\ln\frac{\bar{\mu}^2}{M_W^2} \nonumber \\
&\qquad
+\frac{2}{3}\!\left(4\frac{M_t^2}{M_W^2}-\frac{M_H^2}{M_W^2}\right)F(M_t;M_t,M_H)
+\frac{2}{3}\!\left(1-\frac{M_W^2}{M_t^2}\right)F(M_t;M_t,M_W) \nonumber \\
&\qquad
+\frac{2}{3}\!\left(\frac{M_t^2}{M_W^2}+1-2\frac{M_W^2}{M_t^2}\right)F(M_t;M_W,0)
+4\frac{M_H^2}{M_W^2}\ln\frac{M_H}{M_W}
-32\frac{M_t^4}{M_W^2M_H^2}\ln\frac{M_t}{M_W} \nonumber \\
&\qquad
-\frac{4}{3}\frac{M_t^2(2M_t^2+M_H^2)}{M_W^2(M_t^2-M_H^2)}\ln\frac{M_t}{M_H}
+\frac{128}{3}\frac{g_s^2}{g^2}\ln\frac{M_t}{M_W}
-2
+2\frac{M_t^2}{M_W^2}
-2\frac{M_H^2}{M_W^2} \nonumber \\
&\qquad
-4\frac{M_W^2}{M_H^2}
+16\frac{M_t^4}{M_W^2M_H^2}
-\frac{256}{9}\frac{g_s^2}{g^2}
\Bigg] \ .
\end{align}
In these expressions, we defined the function \(F(m;m_1,m_2)\), which we now introduce \cite{Bohm:1986rj,Kajantie:1995dw}. In the kinematic interval
\[
|m_1-m_2|<k<m_1+m_2 \ ,
\]
it can be written as
\begin{align*}
F(k;m_1,m_2)
&=1-\frac{m_1^2-m_2^2}{k^2}\ln\!\frac{m_1}{m_2}
+\frac{m_1^2+m_2^2}{m_1^2-m_2^2}\ln\!\frac{m_1}{m_2} \\
&\quad
-\frac{2}{k^2}
\sqrt{(m_1+m_2)^2-k^2}\,
\sqrt{k^2-(m_1-m_2)^2}\,
\arctan\!\left(
\frac{\sqrt{k^2-(m_1-m_2)^2}}
{\sqrt{(m_1+m_2)^2-k^2}}
\right) \ .
\end{align*}
Useful limits are
\begin{align*}
F(m_1;m_1,m_2)
&=1-r^2\frac{3-r^2}{1-r^2}\ln r
-2r\sqrt{4-r^2}\,
\arctan\!\sqrt{\frac{2-r}{2+r}} \ ,\\[2mm]
F(m_1;m_2,m_2)
&=2-2\sqrt{4r^2-1}\,
\arctan\!\left(\frac{1}{\sqrt{4r^2-1}}\right) \ ,\\[2mm]
F(m;m,m)
&=2-\frac{\pi}{\sqrt{3}} \ ,
\end{align*}
with
\[
r=\frac{m_2}{m_1} \ .
\]
For the continuation to \(m_2\to 0\), one may use
\[
F(m_1;m_2,0)=1+(r^2-1)\ln\!\left(1-\frac{1}{r^2}\right) \ .
\]

\newpage
\printbibliography

\end{document}